\documentclass[twoside,12pt]{article}
\usepackage{epsfig}

\usepackage{graphicx}
\usepackage{color}
\usepackage{amstext}
\usepackage{comment}

\def\Journal#1#2#3#4{{#1} {#2} (#4) #3}

\def\PRL{\em Phys. Rev. Lett.}

\def\PRA{{\em Phys. Rev.} A}
\def\PRD{{\em Phys. Rev.} D}

\newcommand{\be}{\begin{equation}}
\newcommand{\ee}{\end{equation}}
\newcommand{\bea}{\begin{eqnarray}}
\newcommand{\eea}{\end{eqnarray}}

\begin{document}

\title{ \vspace{1cm} Physics at CERN's Antiproton Decelerator}
\author{M.\ Hori,$^{1,2}$ J.\ Walz$^{3,4}$ 
\\
$^1$Max-Planck-Institut f\"{u}r Quantenoptik, \\ Hans-Kopfermann-Strasse 1, 85748 Garching, Germany\\
$^2$Department of Physics, University of Tokyo, \\ Hongo, Bunkyo-ku, Tokyo 113-0033, Japan \\
$^3$Institut f\"{u}r Physik, Johannes Gutenberg-Universit\"{a}t, \\ D-55099 Mainz, Germany \\
$^4$Helmholtz Institut Mainz \\ D-55099 Mainz, Germany}
\maketitle
\begin{abstract} The Antiproton Decelerator (AD) facility of CERN began operation in 
1999 to serve experiments for studies of $CPT$ invariance by precision laser and microwave spectroscopy of 
antihydrogen ($\overline{\rm H}$) and antiprotonic helium ($\overline{p}{\rm He}^+$) atoms. 
The first 12 years of AD operation saw cold $\overline{\rm H}$ synthesized by
overlapping clouds of positrons ($e^+$) and antiprotons ($\overline{p}$) confined in magnetic Penning traps. 
Cold $\overline{\rm H}$ was also produced in collisions between Rydberg positronium ($Ps$) atoms
and $\overline{p}$. Ground-state $\overline{\rm H}$ was
later trapped for up to $\sim 1000$ s in a magnetic bottle trap, and microwave transitions
excited between its hyperfine levels. In the $\overline{p}{\rm He}^+$ atom,
deep ultraviolet transitions were measured to a fractional precision of 
(2.3--5) $\times$ $10^{-9}$ by sub-Doppler two-photon laser spectroscopy. From this
the antiproton-to-electron mass ratio was determined as 
$M_{\overline{p}}/m_e=$1836.1526736(23), which
agrees with the $p$ value known to a similar precision.
Microwave spectroscopy of $\overline{p}{\rm He}^+$ yielded a 
measurement of the $\overline{p}$ magnetic moment with a precision of $0.3\%$.
More recently, the magnetic moment of a single $\overline{p}$ confined in a Penning 
trap was measured with a higher precision, as $\mu_{\overline{p}}=-2.792845(12)$$\mu_{\rm nucl}$ in 
nuclear magnetons. Other results reviewed here include the first measurements of 
the energy loss ($-dE/dx$) of 1--100 keV $\overline{p}$ traversing
conductor and insulator targets; the cross sections of low-energy ($<10$ keV) $\overline{p}$
ionizing atomic and molecular gas targets; and the cross sections of 5-MeV $\overline{p}$
annihilating on various target foils via nuclear collisions. The biological effectiveness 
of $\overline{p}$ beams destroying cancer cells was measured as a possible method 
for radiological therapy. New experiments under preparation attempt to measure 
the gravitational acceleration of $\overline{\rm H}$ or synthesize $\overline{\rm H}^+$. 
Several other future experiments will also be briefly described. 
\end{abstract}
\eject
\tableofcontents

\newcommand{\aegis}{AE$\bar{\textrm{g}}$IS}

\section{Introduction}

The Antiproton Decelerator (AD) facility of CERN \cite{Maury:97,Pavel:04} began operation in 1999
to carry out high-precision laser spectroscopy of antihydrogen 
($\overline{\rm H}$) and antiprotonic helium ($\overline{p}{\rm He}^+$) atoms. 
It was envisaged that by comparing the characteristic transition frequencies of these atoms with 
the corresponding ones for hydrogen (H) in the $\overline{\rm H}$ case, or quantum electrodynamics (QED) 
calculations in the $\overline{p}{\rm He}^+$ case at the 
highest possible precision, the consistency of $CPT$ invariance could be tested. 
This invariance is deeply engrained within the Standard Model of particle physics, and implies that 
particles and their antiparticle counterparts should have exactly the same mass, and charges and 
magnetic moments of the same values but opposite signs. Atoms should resonate at exactly the same 
frequency as "anti-atoms" made of antiparticles.

Precision laser and microwave spectroscopy of atoms and ions of ordinary matter have been carried out for more than 
50 years, and in recent years have achieved such a high level of sophistication that transition frequencies 
have routinely been measured with an experimental precision of better than $10^{-15}$. This precision
exceeds even the precision by which the international definition of the second can be currently
determined. Some experiments are sensitive to minute
shifts in the frequencies due to the effects of General Relativity. Progress on the
anti-atom side is much more difficult due to the simple fact that cold samples are so difficult to synthesize
in large quantities. The constituent antiprotons ($\overline{p}$) and positrons ($e^+$) can only be produced in very small 
quantities in laboratory nuclear reactions at MeV or GeV energy scales. These particles cannot be directly 
used to form stationary atoms that can be used for precision spectroscopy, 
since their kinetic energy exceeds the eV-scale binding energy by orders of magnitude. 
So first the $\overline{p}$ and $e^+$ must be collected using various experimental techniques, and 
cooled in some cases by 10 orders of magnitude. In fact for achieving the highest possible experimental 
precision, one needs the coldest atoms ($<10$ K) where the Doppler effect on the measured atomic 
frequencies caused by thermal motions is minimized.

The AD is currently the world's only facility where $\overline{p}$ of low ($5.3$ MeV) energy needed
for these experiments can be produced. Important milestones achieved at the AD include the production 
of $\overline{\rm H}$ \cite{ATHENA:2002,ATRAP:Hbar:2002,ATRAP:Cesium:2004,Enomoto:PRL:2010}
in magnetic Penning traps.
The  $\overline{\rm H}$ were later produced at cryogenic temperatures \cite{alpha_cool2010}, and confined
for $\sim 1000$ s in a magnetic bottle trap \cite{ALPHA:Nature:2010,Alpha:NatPhys:2011,ATRAP:TrappedHbar:2012}. 
The first microwave excitations between the ground-state 
hyperfine substates of $\overline{\rm H}$ were recently demonstrated \cite{Alpha:SpinFlips:2012}. 

For $\overline{p}{\rm He}^+$ \cite{hayanorpp}, the atomic transition frequencies were measured by single 
\cite{mhori2001,mhori2003,mhori2006} and sub-Doppler two-photon \cite{hori2011} laser spectroscopy 
to a fractional precision of $\sim 10^{-9}$. By comparing the experimental 
results with QED calculations \cite{korobov2008}, the antiproton-to-electron mass ratio was determined as
$M_{\overline{p}}/m_e=1836.1526736(23)$  \cite{hori2011}. Microwave spectroscopy 
of $\overline{p}{\rm He}^+$ allowed the determination of the $\overline{p}$ magnetic moment with 
a precision of $0.3\%$ \cite{pask2009}.  This agreed with the magnetic moment values derived
from previous X-ray spectroscopy experiments on antiprotonic lead atoms with a similar level of precision \cite{kreissl}.

The magnetic moment of a single $\overline{p}$ confined in a Penning trap was recently measured with
a much higher precision, as $\mu_{\overline{p}}=-2.792845(12)$$\mu_{\rm nucl}$ \cite{atrap_mag} in nuclear magnetons.
This agreed with the known proton value with a fractional precision of $\sim 4\times 10^{-6}$. This experiment
employed the continuous Stern-Gerlach effect, where an inhomogeneous magnetic field of a so-called
"magnetic bottle" was superimposed on the Penning trap. Spin-flips of the $\overline{p}$ resonating with an 
external RF field were then revealed as small shifts in the oscillation frequency of the $\overline{p}$'s axial motion 
in the trap.

Measurements of various atomic \cite{Moller2002,Moller2004,Moller2008,knudsen2008,knudsen2010} 
and nuclear \cite{infn2011} collision cross sections using 
low-energy $\overline{p}$ were also carried out. One experiment studied the biological
effectiveness of $\overline{p}$ beams destroying cancer cells \cite{ace2006}. 
In this paper, we review these experimental results obtained during the first 12 years 
of AD operation. 

The AD was conceived as an economical replacement for the previous Low Energy 
Antiproton Ring (LEAR) facility of CERN, which was shut down in 1996 \cite{Maury:97,Pavel:04,eades}. 
LEAR carried out pioneering studies on $CP$ violation, meson spectroscopy, 
and nuclear reactions using high-intensity beams of $\overline{p}$
circulating inside the storage ring, or extracted as a continuous beam of up to $10^{7}$ s$^{-1}$. 
The high intensity was often needed because of the small cross sections involved in these
particle and nuclear physics experiments. The facility involved, however, a complex
chain of four storage rings (Antiproton Collector (AC), Antiproton Accumulator (AA), Proton Synchrotron (PS), and LEAR)
which handled the sequence of first producing the $\overline{p}$, accumulating and decelerating them
to lower energies, and delivering them to experiments. The AD on the other hand was optimized for precision 
atomic-physics experiments that were assumed to require far fewer $\overline{p}$ per unit time. This allowed 
CERN engineers to devise 
a simplified all-in-one machine where the same sequence was handled by a single storage ring with a cycle time of only 
$\sim 100$ s.  The AD now routinely provides pulsed beams containing 
$\ge 3\times 10^7$ $\overline{p}$ with an emittance of 2--3 $\pi$ mm mrad
and rate of 0.01 Hz.

Much of the experimental work at AD has concentrated on developing the
techniques to decelerate and cool larger numbers of $\overline{p}$. 
The cold $\overline{p}$ were then used as one of the ingredients to
synthesize $\overline{\rm H}$, first by the ATHENA (AnTiHydrogEN Apparatus) \cite{ATHENA:2002}
collaboration followed by the ATRAP \cite{ATRAP:Hbar:2002} 
collaboration in 2002. More recently, ALPHA (Antihydrogen Laser Physics 
Apparatus) and ATRAP have cooled clouds of $10^3$--$10^6$ $\overline{p}$ 
confined in magnetic Penning traps to cryogenic temperatures $T=3.5$--10 K 
\cite{andresencold2010,atrapcold2011}. A small percentage of the $\overline{\rm H}$ 
synthesized using such cold $\overline{p}$ were of sufficiently low ($T<1$ K) temperature that could be subsequently 
confined in a magnetic trap \cite{ALPHA:Nature:2010,Alpha:NatPhys:2011,ATRAP:TrappedHbar:2012}. 
The ASACUSA (Antiproton Spectroscopy And Collisions Using Slow Antiprotons) 
experiment introduced a radiofrequency quadrupole decelerator (RFQD) in collaboration 
with CERN \cite{Bylinsky:RFQD:2000,lombardi2001}. This 3-m-long device decelerated the 
5.3-MeV $\overline{p}$ arriving from AD to $E\sim 60$ keV with a high ($\sim 25\%$) efficiency 
\cite{mhori2003}. The resulting low-energy beam was allowed to come to rest in a helium target, 
thereby synthesizing $\overline{p}{\rm He}^+$ atoms which were studied by 
two-photon laser spectroscopy \cite{hori2011}.
The $\overline{p}$ were also confined in a Penning trap \cite{kuroda2005}, and 
extracted as a continuous beam with an energy of $<1$ keV and average rate 6000--7000 s$^{-1}$. 
The ionization cross sections of $\overline{p}$ colliding with gas targets were measured using this 
beam \cite{knudsen2008,knudsen2010}. Work is also underway to produce high-intensity spin-polarized 
$\overline{\rm H}$ beams in the future, which may 
be used to measure the ground-state hyperfine structure of  $\overline{\rm H}$.
Future experiments on the gravitational acceleration of $\overline{\rm H}$ pursued by the \aegis
(Antihydrogen Experiment: Gravity, Interferometry, 
Spectroscopy) \cite{aegis2010,aegis2011} and GBAR (Gravitational Behavior of Antihydrogen at Rest) 
\cite{gbar2011} collaborations would require $\overline{p}$ beams of even lower (mK-scale) energies.

Following these developments by the AD collaborations, CERN has recently begun 
the construction of a new synchrotron ELENA (Extra-Low ENergy 
Antiproton ring) of 30-m circumference, which would capture the 5.3-MeV $\overline{p}$ 
provided by AD and decelerate them to $E=100$ keV
\cite{elena2010}. Unlike the RFQD described above, ELENA will reduce 
the momentum spread of the $\overline{p}$ using electron cooling techniques, 
thereby achieving a transverse beam emittance of $<5\pi$ mm mrad. The number
of $\overline{p}$ captured in Penning traps per unit time can then be increased by a factor
$100$ for the \aegis, ALPHA, and ATRAP collaborations which now use the 5.3-MeV 
$\overline{p}$ beam.

This article is organized in the following way. In Sect.~\ref{sec:cpt}, we briefly
describe past experimental and theoretical work on $CPT$ symmetry pertaining 
to atomic physics experiments involving $\overline{p}$. The AD facility is
described in Sect.~\ref{sec:ad}. In Sect.~\ref{sec:antihydrogen}, 
progress in $\overline{\rm H}$ production, trapping, and microwave spectroscopy 
made at the AD are outlined. Sect.~\ref{pbarhelium} summarizes the results of 
precision laser and microwave spectroscopy of $\overline{p}{\rm He}^+$.
Measurements of the $\overline{p}$ magnetic moment by inducing
spin-flips of a single $\overline{p}$ confined in a Penning trap carried out by ATRAP
are described in Sect.~\ref{sec:magneticmoment}. Atomic and nuclear collision experiments 
using low-energy $\overline{p}$ are described in Sect.~\ref{collisions}. 
Experiments to measure the biological effectiveness of $\overline{p}$
destroying cancer cells are also presented in the same section.
In Sect.~\ref{sec:future}, future experiments are outlined including, i): $1s-2s$ laser spectroscopy of $\overline{\rm H}$,
ii): higher-precision measurements on the ground-state hyperfine structure of $\overline{\rm H}$,
iii): gravitational acceleration of $\overline{\rm H}$, iv): antiproton-to-electron mass ratio determined by 
sub-ppb-scale laser spectroscopy of $\overline{p}{\rm He}^+$,
v): sub-ppm-scale measurements of the $\overline{p}$ magnetic moment in Penning traps pursued by
ATRAP and a new collaboration BASE (Baryon Antibaryon Symmetry Experiment),
vi): differential cross sections of 100-keV $\overline{p}$ circulating in a new storage ring (ELENA), 
ionizing atoms and molecules contained in a gas jet target. The new ELENA facility, now under construction at CERN, 
is also described in the same section.

\section{$CPT$ symmetry and low-energy antiproton physics}
\label{sec:cpt}

\subsection{\it Antiparticles and symmetries}

The Dirac equation implies that for every variety of fermion observed in nature,
there is a corresponding antifermion with the same mass and opposite electric
charge. Some $E\sim 1$ MeV of energy is needed to produce an $e^-$-$e^+$ pair,
which is readily available in small accelerators. The $e^+$ can also be produced by 
radioactive isotopes that undergo $\beta^+$ decay. When $e^+$ are allowed to come to
rest in matter, they annihilate with atomic $e^-$ via the electromagnetic interaction, which
results in the emission of 2--3 gamma-rays.

The production of $p$-$\overline{p}$ pairs, on the other hand, requires much higher energies of 
$E>2$ GeV; this is typically accomplished by colliding beams of $20-100$ GeV protons on metallic 
targets. Thus there are only a few large-scale synchrotron facilities where $\overline{p}$ can be produced.
A $\overline{p}$ annihilation with a $p$ or neutron ($n$) proceeds 
via strong interaction, and typically results in the emission of several charged and 
neutral pions ($\pi^+$, $\pi^-$, and $\pi^0$) within a picosecond. 
Low-energy $\overline{p}$ and $e^-$, on the other hand, can elastically scatter off each 
other without annihilating, a characteristic which is used to cool $\overline{p}$ confined in 
storage rings and traps.

The properties of particle and antiparticle are related by discrete symmetries \cite{pdg2010} 
in the Standard Model. The substitution of the wavefunction $\left|\Phi\right>$ of a particle 
with its antiparticle $\left|\overline{\Phi}\right>$ can be carried out using the charge conjugation 
operator $\hat{C}\left|\Phi\right>=\left|\overline{\Phi}\right>$. This changes the signs of all 
additive quantum numbers (electric and flavor charges, and baryon and 
lepton numbers), while keeping space-time properties (mass, energy, momentum, 
and spin) as well as multiplicative quantum numbers (parity) unchanged. 
Experimental evidence shows that the laws of electromagnetism, gravity, and strong 
interaction are invariant under this particle-to-antiparticle transformation ($C$-symmetry),
and also preserve the $P$, $T$, and combined $CP$ symmetries. The operator $\hat{P}$ here
reverses the space coordinates from right handed to left handed;
the time reversal operator $\hat{T}$ inverts the flow of time, so that the direction of 
motion and the signs of all time derivatives such as momentum and angular momentum 
are reversed. Violations of the $T$ and $CP$ symmetries in the strong interaction would cause 
$n$ to have a non-zero value of the electric dipole moment (EDM) \cite{dubbers2011}, 
but no EDM has been detected so far to
an upper limit of $2.9\times 10^{-26}$ e$\cdot$cm \cite{baker2006}. 

As discovered in the 1950's, however, the $C$ and $P$ symmetries are maximally violated in weak 
interactions: no antineutrinos with left-handed (LH) chirality, which results from the 
application of either the $\hat{C}$-operator on normal LH neutrinos, 
or the $\hat{P}$-operator on right-handed (RH) antineutrinos, have ever been experimentally detected. 
The Standard Model postulates that, i): the W$^{\pm}$ bosons which mediate the 
charged weak force interact only with LH fermions or RH antifermions, and ii): the 
neutral $Z^0$ boson interacts with charged leptons and quarks of both chiralities with 
different strengths; and only with LH neutrinos or RH antineutrinos, but not their 
respective RH and LH counterparts. 

The weak interaction also violates $CP$ and $T$ symmetries. One of the mass eigenstates
of neutral kaons $K_{\rm long}$ normally decays either semileptonically, or hadronically into
combinations of three neutral and charged pions ($\pi^0\pi^0\pi^0$ or $\pi^+\pi^-\pi^0$) 
which comprise a $CP$ eigenstate of -1. A small fraction ($\sim 10^{-3}$), however, was found to decay 
into $\pi^+\pi^-$ or $\pi^0\pi^0$ pairs \cite{christenson1964,ktev2011,na482007}
with $CP=1$, which violates the $CP$ symmetry. Asymmetries were also recently 
observed in the rates of some decay modes of $B^0$ \cite{babar2001,belle2001} and tentatively $B^0_s$ \cite{bs0} and 
$D^0$ \cite{aaij2012} mesons compared to their antiparticle cases. $CP$ violation is incorporated into the 
Standard Model in the following way: the three generations of quarks are assumed 
to change their flavor only through interactions with $W^{\pm}$ bosons
(e.g., $u^{2/3+}+W^-\rightarrow d^{1/3-}$). Nine such pairings 
$(u,c,t)$ $\leftrightarrow$ $(d,s,b)$ between parent and daughter quark states are 
then possible. The phenomenological 3$\times$3 Cabibbo-Kobayashi-Maskawa 
matrix used to describe the coupling constants of these 9 quark-$W$ pairings 
contains complex components that violate $CP$ symmetry, giving 
rise to the asymmetries in the meson decays. The $CP$ violations measured in the
latest experiments, however, are too small to account for the predominance of matter 
over antimatter in the universe. It is also not understood why the strong interaction 
between the quarks does not exhibit any observable $CP$ violation so far.

On the other hand, the laws of physics are believed to be 
perfectly symmetric under the combined transformations of charge conjugation, 
parity, and time reversal, i.e., $\hat{C}\hat{P}\hat{T}\left|\Phi\right>$$=$$\left|\Phi\right>$. 
In fact, $CPT$ symmetry was
axiomatically proven \cite{pauli1955,luders1957,jost1957,streater1964} to hold for {\it any} relativistic 
quantum field theory under a few basic assumptions, i): Lorentz invariance, 
ii): unitarity, i.e. the sum of all quantum-mechanical probabilities is conserved, iii): 
interactions are local, iv): flat space-time without strong gravitational fields.
This is called the Schwinger-L\"{u}ders-Pauli, or $CPT$, theorem. 

An important consequence of $CPT$ symmetry is that particles and their 
antiparticles have exactly the same mass and lifetimes, and charges and 
magnetic moments of opposite sign and same absolute value. 
A particle propagating in free space can be treated mathematically 
as if it were an antiparticle of exactly the same mass and opposite charge moving backwards 
in space and time ({\it Feynman-Stueckelberg interpretation}). 
The $\hat{C}\hat{P}\hat{T}$ operator relates the scattering matrix $S$
of a physical process to its inverse process $\overline{S}$, with all
particles replaced by antiparticles and 
the spin components reversed. So whereas $CPT$ symmetry
implies that the lifetimes (i.e., sum of all partial decay rates) of
particle and antiparticle should be exactly the same, the {\it partial} decay rate 
of a particle decaying into a certain channel does not necessarily need to 
be equal to its antiparticle case. 

\subsection{\it Theories of possible $CPT$  violation}

Although $CPT$ symmetry remains a fundamental property of relativistic quantum 
field theories, a large number of theoretical and experimental studies have explored its possible breakdown. 
Violation can be introduced by removing some of the assumptions underlying the $CPT$ theorem.
Greenberg \cite{greenberg2002} showed that in any field theory involving local interactions where
unitarity is preserved, $CPT$ violation can only occur if Lorentz invariance is violated as well. The 
converse is not true: Lorentz violation does not necessarily imply $CPT$ violation.
In some theories \cite{chaichian2011} with nonlocal interactions or noncommutative space-time geometry,
$CPT$ violation can occur without Lorentz violation.

Colladay and Kosteleck\'y \cite{Colladay:97,Colladay:PRD_58_116002} 
developed a generalized parameterization of an effective field theory called the Standard Model Extension (SME)
that contains operators that break Lorentz and $CPT$ symmetries. 
This model encompasses the normal Standard Model and general relativity, and retains 
some of their important characteristics such as renormalizability, causality, and invariance 
and covariance under translations and rotations in 
the inertial frame of the observer.  SME additionally assumes the existence of 
a tensor-like "background field", i.e., a preferred direction 
in the vacuum which is frozen and extends over all space and time. Lorentz and 
$CPT$ symmetries are spontaneously broken when two 
identical experiments which are sensitive to this background field are rotated 
or translated relative to each other while being studied by an inertial observer.
The model is expressed by adding several Lorentz and $CPT$-violating terms to 
the Standard-Model Lagrangian. Limits on the sizes of these coefficients can be 
experimentally determined by searching for couplings between the background 
fields and various particle properties such as spin or propagation direction
\cite{kostelecky2011}.  Annual 
and sidereal variations should appear when the Earth rotates and revolves around 
the sun, changing the orientation of the experiment with respect to the field. 
No such unambiguous signal has been detected so far. 

The axiomatic proofs of the $CPT$ theorem are invalid in the highly-curved 
space-time near black holes. Some quantum gravity theories 
\cite{mavromatos2005,mavromatos2010,klinkhamer2004}
involve quantum fluctuations of space-time geometry that are singular, such as 
microscopic black holes with event horizons of Planck-scale  
($10^{-35}$ m) sizes. These backgrounds cause an apparent violation 
of unitarity, since part of the information such as the quantum numbers 
of particles can disappear into the event horizon. Pure ground states 
of quantum gravity thus get mixed and become decoherent as time evolves,
since parts of the quantum states are trapped into the event horizons. 
$CPT$ is obviously violated here since the $S$-matrix of such a process 
cannot be inverted.

Several authors \cite{murayama2001,barenboim2002,chaichian2012} 
have proposed $CPT$-violating models that give
rise to different masses for neutrinos and antineutrinos. These models
attempt to explain anomalies reported in several neutrino oscillation 
experiments described in Sect.~\ref{othercpt}.

\subsection{\it Cyclotron frequency of antiprotons in Penning traps}
\label{trapcyclo}

\begin{figure}[tb]
\epsfysize=9.0cm
\begin{center}
\begin{minipage}[t]{12 cm}
\epsfig{file=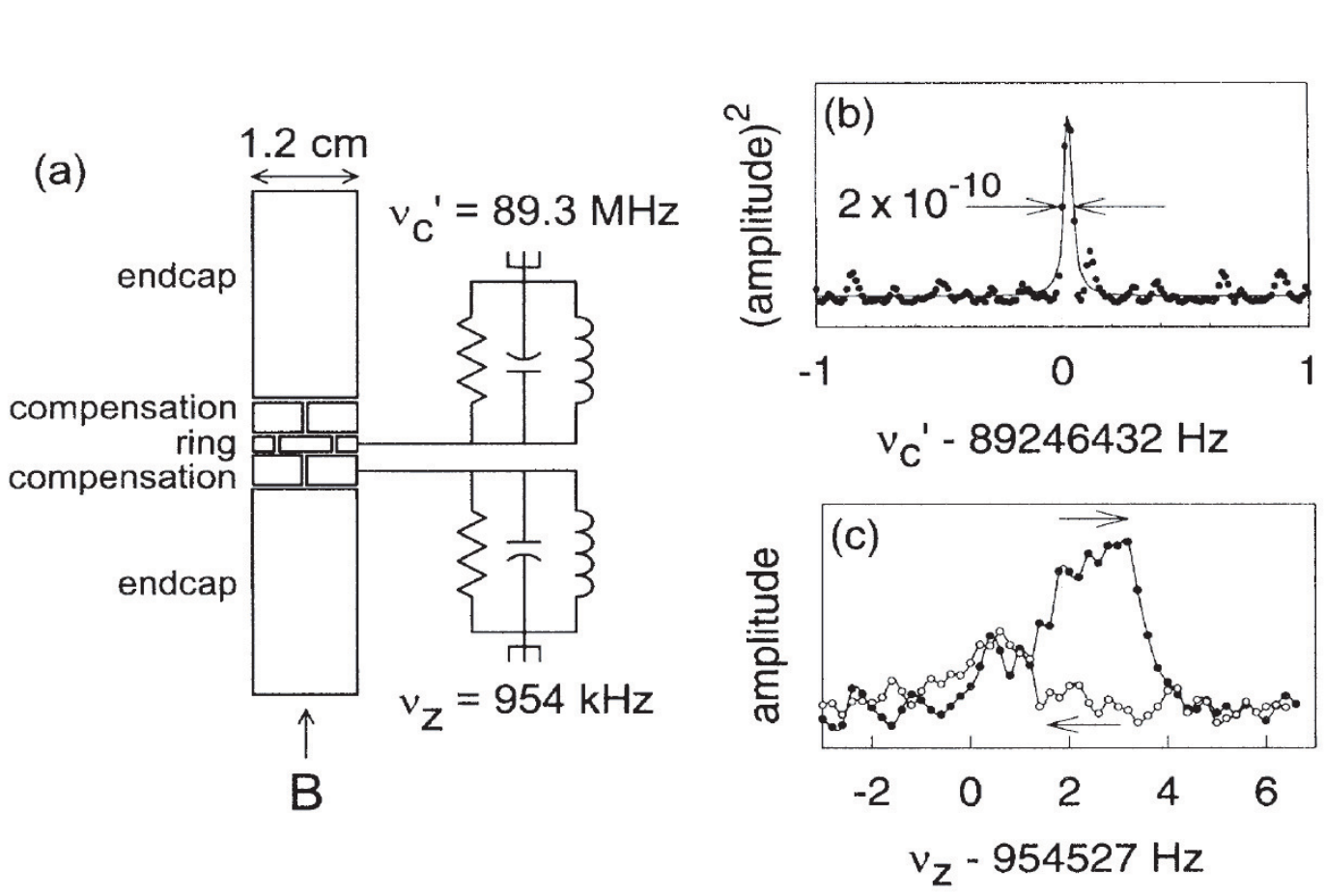,scale=0.8}
\end{minipage}
\begin{minipage}[t]{16.5 cm}
\caption{\label{trapexplain} (a) Open access Penning trap                                                       
electrodes and detection LCR circuits used by the TRAP collaboration 
to confine a single $\overline{p}$,
and measure its cyclotron frequency in a solenoidal magnetic field.
Resonance signals of the (b) modified cyclotron and (c) axial motions 
produced by a single $\overline{p}$. Note extremely narrow $\sim 10^{-10}$
relative width of the cyclotron resonance signal. Figures from Ref.~\cite{gabrielse1995}.}
\end{minipage}
\end{center}
\end{figure}

 \begin{figure}[tb]
\epsfysize=9.0cm
\begin{center}
\begin{minipage}[t]{12 cm}
\epsfig{file=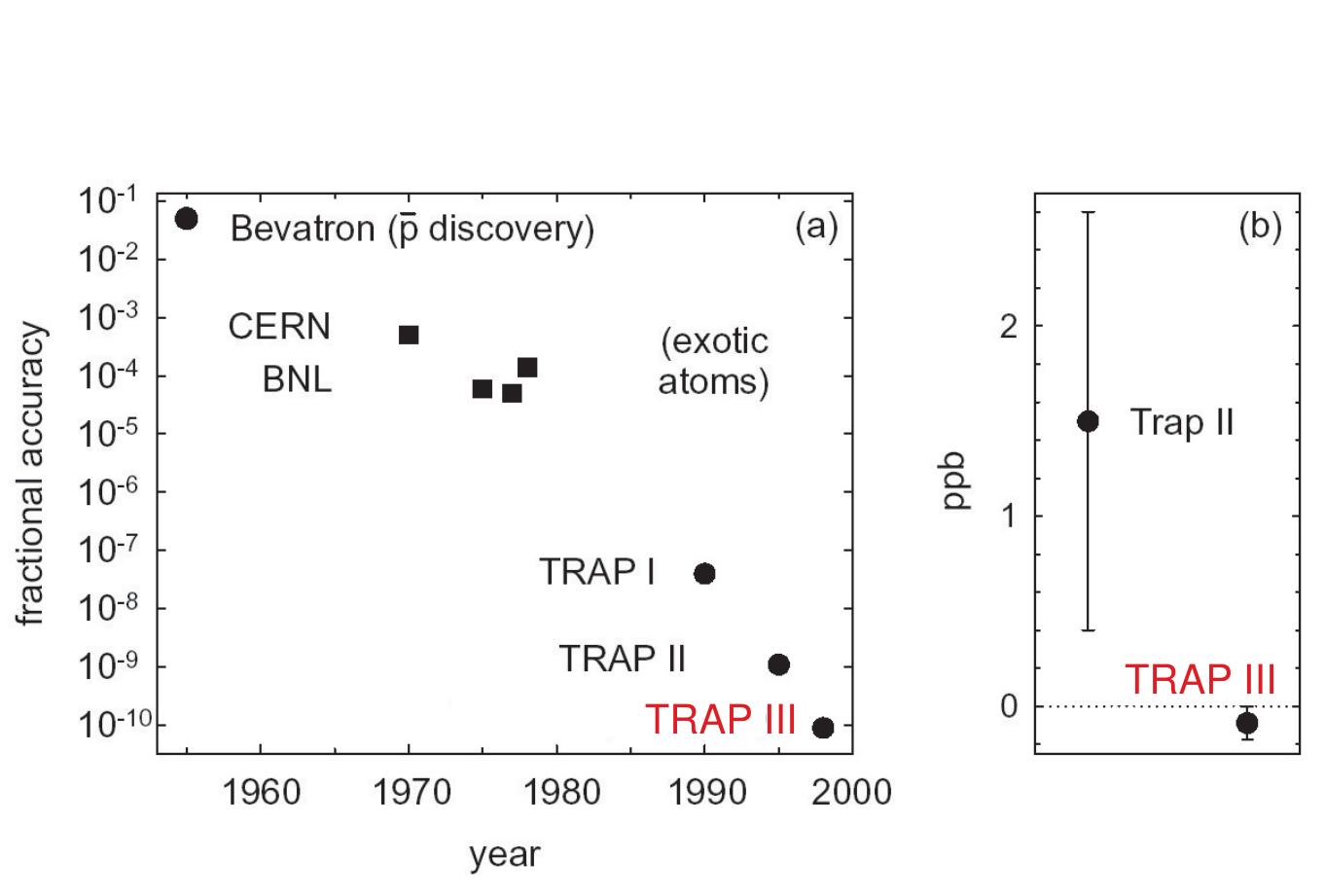,scale=0.9}
\end{minipage}
\begin{minipage}[t]{16.5 cm}
\caption{\label{trapcompare}  (a) Accuracy in comparisons of                                                     
$p$ and $\overline{p}$ as a function of years elapsed. Trap I and II indicate the
results of experiments comparing the cyclotron frequencies of $p$ and $\overline{p}$,
Trap III comparing simultaneously trapped H$^-$ and $\overline{p}$. (b) The difference between $|Q/M|$ of                                                     
$p$ and $\overline{p}$ is shown in ppb units. Figures from Ref.~\cite{gabrielse1999}.
}
\end{minipage}
\end{center}
\end{figure}

Before the construction of LEAR, the mass $M_{\overline{p}}$, 
charge $Q_{\overline{p}}$, and magnetic moment $\mu_{\overline{p}}$ of
$\overline{p}$ were relatively poorly known; the most precise experiments
involved measuring the characteristic X-rays of various types of
antiprotonic atoms \cite{Bamberger,Hu1975,Roberson,Roberts},
and deducing $M_{\overline{p}}$, $Q_{\overline{p}}$, and $\mu_{\overline{p}}$
to a typical precision of around 1 part in $10^3$--$10^4$.

The precision on $Q_{\overline{p}}/M_{\overline{p}}$ was greatly improved
when the TRAP collaboration 
\cite{gabrielse1986,gabrielse1989,gabrielse1990,gabrielse1995,gabrielse1999,Brown:1986} 
confined $\overline{p}$ in a Penning trap (Fig.~\ref{trapexplain} (a))
for the first time. The cyclotron frequency of the $\overline{p}$ was then
determined, which is related to $Q_{\overline{p}}/M_{\overline{p}}$ by,
\begin{equation}
\nu_c(\overline{p})=-\frac{Q_{\overline{p}}B}{2\pi M_{\overline{p}}}.
\label{cycloeq}
\end{equation}

Several types of Penning traps were constructed for these measurements, 
and they typically consisted of two static fields superimposed on each other: 
i): an uniform $B=6$ T magnetic field generated by a superconducting 
solenoidal magnet, and ii): an electrostatic quadrupole field produced 
by cylindrical ring electrodes of inner diameter $d\sim 1$ cm
stacked in series, with voltages of 0.3--20 V
applied to them. The $\overline{p}$ with MeV-scale energies entered the
trap through a thin metallic window,
from which they emerged with keV-scale energies. A small fraction
of the $\overline{p}$ were captured in the electrostatic potential 
well of the trap. They were then cooled to temperatures $T\sim 4$ K by
mixing them with $e^-$ confined simultaneously in the trap. 

In practice, the cyclotron frequency $\nu_c$ is not one of the oscillation
frequencies of the trapped $\overline{p}$, and therefore cannot be directly 
measured. Instead the $\overline{p}$ executed three types of harmonic motion,
i): {\it harmonic axial motion} along the direction 
of the magnetic field at frequency $\nu_z\sim 1$ MHz, 
ii): {\it trap modified cyclotron motion}, a circular motion in a 
perpendicular plane at frequency $\nu_c^{\prime}\sim 90$ MHz, 
iii): {\it magnetron motion}, a low-frequency $\nu_m\sim 5$ kHz circular 
motion occurring in the same plane as the cyclotron motion. 
These motions induce image currents in the trap electrodes,
which can be detected using tuned inductor-capacitor-resistor 
($LCR$) resonance circuits connected to the trap electrodes. The signals were 
amplified by field effect transistors which were cryogenically cooled.
The cyclotron frequency $\nu_c$ was then deduced from the three measured 
eigenfrequencies using the so-called ``invariance theorem'',
\begin{equation}
\left(\nu_c\right)^2=\left(\nu_c^{\prime}\right)^2
+\left(\nu_z\right)^2+\left(\nu_m\right)^2.
\label{invariance}
\end{equation}
This relationship between the four frequencies has been theoretically 
shown to be invariant to leading order, regardless of imperfections in the
electric and magnetic fields in the trap. This fact allows $\nu_c$ to be
determined to extremely high accuracy.

In Fig. ~\ref{trapcompare} (a), the accuracies of experimental comparisons 
between $Q_{\overline{p}}/M_{\overline{p}}$ and $Q_p/M_p$ 
are shown as a function of years elapsed, including values
from previous X-ray spectroscopy experiments of exotic atoms
\cite{Bamberger,Hu1975,Roberson,Roberts}.
The TRAP collaboration initially attained a precision of
$1\times 10^{-9}$ by measuring $\nu_c(\overline{p})$ with a single 
trapped antiproton, then reversing the polarity of the electrostatic 
potential of the trap to confine a single $p$ and measure its frequency 
$\nu_c(p)$, using the techniques described above. This proton-antiproton
comparison eliminated the necessity of measuring the 
magnetic field $B$ with a high absolute accuracy, but systematic errors 
associated with the reversal of the electrostatic potential needed to
trap particles of opposite electric charge limited the experimental precision. 

In later experiments \cite{gabrielse1999}, the precision was improved to 
$9\times 10^{-11}$ by simultaneously trapping a $\overline{p}$ and a ${\rm H^-}$ 
ion in orbits with different cyclotron radii. The
cyclotron frequencies of the two particles $\nu_c(\overline{p})$ and 
$\nu_c({\rm H^-})$ could now be alternately measured without the polarity reversal. 
The measured value $\nu_c({\rm H^-})$ was then converted to the corresponding
proton value $\nu_c(p)$ using the known relationship between the two frequencies,
$\nu_c(p)=1.001089218750(2)\nu_c({\rm H^-})$. Later, it was pointed out that
due to the fact that the two-body H$^-$ ion which undergoes cyclotron motion inside the trap
experiences a Lorentz force and becomes slightly polarized, the $\nu_c({\rm H^-})$ value
is shifted compared to its vacuum value \cite{thompson2004}. From these considerations, 
the charge-to-mass ratios of protons and $\overline{p}$ were experimentally constrained 
\cite{gabrielse1999} as,
\begin{equation}
\frac{Q_{\overline{p}}}{M_{\overline{p}}}/\frac{Q_p}{M_p}+1 = 1.6(9) \times 10^{-10}.
\label{atrapfinal}
\end{equation}
This constitutes one of the most stringent comparisons of particles and antiparticles in the
baryon sector.

\subsection{\it Antiproton lifetime}

$CPT$ symmetry implies that particle and antiparticle decay with the same lifetimes.
The lower limit for the $p$ lifetime is currently $\tau_p>2\times 10^{29}$ y.
This was obtained by the SNO experiment \cite{sno2004} which searched for $\gamma$ rays
emitted from the deexcitation of any residual nucleus that would result from the
decay of a $p$ or $n$ in $^{16}$O nuclei. The TRAP collaboration confined 
some $10^3$ $\overline{p}$ for two months \cite{gabrielse1990} in
a Penning trap, without detecting any sizable annihilation with 
residual gases. By comparing the number of $\overline{p}$ remaining 
after this 2-month period with the initial number loaded into the trap, 
a lower limit,
\begin{equation}
\tau_{\overline{p}}>3.4\ {\rm months}
\end{equation}
was obtained for the $\overline{p}$ lifetime. The lifetime of high-energy $\overline{p}$ 
circulating in a storage ring was also measured at CERN, first in the Initial 
Cooling Experiment facility \cite{bregman1978,bell1979} and later 
at the Antiproton Accumulator \cite{autin1990}.
The APEX collaboration used the Fermilab Antiproton Accumulator 
to search for 13 decay modes of $\overline{p}$ \cite{gear2000}.
A lower limit of $\tau_{\overline{p}}>7\times 10^5$ y was
set for one of the decays, $\overline{p}\rightarrow$$e^-\gamma$.

\subsection{\it Measurements of antiproton magnetic moment prior to AD era}

 \begin{figure}[tb]
\epsfysize=9.0cm
\begin{center}
\begin{minipage}[t]{15 cm}
\epsfig{file=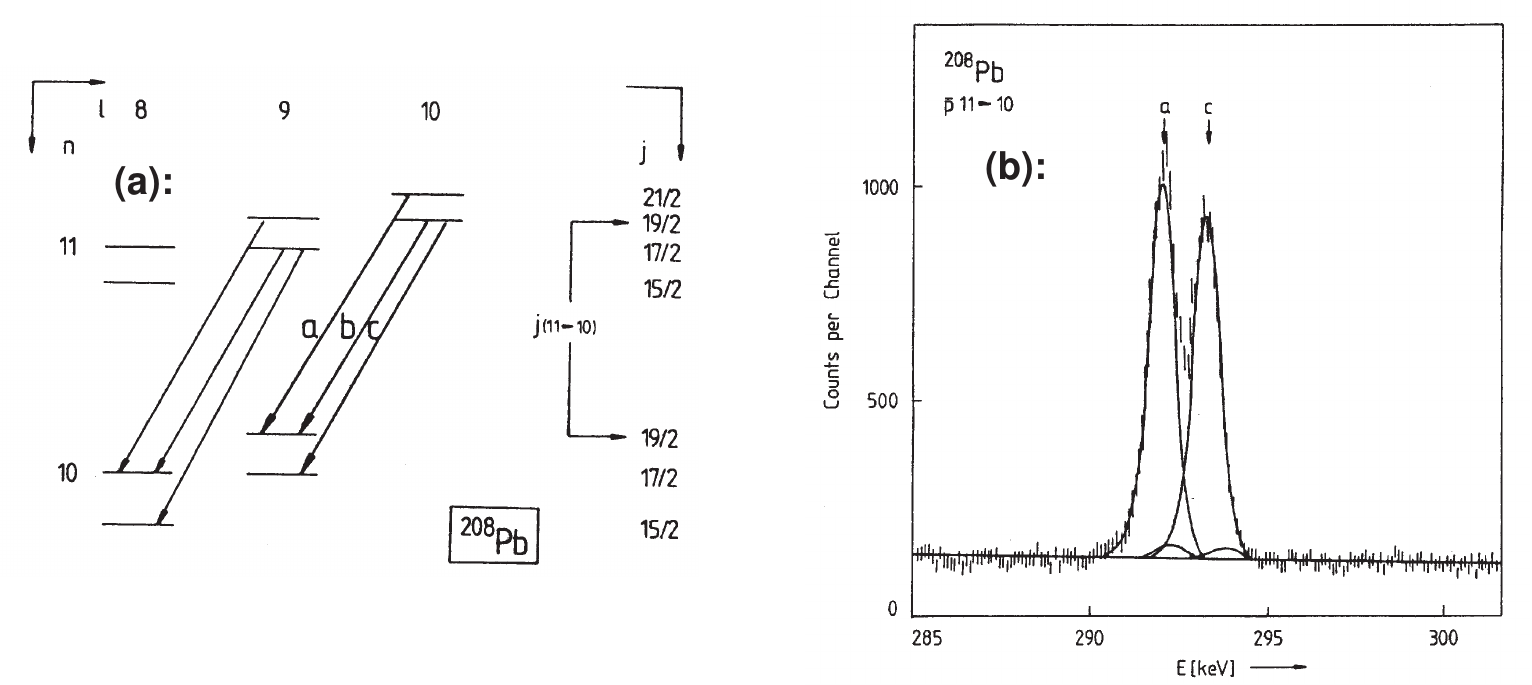,scale=1.0}
\end{minipage}
\begin{minipage}[t]{16.5 cm}
\caption{\label{pbxray} 
(a) Energy level and transition scheme in $\overline{p}{\rm ^{208}Pb}$ atoms. 
(b) X-ray spectrum of the transition $n=11$$\rightarrow$$10$.                                              
The two peaks indicated by ``a'' and ``c'' correspond to the                                                     
two transitions indicated in the level diagram of (a).                                                          
Figures from Ref.~\cite{kreissl}.}
\end{minipage}
\end{center}
\end{figure}

The PS186 collaboration of LEAR measured the magnetic moment $\mu_{\overline{p}}$ of $\overline{p}$
 by X-ray spectroscopy of antiprotonic lead ($\overline{p}{\rm ^{208}Pb}$) atoms
\cite{kreissl}. The experiment involved stopping some $7\times 10^8$ 
$\overline{p}$ of energy $E=20$--50 MeV in an isotopically pure $^{208}$Pb target. 
The resulting $\overline{p}{\rm ^{208}Pb}$ emitted characteristic X-rays as the $\overline{p}$ 
cascaded radiatively through the atomic levels.
X-rays with energies $E<700$ keV were detected by 
germanium semiconductor detectors surrounding the target. The value 
$\mu_{\overline{p}}$ was determined by measuring the "circular" (i.e., those
between states with large principal $n$ and angular momentum $\ell$ quantum numbers) 
transitions $(n,\ell)=(11,10)$$\rightarrow$$(10,9)$ and $(11,9)$$\rightarrow$$(10,8)$
with a transition energy $E=292.5$ keV. The energy level 
diagram is shown in Fig.~\ref{pbxray} (a). Since the spatial overlap between
the $\overline{p}$ and Pb nucleus was negligibly small for these states,
effects due to the strong interaction could be ignored. The measured spectrum 
(Fig.~\ref{pbxray} (b)) had a two-peak structure separated by an 
interval $\Delta E=1199(5)$ eV, which corresponded to the fine structure 
splitting arising from the interaction between the $\overline{p}$ magnetic moment 
and its orbital angular momentum $\ell$. By adjusting the $\mu_{\overline{p}}$-value 
used in theoretical QED calculations and perturbative 
evaluations \cite{borie,bohnert} to reproduce the measured splitting, a value
\begin{equation}
\mu_{\overline{p}}=-2.8005(90)\mu_{\rm nucl},
\end{equation}
was obtained in nuclear magnetons. This agreed with the magnetic moment of $p$ \cite{jpcrd},
\begin{equation}
\mu_{p}=-2.792847356(23)\mu_{\rm nucl},
\label{mup}
\end{equation}
with a precision of $0.3$$\%$. The $p$ value is currently derived
from measurements of the hyperfine splitting of H, using a maser in a variable magnetic
field \cite{Winkler:1972,Karshenboim:PLB566:2003}.  

\subsection{\it Some other tests of $CPT$  symmetry}
\label{othercpt}

The neutral $K^0$ meson oscillates with its antiparticle $\overline{K}^0$
with a frequency of 5 GHz via the weak interaction. The fact that one of the
mass eigenstates of this system $K_S$ decays at a similar rate of $\sim 11$ GHz makes
it possible to experimentally study these oscillations with a high precision 
by detecting the decay products. An analysis \cite{pdg2010} combining the results 
of the CPLEAR, KLOE, KTeV, and NA48 experiments yielded a value,
\begin{equation}
\left|m_{K^0}-m_{\overline{K}^0}\right|<4.0\times 10^{-19} \ {\rm GeV},
\end{equation} 
at a confidence level of 95$\%$. This is commonly considered to be the most precise
test of $CPT$ invariance involving mesons, although this evaluation critically depends 
on some assumptions \cite{kobayashi1992} on the $CP$-violating parameters.
 
 The relative mass difference between $e^-$ and $e^+$ have been constrained to  a precision,
 \begin{equation}
\left|m_{e^+}-m_{e^-}\right|/m_{\rm average}<8\times 10^{-9},
\end{equation}
with a confidence level of 90$\%$. This result was obtained by employing laser spectroscopy
to measure the 1$^3S_1$-2$^3S_1$ interval of positronium ($Ps$) with a precision of $2.6\times 10^{-9}$
\cite{fee93}. The measured atomic transition frequency was then compared with QED calculations 
to derive the above limit.
 
The Liquid Scintillator Neutrino Detector (LSND) collaboration reported an excess of 
$\overline{\nu}_\mathrm{e}$ found in a $\overline{\nu}_\mu$ beam with a statistical
significance of $\sim 3.8\sigma$ \cite{lsnd2001}. The data suggested
$\overline{\nu}_\mu$$\rightarrow$$\overline{\nu}_\mathrm{e}$ flavor oscillations involving a
mass-squared difference of the antineutrino mass eigenstates $\Delta m^2=0.2$--10 eV$^2$. 
This appeared to conflict with observations of atmospheric 
and solar $\nu_\mathrm{e}$ neutrino experiments which imply much smaller
$\Delta m^2$ values of $<2\times 10^{-3}$ eV$^2$.
The Booster Neutrino Experiment (MiniBooNE) also searched for the same oscillations,
and recently reported results \cite{miniboone2010} consistent with LSND, i.e. $\Delta m^2=0.1-1.0$ eV$^2$.
This result had a $2.7\sigma$ confidence level when the data was analyzed in the context of a model
involving the mixing of two neutrinos, but the excess $\overline{\nu}_\mathrm{e}$ events was
also consistent with a null hypothesis at a 3$\%$ confidence level.
Many theoretical groups have attempted to explain these anomalies by suggesting either the existence of
a fourth sterile neutrino that does not weakly interact, or $CPT$ violation
(e.g., that neutrinos and antineutrinos have different masses).
The Main Injector Neutrino Oscillation Search (MINOS) collaboration measured the
disappearance of $\overline{\nu}_\mu$ using an accelerator beam. They initially reported
$\Delta m^2$ and mixing angle values that differed by $40\%$ from the corresponding ${\nu}_\mu$ values \cite{minos2011}. 
This difference was reduced \cite{minos2012} to negligible levels after more statistics was recently collected.
                                                                                                                                                                                                                
\section{Production of low-energy antiprotons}
\label{sec:ad}

\subsection{\it Antiproton Decelerator}
                                                                                                                                                                                 
\begin{figure}[tb]
\epsfysize=9.0cm
\begin{center}
\begin{minipage}[t]{14 cm}
\epsfig{file=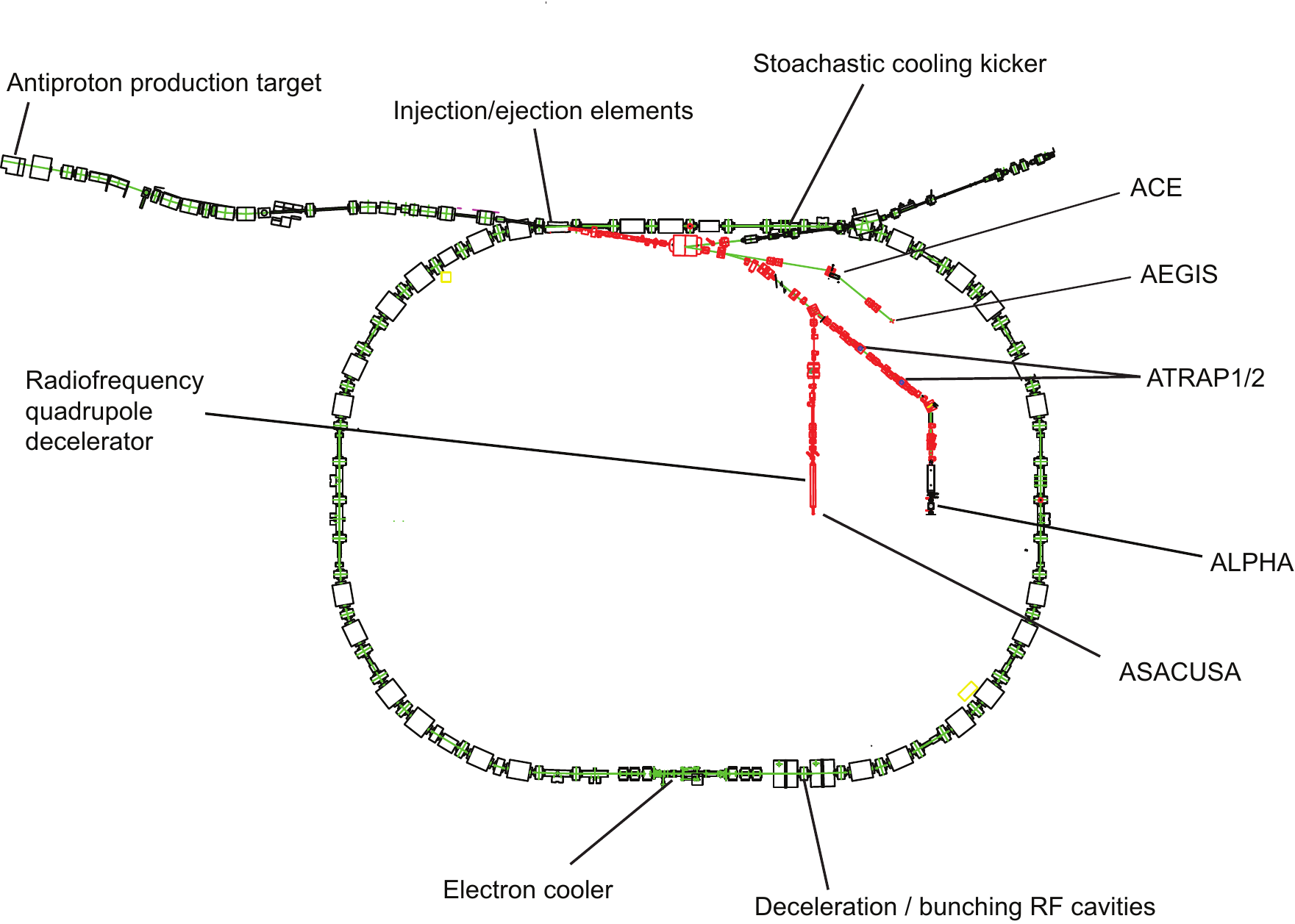,scale=0.8}
\end{minipage}
\begin{minipage}[t]{16.5 cm}
\caption{\label{fig:adall} Schematic diagram of Antiproton Decelerator (AD)
and the positions of experimental installations in 2012.
The $\overline{p}$ emerging from the production target are injected into the 
AD and decelerated to a kinetic energy of 5.3 MeV over a 100-s cycle.
The $\overline{p}$ beam is then transported by magnetic beamlines to the 
experiments ACE, \aegis, ALPHA, ASACUSA and ATRAP. Large numbers of
auxiliary instruments, concrete shielding, and support buildings are not shown for 
clarity.}
\end{minipage}
\end{center}
\end{figure}

The Antiproton Decelerator (AD)
\cite{Maury:97,Pavel:04} is currently the world's only source of                                                                                                            
low energy ($E=5.3$ MeV) $\overline{p}$ (Fig.~\ref{fig:adall}).                                                                                                                                     
The $\overline{p}$ are produced by colliding $p$ beams                                                                                                         
on an Ir target. In a small fraction of the collisions, the following reaction                                                                                                
(or a similar one involving target $n$) occurs under the                                                                                                            
conservation requirements of energy, momentum, and nucleon number,                                                                                                                 
\begin{equation}                                                                                                                                                                   
p{(\rm beam)}+p{(\rm target)}\rightarrow p+p+p+\overline{p}.                                                                                                                           
\end{equation}                                                                                                                                                                     
The minimum kinetic energy of the incoming $p$ needed for this is around
$E\sim 6$ GeV, whereas the $\overline{p}$ and three $p$ emerge with laboratory 
energies $E\sim 1$ GeV.                                                                                   
The PS-AD combination uses higher $p$ energies ($E=26$ GeV) to increase the                                                                                                      
production yield of $\overline{p}$, so that the $\overline{p}$ emerge with a correspondingly
higher energy $E\sim 3.6$ GeV. This energy must be reduced by eight orders of magnitude 
before the $\overline{p}$ can be used for the trap and atomic spectroscopy experiments
described in this paper. A simple deceleration of a cloud containing                                                                                                         
$N$ number of $\overline{p}$, however, would lead to an adiabatic increase in its phase-space                                                                                                    
density $D$ defined as,                                                                                                                                                             
\begin{equation}                                                                                                                                                                   
D=\frac{N}{\sqrt{E_hE_v}L\Delta p/p},  \label{phasespace}                                                                                                                                           
\label{emit}                                                                                                                                                                       
\end{equation}                                                                                                                                                                     
where $E_h$ and $E_v$ denote the horizontal and vertical emittances of the                                                                                                       
cloud, $L$ its longitudinal length, and                                                                                                                                            
$\Delta p/p$ the spread of the $\overline{p}$ momentum distribution \cite{Mohl:97}.                                                                                                                                   

The AD (Fig.~\ref{fig:adall}) is an oval-shaped, 188-m circumference synchrotron. It consists of
four straight sections where the instruments needed for cooling
(i.e., reduce the phase-space and increase the $D$-value) of the beam, 
the RF cavities which decelerates the $\overline{p}$, and diagnostics equipment are 
placed. A series of dipole and quadrupole magnets in the four bending sections compensates 
the dispersion and chromaticity in the beam. Chromaticity here refers to the momentum dependence 
of the frequency of the transverse (e.g. betatron) oscillations in the circulating $\overline{p}$. These 
oscillations may cause the beam to increase in size and strike the inner walls of the synchrotron, 
unless they are compensated.
  
\begin{figure}[tb]
\epsfysize=9.0cm
\begin{center}
\begin{minipage}[t]{10 cm}
\epsfig{file=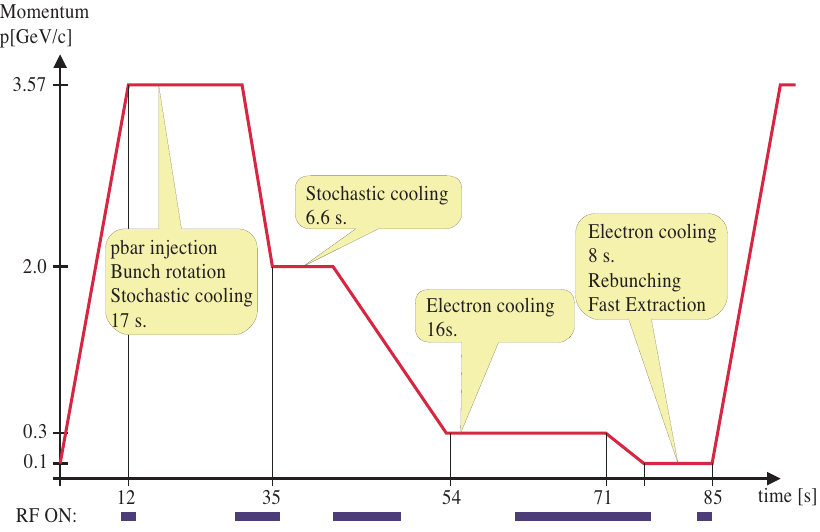,scale=1.2}
\end{minipage}
\begin{minipage}[t]{16.5 cm}
\caption{\label{fig:adcycle} Typical machine cycle of Antiproton Decelerator,
showing the momentum of $\overline{p}$ as a function of time elapsed. 
The timings and durations of the stochastic and electron cooling are indicated.
Figure from Ref.~\cite{Pavel:04}.}
\end{minipage}
\end{center}
\end{figure}

The AD decelerates and cools the $\overline{p}$ \cite{Maury:97,Pavel:04} over a 
100-s cycle in the following way (Fig.~\ref{fig:adcycle}),                                                                                                                                                                                                                                       
i): {\it $\overline{p}$ production and capture,} a pulsed beam containing
$\sim 1.5\times 10^{13}$ protons provided by the CERN Proton Synchrotron (PS) 
is allowed to strike a 50-mm-long Ir target, thereby producing a shower of 
$\overline{p}$ that are focused into a parallel beam by a magnetic horn-type lens. 
This beam containing $>5\times 10^7$ $\overline{p}$ of momentum $p\sim 3.6$ GeV/c, 
transverse emittance $\sim 200$ $\pi$ mm mrad, and momentum spread                                                                                                                  
$\Delta p/p\sim 6\%$ are injected into the AD,  ii): {\it RF bunch rotation}, 
RF fields stretch the pulse lengths of these bunches of $\overline{p}$ from 
$L=30$ m (corresponding to $\Delta t=25$ ns) to 190 m (150 ns).
This stretching in the longitudinal direction conversely reduces the $\Delta p/p$-value of the $\overline{p}$ 
ensemble to $\sim 1.5\%$, since the longitudinal emittance $L\Delta p/p$ is typically 
conserved during such a procedure,
iii): {\it stochastic cooling and deceleration,} so-called ``pickup" electrodes located along 
the circumference of AD detects deviations $\Delta p_i$ and $\Delta x_i$ in the 
momenta and transverse positions of small subgroups of
$\overline{p}$, relative to the mean values of all the orbiting $\overline{p}$. These signals                                                                                                         
are used to correct the orbits of the corresponding subgroups of $\overline{p}$, 
by applying electric pulses to steering electrodes located in the opposite side of the AD. 
Repeated corrections cause the beam to converge to an orbit with an emittance                                                                                                                        
3--4$\pi$ mm mrad and $\Delta p/p\sim 0.07\%$. The $\overline{p}$ are                                                                                                                
then decelerated to $p=2$ GeV/c and similarly cooled,                                                                                     
iv): {\it electron cooling,}  the $\overline{p}$ are decelerated to another intermediate
momentum $p=300$ MeV/c,
and allowed to merge with a 20-mm-diameter $e^-$ beam of current $I\sim 3$ A
in a collinear configuration over a 2-m-long section of the AD. The $e^-$                                                                                                     
and $\overline{p}$ velocities are matched so that in the center-of-mass frame, the 
$\overline{p}$ are bathed in a stationary $e^-$ cloud of low temperature. Coulomb                                                                                                                  
collisions transfer the "heat" of the $\overline{p}$ to the $e^-$.                                                                                                      
The $\overline{p}$ are finally decelerated to $p=100$ MeV/c and                                                                                                                       
electron-cooled to obtain a final emittance of $0.3\pi$ mm mrad and momentum
spread $\Delta p/p\sim 0.01\%$.   At the end of the above 100-s cycle, a 100--200-ns-long 
beam containing $\sim 3\times 10^7$ $\overline{p}$ of energy 5.3 MeV are ejected
from AD. Magnetic beamlines transport the $\overline{p}$ to one                                                                                                                
of four experimental zones located inside the AD (Fig.~\ref{fig:adall}).                                                                                                         

\subsection{\it Radiofrequency quadrupole decelerator}
 \label{rfqdexp}
 
\begin{figure}[tb]
\epsfysize=9.0cm
\begin{center}
\begin{minipage}[t]{15 cm}
\epsfig{file=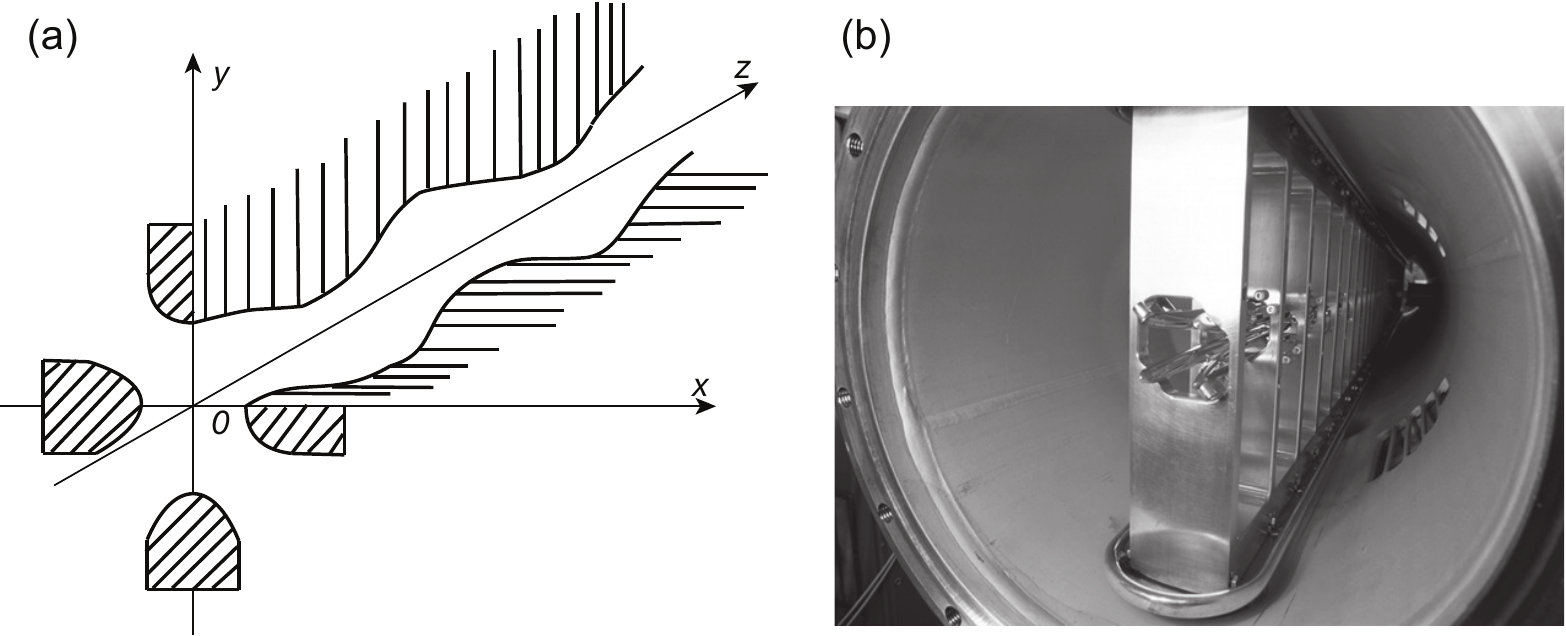,scale=1.0}
\end{minipage}
\begin{minipage}[t]{16.5 cm}
\caption{\label{fig:rfqdelec} (a) Schematic diagram and (b) photo of the quadrupole electrodes in the
radiofrequency quadrupole decelerator. Four 3.4-m-long rod electrodes arranged in a quadrupole
configuration are excited by a quadrupole RF field of 202.5 MHz. This field focuses the
antiproton beam traveling along the z-axis in the transverse direction. Peaks and troughs machined
on the surfaces of the electrode rods give rise to a longitudinal decelerating field.}
\end{minipage}
\end{center}
\end{figure}
    
CERN and the ASACUSA collaboration \cite{lombardi2001}
developed a radiofrequency quadrupole decelerator (RFQD)
which further reduced the 5.3-MeV energy of the $\overline{p}$ arriving from the
AD to even lower energies 10--120 keV needed for atomic physics experiments.
As implied by Eq.~\ref{phasespace}, when particle beams are decelerated 
their physical emittance increases until they can be lost by, e.g., hitting the 
inner walls of the accelerator. To avoid this, the RFQD strongly 
focuses the beam in the transverse direction during deceleration.  The RFQD 
consists of four 3.4-m-long rod electrodes arranged in a quadrupole configuration,
which are excited by a quadrupole RF field of $f\sim 202.5$ MHz 
(Fig.~\ref{fig:rfqdelec}). 
The high field (corresponding to a voltage on the rod electrodes of
$\sim 170$ kV) is achieved by placing the rods in a ladder-shaped 
cavity which resonates in the transverse electric quadrupole (TE210)
mode. The $\overline{p}$ injected axially into the $\sim 1$-cm-diam
aperture between the rods is thus alternately focused and defocused 
in the two transverse planes with a maximum electric field $\sim 30$ MV/m.
This "alternate gradient" focusing provides a net confining effect, which 
ensures that the $\overline{p}$ follows orbits of small diameter that 
oscillate around the RFQD axis. 

A series of peaks and troughs are machined on the surfaces of the electrode rods. The axial positions of the peaks are the same 
for opposing pairs of rod electrodes, but shifted by half a period between neighboring electrodes. This structure deviates a fraction of the transverse electric field into the longitudinal direction, and gives rise to a standing wave along the RFQD axis. This longitudinal component with a strength of a few MV/m decelerates the $\overline{p}$. 
The wavelength $\lambda_r$ of this undulating electrode structure is adjusted to correspond to the flight distance of a
$\overline{p}$ during a single 202.5-MHz RF cycle. The $\lambda_r$-value gradually decreases along the length of the electrode
as the beam is decelerated from $E=5.3$ MeV to $65$ keV along the RFQD.

The RFQD is operated in the following way. The 5.3-MeV $\overline{p}$ extracted from AD first
enters a so-called RF "bunching" cavity excited at $f=202.5$ MHz which is located some 3 m upstream
of the RFQD. This shapes the $\overline{p}$ beam into a train of 30 micropulses with a pulse length 
of $\Delta t=300$ ps. This bunching of the beam is needed to move the $\overline{p}$ into the longitudinal
acceptance of the RFQD defined by the RF phase. Of the $\overline{p}$ that entered the RFQD, some $\sim 25\%$ 
are decelerated to an energy $E=65$ keV. This energy can be varied between $E=10$ and 120 keV 
by biasing the electrodes of the RFQD with a DC potential. An additional energy-correcting 
RF cavity at the input of the RFQD is used to compensate for the changes in the energy of the 
incident beam resulting from this DC biasing, and variations in the energy of the $\overline{p}$ 
extracted from the AD. The majority of the $\overline{p}$ ($\sim 75\%$) misses the longitudinal acceptance 
and emerges with little or no  deceleration. Measurements show that the decelerated $\overline{p}$ beam
has a typical emittance of up to $\sim$100 $\pi$ mm mrad. 
                     
\section{Antihydrogen}
\label{sec:antihydrogen}

The $\overline{\rm H}$ atom is a pure antimatter system which is stable
and electrically neutral. Narrow electromagnetic resonances can be
readily excited between its internal energy levels by laser or microwave
irradiation. These characteristics should in principle allow tests of $CPT$ symmetry (Sect.~\ref{sec:cpt})
to be carried out with unprecedented experimental precision.  Perhaps even more 
fascinating is the prospect of investigating antimatter gravity using $\overline{\rm H}$.  
Both types of future experiments will be discussed in Sect.~\ref{sec:future}.  

\subsection{\it Production of fast antihydrogen atoms}

\begin{figure}[tbh!]
\begin{center}
\includegraphics[width=15cm]{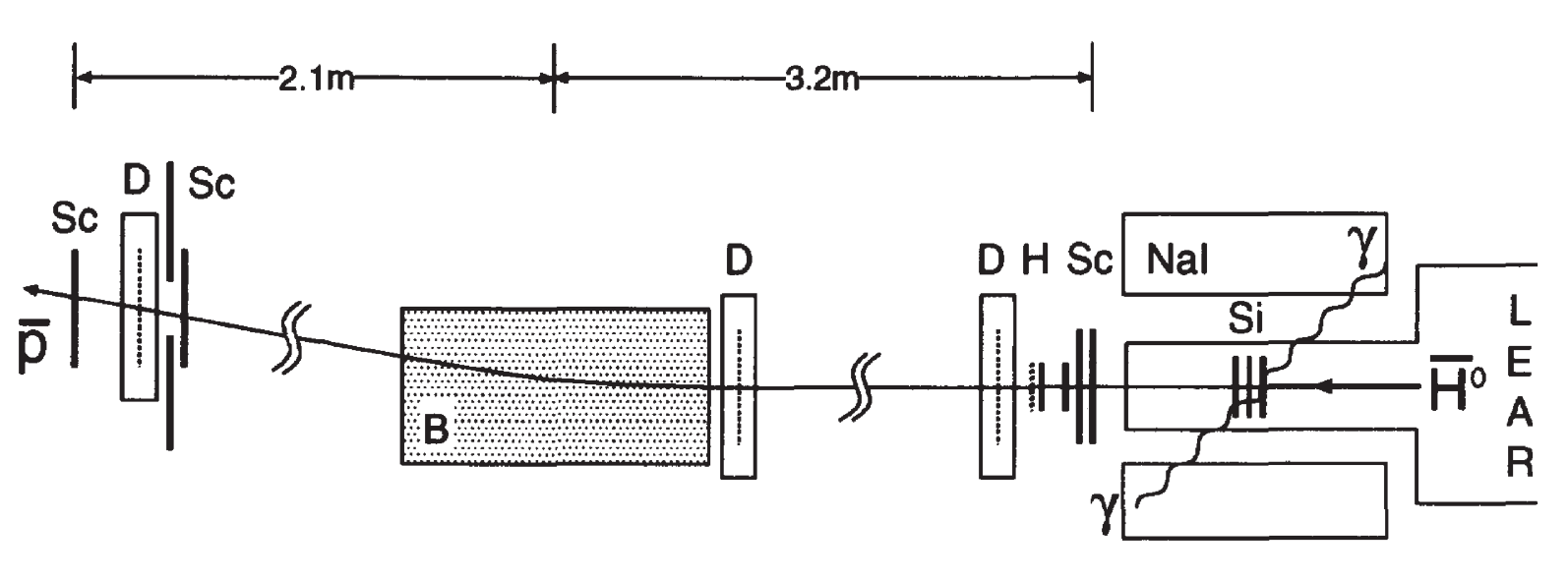}
\begin{minipage}[t]{16.5 cm}
\caption{Experimental layout of the PS210 experiment. 
The $\overline{\rm H}$ emerging from the LEAR storage ring
traversed three silicon detectors, where the $e^+$ was stripped
and allowed to come to rest in the silicon. The two
back-to-back gamma rays emerging from the $e^+$ annihilation
were identified by NaI scintillation counters. The $\overline{p}$
proceeded further along the beamline, traversing several scintillation 
counters and a magnetic spectrometer consisting of a dipole magnet
(B) and three position-sensitive drift chambers (D). The simultaneous
detection of $e^+$ and $\overline{p}$, together with some other
directional, timing, and energy cuts provided the identification of the
$\overline{\rm H}$ event.
Figure from Ref.~\cite{Baur:1996}.
\label{walter_setup}%
}
\end{minipage}
\end{center}
\end{figure}

In the first generation of $\overline{\rm H}$ experiments, 
$\overline{p}$ beams circulating in storage rings were allowed to
repeatedly traverse internal supersonic gas or cluster targets positioned in straight
sections of the rings. Some of the $\overline{p}$ scattered off the strong Coulomb 
field of the target nucleus $Z$, which induced the creation of a $e^+$-$e^-$
pair via the space-like $\gamma\gamma$ production process,
\begin{equation}
\overline{p}+Z\rightarrow\overline{p}+\gamma+\gamma+Z\rightarrow\overline{p}+e^++e^-+Z\rightarrow\overline{\rm H}+e^-+Z.
\end{equation}
In rare cases, the outgoing $\overline{p}$ and $e^+$ had similar velocities so that the two particles combined and formed a 
fast-moving $\overline{\rm H}$ atom \cite{Munger:1994,Bertulani:1998}.  The cross section for this process is extremely small, in the order of 2 pb $\times Z^2$ for a target nucleus of charge $Z$ and depending on the energy of the incoming $\overline{p}$.

In 1995, the PS\,210 collaboration produced $\overline{\rm H}$ 
by circulating a beam of $\sim 10^{10}$ $\overline{p}$ in LEAR at 
a momentum of 1.94 GeV/c, and allowing it to traverse a Xe cluster target 
of density $\sim 1\times 10^{12}$ cm$^2$. At these experimental conditions,
the $\overline{\rm H}$ production cross section was expected to be around 
6000 pb. Once neutral $\overline{\rm H}$ was formed, it was
no longer confined by the magnetic fields of the storage ring, and left LEAR through 
a gap in one of the dipole magnets. A sophisticated set of particle detectors 
(Fig.~\ref{walter_setup}) 
was set up to identify the escaping $\overline{\rm H}$, and distinguish them from 
any background due to, e.g., antineutron ($\overline{n}$) production. The $\overline{\rm H}$ were
directed towards a stack of silicon detectors (indicated by Si in the figure),
where the $e^+$ was stripped away from the $\overline{p}$ and allowed to
stop in one of the detectors. Pairs of 511-keV photons emerging from the resulting
$e^+$ annihilation were identified using NaI scintillation counters. The
$\overline{p}$ emerging from the ionization of $\overline{\rm H}$ continued through
a set of scintillation counters, and a magnetic dipole spectrometer containing
three position-sensitive drift chambers. The $\overline{p}$ was thus identified
by its time-of-flight and magnetic rigidity. A good $\overline{\rm H}$ event 
consisted of a coincidence between the $\overline{p}$ and $e^+$ signals,
together with additional timing, energy, and directional cuts.
Eleven $\overline{\rm H}$ atoms were detected during 15 hours 
of  beamtime \cite{Baur:1996}.  The Fermilab E\,862 experiment later 
reported the detection of 57 $\overline{\rm H}$ atoms \cite{Blanford:1998}
in a similar experiment. 

The velocity of the $\overline{\rm H}$ produced in this way can reach 
$>90$\,\% of the speed of light, whereas reducing the momentum of the
circulating $\overline{p}$ beam in the experiment leads to a rapid reduction of the 
$\overline{\rm H}$ production cross section. Proposals have been put forward 
to measure the energy-level splittings and Lamb shifts of these fast $\overline{\rm H}$ in flight
\cite{Meshkov:1998,Blanford:Lamb:1998,Westig:EPJD57:2010}. 
It is quite natural, however, to expect that even higher precision can be achieved 
using slow $\overline{\rm H}$ beams or stationary atoms confined in a trap.
A second generation of experiments with the goal of producing and
investigating cold $\overline{\rm H}$ atoms was initiated in 1999 at the
AD. 

\subsection{\it Trapping and cooling of antiprotons}
\label{sec:pbartrapping}

\begin{figure}[htbp]
\epsfysize=8.0cm
\begin{center}
\begin{minipage}[t]{16.5 cm}
\begin{center}
\epsfig{file=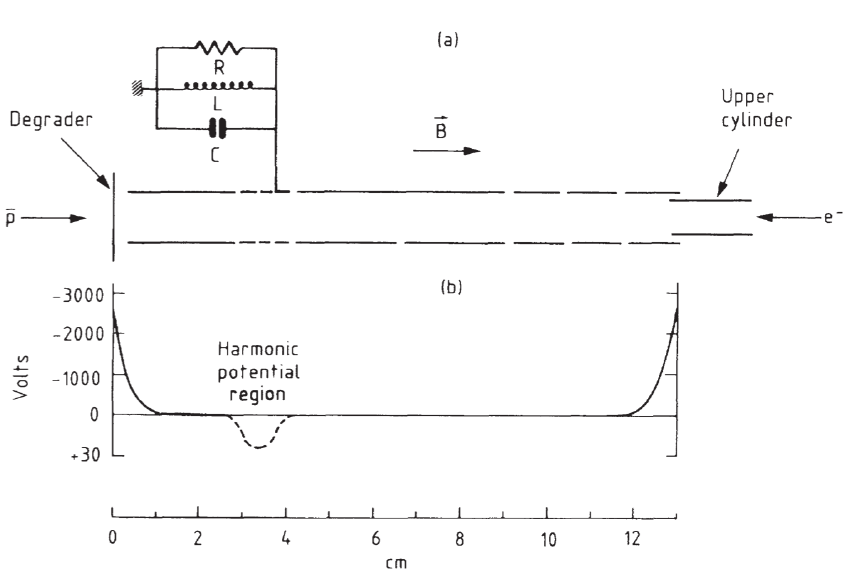,scale=1.5}
\end{center}
\end{minipage}
\begin{minipage}[t]{16.5 cm}
\caption{\label{fig:pbar_setup} 
(a) Schematic layout of a Penning trap used to capture and cool $\overline{p}$
demonstrated by the TRAP collaboration. (b) The $\overline{p}$ were decelerated
in a degrader foil to keV-scale energies before entering through the axis of the
cylindrical electrodes comprising the trap.
The $\overline{p}$ traveled to the end of the trap, and was reflected
by a -3 kV potential applied to one of the cylinder electrodes. Before the $\overline{p}$
could reach the entrance again and leave the trap, however, a second electrode near
the upstream end of the trap was rapidly biased to -3 kV, thereby trapping the $\overline{p}$ 
in a 12-cm-long rectangular potential well. The $\overline{p}$ were cooled by collisions with
a cloud of $e^-$ loaded in the harmonic potential region. 
Figure from Ref.~\cite{gabrielse1989}.}
\end{minipage}
\end{center}
\end{figure}

Before the AD experiments could synthesize $\overline{\rm H}$, the
ingredient $\overline{p}$ had to be trapped and cooled in Penning traps 
\cite{gabrielse1989,Brown:1986,Gabrielse:IJMSIP:1989a}.
As described in Sect.~\ref{trapcyclo}, these traps consist of a stack
of cylindrical electrodes placed in a magnetic solenoid.
The open geometry of the electrodes allows particles to enter along the
trap axis. Similar devices have been used for many years in the Malmberg 
variant of the Penning trap to confine non-neutral plasmas \cite{ONeil:1998}.
The highest kinetic energy of $\overline{p}$ that can be captured by
these traps are roughly equivalent to the electrostatic voltage applied to the
electrodes located at its two ends (Fig.~\ref{fig:pbar_setup}). 
In practice, to avoid electrical 
discharge in the confined space of the superconducting magnetic 
solenoid, this voltage is typically limited 
to a few kV. This is much smaller than the kinetic energy of $\overline{p}$
(5.3 MeV) arriving from AD, so that two alternative methods are currently
employed to first slow down the $\overline{p}$ to keV energies, prior 
to injection into the trap.

The first deceleration method involves passing the 
$\overline{p}$ through a so-called "degrader" foil or a gas cell filled
with, e.g., a SF$_6$/He mixture \cite{Gabrielse:Adv:01} in which they 
are slowed down by atomic collisions. Due to the range-straggling 
effects associated with this energy loss, however, a significant
(typically $\sim 99.9\%$) fraction of the $\overline{p}$ either annihilates in the degrader or emerges
with too high energy ($>10$ keV) to be trapped. It is therefore crucial to 
adjust the degrader thickness to maximize the $\overline{p}$ capture efficiency.
This can be accomplished by either rotating the degrader foil
\cite{Amsler:HydrogenAtom:2001} or by changing the 
gas density in the degrader cell. 
The decelerated $\overline{p}$ travel along the axis of the trap, before encountering 
a negative potential applied to a cylindrical electrode located at the
downstream end of the trap. This reflects the $\overline{p}$ at a 180-degree
angle back towards the trap entrance. Before the $\overline{p}$ can exit
the trap, a negative potential is rapidly pulsed on
to a second electrode located near the trap entrance. This results in the $\overline{p}$ 
being confined between the two electrodes comprising a longitudinal potential well 
of typical length $\sim 100$ mm.

The trapped keV $\overline{p}$ are then cooled by the so-called "sympathetic cooling" 
technique in the following way: clouds of $e^-$ or $e^+$ are first loaded
into the trap, where they undergo cyclotron motion in the strong magnetic field $B$
(Fig.~\ref{fig:pbar_setup}).
They cool down to the temperature ($T\sim 4$--10 K)
of the cryogenic environment by emitting synchrotron radiation \cite{Brown:1986},
provided that possible heating due to space-charge and plasma effects, or induced noise
on the electrodes can be neglected.  The time constant for this synchrotron cooling is 
proportional to $B^{-2}$, and corresponds to around 2.6 s in a 1-T field at an environmental
temperature of $T=4$ K. The cold $e^-$ plasmas are placed
in short (10--30 mm) harmonic potential wells, which are superimposed on the long ($>100$ mm)
well holding the $\overline{p}$. As the $\overline{p}$ elastically scatter off the $e^-$, 
they cool down and collect in the short potential wells containing the $e^-$.  
The long potential well, now empty, is ready to accept another $\overline{p}$ pulse.  
For a review on the early development of these $\overline{p}$ trapping, cooling, and 
stacking techniques, see Ref.~\cite{Gabrielse:Adv:01}.

The radiofrequency quadrupole decelerator (Sect.~\ref{rfqdexp}) 
provides an alternate way to slow down $\overline{p}$ to $\sim 100$ keV with a higher efficiency 
compared to the simple degrader technique. The ASACUSA collaboration allowed the 100-keV
$\overline{p}$ to traverse a thin ($\sim 1$ $\mu{\rm m}$) plastic 
degrader foil, before injecting them into a multiring 
Penning trap called MUSASHI. In this way, $>$1\,$\times$\,10$^7$ $\overline{p}$ 
were trapped out of seven $\overline{p}$ pulses provided by the AD. This constitutes a record 
in overall trapping efficiency of about 5\,\% \cite{Kuroda:PRST:2012}.  The $\overline{p}$ have 
subsequently been radially compressed in this trap \cite{Kuroda:PRL:2008} and extracted as a 
slow beam \cite{Torii:LEAP:2005}.

\subsection{\it Trapping and cooling of positrons}
\label{sec:positrontrapping}

\begin{figure}[tb]
\epsfysize=9.0cm
\begin{center}
\begin{minipage}[t]{15 cm}
\epsfig{file=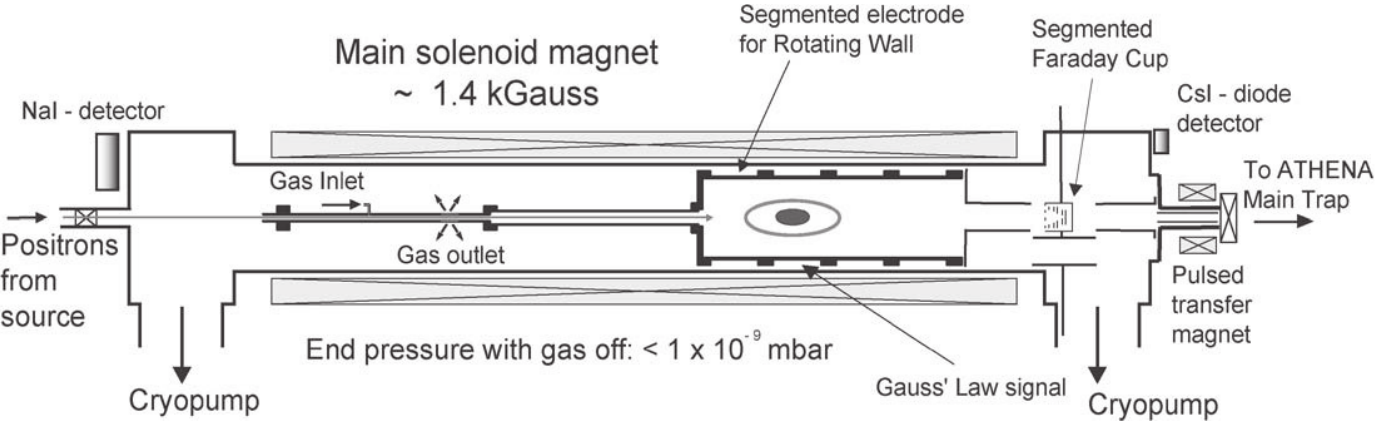,scale=1.1}
\end{minipage}
\begin{minipage}[t]{16.5 cm}
\caption{\label{fig:positron_setup} 
Layout of the $e^+$ accumulation trap used by the ATHENA collaboration.
A continuous beam of $e^+$ emitted by a $^{22}$Na source was moderated 
in a layer of solid Ne frozen on the surface of the cooled source. The $e^+$ was then 
transported into the main solenoid
containing the Penning trap. Differential pumping was used to establish a pressure gradient along
the axis of the trap, so that the $e^+$ passed through a region of high gas pressure
where they were rapidly decelerated, before reaching a low pressure region where they were 
accumulated. The other AD collaborations use similar accumulators based on the same principles.
Figure from Ref.~\cite{ATHENA:Apparatus:2004}.}
\end{minipage}
\end{center}
\end{figure}

The current generation of $\overline{\rm H}$ experiments utilize
radioactive $^{22}$Na sources to harvest $e^+$.
The kinetic energy of $e^+$ emerging from the $\beta^+$ decay in these sources
has a wide distribution with a maximum value of around $\sim 544$ keV.
These fast $e^+$ must be decelerated and cooled via a 
so-called "moderation" process before they can be accumulated.
This moderation is typically accomplished by first allowing the $e^+$ to pass through 
thin foils of solid Ne \cite{mills1986} or W. Solid Ne grown on the surface of a cooled $^{22}$Na
source has been experimentally 
found to be a particularly efficient moderator; indeed, some $\sim 7\times 10^{-3}$ 
of the $e^+$ that traverse these moderators typically emerge 
with a kinetic energy of a few eV. This high efficiency is primarily due to the fact
that solid Ne is an insulator with a large band gap energy $E_b\sim 21$eV.
High-energy $e^+$ traveling through the Ne foil initially lose energy by ionization 
process. When the $e^+$ is slowed down below $E_b$, however, further
energy loss can only proceed via the excitation of low-energy phonons
with a relatively small cross section. The $e^+$ can therefore travel greater distances 
($\sim 0.5$ $\mu$m) in the Ne foil and emerge from the other side,
compared to other metallic moderators. In addition, 
$e^+$ can reflect off the surface of the Ne foil with a high re-emission probability $\sim 90\%$.
By utilizing a conical-shaped Ne foil of carefully adjusted thickness, a moderated beam 
containing several $10^6$ $e^+$ per second can be routinely produced from a 75 mCi 
(i.e., $2.7\times 10^9$ decays per second) source. 

The moderated $e^+$ are then allowed to enter a Penning trap (Fig.~\ref{fig:positron_setup}) filled with 
${\rm N}_2$ gas at pressures between $10^{-1}$ and $10^{-7}$ Pa \cite{Murphy:PRA:1992,Surko:1997,Surko:PP11:2004}. 
An $e^+$ colliding with a ${\rm N}_2$ molecule loses around $9$ eV of kinetic 
energy via the excitation transition, $e^++{\rm N}_2\rightarrow e^++{\rm N}_2^*$.
Since the cross section for this reaction is large, the $e^+$ are rapidly 
slowed down by successive collisions in the trap. Only a small fraction 
of $e^+$ is lost by forming positronium ($Ps$) atoms via the reaction, 
$e^++{\rm N}_2\rightarrow {Ps}+{\rm N^+_2}$ and subsequently annihilates.
Differential pumping is used to establish a pressure gradient along
the axis of the trap, so that the $e^+$ initially passes through a region of high gas pressure
$10^{-1}$ Pa where they are rapidly decelerated, before reaching a low ($10^{-4}-10^{-7}$ Pa) 
pressure region where they are accumulated.
In this way, the trapped $e^+$ typically retain lifetimes of $\sim 100$ s against 
annihilation.  

The description provided in Sect.~\ref{trapcyclo} of particles 
undergoing three types of characteristic oscillations in the harmonic electrostatic
potential of a Penning trap is valid only when the number of confined particles is very low. When large
numbers of $e^+$ are trapped, the space charge of the $e^+$ cloud 
distorts the electrostatic potential, which no longer appears harmonic.
At thermal equilibrium, the density of $e^+$ in a plasma becomes uniform
and the $e^+$ begins to rotate around the symmetry axis of the magnetic field
at a constant angular velocity in a manner resembling a rigid rotor. Numerous
collective modes can be excited in the plasma, which induces changes in its shape.
The radial size for example can be compressed by applying a torque in the form of a 
rotating electric field on the plasma surface, and thereby increasing the angular 
momentum. This is normally carried out by segmenting some of the cylindrical
electrodes of the trap in the azimuthal direction, and applying
oscillating electric potentials of the correct relative phases. The frequency of this 
``rotating wall'' is typically tuned slightly higher than the rotation frequency of 
the plasma, such that it excites the so-called Trivelpiece-Gould plasma
modes \cite{Hollmann:2000}. This important technique increases both the
density, number, and confinement lifetime of $e^+$ in the accumulation trap. The maximum
density $\rho$ of particles of mass $m$ that can be stably confined in the
magnetic field $B$ of the trap, however, is ultimately defined by the Brillouin limit, 
$\rho<B^2/2\mu_0mc^2$, where the vacuum permeability is denoted by 
$\mu_0$ and the speed of light by $c$.

More than $10^8$ $e^+$ have been accumulated within several minutes 
in a Penning trap developed by the ATHENA and ALPHA collaborations \cite{ATHENA:Apparatus:2004,ALPHA:JPCS262:2011}.  
ATRAP and ASACUSA use accumulators of similar design \cite{comeau2012,Imao:HI194:2009}.  
The $e^+$ emerging from all these accumulation traps are then typically injected into 
``nested" Penning traps (see below) to synthesize $\overline{\rm H}$.  Other methods to 
accumulate $e^+$ (see \cite{Gabrielse:AMOP50:2005} for a review) are not presently 
used at the AD. The future GBAR collaboration plans to produce even larger numbers of
$e^+$ by colliding electrons accelerated by an $e^-$ linac on a production target,
as described in Sect.~\ref{sec:gbar}.

\subsection{\it Formation of cold antihydrogen}
\label{sec:hbarformation}

\subsubsection{\it Antihydrogen production methods}
\label{formation}

When trapped clouds of $\overline{p}$ and $e^+$ are brought together in a trap,
they can recombine and form $\overline{\rm H}$ in several possible ways that are enumerated
below,

{\bfseries{Spontaneous radiative recombination:}} In this process,
\begin{equation}
 \bar{p} + e^+ \rightarrow \bar{\rm H} + h \nu,
\end{equation}
$\overline{\rm H}$ forms in binary collisions between $\overline{p}$ and
$e^+$ which is accompanied by the emission of a photon that carries away the excess energy and 
momentum. This reaction resembles the time-reversal of the photoionization process, and tends to produce 
$\overline{\rm H}$ occupying states of small principal quantum numbers \cite{Neumann:1983}. 
The cross section of this process is small because of the slow rate of photon emission 
compared to the typical time scales involved in thermal collisions between $\overline{p}$ and 
$e^+$. The corresponding $p+e^-\rightarrow {\rm H}+h\nu$ process, however, occurs readily in the 
interstellar medium and has been thoroughly investigated in astrophysics \cite{Bates:CaseStudies:1974}.
In the laboratory, this process has been studied by accelerating beams of
protons in storage rings, and allowing them to pass through a co-propagating $e^-$
beam of an electron cooler \cite{Bell:1982,Andersen:1990}.

{\bfseries{Stimulated radiative recombination:}} In this process, a
radiation field is applied to stimulate the radiative recombination, 
\begin{equation}
~\bar{p} + e^+ + h \nu\rightarrow \bar{\rm H} + 2 ~ h \nu~.
\end{equation}
The corresponding reaction for H has been experimentally observed both in a storage ring 
\cite{Schuessler:1995a} and with merged beams of $e^-$ 
and $p$ \cite{Yousif:PRL:1991}.  For a detailed and careful discussion
of possible enhancement factors, see Refs.~\cite{Wolf:NATO:1992,Wolf:97}.

{\bfseries{Pulsed-field recombination:}} This process resembles 
the time-reversal of ionization by the application of a pulsed field.  
An external electric field is applied to a pair of $\overline{p}$ and $e^+$,
so that the sum of the Coulomb field of the $\overline{p}$ and the external
field constitutes
a saddle potential. At the instant when the $e^+$ arrives at the saddle point, 
the external electric field is turned off, thereby forming the 
$\overline{\rm H}$ atom. Pulsed field recombination has been demonstrated 
using Rb$^+$ ion and $e^-$ pairs produced by laser-induced photoionization.  
High recombination rates of 0.3\,\% have been observed
\cite{Noordam:2000,Wesdorp:2001}.  Another experiment observed an
enhancement in the recombination rate of Ca$^+$ ions and $e^-$ using THz
half-cycle laser pulses in the presence of a static electric field
\cite{Zeibel:2002}.

\begin{figure}[tb]
\epsfysize=9.0cm
\begin{center}
\begin{minipage}[t]{15 cm}
\epsfig{file=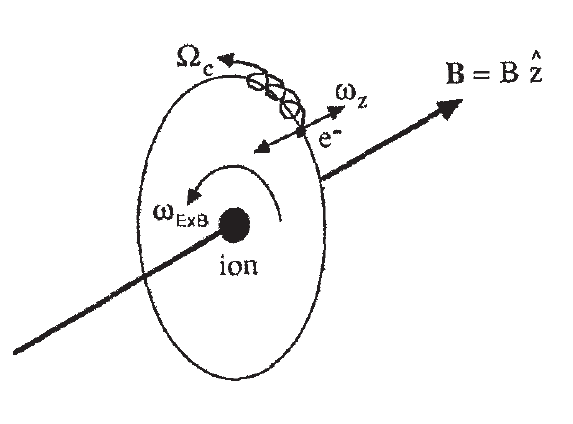,scale=1.1}
\end{minipage}
\begin{minipage}[t]{16.5 cm}
\caption{\label{guiding} 
Drawing of a weakly-bound, guiding-center atom in a strong magnetic
field. The $e^-$ in this example executes three types of motion, i): a 
cyclotron-like motion of frequency $\Omega_c$ in the external
magnetic field $B$, ii): an axial motion
of frequency $\omega_z$ along the magnetic field line while trapped 
in the Coulomb potential of the ion, and iii): a slow $E\times B$ drift
around the ion. Figure from Ref.~\cite{Glinsky:91}.}
\end{minipage}
\end{center}
\end{figure}

{\bfseries{Three-body recombination:}} In simplified form
this process can be denoted as,
\begin{equation}
~\bar{p} + e^+ + e^+ \rightarrow \bar{\rm H} + e^+~.
\end{equation}
Here a $\overline{p}$ collides with two $e^+$ simultaneously,
$\overline{\rm H}$ is formed, and the outgoing $e^+$ carries
away excess energy and momentum \cite{Gabrielse:1988b}.  
Actually this three-body recombination does
{\em{not}} occur in a single step, but involves a complex cascade through
intermediate excited states that arise during the collision. 
The rate of recombination is determined by a kinetic bottleneck which lies at a binding
energy of a few $k_B T$ below the ionization threshold, where the temperature of the 
$e^+$ is denoted by $T$ \cite{Glinsky:91} . Since $k_BT$ is four orders of magnitude
smaller than the Rydberg energy, the collisional dynamics which governs the system is
expected to be classical. 
The recombination rate increases with the square of the $e^+$ density, 
and is proportional to $T^{-9/2}$ \cite{Makin:63}.  
Although theoretical calculations indicate that the strong magnetic fields of 
Penning traps reduces this recombination rate, it is nevertheless expected to be
the dominant $\overline{\rm H}$ production process at low temperatures of the
$e^+$ and $\overline{p}$ clouds \cite{Glinsky:91,Menshikov:95,Fedichev:1997,Robicheaux:PRA73:2006}.
The weakly-bound $\overline{\rm H}$ formed in the magnetic 
field of the trap is theoretically characterized as a so-called ``guiding-center atom".
Here the radius of the cyclotron motion of $e^+$ in the magnetic field at frequency
$\Omega_c$ is much smaller than the scale length in which the interatomic interaction 
potential varies, and so the atom resembles a classical object. The guiding center of 
the $e^+$ orbit (Fig.~\ref{guiding}) simultaneously oscillates in two ways,  i): rapidly 
along the magnetic field line in the Coulomb potential of $\overline{p}$ at frequency $\omega_z$, and
 ii): more slowly executing a $E\times B$ drift around the $\overline{p}$ at frequency
$\omega_{E\times B}$. 

{\bfseries{Resonant charge-exchange:}} The $\overline{\rm H}$ can form in 
collisions between $\overline{p}$ and excited $Ps$ atoms \cite{Humberston:1987},
\begin{equation}
{Ps}^\ast +\bar{p} \rightarrow \bar{\rm H}^\ast + e^-.    
\end{equation}
The corresponding reaction of H formed in collisions of $p$ with
$Ps^{\ast}$ has been experimentally observed \cite{Merrison:1997}. 
The cross section for this should increase with $\propto n^4$ for a 
$Ps^\ast$ occupying excited states with principal quantum number $n$ \cite{Charlton:1990}. 
An elegant method for preparing the excited $Ps^{\ast}$ \cite{Hessels:1998} 
involves first irradiating a thermal beam of alkali atoms with lasers. 
The resulting Rydberg alkaline atoms are next allowed to traverse a cloud of $e^+$ 
confined in a Penning trap. This in turn results in the formation of Rydberg $Ps^\ast$ by charge-exchange 
process. The cross section for this $Ps^\ast$ formation reaches a maximum at
the energy-matching condition where the total binding energies of the initial alkali 
and final $Ps^\ast$ states are similar. The process was experimentally demonstrated
using laser-excited Cs atoms \cite{Speck:2004}. The neutral $Ps^\ast$ atoms are then
allowed to traverse a cloud of $\overline{p}$ confined in an adjacent Penning trap,
thereby producing Rydberg $\overline{\rm H}$ via a second charge-exchange step.

Among these $\overline{\rm H}$ formation processes, 
three-body recombination and resonant charge-exchange have been experimentally demonstrated,
as described in Sects.~\ref{sec:tbr} and \ref{sec:charge_exchange}.
Two other methods have been attempted so far without success:
pulsed-field recombination could not be made to work by the ATRAP collaboration for unknown reasons
(see Ref.~\cite{Gabrielse:AMOP50:2005}, p.~161).  Laser-stimulated radiative recombination 
was attempted by ATHENA \cite{Athena:LaserStim:2006}, but no such event was 
observed so far.  Given the fact that the enhancement to the recombination rate induced by
laser irradiation is understood both theoretically and in experiments involving $p$
in storage rings,  the negative result suggests that spontaneous radiative recombination 
does not contribute appreciably to $\overline{\rm H}$ formation at the experimental conditions 
now used by the AD collaborations.

\subsubsection{\it Interaction of trapped antiprotons and positrons}

\begin{figure}[tbh!]
\begin{center}
\includegraphics[width=12cm]{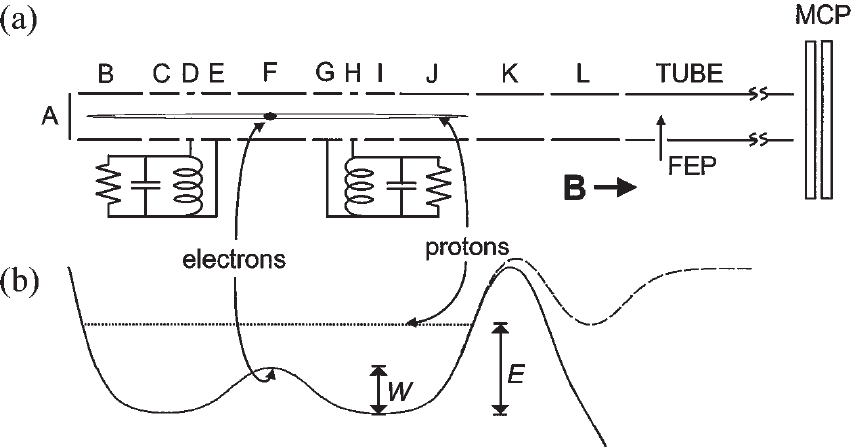}
\begin{minipage}[t]{16.5 cm}
\caption{(a) Example of a nested Penning trap,
including the scale outline of the inner surface
of the electrodes and (b) the electrostatic
potential wells. Here $p$ and $e^-$ are
simultaneously trapped in potential wells of
depth $E$ and $W$. FEP denotes field emission
point, MCP micro-channel plate. Figure from \cite{Hall:1996}.
\label{nested}%
}
\end{minipage}
\end{center}
\end{figure}

Except for resonant charge-exchange, all of the above $\overline{\rm H}$
production methods involve the spatial overlapping of $\overline{p}$ and $e^+$ clouds.
To confine these oppositely-charged particles in close proximity to each other
\cite{Gabrielse:1988b,Hall:1996}, so-called ``nested" Penning traps are employed
in which a stack of cylindrical electrodes generate electrostatic potential wells of alternate 
polarity. In the example of Fig.~\ref{nested}, a positive potential which axially confines $e^-$
is superimposed on a longer negative potential which confines $p$. One disadvantage of this
method is that in thermal equilibrium at low temperatures, the $\overline{p}$ and $e^+$ confined
in this type of trap collect in their respective potential wells, 
i.e., they decouple and separate into two spatially-separated clouds so that
$\overline{\rm H}$ may no longer be produced \cite{Ordonez:1997}.  
Overlap between the two clouds can therefore only be achieved in a non-equilibrium condition, e.g., by
accelerating the $\overline{p}$ into the $e^+$ plasma \cite{ATRAP:Hbar:2002} or
by exciting them using RF fields applied to the trap electrodes
\cite{ATRAP:HbarStates:2002,ALPHA:Autoresonant:2011}.  One milestone
on the way to $\overline{\rm H}$ production was the observation of
sympathetic cooling of $\overline{p}$ with $e^+$ in a nested Penning
trap \cite{Gabrielse:PLB:2001}, which demonstrated their mutual 
interactions at even low relative energies.

Alternate schemes for simultaneous trapping of $\overline{p}$ and $e^+$
involve Paul (RF) traps combined with Penning traps
\cite{Li:89,Walz:95} or two-frequency Paul-traps
\cite{Dehmelt:PhysScr:1995,Widmann:LoI:2003}.  The advantage of these
methods is that overlap can be achieved in thermal equilibrium, but
the drawback of Paul traps is that within clouds of particles, nonlinear
dynamics can couple energy from the trap-driving RF field 
into the motion of particles \cite{Bluemel:89}.  This so-called RF-heating 
process makes it difficult to achieve very low temperatures in the
particle clouds. So far, Paul traps have not been used for $\overline{\rm H}$ 
production, although the ASACUSA collaboration is attempting to develop
particle cooling techniques to try and alleviate the above problems.

\subsubsection{\it Antihydrogen production by three-body recombination}
\label{sec:tbr}

\begin{figure}[tbh!]
\begin{center}
\includegraphics[width=15cm]{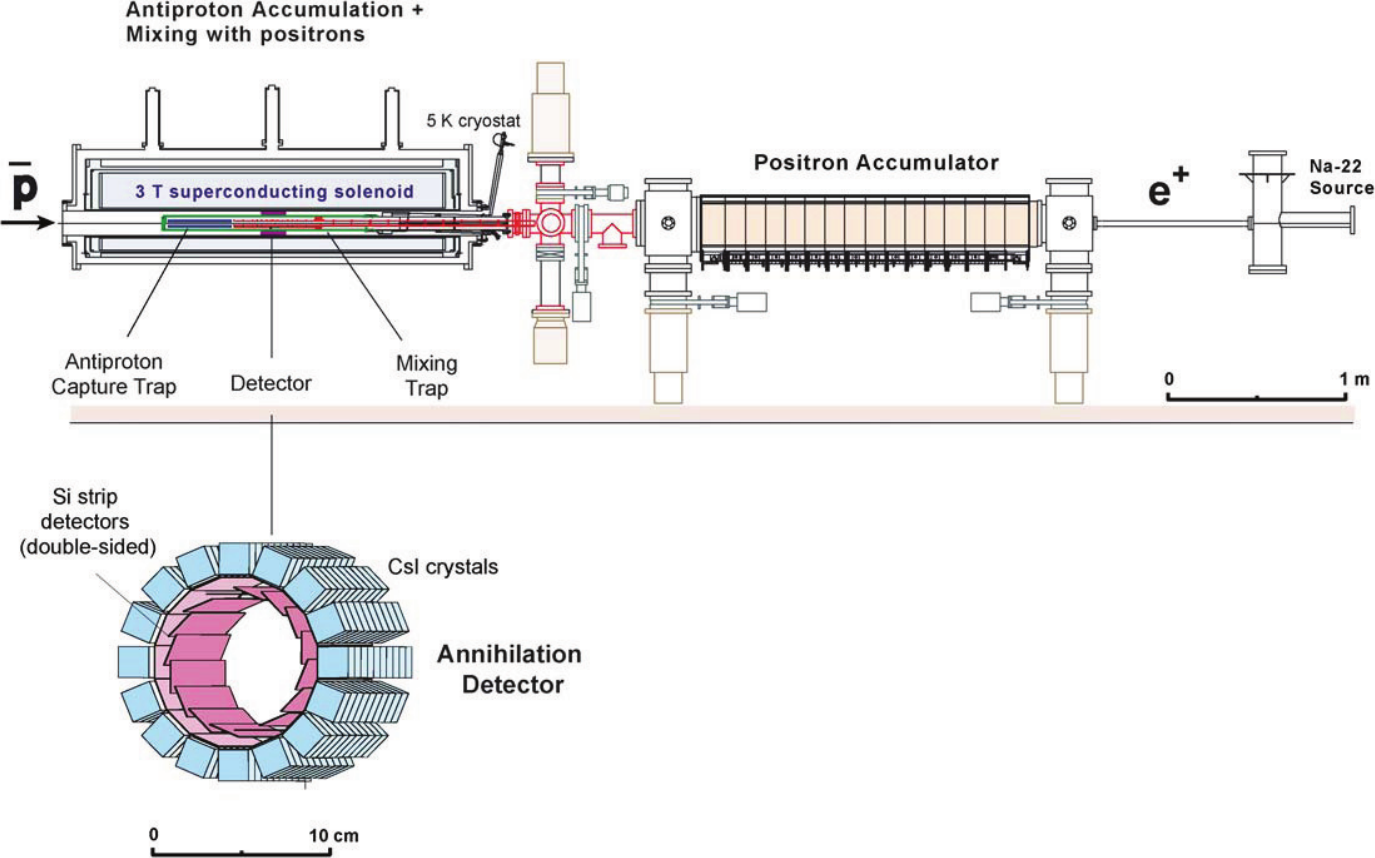}
\begin{minipage}[t]{16.5 cm}
\caption{Experimental layout of the ATHENA experimental apparatus,
consisting of a $^{22}$Na source and $e^+$ accumulator trap, a $\overline{p}$ trap,
and a nested Penning trap where the $\overline{\rm H}$ were produced.
An expanded view of the apparatus used to spatially resolve
the positions of $\overline{\rm H}$ annihilations in the trap is shown
below the main apparatus. Si microstrip detectors measured the tracks
of $\pi^+$ and $\pi^-$ emerging from $\overline{p}$ annihilations, whereas
pairs of 511-keV photons emerging from $e^+$ annihilations were
detected by CsI scintillators. The signature
of a $\overline{\rm H}$ annihilation was a coincidence in space and time of
both annihilation vertices. Figure from Ref.~\cite{ATHENA:Apparatus:2004}.
\label{ATHENA_setup}%
}
\end{minipage}
\end{center}
\end{figure}

\begin{figure}[tbh!]
\begin{center}
\hfill
\includegraphics[width=8cm]{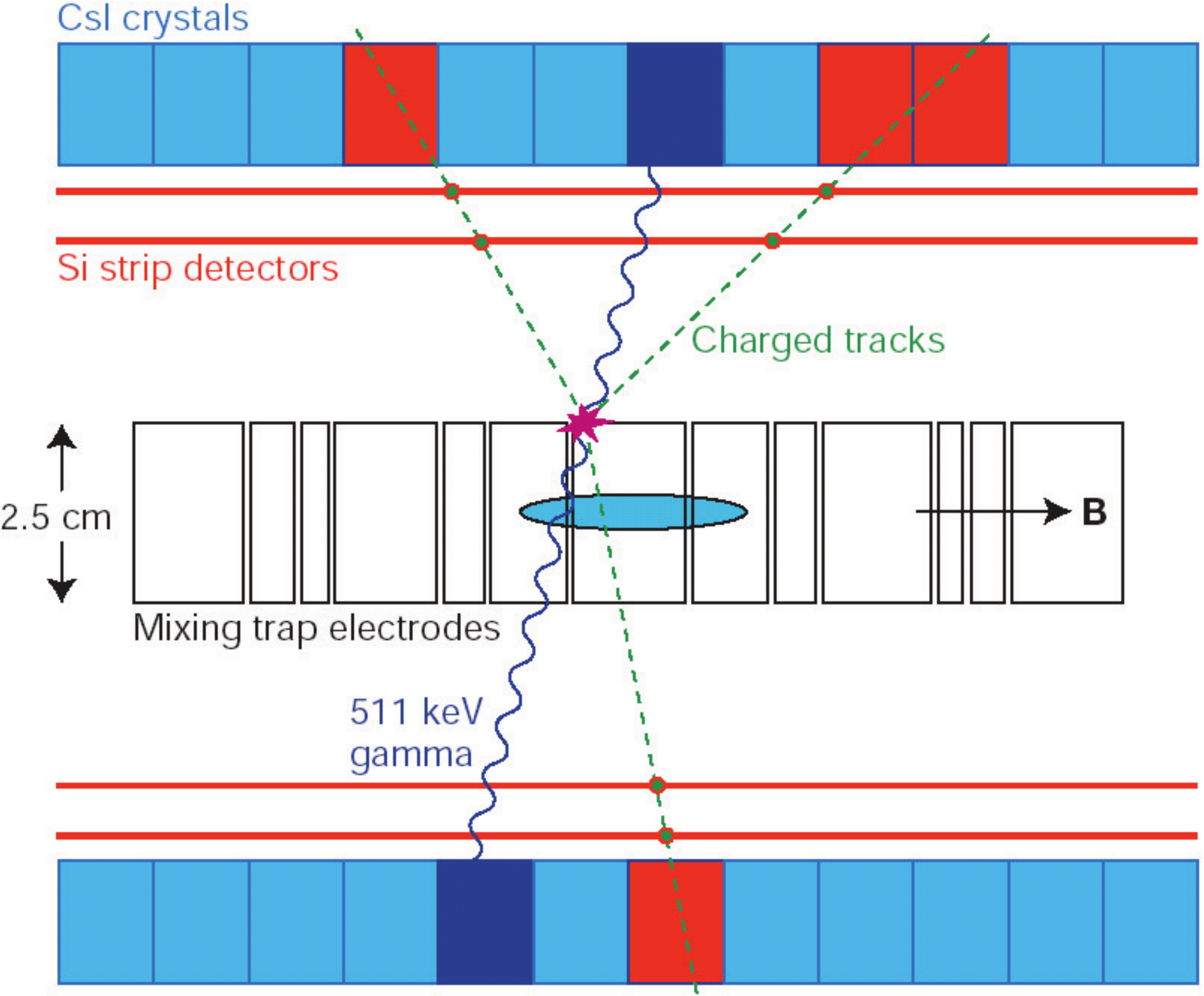}
\hfill
\includegraphics[width=8cm]{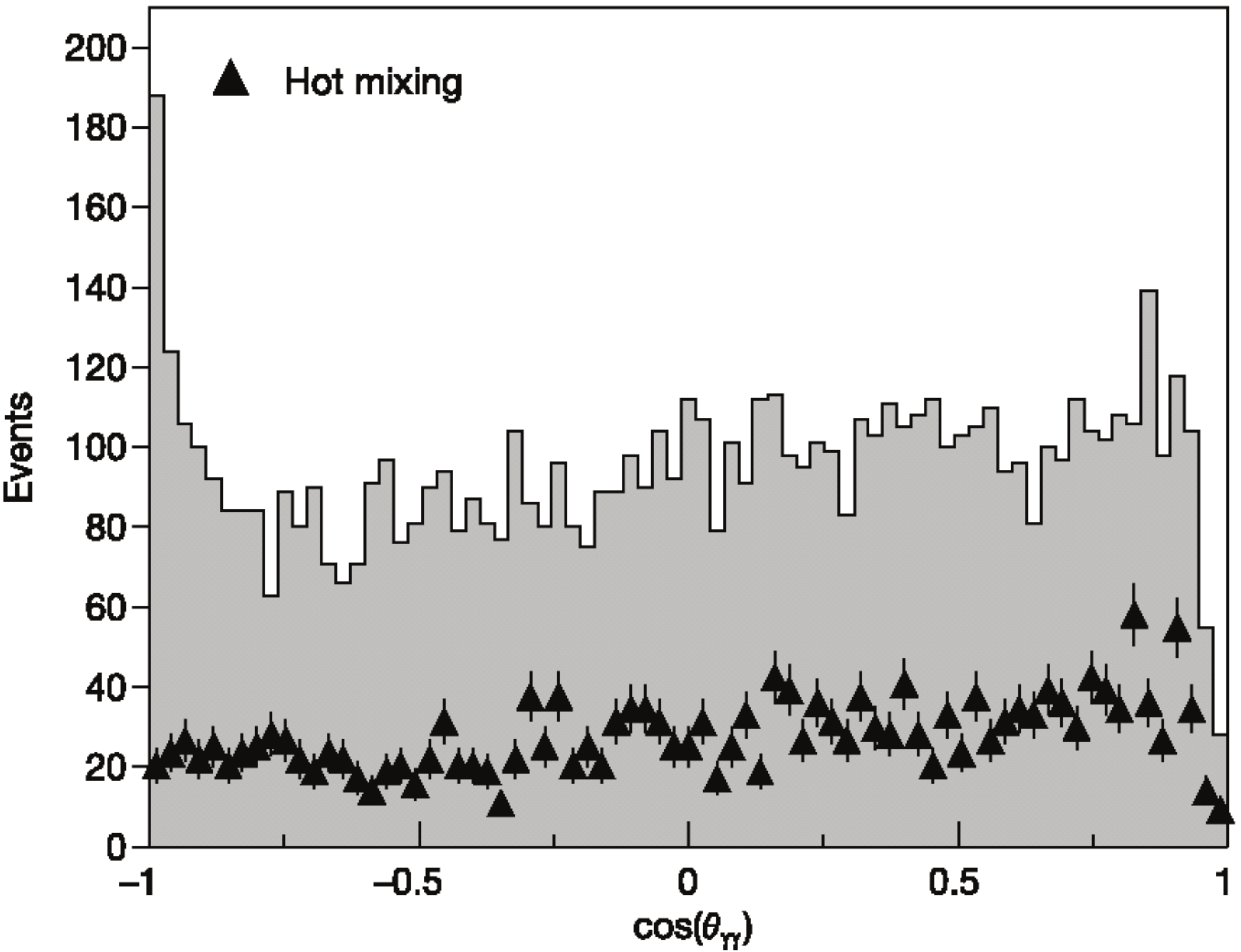}
\hfill$~$
\begin{minipage}[t]{16.5 cm}
\caption{Left panel: Schematic illustration of $\overline{\rm H}$ detection
in the ATHENA experiment using imaging annihilation detectors.  The
stacked cylindrical electrodes of the nested Penning traps are shown
in the center.  Neutral $\overline{\rm H}$ atoms were not confined by the
traps and annihilated on the electrodes.  The dashed
lines show tracks of $\pi^+$ and $\pi^-$ emerging from the $\overline{p}$
annihilation vertex, which were detected by two layers of double-sided
Si-strip detectors.  The wavy lines represent 511\,keV gamma rays
emerging from the $e^+$ annihilation vertex, which were detected by
CsI crystals.  The coincidence of both annihilation vertices signaled
$\overline{\rm H}$ annihilation.
Right panel: The $\overline{\rm H}$ signal.  The
grayed area shows the distribution of events as a function 
of the cosine of the opening angle $\theta_\pi$ of the tracks
of gamma-ray pairs that coincide with a $\overline{p}$ annihilation. 
The peak at $\cos\left(\theta_\pi\right)=-1$ arises from back-to-back gamma
rays emerging from $e^+$ annihilation on the electrode surface. 
The triangles indicate the result from a control experiment ``hot mixing,'' 
in which the $e^+$ plasma was heated using a RF voltage on one of
the trap electrodes, thereby suppressing $\overline{\rm H}$ formation.  
No back-to-back peak appears here. Figures from Ref.~\cite{ATHENA:2002}.
\label{ATHENA_2002}%
}
\end{minipage}
\end{center}
\end{figure}

The first formation of cold $\overline{\rm H}$ by overlapping clouds of
$\overline{p}$ and $e^+$ in nested Penning traps was
reported in 2002, first by the ATHENA collaboration \cite{ATHENA:2002}
and then by the ATRAP collaboration \cite{ATRAP:Hbar:2002}.  ATHENA
detected $\overline{\rm H}$ escaping from their trap by identifying 
its annihilation on the electrode walls (Fig.~\ref{ATHENA_setup}),
as revealed by a coincidence in space and time of $\overline{p}$ and $e^+$ 
annihilation vertices \cite{ATHENA:Apparatus:2004}.
The tracks of $\pi^+$ and $\pi^-$ emerging from the
$\overline{p}$ annihilations were measured by two layers of double-sided
Si microstrip detectors surrounding the cylindrical electrodes of the Penning trap,
with a solid angle of 80$\%$. The apparatus included 8192 detector channels, the 
signals of which were amplified by application-specific integrated circuits 
(ASICs) and recorded by flash analog-to-digital converters at a trigger
rate of $\sim 40$ Hz. Pairs of 511-keV photons emerging from $e^+$
annihilations were detected by 192 CsI scintillation counters of 
size $17\times 17.5\times 13$ mm$^{3}$. These were read out by
avalanche photodiodes (APD's).  The $\overline{\rm H}$
signal is characterized  in Fig.~\ref{ATHENA_2002} by a clear back-to-back 
peak of the two 511-keV photons. This method is insensitive to the
internal quantum states of $\overline{\rm H}$.

\begin{figure}[tbh!]
\begin{center}
\hfill
\includegraphics[width=11cm]{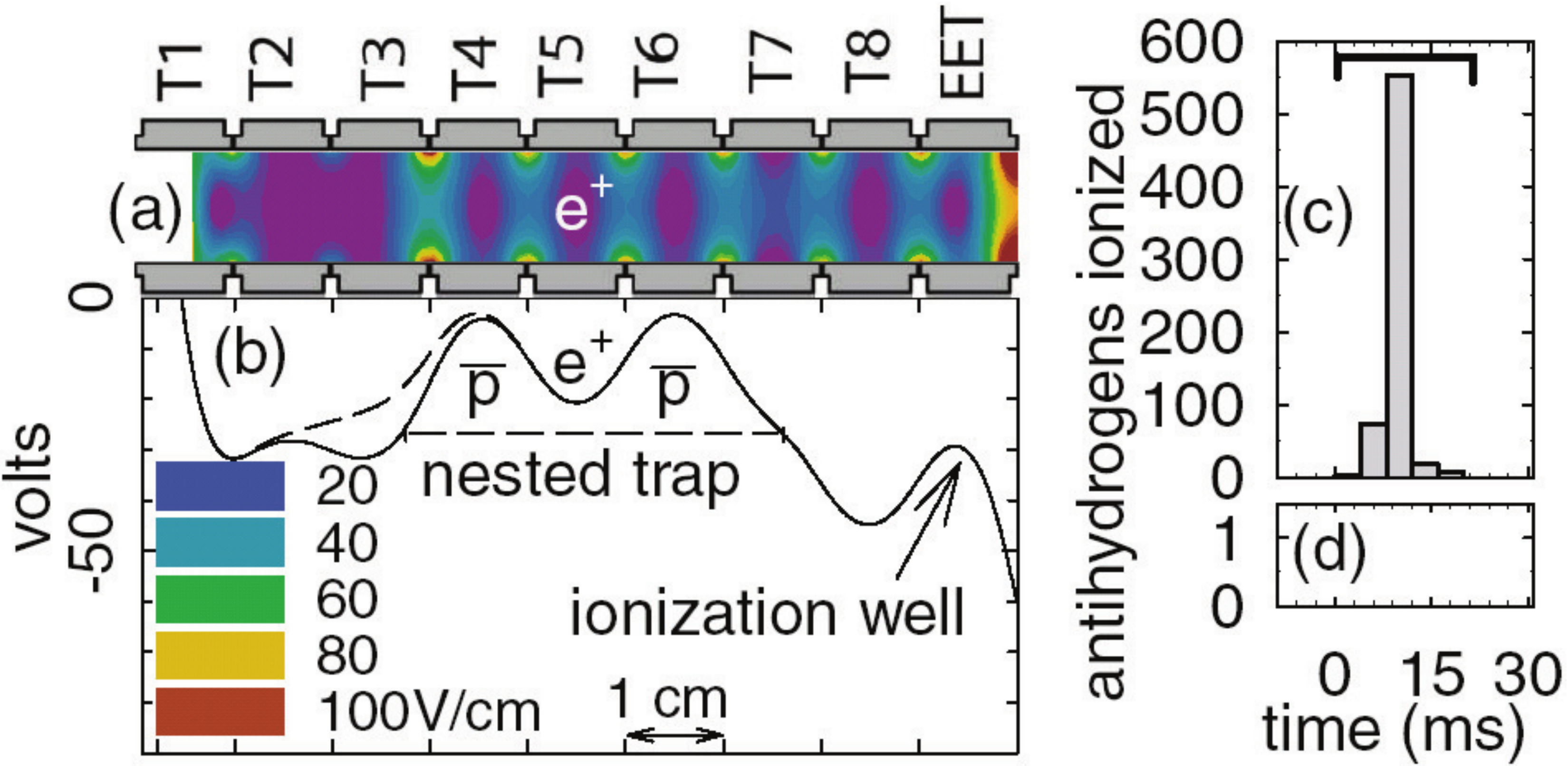}
\hfill$~$
\begin{minipage}[t]{16.5 cm}
\caption{(a) Field-ionization method for detecting $\overline{\rm H}$ used by the
ATRAP collaboration.  The electrodes of a nested Penning trap used for mixing clouds of
$e^+$ and $\overline{p}$ to produce $\overline{\rm H}$, together with a color-coded representation 
of the magnitude of the electric field.  (b) The electrical potential along the trap axis
used for producing $\overline{\rm H}$ indicated by solid lines.  
The dashed line indicates the potential used to launch $\overline{p}$ into the $e^+$ cloud.
(c) Annihilation signals corresponding to $\overline{p}$ that first emerged from the
field-ionization of $\overline{\rm H}$, before being confined in the detection well and
then ejected from it. (d) No $\overline{p}$ signals were detected in a 
control-experiment that was carried out without introducing $e^+$ in the nested Penning trap. 
Figures from \cite{ATRAP:Hbar:2002}.
\label{ATRAP_2002}%
}
\end{minipage}
\end{center}
\end{figure}

A quite different method for $\overline{\rm H}$ detection was used by
the ATRAP collaboration, based on the fact that
both three-body and charge-exchange recombinations
produce $\overline{\rm H}$ that initially occupy Rydberg states \cite{ATRAP:Hbar:2002}. 
Some of these weakly-bound $\overline{\rm H}$ were ionized by an external electric field
which constituted part of an auxiliary (or ``detection") potential well in the trap
(Fig.~\ref{ATRAP_2002}).
The $\overline{p}$ emerging from the $\overline{\rm H}$ ionization were recaptured 
and stored in the auxiliary well, thereby providing a signal of $\overline{\rm H}$ production. 
These signal $\overline{p}$ were detected by opening this well at a later time 
and counting the $\overline{p}$ annihilations on the trap electrodes. A sharp
ms-long signal (Fig.~\ref{ATRAP_2002}) emerged, which could be clearly distinguished from the 
continuous background caused by cosmic rays. To isolate the signal $\overline{p}$ from the 
background caused by any other $\overline{p}$ that was not involved in the $\overline{\rm H}$ 
ionization, a positive electric potential was applied between the $\overline{\rm H}$ formation and 
detection regions. This allowed the passage of neutral $\overline{\rm H}$ between
the two regions, whereas the background $\overline{p}$ were reflected. 
Any other background caused by the loss of $\overline{p}$ or $e^+$ from the
nested Penning trap during the mixing of the two particle clouds was rejected,
by delaying the opening of the auxiliary well until after the mixing phase.

The parameters of the $\overline{p}$ and $e^+$ clouds during $\overline{\rm H}$
production were systematically studied 
\cite{ATHENA:Diagnostic:2003,ATHENA:PRL:2003,ATRAP:ApertureMethod:2004}.
By using the vertex reconstruction and imaging techniques described above, ATHENA
found that $\overline{p}$ escaping from the Penning trap preferentially 
annihilate on localized ``hot spots'' along the surface of the trap electrodes, 
possibly due to small off-axis displacements of adjacent electrodes in the trap.
By contrast, the spatial distribution of $\overline{\rm H}$ annihilations on the electrodes
were found to be radially symmetric \cite{ATHENA:Imaging:2004}.  
It was also discovered that $\overline{\rm H}$ production initiates only after the
$\overline{p}$ has been cooled close to thermal equilibrium, which occurs on 
a time scale of $\sim 10\,\text{ms}$ \cite{ATHENA:Dynamics:2004}.

The efficiency of $\overline{\rm H}$ production in these experiments
is remarkable: some $17 \pm 2\,\%$ \cite{ATHENA:HighRate:2004} and 
$> 11\,\%$ \cite{ATRAP:Hbar:2002} of the trapped $\overline{p}$ were found to 
form $\overline{\rm H}$ in respectively ATHENA and ATRAP. 
Such a large $\overline{\rm H}$ production rate is 
incompatible with the expected values for spontaneous radiative recombination, and 
strongly suggests that three-body recombination plays an important role. 
The rate of three-body recombination in the
equilibrium state, however, is theoretically expected to have a steep $T^{-9/2}$
dependence on $e^+$ temperature, whereas the experiments have measured 
a less pronounced $T^{-1.1 \pm 0.5}$ scaling
\cite{ATHENA:temperature:2004,ATHENA:Fujiwara:PRL:2008}.  Although
many insights into $\overline{\rm H}$ formation have been gained from simulations as
discussed in a recent review \cite{Robicheaux:JPB41:2008}, there is
no clear explanation for the apparent discrepancy in this temperature scaling 
\cite{Jonsell:JPB42:2009}.

The state distributions of the Rydberg $\overline{\rm H}$ atoms was studied
by measuring the $\overline{p}$ annihilation originating from the ionization of
$\overline{\rm H}$, as a function of the strength of an analysis field
located between the nested trap and the ionization well of Fig.~\ref{ATRAP_2002} 
\cite{ATRAP:HbarStates:2002,Pohl:PRL:2006}.  The velocity distribution of
$\overline{\rm H}$ was also studied, by allowing the 
$\overline{\rm H}$ to pass through an oscillating electric field generated by 
one of the trap electrodes, before reaching the ionization well. The fraction
of $\overline{\rm H}$ that passed through this field without ionizing was 
measured as a function of the frequency of the oscillating field
\cite{ATRAP:Velocity:2004,Pohl:PRL:2006}. These initial experiments detected
weakly-bound $\overline{\rm H}$ traveling along the magnetic field direction
in the trap with relatively high velocities, corresponding to an energy $E=200$ meV
and temperature $T=2400$ K. 
The interpretation of the results is complicated \cite{Pohl:PRL:2006} due to
the complex way in which guiding-center $\overline{\rm H}$ atoms can interact 
with the magnetic field and with other $\overline{p}$ confined in the trap.
It has been suggested that although $\overline{\rm H}$ with initial energies 
of 1--10 meV may be produced in the experiment, they are converted into 
higher-energy $\overline{\rm H}$ by undergoing charge-exchange collisions 
with fast $\overline{p}$ in the trap.

An interesting byproduct of these experiments was the observation of evidence
by the ATHENA collaboration of cold protonium ($Pn\equiv\overline{p}$-$p$) atoms of 
temperature 400-700 meV being produced in the trap. The data was inferred from
the observed axial and radial distributions of $\overline{p}$ annihilation vertices
occurring in the trap. This atom is believed
to be formed when H$_2^+$ ions confined in the nested Penning trap collide
with $\overline{p}$ via the reaction \cite{zurlo},
\begin{equation}
\overline{p}+{\rm H}_2^+\rightarrow {Pn}(n,\ell)+{\rm H}.
\end{equation} 

\subsubsection{\it Antihydrogen production using resonant charge exchange}
\label{sec:charge_exchange}

\begin{figure}[tbh!]
\begin{center}
\hfill
\includegraphics[width=10cm]{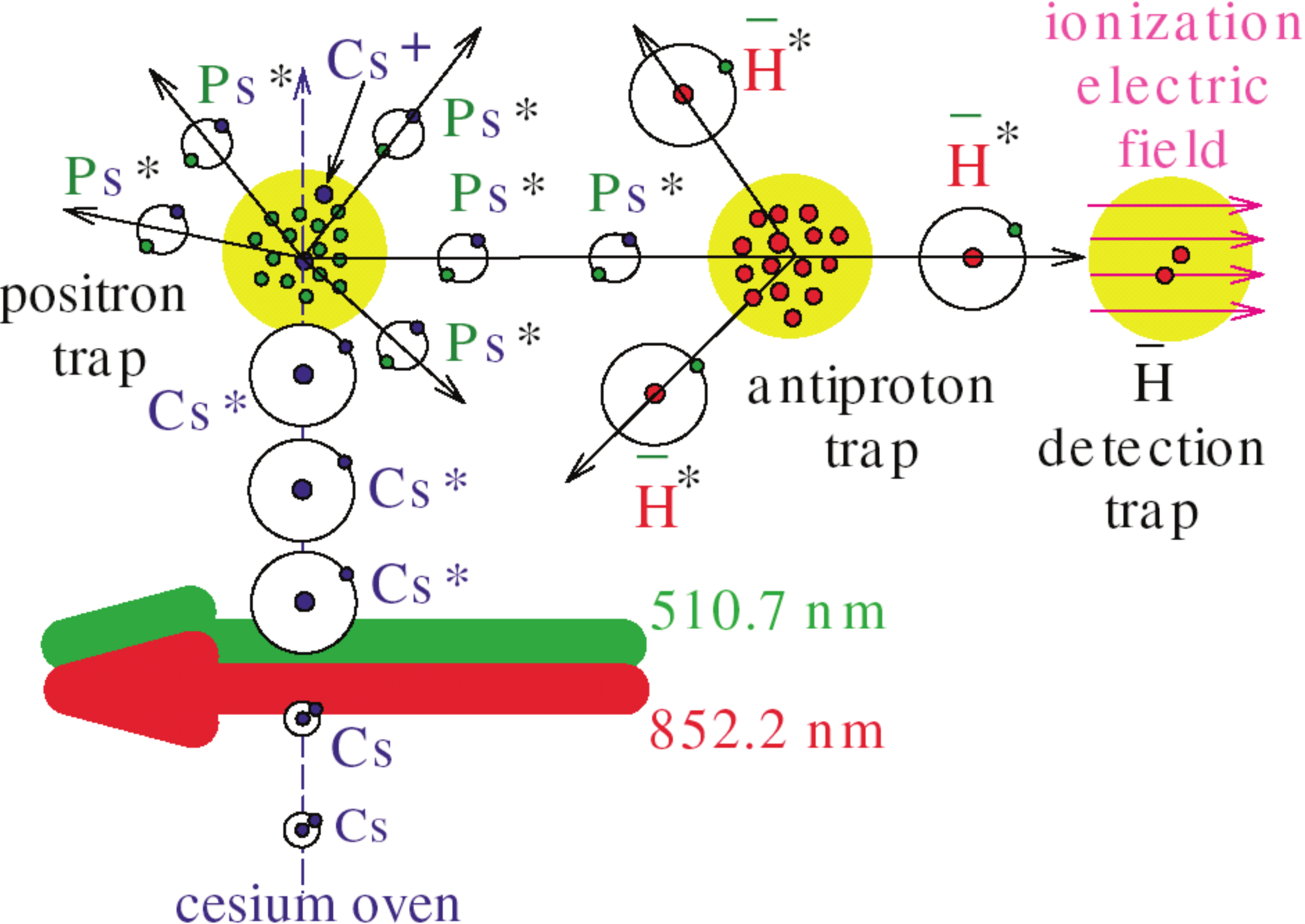}
\hfill$~$
\begin{minipage}[t]{16.5 cm}
\caption{Schematic layout of the experiment which produced $\overline{\rm H}$
by resonant charge-exchange. Laser beams of wavelengths 852.2\,nm and 510.7\,nm 
excited Cs into Rydberg states with principal quantum number $n \simeq 37$.  
These Cs$^\ast$ traversed a cloud of $e^+$ confined in a Penning trap,
thereby forming Rydberg $Ps$ via the reaction,
$~{\rm Cs}^\ast + e^+ \rightarrow {Ps}^\ast + {\rm Cs}^+$.  Some of these
atoms traveled through a cloud of $\overline{p}$ in an adjacent Penning
trap and formed $\overline{\rm H}$ in a second charge-exchange step,
${Ps}^\ast + \bar{p} \rightarrow \bar{\rm H}^\ast + e^-$.  These 
Rydberg $\overline{\rm H}$ atoms traversed a detection region that contained
an ionizing electric field. Figure from Ref.~\cite{ATRAP:Cesium:2004}.
\label{ATRAP_Cesium_2004}%
}
\end{minipage}
\end{center}
\end{figure}

\begin{figure}[tbh!]
\begin{center}
\hfill
\includegraphics[width=11.5cm]{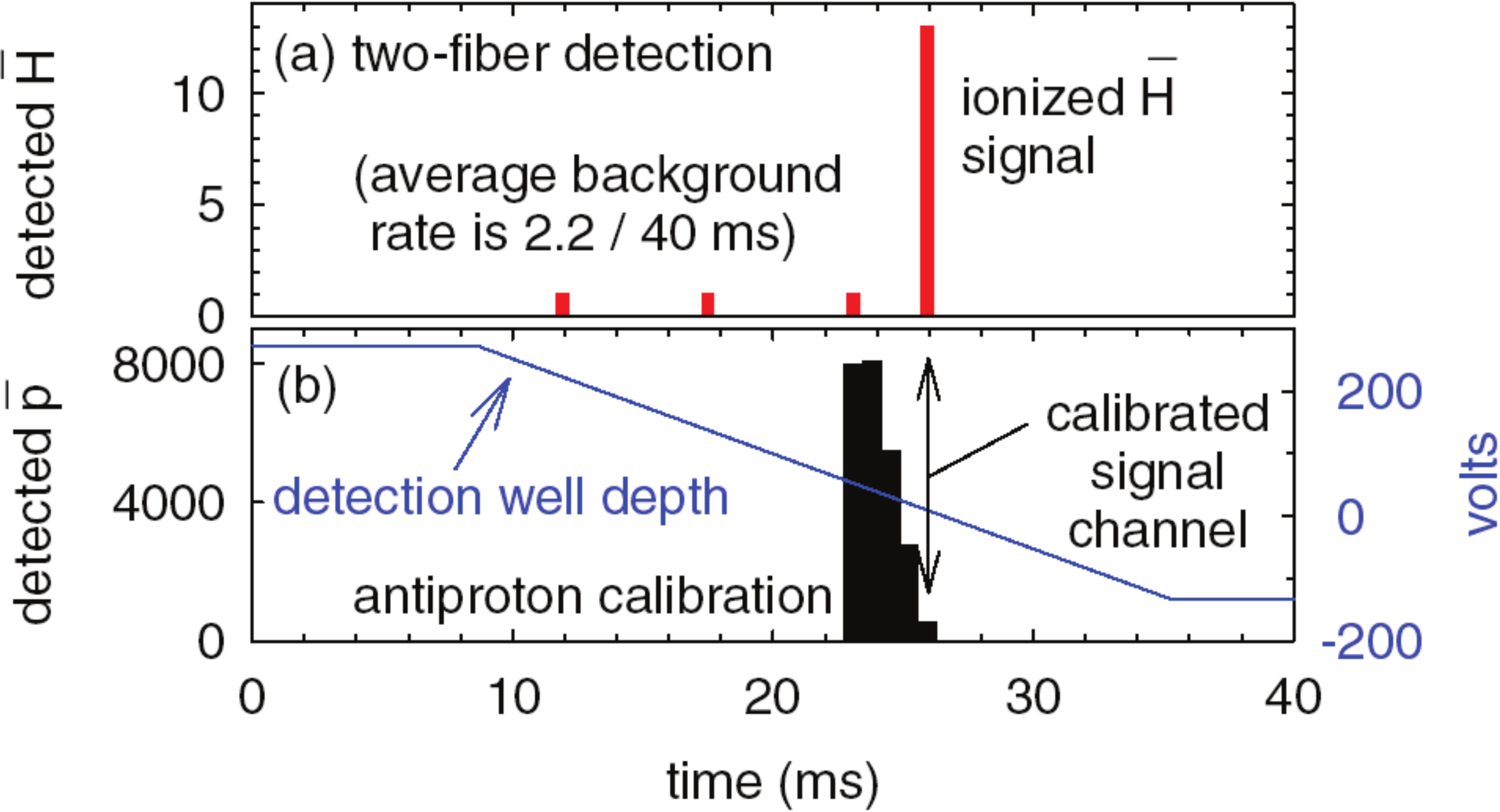}
\hfill$~$
\begin{minipage}[t]{16.5 cm}
\caption{(a) Annihilation signal corresponding to $\overline{p}$ emerging from
field-ionized $\overline{\rm H}$ and being captured by a so-called ``detection well"
constituting an auxiliary Penning trap.
The peak corresponding to the ionized $\overline{\rm H}$ was detected by allowing the 
electrostatic potential of the detection well to be ramped down. Under the assumption 
that the velocity of these $\overline{\rm H}$ formed by resonant charge-exchange
 is isotropic, this signal corresponds 
to 100--200 atoms produced in 6 trials. (b) Signal for calibrating the efficiency of the
annihilation detector.  Figure from Ref.~\cite{ATRAP:Cesium:2004}.
\label{ATRAP_Cesium_2004_2}%
}
\end{minipage}
\end{center}
\end{figure}

The resonant charge-exchange method for $\overline{\rm H}$
production (Fig.~\ref{ATRAP_Cesium_2004}) is quite complex to implement, 
as a thermal alkali beam and multiple laser beams must be incorporated into the experimental apparatus 
together with the nested Penning traps, which are at
cryogenic temperatures and in a strong magnetic field.
The ATRAP collaboration excited a thermal beam of Cs atoms to states of
$n=37$ using a continuous-wave (cw) diode laser of wavelength 852 nm, and a pulsed 
copper vapor laser of wavelength 511 nm.  The resulting Rydberg Cs$^\ast$ atoms 
then entered the electrode stack comprising the Penning trap through a 0.3-mm diameter hole.
They were allowed to interact with a cloud of $\sim 1.4\times 10^6$
$e^+$ cooled to temperature $T\sim 4$ K, thereby producing Rydberg
$Ps^\ast$. These then passed through a second potential well containing
$2.4\times 10^5$ $\overline{p}$. In the initial demonstration experiment,
ATRAP detected 14$\pm$4 $\overline{\rm H}$ events produced in this way
(Fig.~\ref{ATRAP_Cesium_2004_2}).

An important advantage of this method 
is that the $n$-value of the formed $\overline{\rm H}$ can be selected 
by simply tuning the wavelength of the lasers used to excite the alkali atoms to a Rydberg state,
prior to the first charge-exchange collision that produces Rydberg $Ps^\ast$ (see Sect.~\ref{formation}).
Moreover the $\overline{\rm H}$ is expected to be formed at significantly lower temperature than
those produced by, e.g., three-body recombination in nested Penning traps. This is related to the 
fact that the velocity of $\overline{\rm H}$ most likely corresponds to the velocity of the ingredient 
$\overline{p}$ prior to recombination. Whereas in nested 
Penning traps the $\overline{p}$ and $e^+$ clouds must be accelerated towards each other
to overlap which can potentially increase the $\overline{\rm H}$ temperature, in the resonant 
charge-exchange case the target $\overline{p}$ can be kept static and cold in the Penning
trap, waiting to collide with the cold $Ps^{\ast}$ and form $\overline{\rm H}$ of the same
temperature. These features may be important for future experiments on antimatter
gravity (see Sec.~\ref{sec:antimatter_gravity}).

\subsection{\it Antihydrogen trapping}
\label{sec:hbartrapping}

\subsubsection{\it Magnetic traps}
\begin{figure}[tbh!]
\begin{center}
\includegraphics[height=7cm]{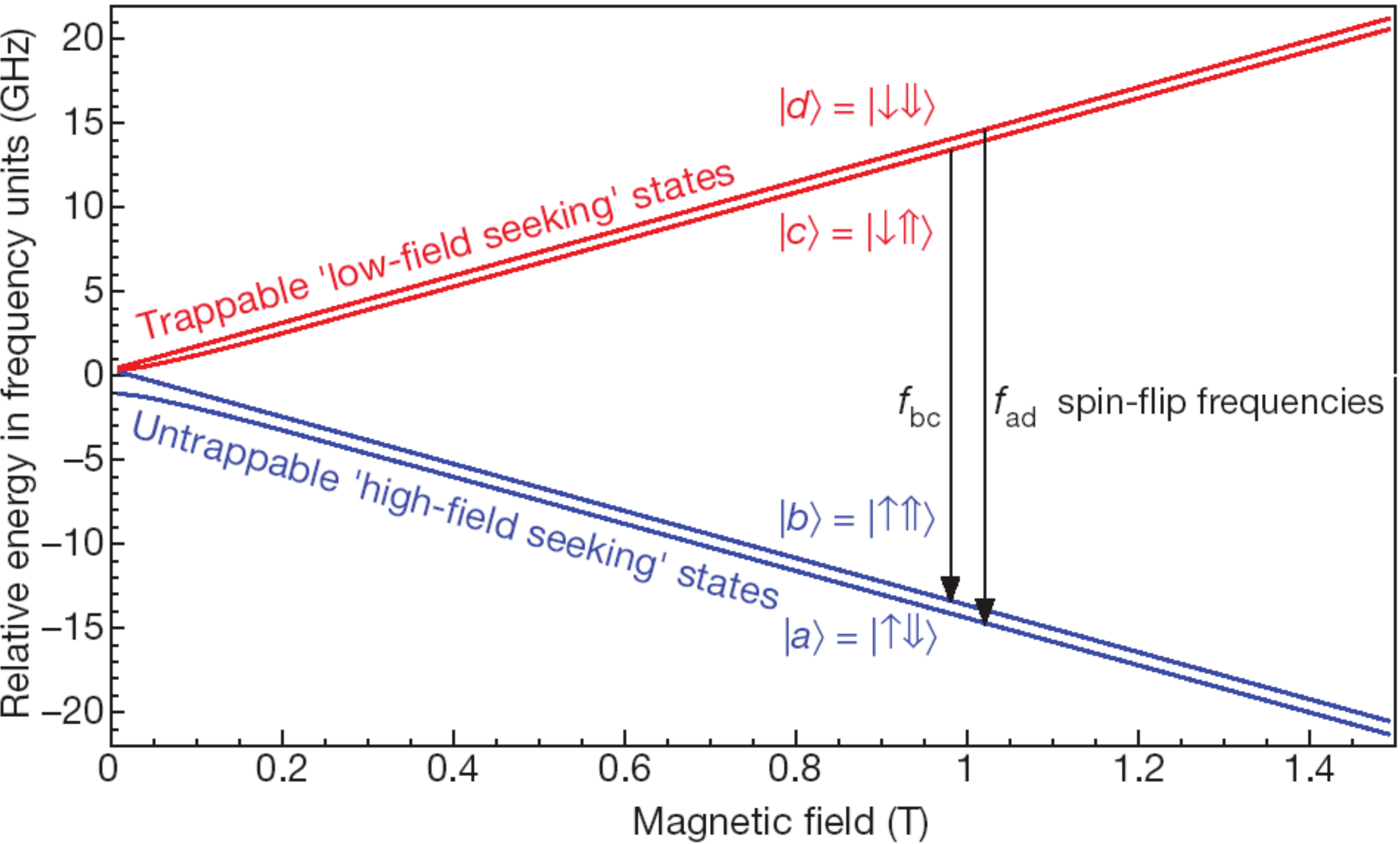}
\begin{minipage}[t]{16.5 cm}
\caption{Hyperfine energy levels of $\overline{\rm H}$ in
the $1s$ ground state as a function of the strength of an external magnetic field
(Breit-Rabi diagram).  In the state vectors shown, the single arrow
refers to the $e^+$ spin and the double arrow refers to the
$\overline{p}$ spin. Figure from Ref.~\cite{Alpha:SpinFlips:2012}.
\label{hydrogen_levels}
}
\end{minipage}
\end{center}
\end{figure}

Experiments with $\overline{\rm H}$ have important advantages compared to
the corresponding ones for H, since even a few $\overline{\rm H}$ can be detected
with high sensitivity by measuring their annihilation in the experimental apparatus. 
The very low numbers of $\overline{\rm H}$ now available to experimentalists, however, also 
represents a challenge. Whereas precision laser spectroscopy experiments of H \cite{Parthey:PRL107:2011} 
typically employ cold beams with a flux of 10$^{17}$ s$^{-1}$ \cite{Walraven:1982},  the 
corresponding flux in the $\overline{\rm H}$ case is in principle limited among many 
things by the production rate of $\overline{p}$ at the AD of $3\times 10^7$ every 
90--100 s. Trapping of $\overline{\rm H}$ thus suggests itself as an obvious means to 
make efficient use of the rare atoms.

Neutral atom traps confine $\overline{\rm H}$ using the fact that those atoms whose 
magnetic moments are aligned opposite an external magnetic field $B$ will have lower 
energies in a lower field. The Breit-Rabi diagram of Fig.~\ref{hydrogen_levels} 
schematically shows the binding energies of the hyperfine sublevels of the $\overline{\rm H}$
$1s$ ground state, as a function of $B$.  Since the energies of the so-called ``low-field-seeking" 
states $c$ and $d$ increase with $B$, $\overline{\rm H}$ that populate these states,
when placed in the inhomogeneous magnetic field of the trap, tend to drift towards and collect 
at the center where the field is minimum. Conversely, the ``high-field-seeking" states
$a$ and $b$ are repelled from the field minimum and can be ejected from the trap.

In practice there are several possible field configurations: Ioffe traps \cite{Bergeman:1987} employ a quadrupole or 
octupole magnetic field for confinement of atoms in the radial direction, 
and ``pinch'' coils for axial confinement.  These traps have been used to confine 
H atoms \cite{Hess:1987,Roijen:1988} and achieve Bose-Einstein condensation 
\cite{Fried:1998}. An alternative geometry involves the use of 
anti-Helmholtz coils which create a quadrupole trap with an axial
symmetry \cite{Bergeman:1987,Mohri:2003}.  

Typical values of the magnetic field gradients in Ioffe traps that can be achieved by
current technology is $\Delta B\sim 1$ T between the center and walls of the trap.
This corresponds to a potential well depth expressed in temperature units of 
$\mu_\text{Bohr}\Delta B/k_B \simeq 0.7\,\textrm{K}$,
implying that only $\overline{\rm H}$ with mK-scale temperatures can be trapped.
This is much smaller than the typical well depths (few kV) that can be achieved for 
charged particles in Penning traps.

\subsubsection{\it Compatibility of traps for neutral atoms and charged particles}

Neutral H are normally loaded into atom traps by either decelerating them via atomic 
collisions, or by allowing them to interact with cryogenic surfaces within
the trap which are covered with liquid He. So far there are no corresponding techniques 
to load such a trap with $\overline{\rm H}$ arriving from the outside
(a possible exception is the coil gun technique \cite{Raizen:Coilgun:2009}).
The $\overline{\rm H}$ atoms must instead be produced directly within the magnetic fields of the 
trap, before they can be successfully captured. This implies the use of an
atom trap which is superimposed on some Penning traps, for simultaneous 
confinement of $\overline{\rm H}$, $\overline{p}$, and $e^+$.
One such design employs a Ioffe-trap with a non-zero magnetic field at the center. 
This so-called ``bias field'' serves as the magnetic confinement field of a nested 
Penning trap for charged particles.

There were previous concerns, however, with the feasibility of this design,  
since the Ioffe trap includes a radial magnetic multipole field which breaks the
cylindrical symmetry. This seemed to be at odds with the confinement theorem for stably trapping 
non-neutral plasmas in Penning or Penning-Malmberg traps, which imposes a strong requirement 
for preferring a cylindrical symmetry \cite{ONeil:PF23:22161980}.  In fact, experiments have 
shown that when magnetic quadrupole fields are superimposed on Penning traps,
the radial diffusion of the trapped $e^-$ plasmas is significantly increased 
\cite{Gilson:2003}, and the confinement properties can even be destroyed \cite{Fajans:2005}.  
Atom traps with higher-order radial multipole \cite{Fajans:2004} and other field configurations 
\cite{Dubin:2001} have been suggested to alleviate this problem.  On the other hand, it has 
been shown that single charged particles confined in Penning-Ioffe traps with a radial quadrupole 
field \cite{Squires:2001} can follow stable trajectories. This may imply that relatively low-density 
clouds of charged particles can still be stably trapped, even if high-density non-neutral plasmas cannot.

It was to later great relief that further experiments demonstrated $\overline{p}$ and $e^+$ 
confinement in nested Penning traps superimposed on Ioffe traps, with either radial octupole
\cite{ALPHA:2007a} or quadrupole \cite{Gabrielse:PenningIoffe:2007} magnetic fields.  
Two additional intermediate goals were soon achieved on the path towards
$\overline{\rm H}$ trapping, i): production of $\overline{\rm H}$ in a Penning trap
of comparably low (1 T) magnetic field \cite{Andresen:JPB41:2008} compatible with the 
magnetic bias field of an Ioffe trap, and ii): $\overline{\rm H}$ production in a Penning-Ioffe 
trap with a quadrupole magnetic field \cite{PenningIoffe:PRL:2008}.

\begin{figure}[tbh!]
\begin{center}
\hfill
\includegraphics[height=6cm]{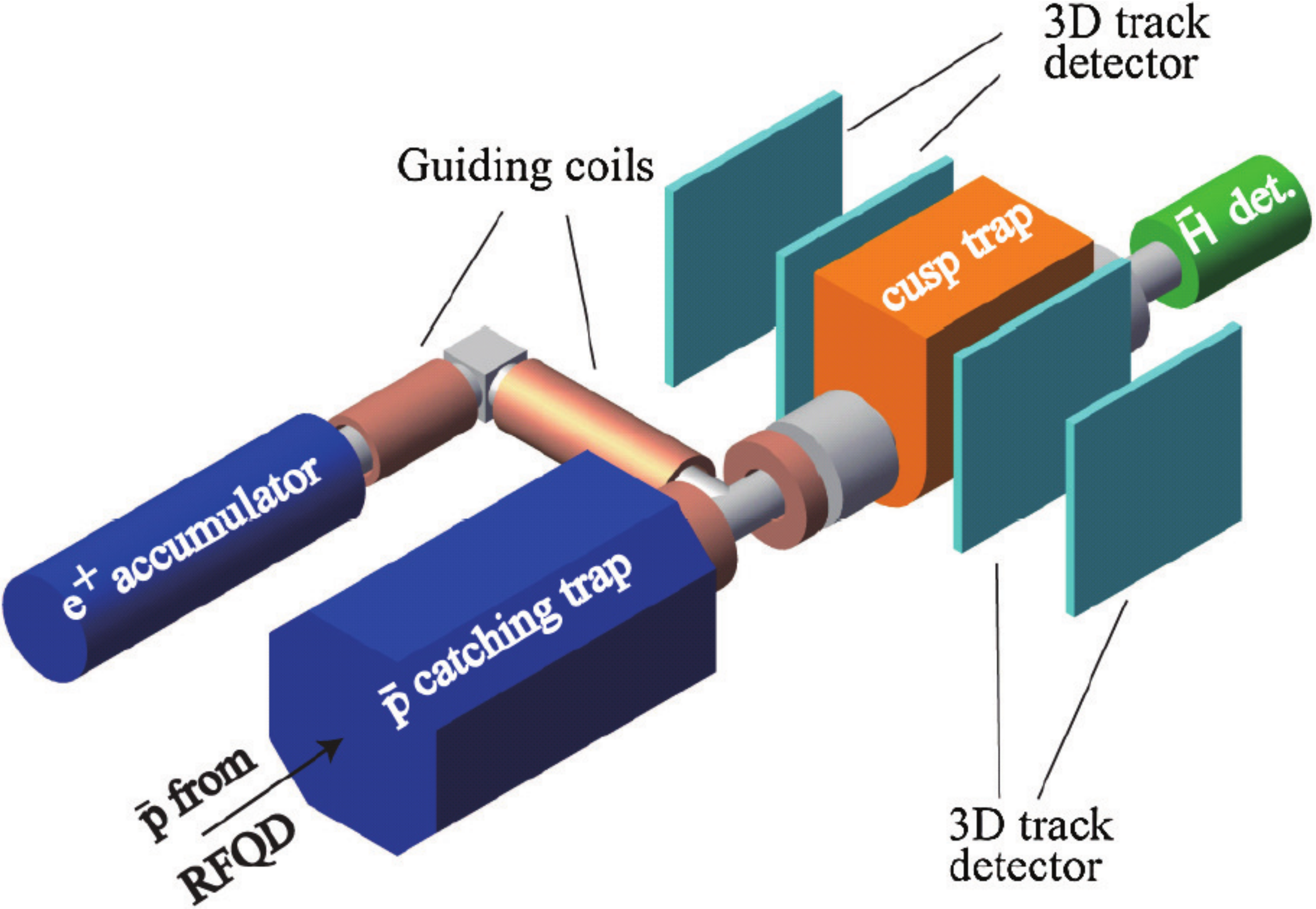}
\hfill
\includegraphics[height=6.3cm]{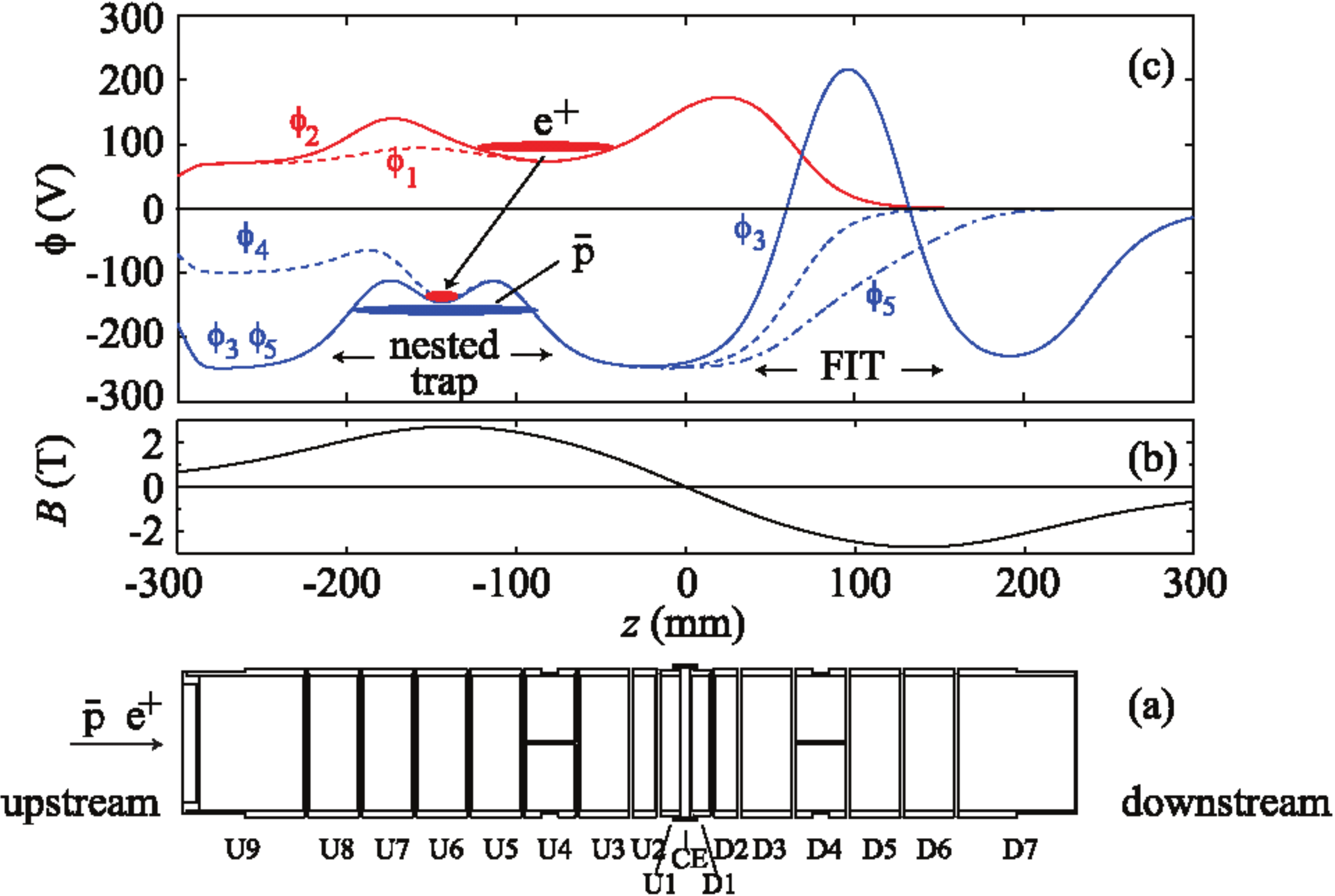}
\hfill
\begin{minipage}[t]{16.5 cm}
\caption{Left panel: Schematic layout of the cusp trap for mixing $\overline{p}$ decelerated by the ASACUSA radiofrequency quadrupole decelerator with $e^+$ accumulated in a Penning trap. 
 \newline
Right panel: Schematic drawings of the (a) trap electrodes, (b) magnetic field along 
the trap axis, and (c) electrostatic potential on the axis.  The $e^+$ and $\overline{p}$ were
mixed in a nested Penning trap located in the upstream part of the apparatus.  
The $\overline{\rm H}$ formed here traversed the magnetic cusp field (note the zero-crossing of the
magnetic field in (b) at $z = 0\,\textrm{mm}$) and were field-ionized downstream in the field-ionization trap (FIT). 
Figures from Ref.~\cite{Enomoto:PRL:2010}.
\label{cusp_2010}
}
\end{minipage}
\end{center}
\end{figure}

A different proposal for superimposing neutral atom and charged particle
traps is the so-called ``cusp trap'' \ref{cusp_2010}, which attempts to employ a magnetic 
quadrupole field with cylindrical symmetry \cite{Mohri:2003}.  Here charged 
particles can escape along the magnetic field lines which constitutes a ``loss cone'' of
plasma confinement in a magnetic mirror. This loss, however, can be prevented by
overlapping an electric octupole field with a cylindrical symmetry
\cite{Mohri:2003}.  This so-called ``magnetic cusp and electric octupole
(MCEO)'' idea can be seen as a higher-order multipole
generalization of the Penning trap principle.  Confinement of
non-neutral plasmas in such a MECO trap has been demonstrated.  The
density profile, however, shows a crater near the trap center
\cite{Mohri:AIP793:2005}.  Simulations show that one distinct feature
of the cusp trap is the possibility to extract a polarized beam of
$\overline{\rm H}$ \cite{Mohri:2003}, which may be important for 
carrying out future measurements of $\overline{\rm H}$ hyperfine splitting 
using such a beam (Sect.~\ref{sec:antihydrogen_hfs}).  
Other simulations show that the cusp trap may have advantages 
in a cooling process which occurs as Rydberg $\overline{\rm H}$ 
spontaneously decay inside a magnetic trapping field 
\cite{Pohl:PRL97:2006}.  Some $\overline{\rm H}$ have recently 
been produced using conventional nested Penning traps in the
cusp trap apparatus, and detected on the other side of the
magnetic cusp field \cite{Enomoto:PRL:2010}. 

\subsubsection{\it Magnetic trapping of antihydrogen}
\label{hbarHFS}

\begin{figure}[tbh!]
\begin{center}
\includegraphics[height=5cm]{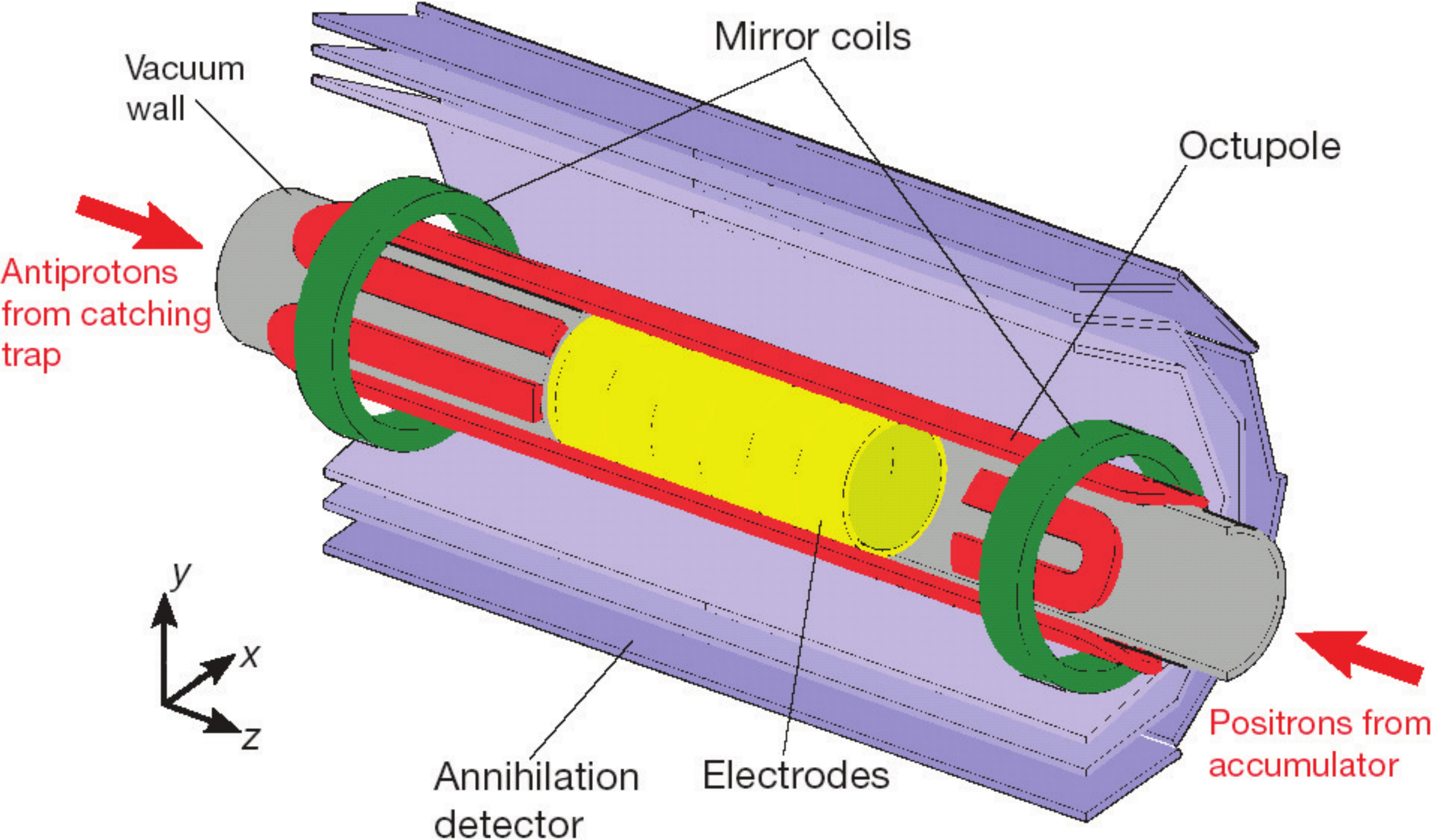}
\includegraphics[height=7cm]{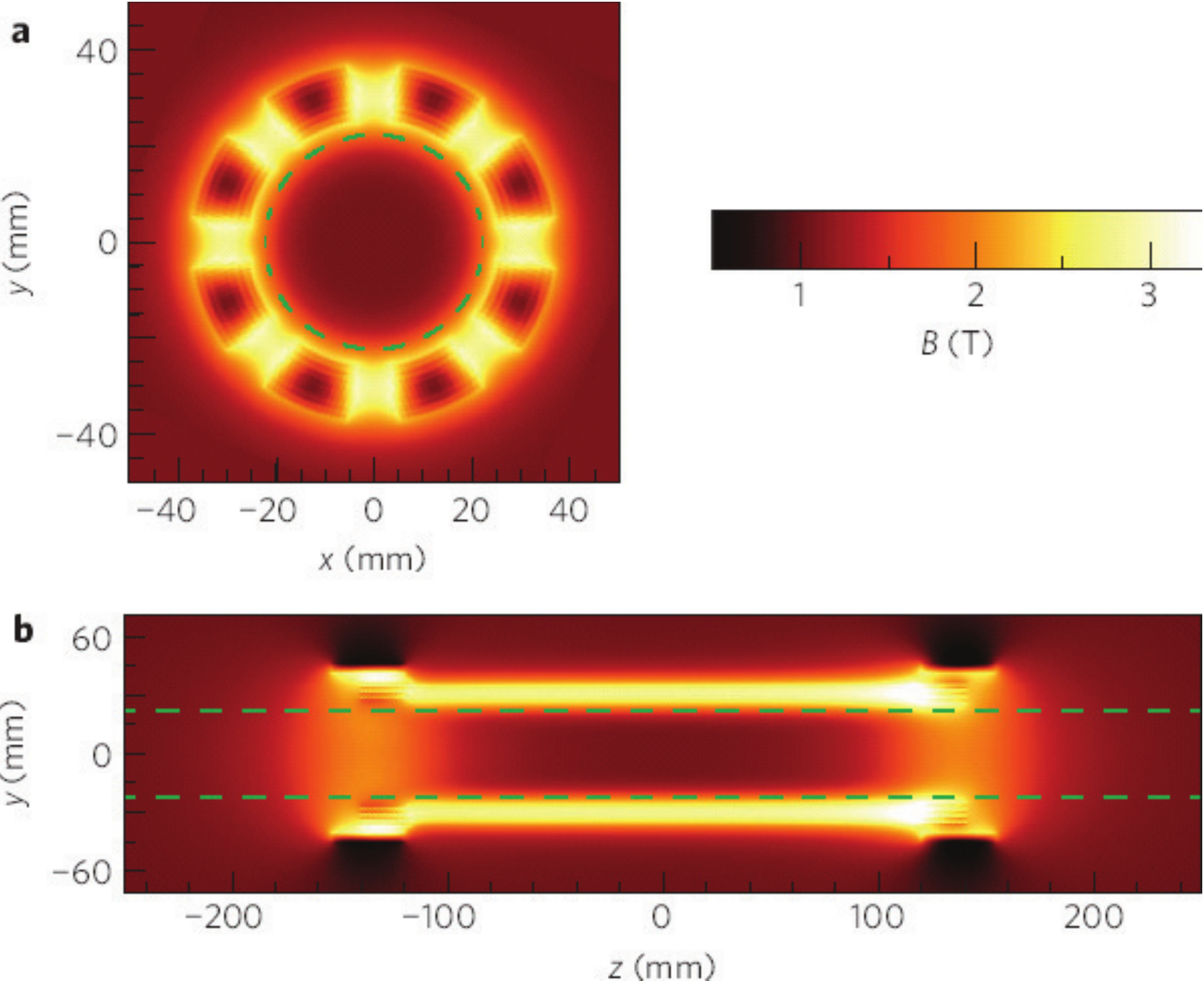}
\begin{minipage}[t]{16.5 cm}
\caption{Left panel: Schematic of the ALPHA apparatus showing the
electrodes of nested Penning traps at the center.  The neutral
$\overline{\rm H}$ atom trap utilized mirror coils for axial confinement and
octupole coils for radial confinement.
Right panel: (a) Cross-sectional drawing of the $\overline{\rm H}$ trap
showing the distribution of the radial octupole magnetic field along
the $xy$-plane. (b) Cross-sectional drawing of the magnetic field strength in the y-z-plane 
which include the trap axis. The effect of the mirror coils for axial
confinement can clearly be seen.  Green dashed lines depict the locations of the inner
walls of the electrodes.  A bias magnetic field at the trap center of
1\,T was produced by an external solenoid (not shown) and
used for trapping the charged constituents $\overline{p}$ and $e^+$
in nested Penning traps. Figures from Ref.~\cite{ALPHA:Nature:2010}.
\label{ALPHA_trapping}%
}
\end{minipage}
\end{center}
\end{figure}

In 2010, the ALPHA collaboration demonstrated a major breakthrough of
magnetically trapping cold $\overline{\rm H}$
\cite{ALPHA:Nature:2010}.  A superconducting Ioffe trap with a radial
octupole field having a potential depth of about 0.5\,K was used in this experiment, the
magnetic field distribution of which is shown in Fig.~\ref{ALPHA_trapping}. 
By mixing $3\times 10^4$ $\overline{p}$ and 
$2\times 10^6$ $e^+$ in nested Penning traps located within the bias field of the 
Ioffe trap, several thousand $\overline{\rm H}$ were produced.  
On average a single $\overline{\rm H}$ in several such trials was cold enough to remain trapped.  
The magnetic trapping field was then ramped down, releasing the 
$\overline{\rm H}$ within a well-defined time window.  An imaging 
silicon detector recorded the vertices of the resulting annihilation events.

\begin{figure}[tbh!]
\begin{center}
\includegraphics[height=7cm]{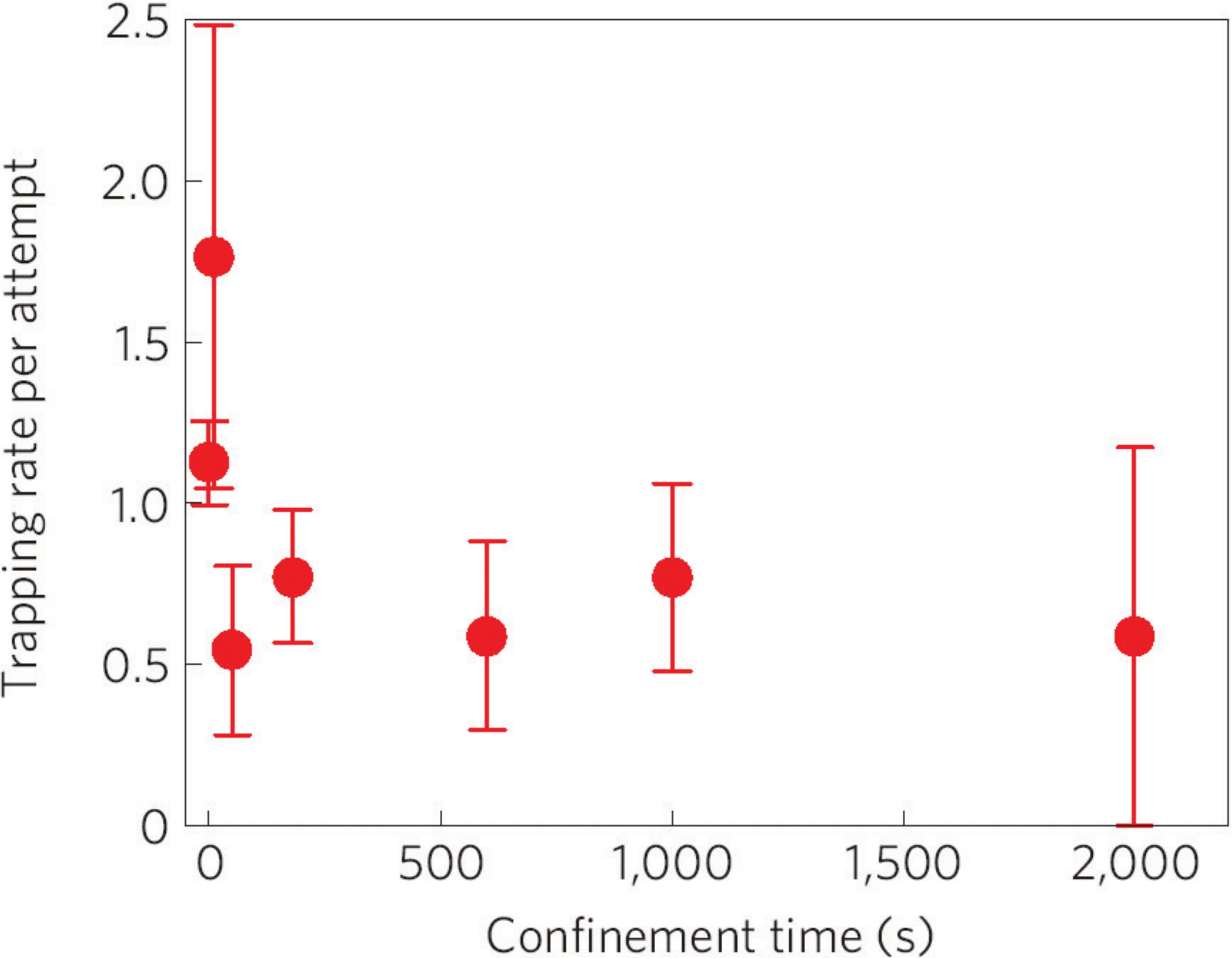}
\begin{minipage}[t]{16.5 cm}
\caption{
The $\overline{\rm H}$ trapping rate as a function of confinement
time measured by the ALPHA collaboration.  These $\overline{\rm H}$ have decayed down to the
$1s$ ground state. Figure from Ref.~\cite{Alpha:NatPhys:2011}.
\label{ALPHA_trapping2}%
}
\end{minipage}
\end{center}
\end{figure}

Shortly after this initial demonstration, ALPHA
extended the $\overline{\rm H}$ trapping time to $>1000$ s.  
Fig.~\ref{ALPHA_trapping2} shows the observed rate of $\overline{\rm H}$ trapping 
as a function of confinement time.  Whereas three-body recombination initially produced 
Rydberg $\overline{\rm H}$, the long confinement times ensured that they
cascade down to the $1s$ ground state. Theoretical calculations indicate that the time scale for this to occur
is in the ms regime, for initial states near $n=25-30$ \cite{Taylor:JPB4945:2006}. 

\begin{figure}[tbh!]
\begin{center}
\includegraphics[height=7cm]{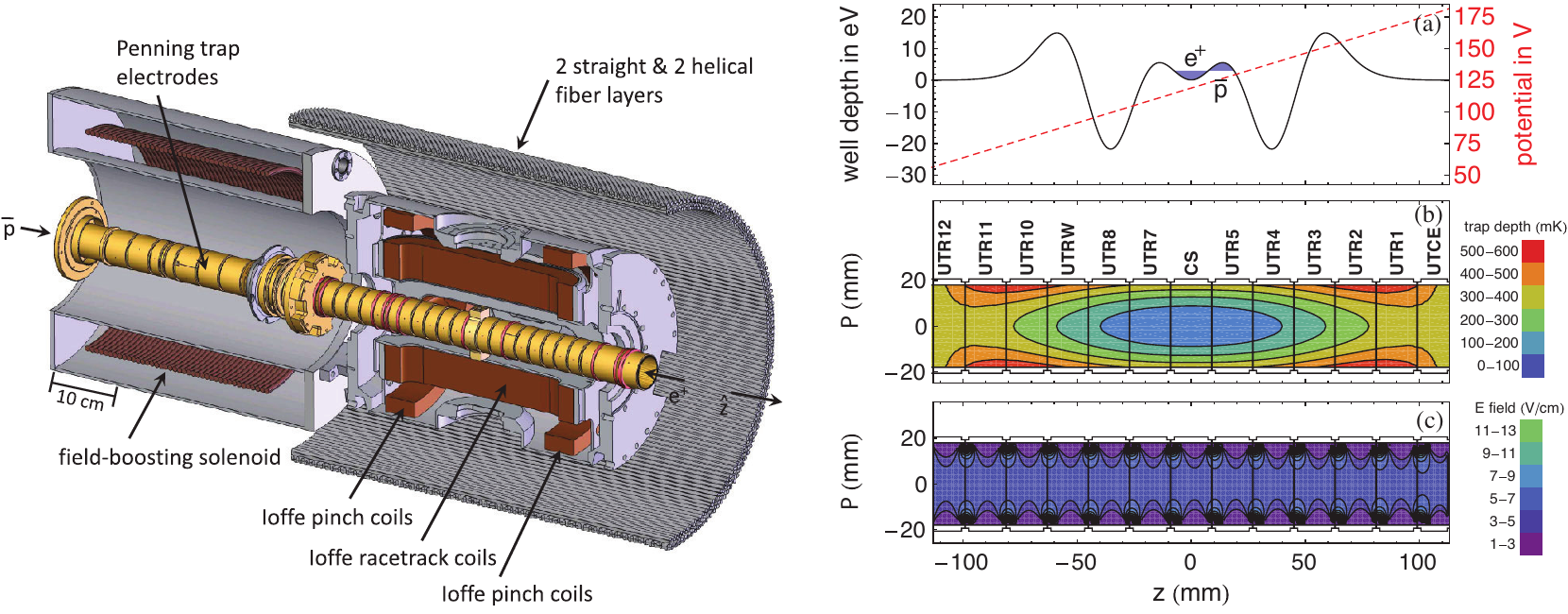}
\begin{minipage}[t]{16.5 cm}
\caption{Left panel: Schematic layout of the ATRAP apparatus showing the
nested Penning and Ioffe traps. An external solenoid (not shown)
added a 1-T magnetic field along the trap axis $\hat{z}$.
The experimental apparatus was actually vertical, but shown horizontally
in this figure for clarity.
Right panel: (a) Electrostatic potentials along the
central axis of the trap used to contain
(solid line) and remove (dashed line) charged particles.
(b) Electrode cross sections with equipotential energy
contours for low-field-seeking, ground-state $\overline{\rm H}$
confined in the Ioffe trap. (c) Axial electric field contours
used to clear $\overline{p}$ and $e^+$ before trapped
$\overline{\rm H}$ are detected. Figures from Ref.~\cite{ATRAP:TrappedHbar:2012}.
\label{ATRAP:TrappedHbar:2012}%
}
\end{minipage}
\end{center}
\end{figure}

The ATRAP collaboration also observed $\overline{\rm H}$ confined for 15--1000 s in a slightly different type of Ioffe trap 
(Fig.~\ref{ATRAP:TrappedHbar:2012})
with a radial quadrupole field \cite{ATRAP:TrappedHbar:2012}.  More ($10^6$) $\overline{p}$ were used in this experiment, 
and a high $\overline{\rm H}$ trapping rate of $5 \pm 1$ per trial was observed. 
The trapping of ground-state $\overline{\rm H}$ constituted another crucial step towards $\overline{\rm H}$ spectroscopy.
High precision experiments will presumably require a large number of trapped $\overline{\rm H}$, and so both collaborations 
are working towards this goal.

\subsection{\it Microwave spectroscopy of the ground-state hyperfine structure of antihydrogen}
\label{antihydrogenhyperfine}

\begin{figure}[tbh!]
\begin{center}
\includegraphics[height=9cm]{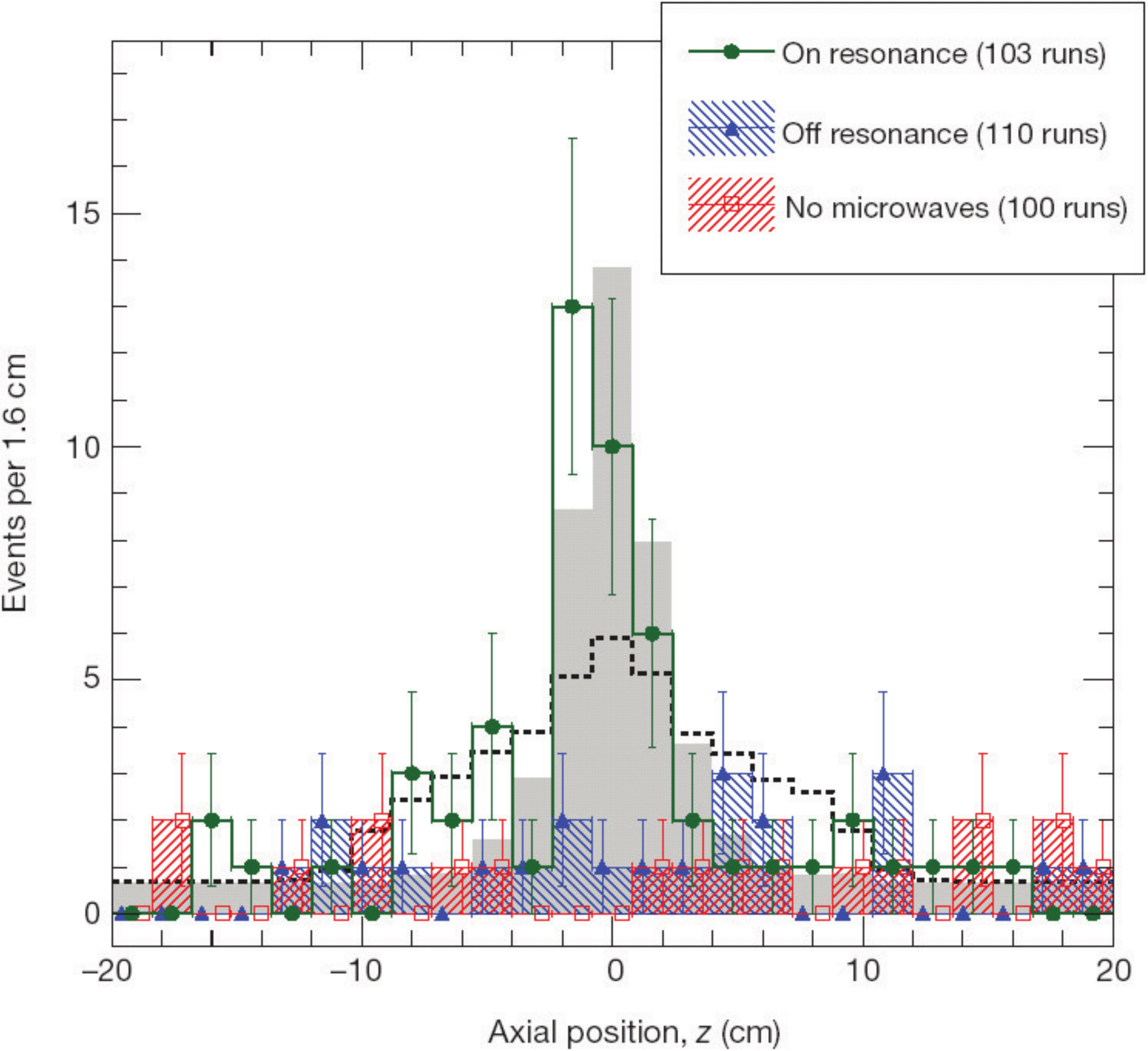}
\begin{minipage}[t]{16.5 cm}
\caption{
Detected $\overline{\rm H}$ vertices for a 30\,s time interval
as a function of position along the trap axis in the first microwave spectroscopy
experiment of $\overline{\rm H}$ carried out by the ALPHA collaboration. 
An annihilation peak
appears near the center of the Ioffe trap when the microwaves injected
into the apparatus are resonant with the spin-flip transition.  No
such signal appears both when the microwaves are off resonance and
when no microwaves are injected at all.  The signal is consistent with
a simulation, gray histogram.  The dashed black line is the result of
a simulation of trapped $\overline{\rm H}$ annihilating on residual gas.
This is not compatible with the observed signal.  Figure from \cite{Alpha:SpinFlips:2012}.
\label{ALPHA_spinflip}%
}
\end{minipage}
\end{center}
\end{figure}

The $1s$-ground state of $\overline{\rm H}$ contains four magnetic substates characterized by
the total angular momentum and magnetic quantum numbers $(F,M) = (0, 0)$, $(1,1)$,
$(1, 0)$ and $(1, 1)$ (Fig.~\ref{hydrogen_levels}). The ground-state hyperfine splittings 
between the $F=0$ (singlet) and $F=1$ (triplet) states of H has been measured 
\cite{hellwig1970,essen1971} to a precision of $\sim 10^{-12}$ using a H maser,
\begin{equation}
\nu_{\rm HFS}= 1\ 420\ 405\ 751. 766\ 7(9) \ {\rm Hz}.
\end{equation}
This value is to leading order determined by the Fermi contact interaction
and is proportional to the $\overline{p}$ magnetic moment $\mu_{p}$,
\begin{equation}
\nu_{\rm HFS}=\frac{16}{3}\left(\frac{M_p}{M_p+m_e}\right)^3\frac{m_e}{M_p}\frac{\mu_p}{\mu_{\rm nucl}}\alpha^2c
R_{\infty}(1+\Delta).
\label{hfsequation}
\end{equation}
The term $\Delta$ contains QED and other corrections of 
relative size $\sim 1.1\times 10^{-3}$. Since the $\mu_{\overline{p}}$ value has recently been measured by
a separate Penning trap experiment \cite{atrap_mag} with a precision of $\sim 10^{-6}$ 
(see Sect.~\ref{sec:magneticmoment}), 
a measurement of $\nu_{\rm HFS}$ of $\overline{\rm H}$ to a precision of $10^{-6}$ would 
provide information on the magnetic form factors of $\overline{p}$ \cite{Brodsky:2005,cjp:2006:429}
i.e., the non-relativistic magnetic size (Zemach) radius which contributes to a
$\sim 30$-ppm shift in $\nu_{\rm HFS}$. The proton polarizability contributes a
further shift of $<4$ ppm.
 
The ALPHA collaboration recently succeeded in inducing microwave 
transitions between these hyperfine levels in trapped $\overline{\rm H}$.
In the energy level diagram of Fig.~\ref{hydrogen_levels}, the proton spin resonance (PSR) transitions 
$\left|c\right>\rightarrow\left|a\right>$ and $\left|d\right>\rightarrow\left|b\right>$ 
were measured. For this, microwave radiation at a frequency of $\sim 29$\,GHz was injected
along the trap axis using a horn antenna. The spin-flip transitions excited states 
$\left|c\right>$ and $\left|d\right>$ of trapped atoms into high-field seeking states, which led to 
$\overline{\rm H}$ expulsion from the trap. When the magnetic field of the neutral atom trap was ramped 
down after exposing the 
$\overline{\rm H}$ to resonant microwaves, those atoms which were
expelled from the trap were missing; this constituted the so-called
``disappearance mode'' signal.  Another method of observing resonant
transitions was to detect annihilation events during the injection of
microwaves, which constituted an ``appearance mode'' signal.  Both types of
signals have been clearly observed in the experiment, as shown in 
Fig.~\ref{ALPHA_spinflip}. This observation of induced spin-flip
transitions in ground-state atoms marks the advent of $\overline{\rm H}$ spectroscopy.

\section{Antiprotonic helium}
\label{pbarhelium}

\subsection{\it Metastable antiprotonic helium atoms}

\begin{figure}[tb]
\epsfysize=9.0cm
\begin{center}
\begin{minipage}[t]{16 cm}
\epsfig{file=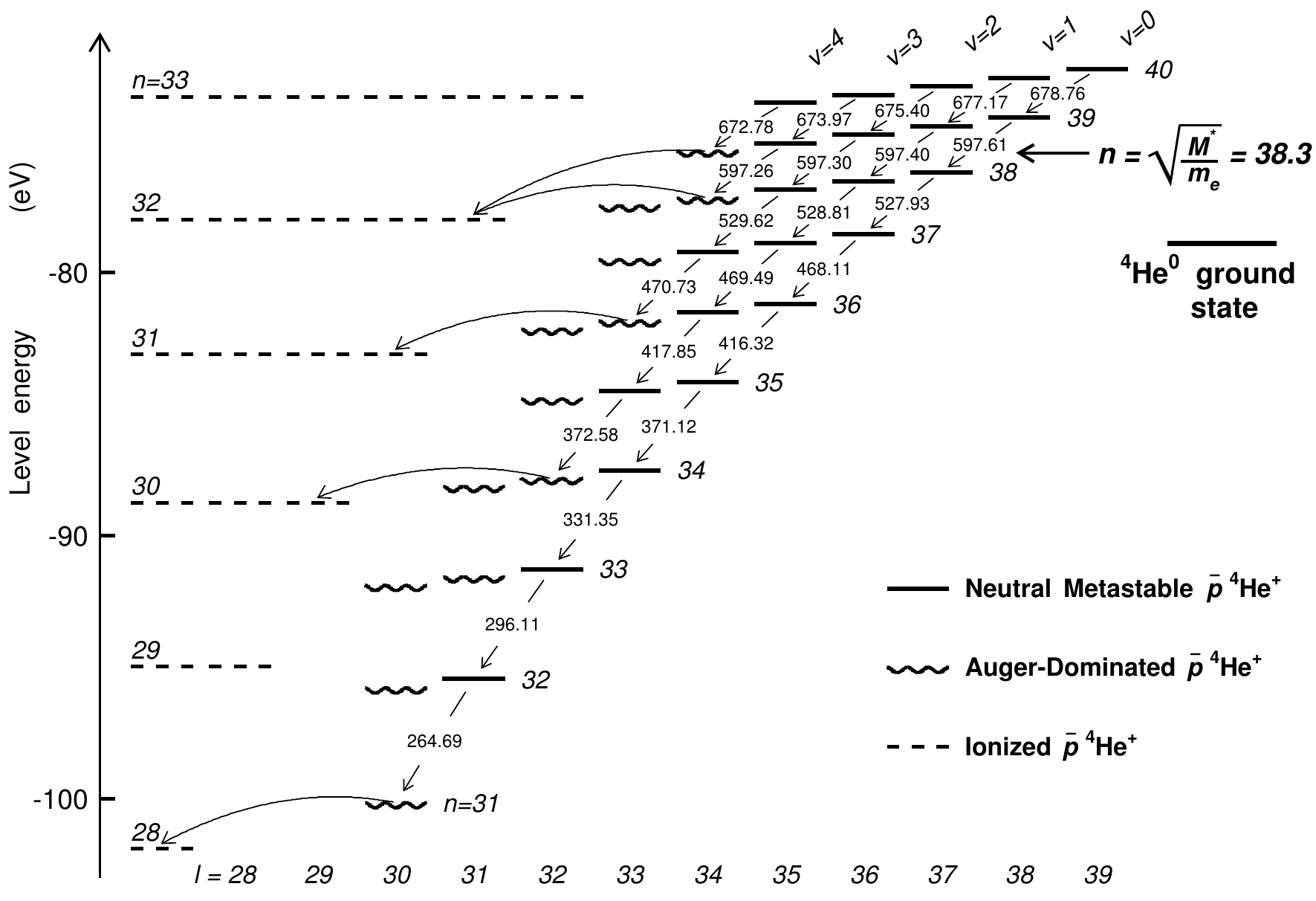,scale=0.9}
\end{minipage}
\begin{minipage}[t]{16.5 cm}
\caption{\label{fig:energylevel} 
Level structure of the $\overline{p}^4$He$^+$ atom.
The solid lines indicate the radiation-dominated metastable levels with 
lifetimes of 1--2 $\mu{\rm s}$, the wavy lines Auger-dominated
short-lived states. The broken lines show the final 
$\overline{p}{\rm He}^{2+}$ 
ionic states formed after Auger $e^-$ emission. The curved arrows
indicate Auger transitions with minimum $|\Delta\ell|$. 
On the left-hand scale the theoretical absolute energy of each 
state $(n,\ell )$ is plotted relative to the three-body-breakup threshold.
The calculated resonance wavelengths of radiative transitions following the 
constant-$v$ propensity rule are given in nanometers. Figure from Ref.~\cite{mhori2004}.}
\end{minipage}
\end{center}
\end{figure}

Antiprotonic helium atoms ($\overline{p}{\rm He}^+\equiv \overline{p}+{\rm He}^{2+}+e^-$) 
\cite{hayanorpp,yamazaki2002,iwasaki1991} are three-body Coulomb systems composed 
of a helium nucleus, an $e^-$ in the 
1s-ground state, and a $\overline{p}$ populating a Rydberg state with principal and angular momentum 
quantum numbers of $n\sim \ell+1\sim 38$. The energy level diagram of the $\overline{p}{\rm ^4He}^+$ 
isotope in the region $n=31-40$ and $\ell=30$--39 is shown in Fig.~\ref{fig:energylevel}.
Precise measurements on the transition frequencies of 
$\overline{p}{\rm He}^+$ can be used to determine the antiproton-to-electron mass ratio
\cite{mhori2006,hori2011}, and
constrain the equality of the $\overline{p}$ and $p$ charges and masses \cite{mhori2001,mhori2003,hori2011}.

Antiprotonic atoms (denoted  $\overline{p}{\rm X}^+$) can be readily synthesized for a given element 
${\rm X}$ by replacing one or more atomic $e^-$ with a negatively-charged $\overline{p}$ \cite{briggs1999,cohen2004,hesse,ovchinnikov,tong08,genkin,sakimoto2007,sakimoto2009}. 
The substitution takes place spontaneously, when $\overline{p}$ are brought to rest in the substance concerned.
These exotic atoms, however, are usually destroyed within picoseconds by electromagnetic 
cascade mechanisms that result in the rapid deexcitation of the $\overline{p}$ and its annihilation 
in the nucleus of $X$ via strong interaction.

The  $\overline{p}{\rm He}^+$ atom alone retains $\mu{\rm s}$-scale lifetimes against $\overline{p}$ annihilation
in the nucleus, even in dense helium targets \cite{iwasaki1991}. 
The extreme longevity is due to the fact that the $\overline{p}$ trapped in the Rydberg states have 
almost no overlap with the nucleus, and furthermore cannot easily deexcite by Auger emission 
\cite{kor97,revai97,kartavtsev2000,yamaguchi2002,yamaguchi2004} of the remaining
$e^-$ owing to its large ($I\sim 26$ eV) binding energy and the large multipolarity ($\Delta\ell\ge 3$) of the 
necessary transition. This $1s$ $e^-$ protects the $\overline{p}$ against Stark mixing with low-$\ell$ states
which overlap with the nucleus, during collisions \cite{mhori1998,mhori2004} with other helium atoms.
The atoms can be synthesized via the reaction,
\begin{equation}
\overline{p} + {\rm He} \rightarrow \overline{p}{\rm He}^+_{(n,\ell)}+e^-.
\label{eq}
\end{equation}
Some of the $\overline{p}$ are then captured \cite{tokesi,revai06,mhori2002} into states with $n$-values of around,
\begin{equation}
n\sim n_0=\sqrt{M^*/m_e},
\label{eq2}
\end{equation}
where $M^*$  denotes the reduced mass of the atom, and $m_e$ the $e^-$ mass. This corresponds to the $\overline{p}$ orbital with the same radius and binding energy as that of the displaced $1s$ $e^-$ in Eq.~\ref{eq}. The $n_0$ values for the $\overline{p}{\rm ^4He}^+$ and $\overline{p}{\rm ^3He}^+$ isotopes are respectively $38.3$ and $37.1$. The relative ease in synthesizing large numbers of $\overline{p}{\rm He}^+$ and its long lifetime make this atom amenable to high precision laser spectroscopy.

\subsection{\it Theoretical calculations}
The $\overline{p}{\rm He}^+$ energy levels have been calculated by three-body QED calculations 
to relative precision of 1 part in $10^9$ \cite{korobov2008}. This involved first solving 
\cite{Shimamura:92,Yamazaki:92,Greenland:93,korobov1996,Kartavtsev:96,Elander:97,Korobov:97a,andersson1998,Kino:99,Korobov:99c,kino2001hpc,Korobov:00,Korobov:03,kino2004} the non-relativistic Hamiltonian expressed in 
natural units of $m_e=\hbar=e=1$,
\begin{equation}
E_{\rm nr} =
   \left\langle
      -\frac{1}{2m_{13}}{\bf \nabla}^2_{\mathbf{r}_1}
      -\frac{1}{2m_{23}}{\bf \nabla}^2_{\mathbf{r}_2}
      -\frac{1}{m_3}{\bf \nabla}_{\mathbf{r}_1}
                    {\bf\nabla}_{\mathbf{r}_2}
      +\frac{Z_1Z_3}{r_1}+\frac{Z_2Z_3}{r_2}+\frac{Z_1Z_2}{R},
   \right\rangle.
\end{equation}
using numerical variational methods. Here the indices 1, 2, and 3 correspond respectively to the He 
nucleus, $\overline{p}$, and $e^-$. In fact, in the case of the ground state of normal He atoms, the non-relativistic 
binding energy $E_{\rm nr}=-2.903724377034119598296$ a.u. has been calculated to 19--22 digits of precision 
by several theoretical groups \cite{Korobov:00,goldman1998,drake1999} using a variety of trial functions containing 
a few thousand basis sets.  The case of 
$\overline{p}{\rm He}^+$ is in principle more difficult since the states are resonances that can decay by Auger 
emission of the $e^-$. Nevertheless for some $\overline{p}{\rm He}^+$ states with $\mu{\rm s}$-scale 
lifetimes against Auger and radiative decay, the $E_{\rm nl}$ values were calculated with a claimed precision 
of $\sim 15$ digits using the Feshbach formalism \cite{Korobov:00}. The Hamiltonian was projected onto
the subspace of closed 
channels which provided a sufficiently accurate approximation of the wave function. States with ns-scale
Auger lifetimes, on the other hand, were calculated to $\le 12$ digits of precision \cite{korobov2008,Korobov:03} 
using the complex-coordinate rotation (CCR) method \cite{kino2001hpc}, which takes into account the resonance 
nature of $\overline{p}{\rm He}^+$. 

In addition to $E_{\rm nr}$, perturbative calculations \cite{korobov2008} were carried out to determine the 
relativistic corrections of the $1s$ $e^-$ ($E_{\rm rc}$), the nucleus and $\overline{p}$ ($E_{\rm kin}$), and
the anomalous magnetic moment of the $e^-$ ($E_{a_e}$). Also calculated were QED corrections due to the 
one-transverse-photon exchange ($E_{\rm exch}$); the one-loop self-energy ($E_{\rm se}$) and
vacuum polarization ($E^{(3)}_{\rm vp}$); the recoil correction 
($E^{(3)}_{\rm recoil}$) of order $R_{\infty}\alpha^3 m_e/M_{\overline{p}}$; 
and one- and two-loop corrections ($E_{\alpha^4}$, $E_{\alpha^5}$)
of orders $R_{\infty}\alpha^4$ and $R_{\infty}\alpha^5$.   
The charge radii of the $^3$He and $^4$He nuclei give corrections 
of $E_{\rm nuc}=4-7$ MHz, whereas that of the $\overline{p}$ is much smaller ($<1$ MHz) owing 
to the large $\ell$-value of the states. The values for the $\overline{p}{\rm ^4He}^+$ 
transition $(36,34)\to(34,32)$ are shown in Table~\ref{tabmcorrections} (Ref.~\cite{korobovprivate}).

The latest calculation \cite{korobov2008} uses fundamental constants compiled in CODATA 2002 \cite{codata2002}, 
including the $^3$He- and $^4$He-to-electron mass ratios, the Bohr radius, and Rydberg constant. 
To preserve independence the more recent CODATA 2010 values \cite{jpcrd} were not used, which include results 
from previous experiments \cite{mhori2006} and three-body QED calculations on $\overline{p}{\rm He}^+$.
Similar calculations have been carried out on HD$^+$ and H$_2^+$ molecular ions
\cite{hannemann,hilico,koelemeij}, and the results agree with laser spectroscopy experiments to a 
fractional precision of $\sim 10^{-9}$.

\begin{table}[hbt]
\begin{center}
\begin{minipage}[t]{8.5 cm}
\begin{tabular}{l@{\,\,}c@{\,\,}r@{}l@{}}\hline\hline
$\Delta E_{nr}$        &=&~1\,522\,150\,208.3& \\
$\Delta E_{rc}$        &=&       $-$50\,800.9& \\
$\Delta E_{a_e}$       &=&              454.9& \\
$\Delta E_{exch}$      &=&            $-$84.9& \\
$\Delta E_{kin}$       &=&              105.7& \\
$\Delta E_{nuc}$       &=&                4.7& \\
$\Delta E_{se}^{(3)}$  &=&           7\,311.0& \\
$\Delta E_{vp}^{(3)}$  &=&           $-$243.0& \\
$\Delta E_{recoil}^{(3)}$&=&              1.4& \\
$\Delta E_{\alpha^4}$  &=&              113.1& \\
$\Delta E_{\alpha^5}$  &=&            $-$11.5& \\
\hline\\[-3.5mm]
$\Delta E_{total}$     &=& 1\,522\,107\,058.9&(2.1)(0.3)\\
\hline\hline
\end{tabular}
\end{minipage}
\begin{minipage}[t]{16.5 cm}
\caption{\label{tabmcorrections}
Contributions from various relativistic and QED corrections to
the transition frequency of the $\overline{p}{\rm ^4He}^+$ 
transition $(n,\ell)=(36,34)\rightarrow (34,32)$. From Ref.~\cite{korobovprivate}.} 
\end{minipage}
\end{center}
\end{table}

\subsection{\it Single-photon laser spectroscopy}

Initially, all laser spectroscopy experiments of $\overline{p}{\rm He}^+$ were carried out by inducing single-photon 
laser transitions \cite{Mor93,Mor94} from metastable states (indicated by solid lines in Fig.~\ref{fig:energylevel}) to states with 
ns-scale lifetimes (wavy lines) against Auger emission of the $e^-$. The transitions were excited 
using simple, ns pulsed dye lasers \cite{mhorioptics}.
Two-body $\overline{p}{\rm He}^{2+}$ ions \cite{mhori2005} remained after Auger decay. Since the
$\ell$-substates of the ion are highly degenerate, collisions with other He atoms in the experimental 
target caused Stark mixing with  S, P, and D states at high $n$ that have a large overlap with the He nucleus.
The $\overline{p}$ was absorbed in the He nucleus within picoseconds, resulting in annihilation and the emission of several 
$\pi^+$ and $\pi^-$. The resonance condition between the laser and the atom was detected by measuring the count rate of 
these $\pi^+$ and $\pi^-$ as a function of laser frequency, using acrylic Cherenkov 
counters \cite{cherenkov} surrounding the experimental target.

\begin{figure}[tb]
\epsfysize=9.0cm
\begin{center}
\begin{minipage}[t]{16.5 cm}
\epsfig{file=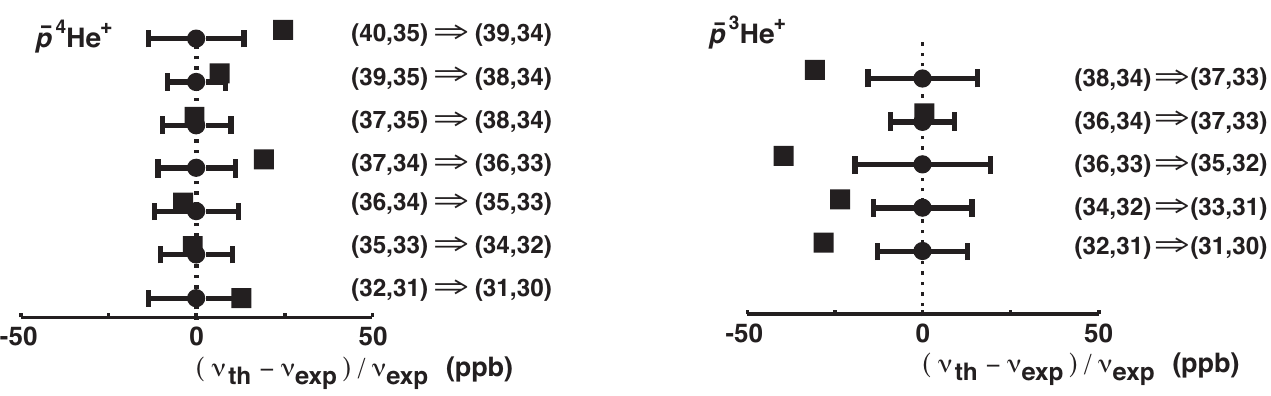,scale=1.30}
\end{minipage}
\begin{minipage}[t]{16.5 cm}
\caption{\label{fig:ppm2004} 
Fractional deviations of experimental (filled circles with error bars) and theoretical (squares) transition frequencies
of $\overline{p}{\rm ^4He}^+$ and $\overline{p}{\rm ^3He}^+$. Figure from Ref.~\cite{mhori2006}.}
\end{minipage}
\end{center}
\end{figure}

Twelve transition frequencies $\nu_{\rm exp}$ in the $\overline{p}{\rm ^4He}^+$ and $\overline{p}{\rm ^3He}^+$ 
isotopes were measured \cite{mhori2003,mhori2006} by single photon laser spectroscopy in this way, with a fractional precision of 
$(9-16)\times 10^{-9}$. In Fig.~\ref{fig:ppm2004}, the $\nu_{\rm exp}$ values
(indicated by filled circles with error bars) are compared with three-body QED calculations $\nu_{\rm th}$
(squares) of Ref.~\cite{korobov2008}. The four highest-precision measurements in
$\overline{p}{\rm ^4He}^+$, and $(36,34)$$\rightarrow$$(37,33)$ in $\overline{p}{\rm ^3He}^+$
agreed with $\nu_{\rm th}$ within $<1\times 10^{8}$. Four $\nu_{\rm exp}$ frequencies for $\overline{p}{\rm ^3He}^+$
were 2$\sigma$ above the $\nu_{\rm th}$ values. 

\subsection{\it Sub-Doppler two-photon laser spectroscopy}

\begin{figure}[tb]
\epsfysize=9.0cm
\begin{center}
\begin{minipage}[t]{16 cm}
\epsfig{file=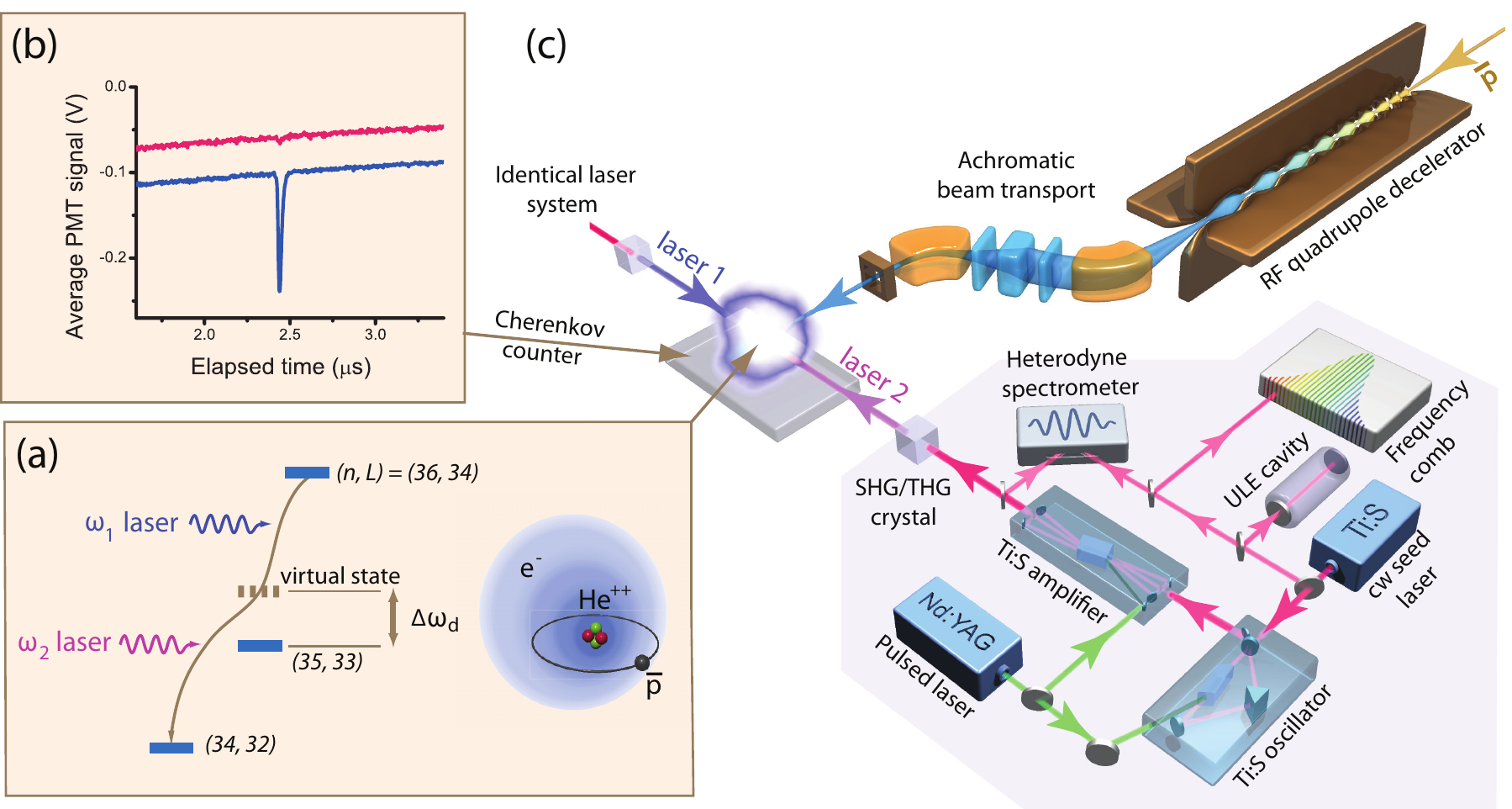,scale=0.9}
\end{minipage}
\begin{minipage}[t]{16.5 cm}
\caption{\label{fig:figure1_layout} 
(a) Energy levels, (b) Cherenkov detector signals, and (c) experimental
 layout for two-photon laser spectroscopy of
$\overline{p}{\rm He}^+$.
(a) Two counterpropagating laser beams induced the two-photon transition 
$(n,\ell)=(36, 34)$$\rightarrow$$(34,32)$ in $\overline{p}{\rm ^4He}^+$
via a virtual intermediate state of the $\overline{p}$ tuned close to the real
state $(35,33)$. (b) Cherenkov detectors revealed the annihilation of
$\overline{p}{\rm ^4He}^+$
following the nonlinear two-photon resonance induced at $t=2.4$ $\mu{\rm s}$ (blue).
When one of the lasers was detuned from resonance frequency by $\sim 500$ MHz,
the two-photon signal abruptly disappeared (red). 
(c) The $\overline{p}{\rm He}^+$ were synthesized by decelerating a beam of $\overline{p}$
using a radiofrequency quadrupole, and allowing them to stop in a cryogenic
helium target. Two Ti:sapphire pulsed lasers whose optical frequencies were
stabilized to a femtosecond frequency comb were used to carry out the
spectroscopy. CW, continuous wave; RF, radio frequency; SHG, second harmonic
generation; THG, third-harmonic generation; ULE, ultralow expansion;
PMT, photomultiplier tube. Figures from Ref.~\cite{hori2011}.}
\end{minipage}
\end{center}
\end{figure}

\begin{figure}[tb]
\epsfysize=10.0cm
\begin{center}
\begin{minipage}[t]{9 cm}
\epsfig{file=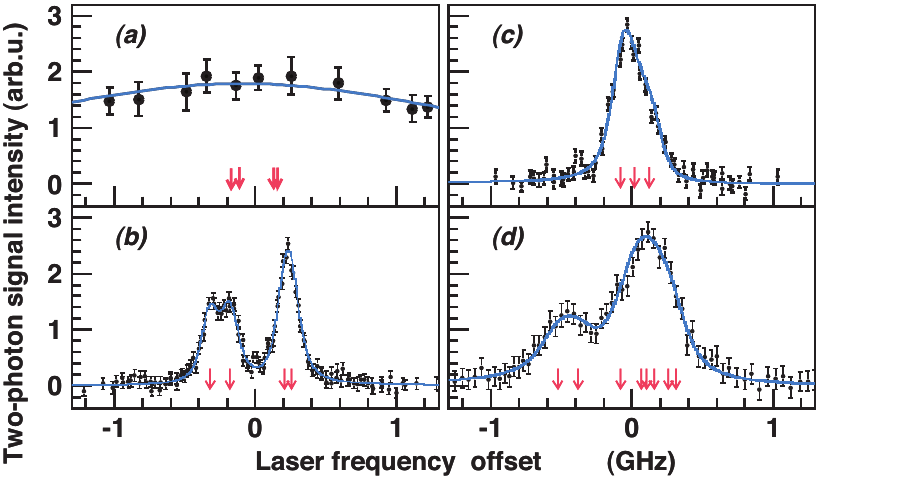,scale=1.3}
\end{minipage}
\begin{minipage}[t]{16.5 cm}
\caption{\label{fig:figure2_twoprof} 
Profiles of sub-Doppler two-photon laser resonances of $\overline{p}{\rm He}^+$. (a) Doppler- and
power-broadened profile of the single-photon resonance $(n,\ell)=(36, 34)$$\rightarrow$$(35,33)$ of
$\overline{p}{\rm ^4He}^+$. (b) Sub-Doppler two-photon profile of $(36, 34)$$\rightarrow$$(34, 32)$ 
involving the same parent state. (c,d) Profiles of $(33,32)$$\rightarrow$$(31,30)$ 
of $\overline{p}{\rm ^4He}^+$ and $(35,33)$$\rightarrow$$(33, 31)$ of 
$\overline{p}{\rm ^3He}^+$. Black filled circles indicate experimental data
points, blue lines are best fits of theoretical line profiles (see text) and partly 
overlapping arrows indicate positions of the hyperfine lines. 
Figures from Ref.~\cite{hori2011}.}
\end{minipage}
\end{center}
\end{figure}

\begin{figure}[tb]
\epsfysize=9.0cm
\begin{center}
\begin{minipage}[t]{16.5 cm}
\epsfig{file=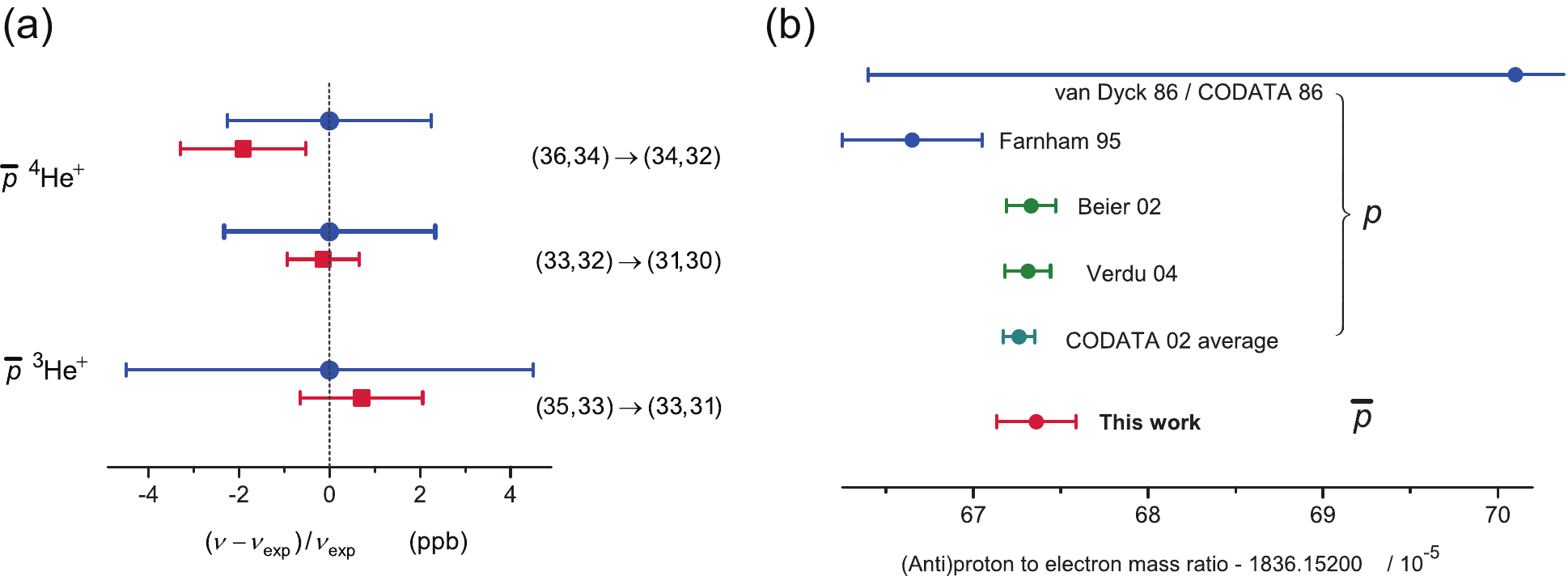,scale=0.86}
\end{minipage}
\begin{minipage}[t]{16.5 cm}
\caption{\label{fig:figure34_pbareratio} 
(a) Fractional deviations between the two-photon transition frequencies of 
$\overline{p}{\rm He}^+$: experimental 
($\nu_{\rm exp}$, blue circles) and theoretical values ($\nu_{\rm th}$, red squares).
(b) Antiproton-to-electron and proton-to-electron mass ratios. The
antiproton-to-electron mass ratio determined from the $\overline{p}{\rm He}^+$
data agrees to within a
fractional precision of $\sim 1.3\times 10^{-9}$ with the proton-to-electron values measured
in previous experiments and the CODATA 2002 recommended value
obtained by averaging them. Figures from Ref.~\cite{hori2011}.}
\end{minipage}
\end{center}
\end{figure}

The precision of the single-photon spectroscopy experiments above was 
limited to $10^{-7}$--$10^{-8}$ due to the Doppler broadening effect. As in normal atoms, the thermal motion of  
$\overline{p}{\rm He}^+$ at temperature $T$ strongly broadens the measured widths of the laser resonances by
$\nu\sqrt{8k_BT\log(2)/Mc^2}$, wherein $\nu$ denotes the transition frequency, $k_B$ the Boltzmann constant, $M$ 
the atom's mass, and $c$ the speed of light. One way to reach an experimental precision beyond this Doppler limit 
is provided by nonlinear two-photon spectroscopy. As described in Sect.~\ref{sec:future}, the $1s-2s$ transition frequency in atomic H has been measured to a precision of  $10^{-14}$ by irradiating the atom with two counterpropagating laser beams, each with a frequency corresponding to half the $1s-2s$ value. This arrangement cancels the Doppler broadening to first order because the red shift in the frequency of one of the lasers seen by the $\overline{p}{\rm He}^+$ is exactly canceled by a corresponding blue shift in the other laser. It is normally difficult, however, to apply this to the $\overline{p}{\rm He}^+$ case, because of the small probabilities involved in the nonlinear transitions of the massive antiproton.

Two-photon transitions of the type $(n,l)\rightarrow(n-2,\ell-2)$ however, were successfully induced 
(Fig.~\ref{fig:figure1_layout}(a)) by exciting $\overline{p}$ between 
the parent and daughter states through a so-called "virtual" intermediate state. This is a dressed-atom 
state that temporarily arises due to the interaction of the $\overline{p}{\rm He}^+$ with the laser field. 
If the frequencies $\nu_1$ and $\nu_2$ of the counterpropagating laser beams are tuned
so that the virtual state lies near (e.g., within $\Delta\nu_d\sim 10$ GHz) to a real state $(n-1,\ell-1)$, the overlap
of their wavefunctions becomes large and so the probability of the two-photon transition $(n,l)\rightarrow(n-2,\ell-2)$
is enhanced by a factor of $>10^5$ \cite{mhori2010}. Note that the population in the state $(n-1,\ell-1)$ is unaffected by this; 
the state only serves to enhance the transition probability. The first-order Doppler width is then reduced by a factor 
$\left|\nu_1-\nu_2\right|/(\nu_1+\nu_2)$.
 
Even under these conditions of enhanced transition probability, MW-scale laser pulses are needed to excite the two-photon transitions. Lasers with high spectral purity and low phase noise are needed to avoid rapid dephasing in the amplitude of the two-photon transition. For this (Fig.~\ref{fig:figure1_layout} (c)) two sets of Ti:Sapphire lasers of pulse length 30-100 ns and linewidth $\sim 6$ MHz were developed \cite{chirp2009}.
They were based on continuous-wave (cw) lasers of wavelengths 728--940 nm whose frequencies were measured to a precision of $<10^{-10}$ using a femtosecond optical comb \cite{udem} locked to a global-positioning-system disciplined, quartz oscillator. This seed beam was pulse-amplified to the 1-MW peak power needed to excite the two-photon transitions, using a Ti:Sapphire pulsed oscillator and amplifier. Spurious modulations in the pulsed laser frequency or "chirp" induced during this amplification are an important source of systematic error, and were measured using a heterodyne spectrometer \cite{eikema1997,meyer}.
The precision of this laser system of $<1.4\times 10^{-9}$ was verified \cite{chirp2009} by measuring some two-photon transition frequencies in Rb and Cs \cite{fendel} at wavelengths of $\lambda=778$ nm and 822 nm.

It was essential to use cryogenic He targets of low enough density for the relaxations caused by collisions between $\overline{p}{\rm He}^+$ and other He atoms that could inhibit the two-photon transition to remain small. This implied the use of $\overline{p}$ beams of low enough energy ($E\sim 70$ keV) to be stopped in such targets within the volume irradiated by the 20-mm-diameter laser beams. Pulsed beams containing $7\times 10^6$ $\overline{p}$ were provided by the RFQD (Fig.~\ref{fig:figure1_layout} (c)). The beam was 
transported \cite{horiphoto,parallel} by an achromatic magnetic beamline to the target chamber filled with 
$^4$He or $^3$He gas at temperature $T\sim 15$ K and pressure $p=0.8-3$ mbar. At a time $\sim 2$ $\mu{\rm s}$ after the
formation of $\overline{p}{\rm He}^+$, horizontally-polarized laser beams of energy density 
$\sim 1$ mJ/cm$^2$ were simultaneously fired through the target in a perpendicular direction to the $\overline{p}$ beam. 
Fig.~\ref{fig:figure1_layout} (b) shows the Cherenkov signal (indicated in blue solid line) as a function of time elapsed since the arrival of $\overline{p}$ pulses at the target. Lasers of wavelengths $c/\nu_1=417$ and $c/\nu_2=372$ nm were tuned to the two-photon transition $(36,34)\rightarrow (34,32)$ so that the virtual intermediate state lay 6 GHz away from the real state $(35,33)$. The above-mentioned annihilation spike corresponding to the two-photon transition can be seen at $t=2.4 \mu{\rm s}$. When the 417-nm laser alone was tuned slightly (by 0.5 GHz) off the two-photon resonance condition (red line), the signal abruptly disappeared, which indicated that the background from any Doppler-broadened, single-photon transitions is very small.

Fig.~\ref{fig:figure2_twoprof} (b) shows the resonance profile measured by detuning the $\nu_1$ laser to $\Delta\nu_d=-6$ GHz, whereas $\nu_2$ was scanned between -1 and 1 GHz around the two-photon resonance defined by $\nu_1+\nu_2$. The measured linewidth ($\sim 200$ MHz) is more than an order of magnitude smaller than the profile of the corresponding single-photon resonance $(36,34)\rightarrow (35,33)$ (Fig.~\ref{fig:figure2_twoprof} (a)). This sharp line allowed the determination of the atomic transition frequency with a correspondingly higher precision. The two-peak structure with a frequency interval of 500 MHz arises from the dominant interaction between the $e^-$ spin and the orbital angular momentum of the $\overline{p}$ (Sect.~\ref{lasermicro}). Each peak is a superposition of two hyperfine lines caused by a further interaction between the $\overline{p}$ and $e^-$ spins. The asymmetric structure is reproduced by lineshape calculations and is due to asymmetric spacings between the hyperfine components.

Fig.~\ref{fig:figure2_twoprof} (c) shows the $(33,32)\rightarrow (31,30)$ resonance at wavelength $\lambda=139.8$ nm with the lowest n-values among the two-photon transitions. All hyperfine lines are much closer together (at intervals of $\sim 100$ MHz). 
The $\overline{p}^3{\rm He}^+$ resonance $(35,33)\rightarrow (33,31)$ (Fig.~\ref{fig:figure2_twoprof} (d)) contains eight partially-overlapping hyperfine lines arising from the spin-spin interactions of the $^3{\rm He}$ nucleus, $e^-$, and $\overline{p}$.
The spin-independent transition frequencies $\nu_{\rm exp}$ were obtained by fitting each profile with a theoretical lineshape (indicated by solid lines) which was determined by numerically solving the nonlinear rate equations of the two-photon process. Various sources of statistical and systematic errors such as the AC Stark shifts \cite{mhori2010,Haas2006}, Zeeman shifts, and frequency chirp were evaluated.

The $\nu_{\rm exp}$ values (indicated by filled circles with error bars in Fig.~\ref{fig:figure34_pbareratio} (a)) agree with theoretical $\nu_{\rm th}$ values (squares) within $(2.3-5)\times 10^{-9}$. This agreement is a factor 5--10 times better than in the single-photon experiments described in the previous section. 

When the antiproton-to-electron mass ratio $M_{\overline{p}}/m_e$ in these calculations was changed by $10^{-9}$, 
the $\nu_{\rm th}$-value changed by 2.3--2.8 MHz. The best agreement between the experimental and calculated frequencies
were obtained with a mass ratio,
\begin{equation}
M_{\overline{p}}/m_e=1836.1526736(23).
\end{equation}
The uncertainty of $23\times 10^{-7}$ includes the statistical and systematic experimental, and theoretical contributions of 
$18\times 10^{-7}$, $12\times 10^{-7}$, and $10\times 10^{-7}$. 

This is in good agreement with the three previous measurements 
\cite{farnham,beier,verdu} of the proton-to-electron mass ratio  (Fig.~\ref{fig:figure34_pbareratio}(b)) with a similar experimental precision. The most precise value for $p$ is currently obtained by comparing the g-factors of hydrogen-like $^{12}$C$^{5+}$ and $^{16}$O$^{7+}$ ions measured by the GSI-Mainz collaboration with high-field QED calculations. The CODATA recommended value \cite{jpcrd,codata2002}
for $M_p/m_e$  is taken as the average of these experiments. By assuming $CPT$ invariance and using the CODATA
recommended value for the $p$ mass, $M_{\overline{p}}=M_p=1.00727646677(10)$ u, one can further derive a value for the $e^-$ mass,
\begin{equation}
 m_e=0.0005485799091(7) {\rm u},
\end{equation}
from the $\overline{p}{\rm He}^+$ result. 

Hughes and Deutch \cite{hughes} constrained the equality between the $\overline{p}$ and $p$ charges and masses 
$\delta_Q=(Q_p-Q_{\overline{p}} )/Q_p$ and $\delta_M=(M_p-M_{\overline{p}})/M_p$ to better than $2\times 10^{-5}$. For this they combined X-ray spectroscopic data of antiprotonic atoms (which is proportional to $Q^2_{\overline{p}}M_{\overline{p}}$) and the cyclotron frequency ($\propto Q_{\overline{p}}/M_{\overline{p}}$) of $\overline{p}$ confined in Penning traps measured to a higher precision. One can improve this limit by factor $>10^4$ using the linear dependence of $\delta_M$ and $\delta_Q$ on the $\nu_{\rm th}$-values of $\overline{p}{\rm He}^+$, i.e., $\delta_M\kappa_M+\delta_Q\kappa_Q\le\left|\nu_{\rm exp}-\nu_{\rm th}\right|/\nu_{\rm exp}$. For the three transitions, the constants were estimated as $\kappa_M=2.3$--2.8 and $\kappa_Q=2.7$--3.4, whereas the right side of this equation was evaluated by averaging over the three transitions as, $<(8\pm 15)\times 10^{-10}$. Meanwhile the constraint of Eq.~\ref{atrapfinal} from the TRAP experiment implies that $\delta_Q\sim \delta_M$.  From this it was concluded that any deviation between the charges and masses are $<7\times 10^{-10}$ at 90$\%$ confidence level \cite{hori2011}.

\subsection{\it Laser-microwave-laser triple resonance spectroscopy}
\label{lasermicro}

\begin{figure}[tb]
\epsfysize=9.0cm
\begin{center}
\begin{minipage}[t]{16.5 cm}
\epsfig{file=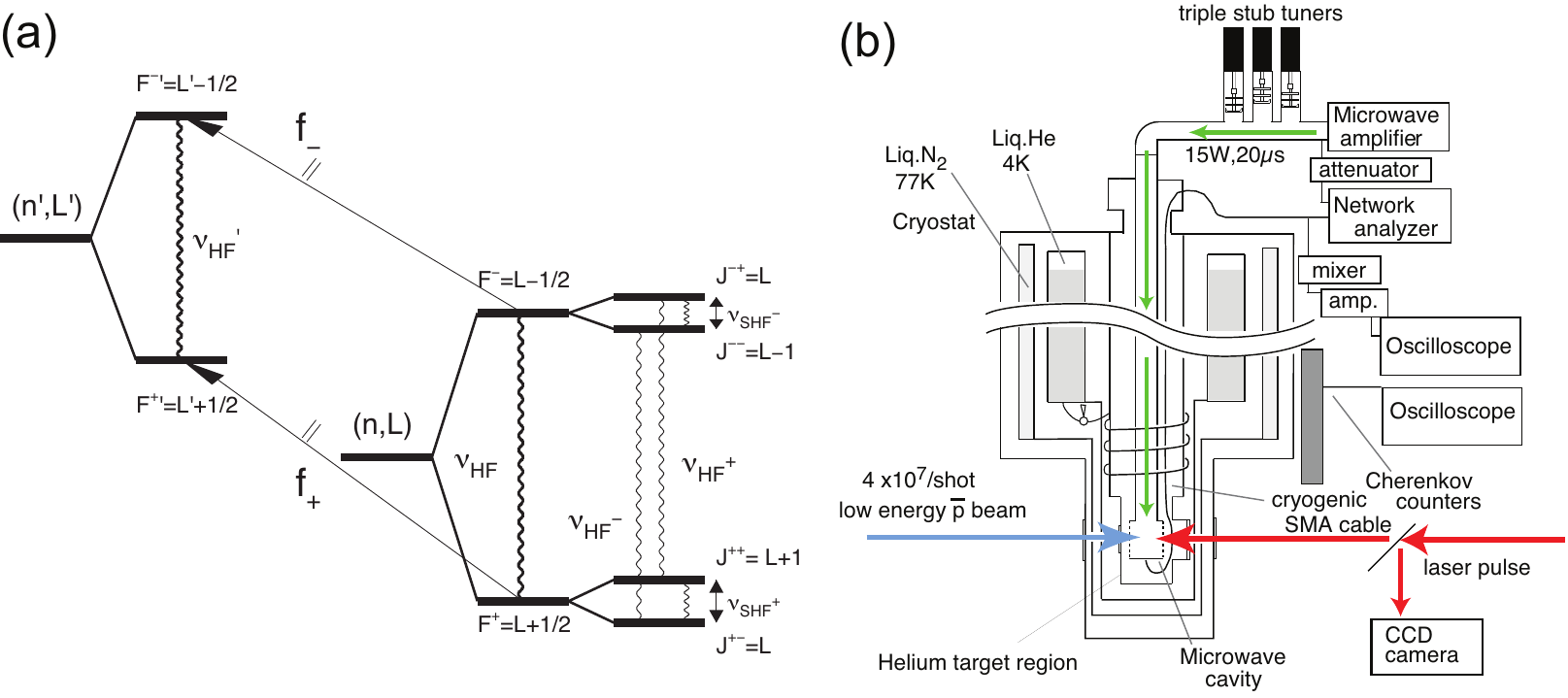,scale=1.05}
\end{minipage}
\begin{minipage}[t]{16.5 cm}
\caption{\label{fig:hfs} 
(a) Energy level diagram indicating the splitting of a                                                                        
$\overline{p}{\rm ^4He}^+$ state and observable                                                                   
laser transitions $f_-$ and $f_+$ from the atomic                                                               
state $(n,L)$ to a daughter state $(n^{\prime},L^{\prime})$.                                                    
Wavy lines denote allowed magnetic transitions                                                                  
with frequencies $\nu_{\rm HF}^+$ and $\nu_{\rm HF}^-$                                                          
associated with an $e^-$ spin flip.                                                                          
(b) Schematic layout of microwave spectroscopy experiment.
}
\end{minipage}
\end{center}
\end{figure}

The microwave transition frequencies between the hyperfine sublevels of 
$\overline{p}{\rm He}^+$ was measured by 
laser-microwave-laser triple resonance spectroscopy \cite{yamazaki2002,widmann2002,sakaguchi2004,pask2008}. 
The spin-spin and spin-orbit interactions between the $\overline{p}$ and $e^-$ 
in $\overline{p}{\rm ^4He}^+$ cause each metastable state to split into four 
magnetic substates, denoted by $J^{-+}$, $J^{--}$, $J^{++}$, and $J^{+-}$
as shown in Fig.~\ref{fig:hfs} (a). The dominant splitting corresponding to a frequency interval 
$\Delta\nu_{HF}=10$--15 GHz arises from the interaction between the $e^-$ spin and the orbital angular momentum 
of the $\overline{p}$. The size of $\Delta\nu_{HF}$ is primarily sensitive to the magnetic moment of the
$e^-$, rather than the $\overline{p}$. A smaller spitting of $\Delta\nu_{SHF}=150$--300 MHz is caused by 
the interaction between the $\overline{p}$ spin and its orbital angular momentum. This frequency 
$\Delta\nu_{SHF}$ is in principle more interesting for $CPT$ consistency tests since it is roughly 
proportional to the $\overline{p}$ magnetic moment \cite{bakahfs}, but the short lifetime of $\overline{p}{\rm He}^+$ and 
other experimental limitations makes it unfortunately difficult to directly measure this with a high 
precision. All measurements therefore concentrated on the microwave $\Delta\nu_{HF}$ transitions.

In the experiment (Fig.~\ref{fig:hfs} (b)), $\overline{p}{\rm ^4He}^+$ were first
synthesized by allowing a beam of 5.3-MeV $\overline{p}$ to come to rest in a microwave
cavity filled with cryogenic helium gas of typical atomic density
$\rho\sim 3\times 10^{20}$ cm$^{-3}$. The stainless-steel cavity \cite{sakaguchi2004} was cylindrical 
with a loaded quality factor (i.e., the ratio between the cavity bandwidth and resonance frequency) 
of $\sim 100$ and a central frequency of $f\sim 12.91$ GHz. The atoms were first irradiated with 
a laser pulse tuned to the transition $(n,\ell)=(37,35)$$\rightarrow$$(38,34)$ 
at wavelength $\lambda=726.1$ nm, which stimulated the transition 
denoted by $f_-$ in Fig.~\ref{fig:hfs} (a). This selectively depopulated the $\overline{p}$ occupying
the two states $J^{-+}$ and $J^{--}$, while those in the states
$J^{++}$ and $J^{+-}$ were unaffected by the laser beam. 

A microwave pulse of frequency $f=12.9$ GHz and typical power $\sim 4$ W was 
admitted into the cavity through a waveguide, thus generating a 
standing wave in the cavity. A triple-stub-tuner was inserted into the waveguide circuit
outside the cryostat, which allowed the impedance of the transmission 
line to be matched to the cavity and the central frequency chosen. 
By this method the cavity was tunable across a frequency range
of 100 MHz, while achieving at each frequency point a resonance condition
with a quality factor close to $\sim 2700$.  The microwave stimulated electron spin-flip transitions between the state 
$J^{-+}$ and $J^{++}$ at frequency $\nu_{\rm HF}^+$, and between $J^{--}$ 
and $J^{+-}$ at $\nu_{\rm HF}^-$. The resulting change in the populations in 
$J^{-+}$ and $J^{--}$ were detected by a second laser pulse, again 
tuned to the $f_-$ transition. 

\begin{figure}[tb]
\epsfysize=9.0cm
\begin{center}
\begin{minipage}[t]{16.5 cm}
\epsfig{file=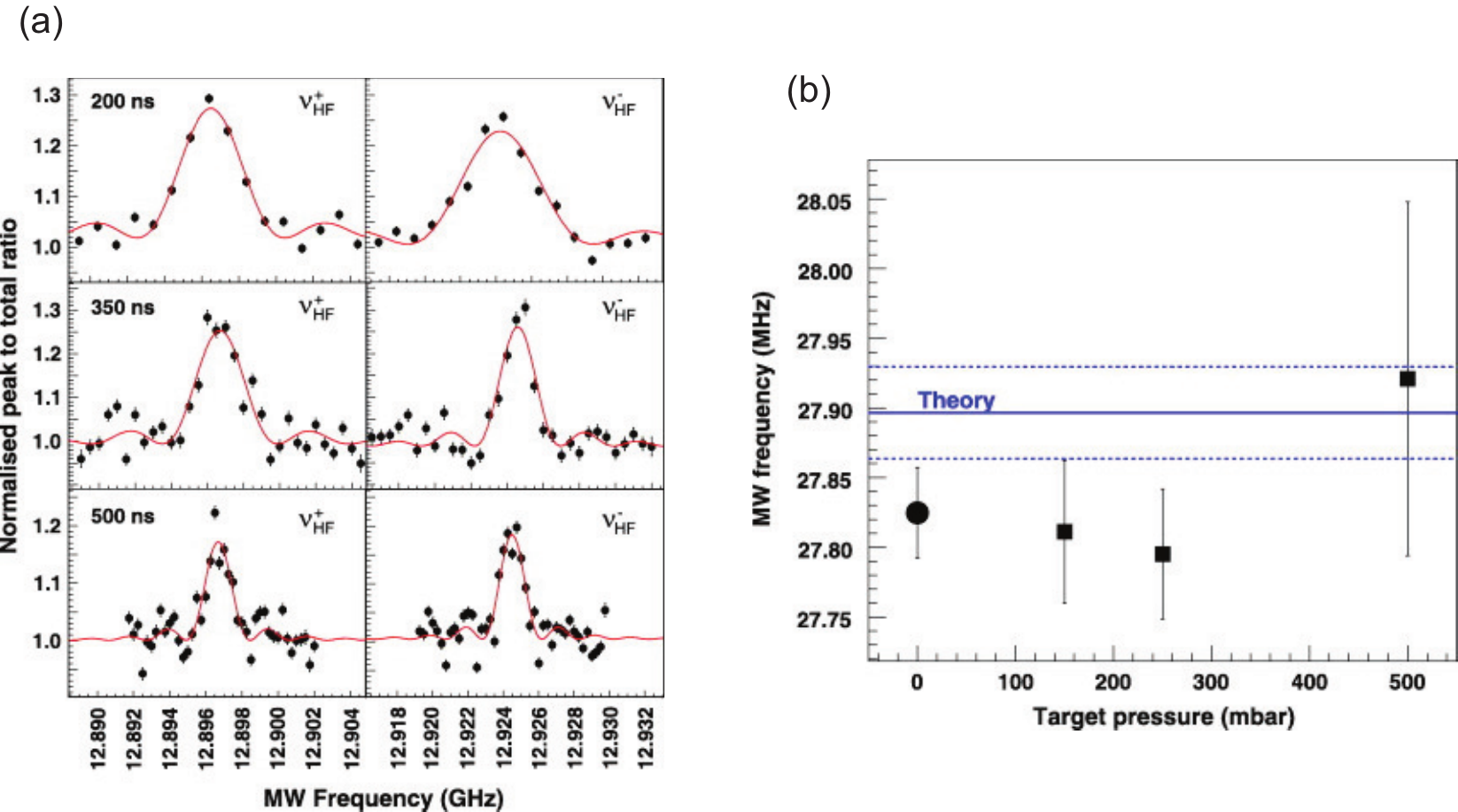,scale=1}
\end{minipage}
\begin{minipage}[t]{16.5 cm}
\caption{\label{fig:hfs_result} 
(a) Laser-microwave-laser triple resonance signal involving the 
$\overline{p}{\rm ^4He}^+$ state
$(n,\ell)=(37,35)$. The delay between the first and
second laser pulses which respectively creates a population
asymmetry between the hyperfine states, and detects the
microwave transition is varied between 200--500 ns. 
(b) Frequency difference $\nu_{\rm HF}^--\nu_{\rm HF}^+$ measured at
target pressures between 150--500 mb (filled squares) and the average
(circles) of the three experimental data points.
Figures from Ref.~\cite{pask2009}.}
\end{minipage}
\end{center}
\end{figure}

In Fig.~\ref{fig:hfs_result} (a), the resonance signals detected in this way are 
shown as a function of microwave frequency. The expected resonances at 
$\nu_{\rm HF}^+$ and $\nu_{\rm HF}^-$ were clearly observed. By increasing
the timing delay between the two $f_-$ lasers between 200 and 500 ns, the
resolution was increased. The frequency difference $\nu_{\rm HF}^--\nu_{\rm HF}^+$ MHz
(Fig.~\ref{fig:hfs_result} (b)) is especially sensitive to the $\overline{p}$ magnetic moment.
By averaging the results measured at four target pressures, a value
$\nu_{\rm HF}^--\nu_{\rm HF}^+=27.825$(33) MHz was obtained \cite{pask2009,pask2008}.
By comparing this with the results of three-body QED calculations \cite{bakahfs,yam01,kinohfs,Kor06,korobov2009},
the magnetic moment was obtained as $\mu_{\overline{p}}=-2.7862(83)\mu_{\rm nucl}$.
This determination with a precision of $0.3\%$ is in good agreement with the 
value measured by X-ray spectroscopy of $\overline{p}$Pb atoms 
to a similar precision. The precision on $\mu_{\overline{p}}$ was later
improved by the Penning trap experiment described in Sect.~\ref{sec:magneticmoment}.
Two microwave transitions between the hyperfine sublevels of 
state $(n,\ell)=(36,34)$ in $\overline{p}{\rm ^3He}^+$ were similarly 
detected \cite{friedreich11} at frequencies of $11.12559(14)$ GHz and $11.15839(18)$ GHz. 

\subsection{\it Chemical physics}

A variety of systematic studies was also carried out on the chemical-physics properties of $\overline{p}{\rm He}^+$.
The numbers of $\overline{p}$ populating the metastable states $(n,\ell)$ were studied by 
measuring the intensity of the laser resonance involved in each state \cite{mhori2002}.
Nearly all the $\overline{p}{\rm He}^+$ were found to lie in the region $n=37$--40 immediately after the
formation of the atom, with the $n=38$ and 37 states having the 
largest population. This appears to support the estimation given in Eq.~\ref{eq2}. On the other hand,
theoretical calculations predicted sizable populations in the $n>40$ states, but experiments detected very 
few $\overline{p}$ in them. This may be due to collisions between $\overline{p}{\rm He}^+$ and other 
helium atoms in the target which destroy the populations in these states. 

The Auger rates of many $\overline{p}{\rm ^3He}^+$ and $\overline{p}{\rm ^4He}^+$ states were measured in 
Refs.~\cite{yamaguchi2002,yamaguchi2004}. Most of the results agreed with theoretical 
calculations, but state $(37,33)$ in $\overline{p}{\rm ^4He}^+$ revealed decay rates which 
were orders of magnitude larger than the theoretical values.
Calculations \cite{kartavtsev2000} indicated that such a short lifetime is caused by a strong coupling to an electronically 
excited $\overline{p}{\rm ^4He}^+$ state, where the electron
occupies the $3d$ orbital, and the $\overline{p}$ the state $(n,\ell)=(32,31)$ \cite{mhori2001}.

The radiative lifetimes of $\overline{p}{\rm ^4He}^+$ states were measured \cite{mhori2004,mhori1998} 
as a function of the atomic density $\rho$ of the helium target. One state $(n,\ell)=(39,35)$ retained a 
lifetime $\tau\sim 1.5$ $\mu{\rm s}$ at even liquid helium densities \cite{mhori1998}, whereas other states 
became dramatically short-lived. For example, the lifetime of $(37,34)$ decreased from 
$\tau=1.2$ $\mu{\rm s}$ to $130$ ns, as $\rho$ was increased from $1\times 10^{20}$ to $6\times 10^{21}$
${\rm cm}^{-3}$. Theoretical calculations \cite{obreshkov2004,korenman07} have been unable to qualitatively 
explain the reason for this.

The antiprotonic helium ion ($\overline{p}{\rm He}^{2+}$)
is a singly-charged, two-body system composed of an antiproton
and helium nucleus. Cold (temperature $T\sim 10$ K)
$\overline{p}{\rm ^4He}^{2+}$ and $\overline{p}{\rm ^3He}^{2+}$ ions with lifetimes 
$\tau_i\sim 100$ ns against annihilation were produced \cite{mhori2005}. These states had
principal and angular momentum quantum numbers $n_i=28$--32 and $\ell_i=n_i-1$, 
and constituted ideal semiclassical Bohr systems. Their spin-independent 
parts of the energy levels (left side of Fig.~\ref{fig:energylevel}) 
can be calculated to very high precision ($\sim 10^{-8}$) using the simple Bohr formula,
\begin{equation}
E_n=-\frac{4R_{\infty}hc}{n_i^2}\frac{M}{m_e}\frac{Q^2_{\overline{p}}}{e^2}.
\end{equation}
Owing to this simple structure, the ion may be a candidate for future precision laser spectroscopy experiments.

When small (10--100 ppm) admixtures of H$_2$ and D$_2$ impurity gases were 
mixed in the target He, the resulting chemical reactions with 
$\overline{p}{\rm ^4He}^+$ caused the state lifetimes to shorten to ns scales \cite{juhasz2002,juhasz2003,juhasz2006}. 
The cross section for this reaction involving state $(n,\ell)=(39,35)$ at target temperatures $T=10$--60 K 
was measured to be around $\sim 3\times 10^{-15}$ cm$^2$. This roughly corresponds to the geometrical cross section.
The cross section for $(37,34)$ on the other hand was much smaller, and decreased from $5\times 10^{-16}$ to 
$1\times 10^{-16}$ cm$^2$ as the temperature was reduced from 300 K to 30 K, then leveled off below 30 K. This 
behavior was interpreted to indicate the presence of a quantum-tunneling effect with a small activation barrier at 
low temperature \cite{juhasz2006}.

\section{Antiproton magnetic moment measured in a Penning trap}
\label{sec:magneticmoment}

\begin{figure}[tbh!]
\begin{center}
\includegraphics[height=8.5cm]{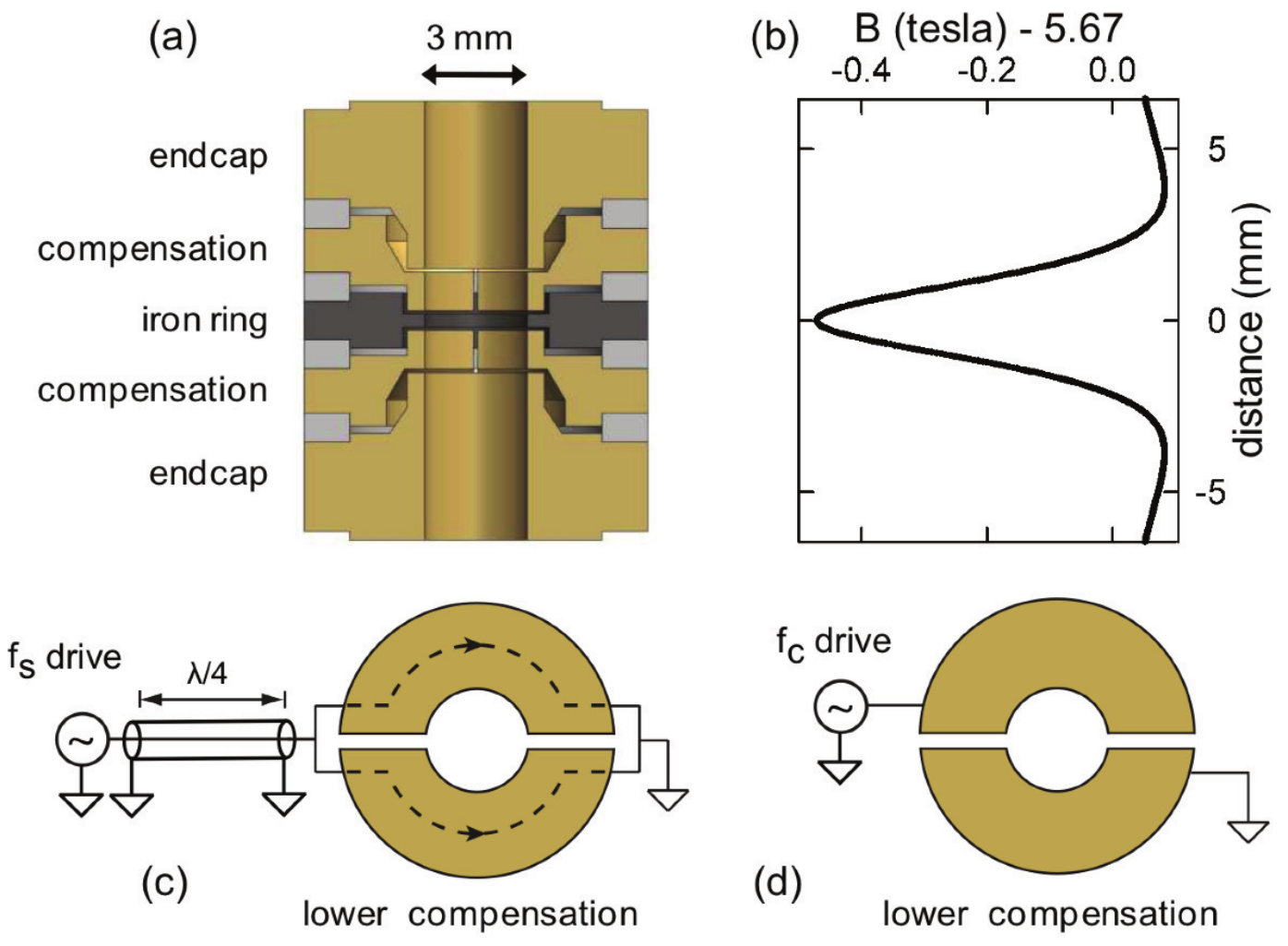}
\begin{minipage}[t]{16.5 cm}
\caption{ (a) Cross-sectional view of the analysis Penning trap
used by the ATRAP collaboration to measure the magnetic moment of $\overline{p}$.
The electrodes were made of copper, with an iron ring which
introduced an inhomogeneous magnetic field superimposed
on the solenoidal one of the Penning trap. (b) The B-field along
the trap axis. (c) Top view of the paths of the oscillating current
which flipped the $\overline{p}$ spin. (d) An oscillating electric
field introduced on the electrodes drove the cyclotron motion of
$\overline{p}$.
Figures from Ref.~\cite{atrap_mag}.
\label{disciacca:fig2}
}
\end{minipage}
\end{center}
\end{figure}

Part of the ATRAP collaboration recently measured the magnetic moment of a 
single $\overline{p}$ confined in a Penning trap as \cite{atrap_mag},
\begin{equation}
\mu_{\overline{p}}=-2.792845(12)\mu_{\rm nucl}.
\end{equation}
A comparison to the proton value,
\begin{equation}
 \mu_p=2.792846(7)\mu_{\rm nucl},
 \end{equation}
which was measured \cite{DiSciacca:Proton:2012} using the same method and trap electrodes resulted in the determination of the constraint,
\begin{equation}
 \mu_{\overline{p}}/\mu_p =-1.000000(5),
 \end{equation}
to a precision of $5\times 10^{-6}$. In this experiment, the spin of the $\overline{p}$ confined 
in the static magnetic field $B$ of a Penning
trap (Fig.~\ref{disciacca:fig2}) was flipped by applying an oscillating magnetic RF field close to the Larmor frequency,
\begin{equation}
\nu_L = -\frac{\mu_{\overline{p}}}{\mu_{\rm nucl}}\frac{Q_{\overline{p}} B}{ 2 \pi M_{\overline{p}}}.
\end{equation} 
By measuring the rate of observed spin-flips as a function of the applied drive frequency,
a resonance curve emerged from which the $\overline{p}$ Larmor frequency
was extracted. The value of the field $B$ was then determined from the
cyclotron frequency of the trapped $\overline{p}$ using Eqs.~\ref{cycloeq} and \ref{invariance}. 
The magnetic moment was derived from the ratio of the two measured frequencies,
\begin{equation}
\nu_L/\nu_c = \mu_{\overline{p}}/\mu_{\rm nucl}.
\end{equation}

\begin{figure}[tbh!]
\begin{center}
\includegraphics[height=6cm]{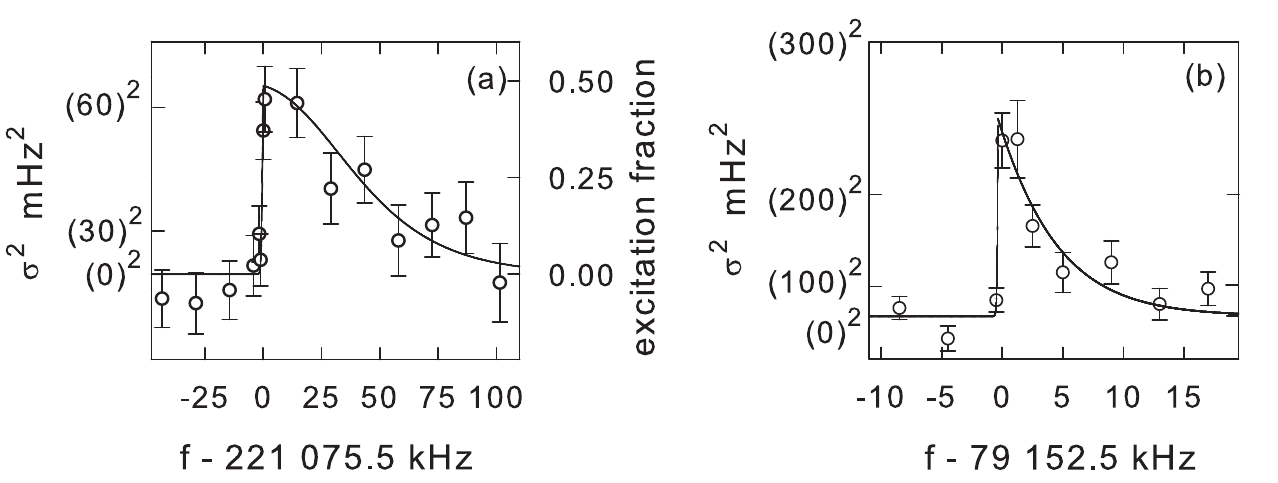}
\begin{minipage}[t]{16.5 cm}
\caption{Time-averaged signals indicating 
spin-flips induced in a single $\overline{p}$ confined in a Penning trap
measured by ATRAP collaborators. The signal
was observed as a sharp excursion in the eigenfrequency
of the axial motion, when a drive RF was applied
to the electrodes to excite quantum transitions in either (a) the 
$\overline{p}$ spin-flip, or (b) cyclotron, degrees of freedom.
Figures from Ref.~\cite{atrap_mag}.
\label{disciacca:fig}
}
\end{minipage}
\end{center}
\end{figure}

The experimental challenge lay in the detection of this spin flip (Fig.~\ref{ulmer:fig}). 
This was carried out by applying the continuous Stern-Gerlach effect,
where an inhomogeneous magnetic field of a so-called ``magnetic bottle'' was superimposed on the Penning trap. 
This inhomogeneous field caused a small shift in the axial oscillation frequency $\nu_z$ of the trapped $\overline{p}$ 
depending on its spin orientation (Fig.~\ref{disciacca:fig}).  The continuous Stern-Gerlach effect was also 
used in the famous experiments which measured the $g-2$ of the $e^-$ and $e^+$ \cite{Dehmelt:PNAS:1986a}. 
The latest incarnation of this experiment \cite{odom,hanneke} determined the $g-2$ of $e^-$ as,
\begin{equation}
g/2=1.00115965218073(28),
\end{equation}
by detecting the individual quantum jumps associated with a single $e^-$ flipping its spin in a Penning trap.
The sensitivity of this effect for $\overline{p}$, however, scales proportionally 
to $\mu_{\overline{p}}/M_{\overline{p}}$, the size of which is $10^{-6}$ of the $e^-$ and $e^+$ cases.
At the experimental conditions of the $p$ work of Ref.~\cite{Ulmer:ProtonSpinFlips:2011} for example,
a $p$ spin-flip shifted the eigenfrequency 674\,kHz of the axial motion by 190 mHz.  In addition to this
coupling to the spin magnetic moment, the magnetic bottle also
couples to the magnetic moment arising from the orbital angular moment of the
particle motion to the axial frequency.  This second coupling causes background
baseline shifts and fluctuations in the axial frequency which are difficult to control, presenting
a considerable experimental challenge. Single spin-flip events have not been
detected so far due to insufficient signal-to-noise ratios. 
Nevertheless, the time-averaged spin-flips of a single $\overline{p}$ \cite{atrap_mag} as well as a 
$p$ \cite{DiSciacca:Proton:2012,Ulmer:ProtonSpinFlips:2011,Rodegheri:NJP:2012} 
have been detected using a statistical method, where
the signals from several spin-flips were added together to enhance the signal-to-noise ratio.

\section{Atomic and nuclear collisions, and applications}
\label{collisions}
\subsection{\it Stopping powers}
\label{stoppingpowers}
The stopping powers $-dE/dx$ of $\overline{p}$ with kinetic energies 1--100 keV in various conductor (C, Al, Ni, Au) and
insulator (LiF) targets were systematically measured using the decelerated beam emerging from the RFQD \cite{Moller2002,Moller2004,Moller2008}. The $\overline{p}$ at such low velocities $v_{\overline{p}}$
lose their energy predominantly by inducing some electronic excitations in the target material, whereas the contribution 
from collisions with the atomic nuclei is negligible. Fermi and Teller \cite{fermi1947} predicted that $-dE/dx$
would then be roughly proportional to $v_{\overline{p}}$ using the following simple argument:
the conduction $e^-$ in metals can be treated as a free degenerate gas with a thermal distribution of velocities 
having a maximum value $v_{e}$ which is larger than $v_{\overline{p}}$. 
Only the fast classes of $e^-$ having velocities close to $v_e-v_{\overline{p}}$ can then collide with the $\overline{p}$, 
since the Pauli principle prevents slower $e^-$ from 
scattering into the final states that are already occupied in the conductor. The number of such high-speed $e^-$ per 
unit volume in the material is of order $n\sim m_e^3v_e^2v_{\overline{p}}/\hbar^3$, whereas the collision cross section is
around $\sigma\sim (e^2/m_ev_e^2)^2$ and the energy transfer per collision $\Delta E\sim m_e v_e v_{\overline{p}}$.
The stopping power can then be estimated as,
 $-dE/dx\sim \Delta E\sigma nv_{\overline{p}}/v_{\overline{p}}\sim m_e^2 e^4 v_{\overline{p}}/\hbar^3$,
which implies its proportionality against the $\overline{p}$ velocity.

\begin{figure}[htbp]
\epsfysize=8.0cm
\begin{center}
\begin{minipage}[htbp]{15 cm}
\epsfig{file=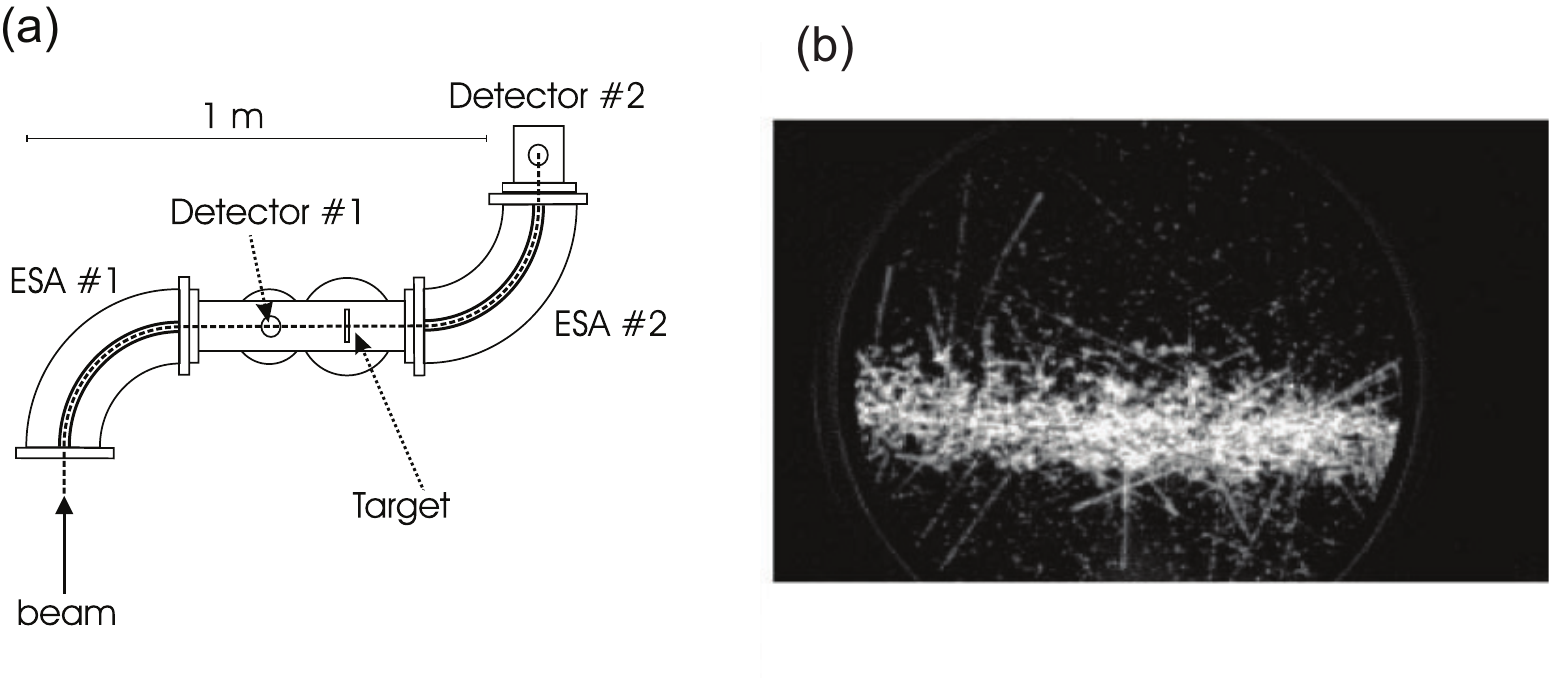,scale=0.95}
\end{minipage}
\begin{minipage}[t]{16.5 cm}
\caption{\label{fig:dedx_apparatus} 
(a) Schematic diagram of the stopping-power experiment of $\overline{p}$ in various foil targets. 
The positions of the micro-channel plates are indicated as Detectors \#1 and \#2. (b) Image of $
\overline{p}$ striking a micro-channel plate, the multiplied $e^-$ of which are imaged by a phosphor 
screen. The tracks visible in the image are presumably caused by ions emerging from the 
$\overline{p}$ annihilations. Figures from Refs.~\cite{Moller2002,andersen2002}.}
\end{minipage}
\end{center}
\end{figure}

Stopping powers linear to the projectile velocity have indeed been observed
in most experiments involving positive ion beams. Deviations from this proportionality
have only been observed in He \cite{golser1991} and Ne \cite{schief1993} targets. 
The theoretical interpretation of these results 
was greatly complicated by charge-exchange effects in which the projectile ions 
can capture $e^-$ from the target atom. No such capture is possible
in the $\overline{p}$ case, and so the velocity-proportional $-dE/dx$ can be directly studied. 

The measurements were carried out by allowing $\overline{p}$ beams of energy
18--63 keV to enter a pair of 90-degree electrostatic 
spherical analyzers (Fig.~\ref{fig:dedx_apparatus} (a)).
The first analyzer \cite{andersen2002} selected $\overline{p}$ with energy 
$E_1$. The beam then traversed a target
foil of thickness $\Delta x=20$--40 $\mu{\rm m}$, and a second analyzer 
which measured the energy of the emerging beam centered around
$E_2$. The stopping power at an average energy $(E_1+E_2)/2$
was then determined as $-dE/dx=(E_1-E_2)/\Delta x$. By 
suitably biasing the foil with a DC potential, the $-dE/dx$ values at various 
$\overline{p}$ energies could be rapidly measured.
The $\overline{p}$ that traversed the two analyzers struck a micro-channel 
plate (MCP) read out by a charged-coupled device (CCD) detector located at the end of the experiment.
The position resolution of $\sim 1$ mm on the MCP yielded an energy
resolution of $\pm 0.2\%$ as shown in Fig.~\ref{fig:dedx_apparatus} (b).
The tracks seen here are produced by heavy ions recoiling from $\overline{p}$
annihilations on the surface of the MCP \cite{andersen2002}.

\begin{figure}[htbp]
\epsfysize=8.0cm
\begin{center}
\begin{minipage}[htbp]{16 cm}
\epsfig{file=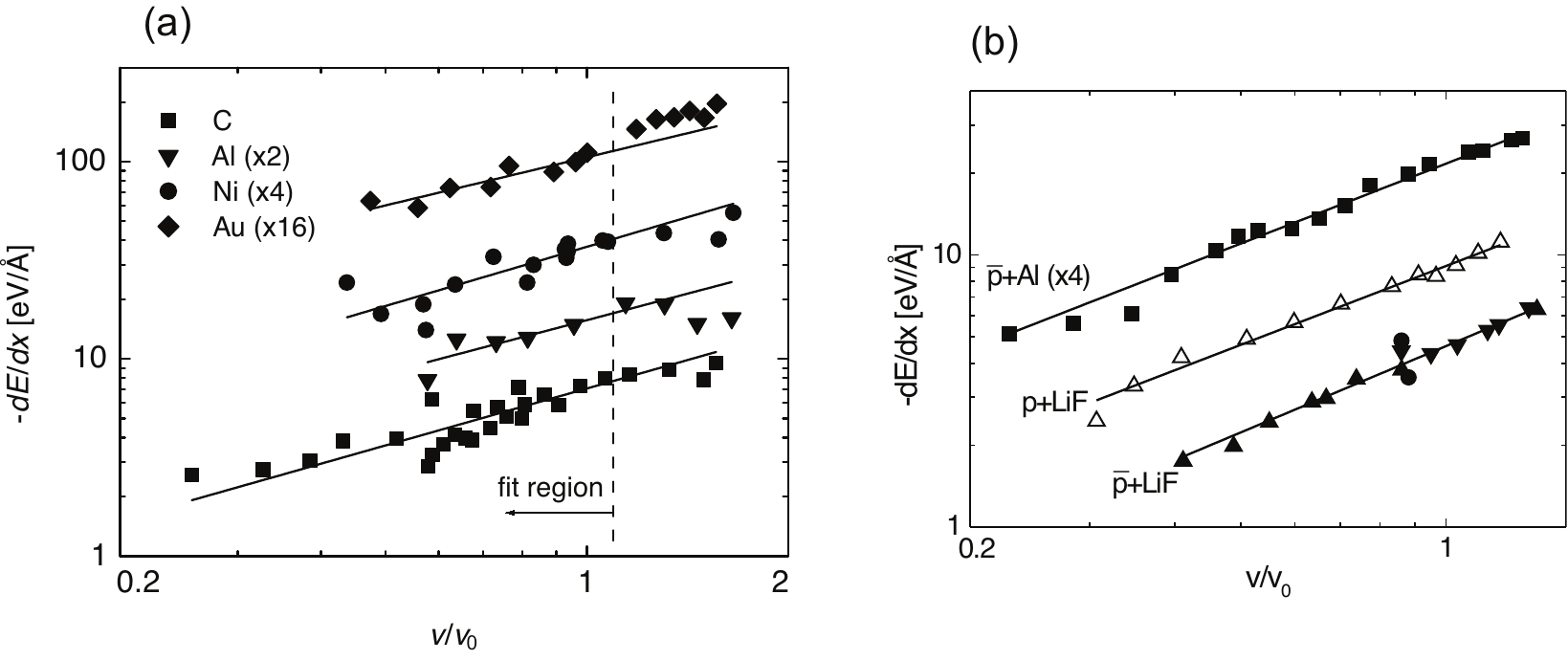,scale=0.95}
\end{minipage}
\begin{minipage}[t]{16.5 cm}
\caption{\label{fig:stopping} 
(a) Stopping powers of $\overline{p}$ traversing C, Al, Ni, and Au targets
as a function of $\overline{p}$ velocity normalized to the
Bohr velocity, $v/v_0$.
(b) Stopping powers of $\overline{p}$ and $p$ on Al and LiF targets
as a function of velocity. Lines indicate
velocity-proportional linear fits to the data. Note logarithmic axes. 
Figures from Refs.~\cite{Moller2002,Moller2004}.}
\end{minipage}
\end{center}
\end{figure}

Fig~\ref{fig:stopping} (a) shows the measured $-dE/dx$ values for C, Al, Ni, and Au targets
as a function of $\overline{p}$ velocity $v_{\overline{p}}$ normalized to the Bohr velocity $v_0$. 
As expected, a linear dependence was observed below $v_{\overline{p}}/v_0\sim 1$. 
The experimental results agreed for some targets and velocity ranges
with theoretical calculations based on the free electron gas
\cite{arista2002,sorensen1990}, quantum mechanical harmonic oscillator
\cite{mikkelsen1989}, and classical binary scattering models \cite{sigmund2001}. 
The $\overline{p}$ stopping powers were $50-60\%$ of the $p$ ones due to the
Barkas effect, in which the polarization of the target $e^-$ induced by the passage 
of $\overline{p}$ causes a reduction in $-dE/dx$ \cite{barkas1956}.

In the case of wide-band gap insulators, it was theoretically predicted that the $-dE/dx$
values would strongly deviate from a linear dependence and possibly exhibit a threshold effect
\cite{schief1993}. This is because when the projectiles ions are very slow, the $e^-$ in
the target cannot easily excite over the band gap. Surprisingly however, the measured 
stopping powers for both $p$ \cite{eder1997} and $\overline{p}$ were found \cite{Moller2004} 
to have a linear dependence between energies of 2 and 50 keV (Fig.~\ref{fig:stopping} (b)). The 
reason for this is not understood.

\subsection{\it Atomic and molecular ionization cross sections}

\begin{figure}[htbp]
\epsfysize=9.0cm
\begin{center}
\begin{minipage}[htbp]{12 cm}
\epsfig{file=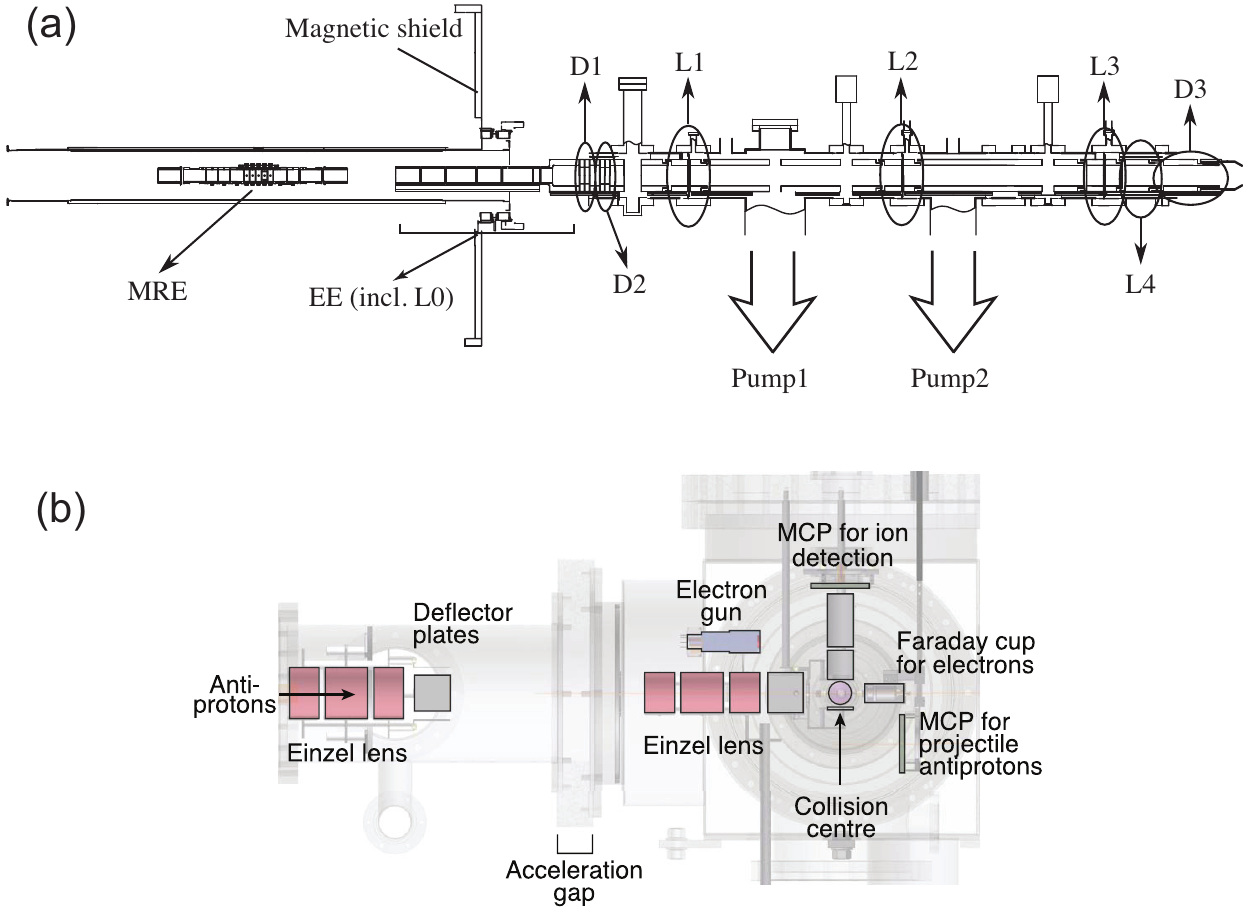,scale=1.0}
\end{minipage}
\begin{minipage}[t]{16.5 cm}
\caption{\label{fig:ionization} 
(a) Experimental layout where a $\overline{p}$ beam with keV-scale kinetic energy was extracted from a Penning trap (labeled MRE)
and transported by an electrostatic beamline. The beamline included an extractor electrode (EE), several electrostatic lenses (L0--L4), and steering deflectors (D1--D3). (b) Experimental setup for measuring the total cross section of $\overline{p}$
ionizing various gas targets. A beam of $\overline{p}$ was
focused by Einzel lenses onto a gas jet target (collision center),
and the emerging $e^-$, ions, and $\overline{p}$ were detected.
Figures from Refs.~\cite{knudsen2008,kuroda2005}.}
\end{minipage}
\end{center}
\end{figure}

The total cross sections for $\overline{p}$ of kinetic energy $E=2$--25 keV
ionizing He and Ar targets were also measured \cite{knudsen2008,knudsen2009}. 
The ionization processes of atoms by slow charged particles are not fully understood 
\cite{knudsen2011}, because the correlations between the many $e^-$ in the system make it difficult 
to theoretically treat this dynamic problem. As in the $-dE/dx$ measurements described in
Sect.~\ref{stoppingpowers}, the $\overline{p}$ is an ideal projectile to study this since there is no 
complication from charge transfer. 

This experiment involved first trapping some $(6-7)\times 10^{5}$ $\overline{p}$ in a Penning 
trap (Fig.~\ref{fig:ionization} (a)), and cooling them to sub-eV energies by mixing the
$\overline{p}$ with a cloud of $e^-$ also confined in the same trap \cite{kuroda2005}. The 
diameter of the $\overline{p}$ cloud was compressed to a few millimeters by applying a rotating electric field on it 
\cite{higaki2002,Kuroda:PRL:2008,Kuroda:PRST:2012}. The $\overline{p}$ were then extracted
from the trap as a continuous beam, and transported through an electrostatic beam transport
line \cite{yoshiki2003} to the experimental target. The beam passed through three apertures 
(indicated by L1--L3 in Fig.~\ref{fig:ionization} (a))
which separated the ultrahigh vacuum in the trap from contamination gases originating from 
the target. The beam was accelerated to energy 2--25 keV and steered through a gas jet target
(Fig.~\ref{fig:ionization} (b)) consisting of a mixture of $90\%$ helium and $10\%$ argon.
The $\overline{p}$ emerging from the target were detected by a MCP detector,
whereas the ions were extracted by a 333 V$\cdot$cm$^{-1}$
electric field in a perpendicular direction to the $\overline{p}$ beam, and 
focused onto a second MCP. 

The He and Ar ionization events were isolated using time-of-flight methods,
by recording the timing difference between the arrivals of the $\overline{p}$ and ion. 
The cross sections $\sigma$ were obtained using the relation, 
$N_{\rm ion}=N_{\overline{p}}\sigma n_tl_t\varepsilon$,
where $N_{\rm ion}$ and $N_{\overline{p}}$ denote the
number of ions and $\overline{p}$ events, $n_tl_t$ the integral of the gas density 
along the projectile path, and $\varepsilon$ the efficiency of detecting the ions. The
value $n_tl_t\varepsilon$ was calibrated in a separate experiment
involving the ionization of the gas by a 3-keV $e^-$ beam.

\begin{figure}[htbp]
\epsfysize=9.0cm
\begin{center}
\begin{minipage}[htbp]{15 cm}
\epsfig{file=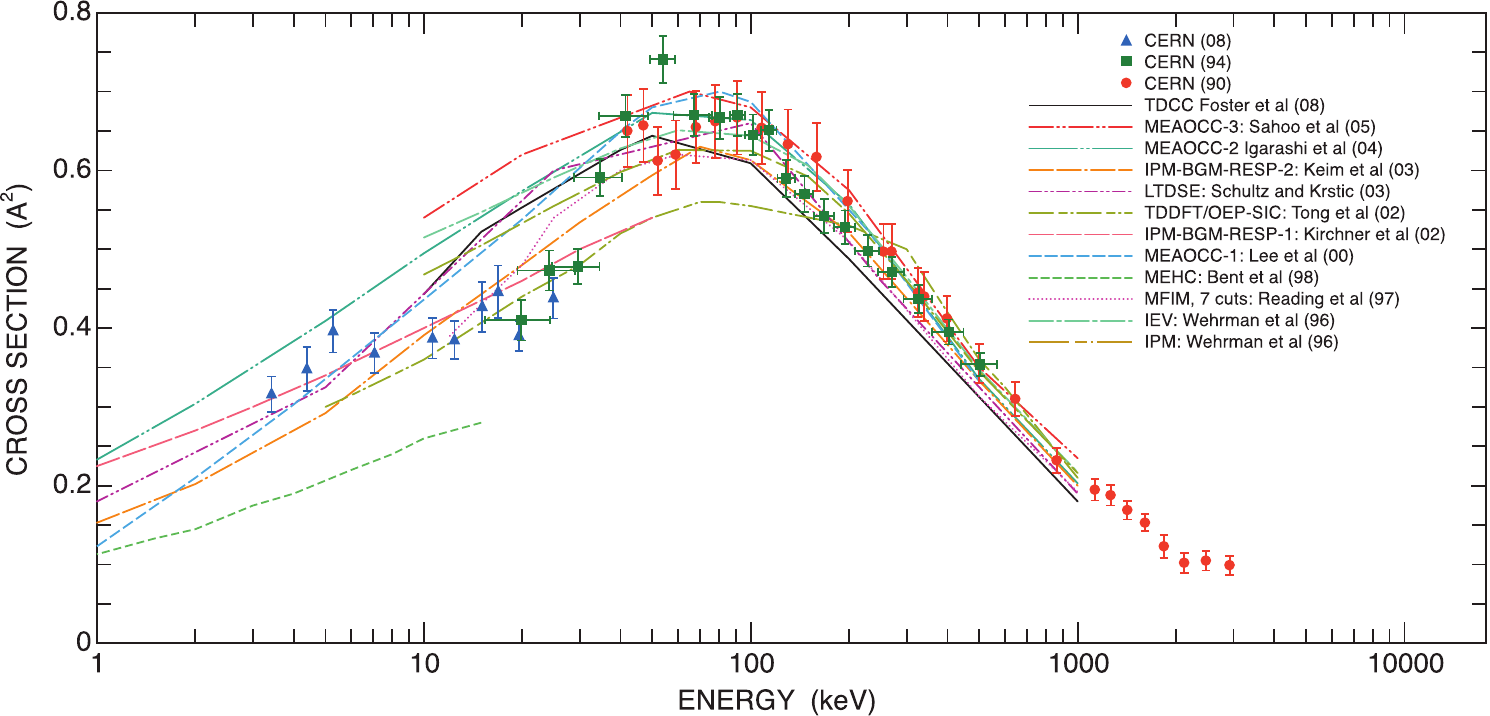,scale=1.0}
\end{minipage}
\begin{minipage}[t]{16.5 cm}
\caption{\label{fig:knudsen_he} 
Total cross section of single ionization of He atoms at $\overline{p}$ energies
$E=3$--3000 keV measured in Refs.~\cite{knudsen2008,andersen1990,hvelplund1994}. Results of several theoretical calculations \cite{wehrman1996,reading1997,bent1998,lee2000,igarashi2000,igarashi2004,sahoo2005,schultz2003,tong2002,kirchner2002,keim2003,foster2008} are shown superimposed. Figure from Ref.~\cite{knudsen2008}.}
\end{minipage}
\end{center}
\end{figure}

The single-ionization cross sections for He measured
at $\overline{p}$ energies $E=3$--30 keV are plotted in
Fig.~\ref{fig:knudsen_he} (indicted by filled triangles \cite{knudsen2008}), together with past experimental results
measured at the higher beam energies of LEAR (squares \cite{andersen1990} and circles \cite{hvelplund1994}).
They are compared with the results of numerous theoretical calculations 
\cite{wehrman1996,reading1997,bent1998,lee2000,igarashi2000,igarashi2004,sahoo2005,schultz2003,tong2002,kirchner2002,keim2003,foster2008}. The results of a time-dependent density functional theory with an optimized effective potential
and self-interaction correction \cite{keim2003} showed the best overall agreement with the experimental data, including
the region around the maximum of the cross section.

\begin{figure}[htbp]
\epsfysize=9.0cm
\begin{center}
\begin{minipage}[htbp]{14 cm}
\epsfig{file=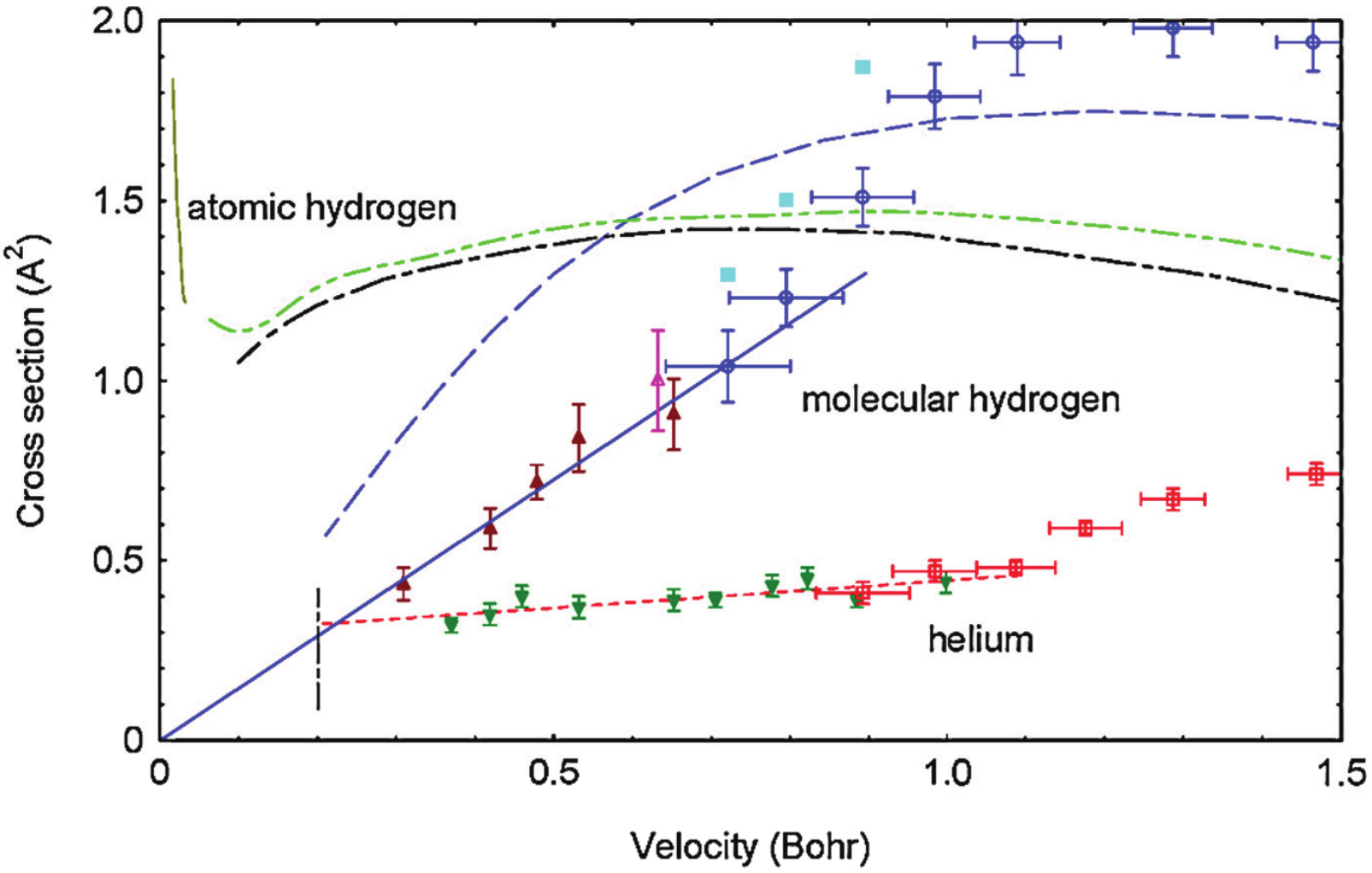,scale=0.8}
\end{minipage}
\begin{minipage}[t]{16.5 cm}
\caption{\label{fig:knudsen_h2} 
Cross sections for single ionization by $\overline{p}$ impact on molecular H$_2$ and He as a function of the 
laboratory velocity of the incoming $\overline{p}$. For atomic H, calculations of Refs. \cite{igarashi2004} 
(dashed-dotted black curves),
\cite{luhr2008} (green dash-dot-dot), and \cite{cohen1997} (brown solid curve at low velocity) are shown. 
For H$_2$, experimental data of Refs.~\cite{knudsen2010} (triangles) and \cite{hvelplund1994} (open circles) are plotted together with a solid line to indicate a linear fit below 1 a.u., the sum of the cross sections for nondissociative and dissociative ionization (filled squares), and theoretical calculations of Ref.~\cite{luhr2010} (long-dashed dark blue curve). 
The vertical line indicates the projectile energy above which more than 90$\%$ of the $D_2^+$ ions and projectiles
emerging from the collisions are collected by the experimental apparatus. For He, the experimental data of Refs.~\cite{knudsen2008} (inverted filled green triangle) and \cite{hvelplund1994} (open squares) are shown. Figure from Ref.~\cite{knudsen2010}.}
\end{minipage}
\end{center}
\end{figure}

Ionization measurements were also carried out for molecular D$_2$ targets, by detecting the
D$_2^+$ ions emerging from single, nondissociative ionizing collisions with $\overline{p}$ \cite{knudsen2010}.
In Fig.~\ref{fig:knudsen_h2}, the measured cross sections $\sigma$ for D$_2$ \cite{knudsen2010} and He \cite{knudsen2008} 
targets are compared for atomic Bohr velocities $v_{\overline{p}}$ of the $\overline{p}$ between 0.3 and 1.5 a.u. 
The $\sigma$-value for He is roughly constant between 0.4--1 a.u., with a slightly
decreasing tendency towards lower velocities. The cross sections for atomic H for this velocity region
has never been measured, but a similar behavior is predicted by two theoretical calculations, the results
of which are indicated by green \cite{luhr2008}, brown \cite{cohen1997}, and black \cite{igarashi2004} 
dashed-dotted lines. Surprisingly, the values for 
molecular D$_2$ were roughly linear to $v_{\overline{p}}$ in the region 0.3--1.0 a.u., in contrast to 
the behavior of atomic targets. The experimental data also disagrees with the results of two-center
atomic orbital close coupling calculations \cite{luhr2010}. Intensive theoretical studies are now underway
to understand these issues.

\subsection{\it Nuclear annihilation cross sections}

Numerous experimental groups have used the LEAR facility to measure
the cross sections $\sigma_{\rm anni}$ of $\overline{p}$ with kinetic energy $E>1$ MeV 
colliding with various target nuclei, and undergoing annihilation. In the semiclassical regime at relatively
high energy, where the de-Broglie wavelength of the $\overline{p}$ is small compared to the radius $R$ of the
target nucleus of mass number $A$, $\sigma_{\rm anni}$ is na\"{i}vely assumed to be independent 
of $E$ and roughly equal to the geometric cross section, $\pi R^2\propto A^{2/3}$. In such a model
the target nucleus resembles a simple black disk. 
Measurements with antineutron ($\overline{n}$) beams \cite{iazzi2000} of energy $\sim 2$ MeV annihilating on 
various targets have indeed shown a $\sim A^{2/3}$ behavior.

\begin{figure}[htbp]
\epsfysize=9.0cm
\begin{center}
\begin{minipage}[htbp]{16 cm}
\epsfig{file=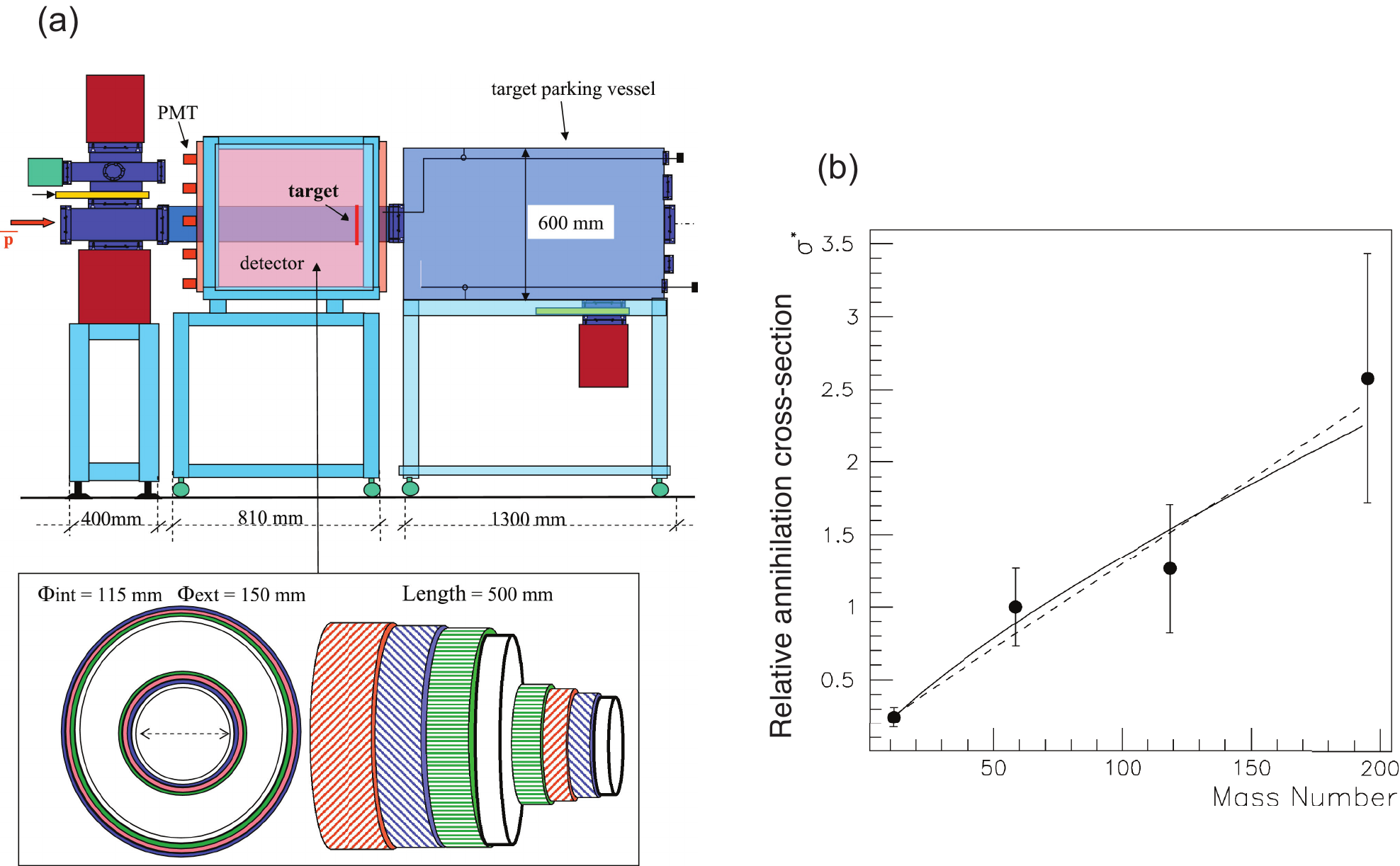,scale=0.8}
\end{minipage}
\begin{minipage}[t]{16.5 cm}
\caption{\label{fig:nuclearsetup} 
(a) Experimental setup for measuring cross sections of $\overline{p}$ annihilations at 5.3 MeV. The $\overline{p}$ annihilations
on the target foil were measured by layers of scintillation fibers surrounding the target. 
(b) Relative values of annihilation cross sections measured using Mylar, Ni, Sn, and Pt targets. The results of
best fits with
functions $CA^{\alpha}$ (solid line) and $K\left[1+\frac{Ze^2(m+M)}{4\pi\varepsilon_0ERM}\right]$ (dashed line)
are shown superimposed. The nuclear radius $R$ is parameterized as 1.840+1.120$A^{1/3}$ fm.
Figures from Ref.~\cite{infn2011}.}
\end{minipage}
\end{center}
\end{figure}

The annihilation cross sections of negatively-charged $\overline{p}$ at kinetic energies 
$E<5$ MeV are theoretically assumed to be enhanced by the Coulomb force, which attracts
the trajectory of the $\overline{p}$ towards the nucleus of charge
$Z$ and mass $M$. The sum of this Coulomb focusing and the black disk \cite{batty2001} then yields,
\begin{equation}
\sigma_{\rm anni}\sim\pi R^2\left(1+\frac{Ze^2\left(m_{\overline{p}}+M\right)}{4\pi\varepsilon_0 ERM}\right) ,
\label{crosssect}
\end{equation}
where $\varepsilon_0$ denotes the dielectric constant of vacuum. Eq.~\ref{crosssect} implies
that at very low energies the dependence on the mass number scales as 
$\sigma_{\rm anni}\propto ZR\propto ZA^{1/3}$. This has never been experimentally verified
due to the fact that such low energy, high intensity beams needed to measure $\sigma_{\rm anni}$ were 
not available at LEAR. In fact, past LEAR experiments in the region $E<100$ MeV 
\cite{bruckner1990,bertin1996,benedettini1997,zenoni1999,kalogeropoulous1980,bizzarri1974,balestra1989,balestra1984,bianconi2000} have
been generally limited to gas (e.g., H$_2$, D$_2$, $^3$He, $^4$He, and Ne) targets, whereas solid targets have not been studied.

\begin{figure}[htbp]
\epsfysize=9.0cm
\begin{center}
\begin{minipage}[t]{12 cm}
\epsfig{file=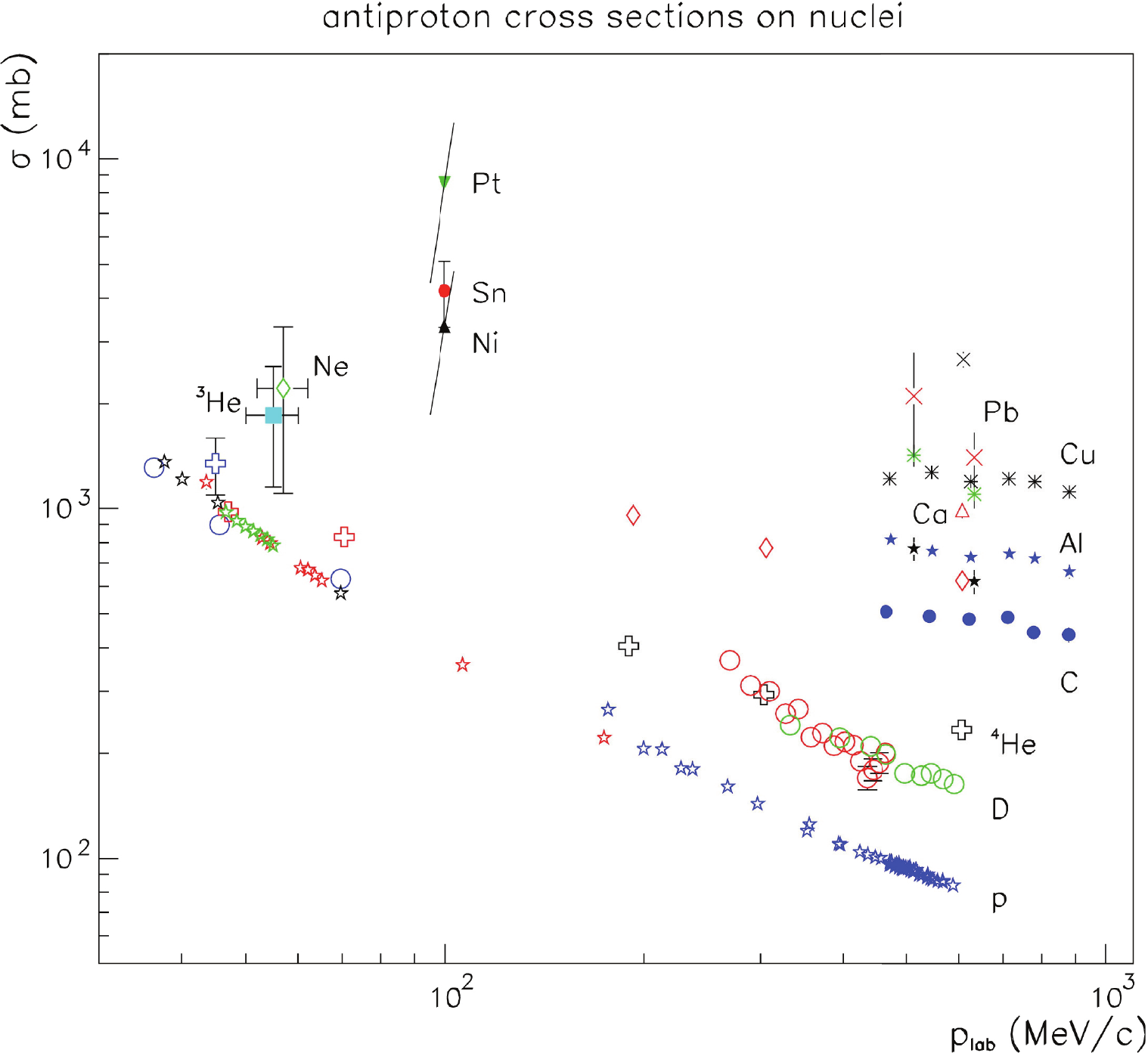,scale=0.8}
\end{minipage}
\begin{minipage}[t]{16.5 cm}
\caption{\label{fig:nucl_cross} 
Nuclear reaction cross sections measured by various experiments for $\overline{p}$ beams striking H (blue \cite{bruckner1990}, red \cite{bertin1996}, green \cite{benedettini1997}, and black \cite{zenoni1999} open stars), D (blue \cite{zenoni1999}, red \cite{kalogeropoulous1980}, green \cite{bizzarri1974} open circles), $^4$He (blue \cite{balestra1989}, red \cite{zenoni1999}, black \cite{balestra1984} open crosses), $^3$He 
(filled squares \cite{bianconi2000}), C (filled circles \cite{nakamura1984}), Ne (green \cite{bianconi2000a}, red \cite{balestra1986} open diamonds), Al (blue \cite{nakamura1984}, black \cite{ashford1985} filled stars), Ca (open triangles \cite{garreta1984}), Cu (black \cite{nakamura1984}, green \cite{ashford1985} eight-pointed asterisk), Pb (red \cite{ashford1985} and black \cite{garreta1984} crosses), and Ni, Sn, and Pt \cite{infn2011} (indicated) targets.
Figure from Ref.~\cite{infn2011}.}
\end{minipage}
\end{center}
\end{figure}

The $\sigma_{\rm anni}$ values at $E\sim 5.3$ MeV were recently measured \cite{infn2011} using
Mylar, Ni, Sn, and Pt targets. The pulsed $\overline{p}$ beam arriving from the AD normally resulted 
in such a high instantaneous rate of annihilations on the experimental target, that it would be difficult to 
resolve and count the individual events needed to determine $\sigma_{\rm anni}$. This problem was solved by using annihilation 
detectors of high spatial granularity \cite{corradini2013} to isolate the individual annihilations, and by changing the 
time structure of the $\overline{p}$ beam to reduce the instantaneous rate.
The $3\times 10^7$ $\overline{p}$ circulating in the AD were divided into six pulses which 
were distributed equidistantly around its 190-m-circumference. Each pulse of length 
40--50 ns was then sequentially extracted to the experiment at intervals of $\sim 2.4$ s.
The experimental apparatus (Fig.~\ref{fig:nuclearsetup} (a)) consisted of a 1-m-long, 150-mm-diameter
vacuum chamber that contained 0.9-$\mu{\rm m}$-thick Mylar target foils. 
Ni, Sn, and Pt layers were sputtered on the Mylar surface, their thicknesses calibrated to a 
precision of 5--40 nm by separate Rutherford backscattering measurements using an $\alpha$ source.
The target was surrounded by three layers of scintillating fibers that reconstructed the tracks 
of the $\pi^+$ and $\pi^-$ emerging from the annihilations with a spatial resolution of 3--5 mm.

For each target, the number of reconstructed vertices $N_{\rm ev}$ representing antiproton
annihilations in the foil were counted. The annihilation cross section was determined from
this data using the formula, $\sigma_{\rm anni}\sim N_{\rm ev}M_A/N_{\overline{p}}N_A\rho l_t\varepsilon$,
where $M_A$, $\rho$, and $l_t$ denote the atomic weight, density, and thickness of the target, 
$N_A$ the Avogadro's number,  and $\varepsilon$ the efficiency of detecting an annihilation vertex. 
It was difficult to precisely determine the number $N_{\overline{p}}$ of $\overline{p}$  contained in the
pulsed beam, but approximate values were estimated by counting the small fraction of $\overline{p}$ 
undergoing Rutherford backscattering in the target, and annihilating in the lateral walls of the chamber.
This introduced a relatively large uncertainty on the absolute normalization of $\sigma_{\rm anni}$.
Experimental values of $\sigma_{\rm anni}=(3.3\pm 1.5)$, $(4.2\pm 0.9)$, and $(8.6\pm 4.1)$ b
were obtained for the Ni, Sn, and Pt targets. 
This was consistent with the theoretical value 4.9 b
for Sn, obtained from a black-disk model with Coulomb corrections \cite{batty2001}. In Fig.~\ref{fig:nuclearsetup} (b), the 
relative cross sections are shown as a function of $A$. The results were consistent with the expected
${\rm A}^{2/3}$ dependence within the relatively large error bars. 
In Fig.~\ref{fig:nucl_cross}, the results are compared with past cross sections measured for $\overline{p}$ traversing various gas and solid foil targets \cite{bruckner1990,bertin1996,benedettini1997,zenoni1999,kalogeropoulous1980,bizzarri1974,balestra1989,balestra1984,bianconi2000,nakamura1984,bianconi2000a,balestra1986,ashford1985,garreta1984}.

Recently, $\overline{p}$ beams of energy 130 keV decelerated by the RFQD 
were directed \cite{todoroki2012} through 70-nm-thick foils of C, and other targets
with 5--19 nm of Pd and Pt evaporated on the C foil surface \cite{aghai2012}. Whereas the majority
($>99.99\%$) of the $\overline{p}$ passed through the thin foils, a small amount
annihilated by undergoing in-flight reactions with the target nuclei, or underwent Rutherford scattering. 
These foil annihilations were measured by plastic scintillation counters surrounding the target. This may 
lead to quantitative measurements of  $\sigma_{\rm anni}$ at energies $\sim 130$ keV.

\subsection{\it Cancer therapy using antiproton beams}

\begin{figure}[htbp]
\epsfysize=9.0cm
\begin{center}
\begin{minipage}[t]{14 cm}
\epsfig{file=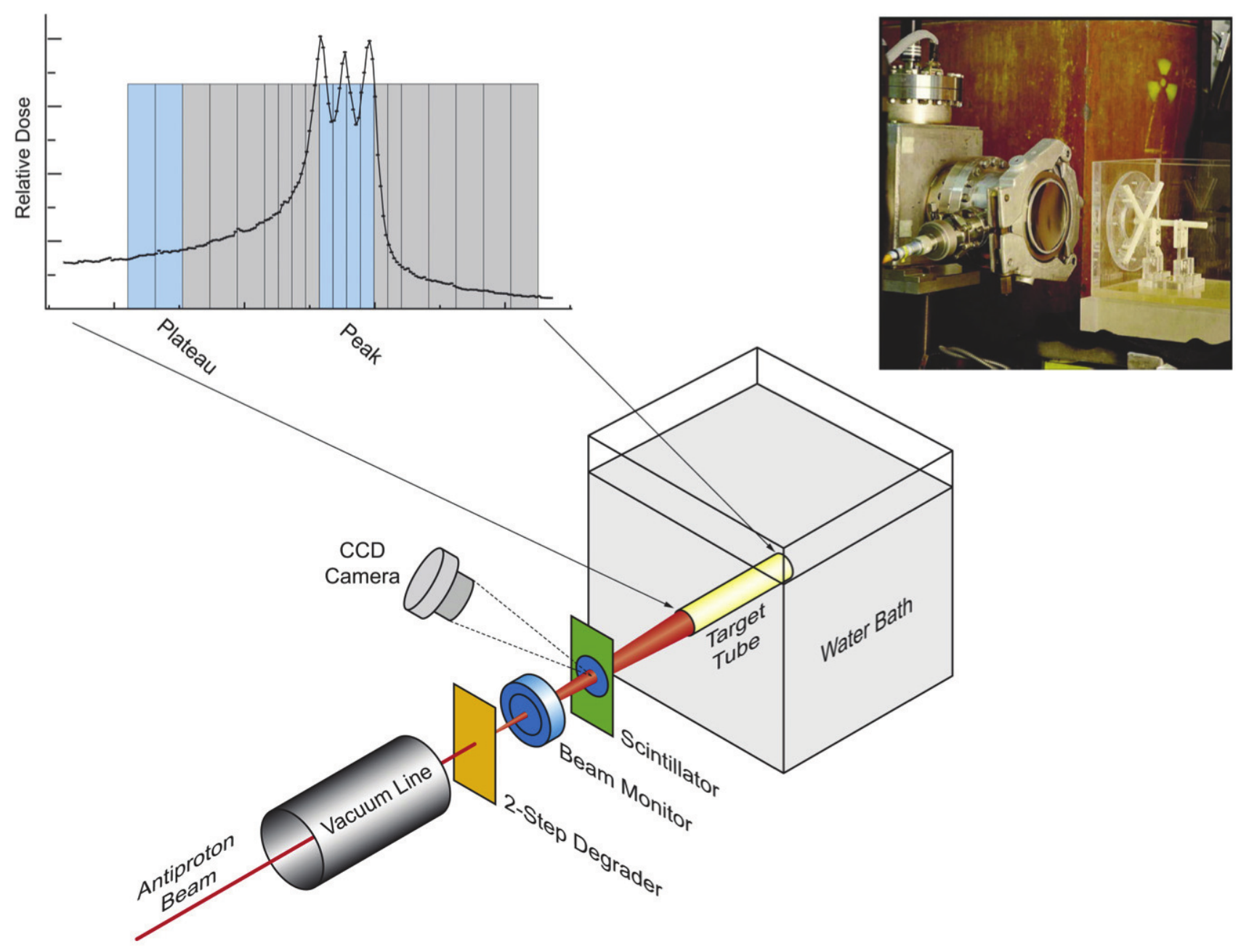,scale=0.9}
\end{minipage}
\begin{minipage}[t]{16.5 cm}
\caption{\label{fig:acesetup} 
 Layout of the ACE experiment. A $\overline{p}$ beam of energy 50 MeV left the vacuum tube of the beamline
 through a Ti window, traversed a 2-step degrader, a beam current monitor and a scintillator before
 entering a plexiglas tank containing a glycol-water mixture and the biological sample (photo insert upper right).
 The $\overline{p}$ dose profile and the slicing protocol for extracting the cell clonological survival data is shown 
 at upper left. Figure from Ref.~\cite{ace2006}.}
\end{minipage}
\end{center}
\end{figure}

\begin{figure}[htbp]
\epsfysize=9.0cm
\begin{center}
\begin{minipage}[t]{12 cm}
\epsfig{file=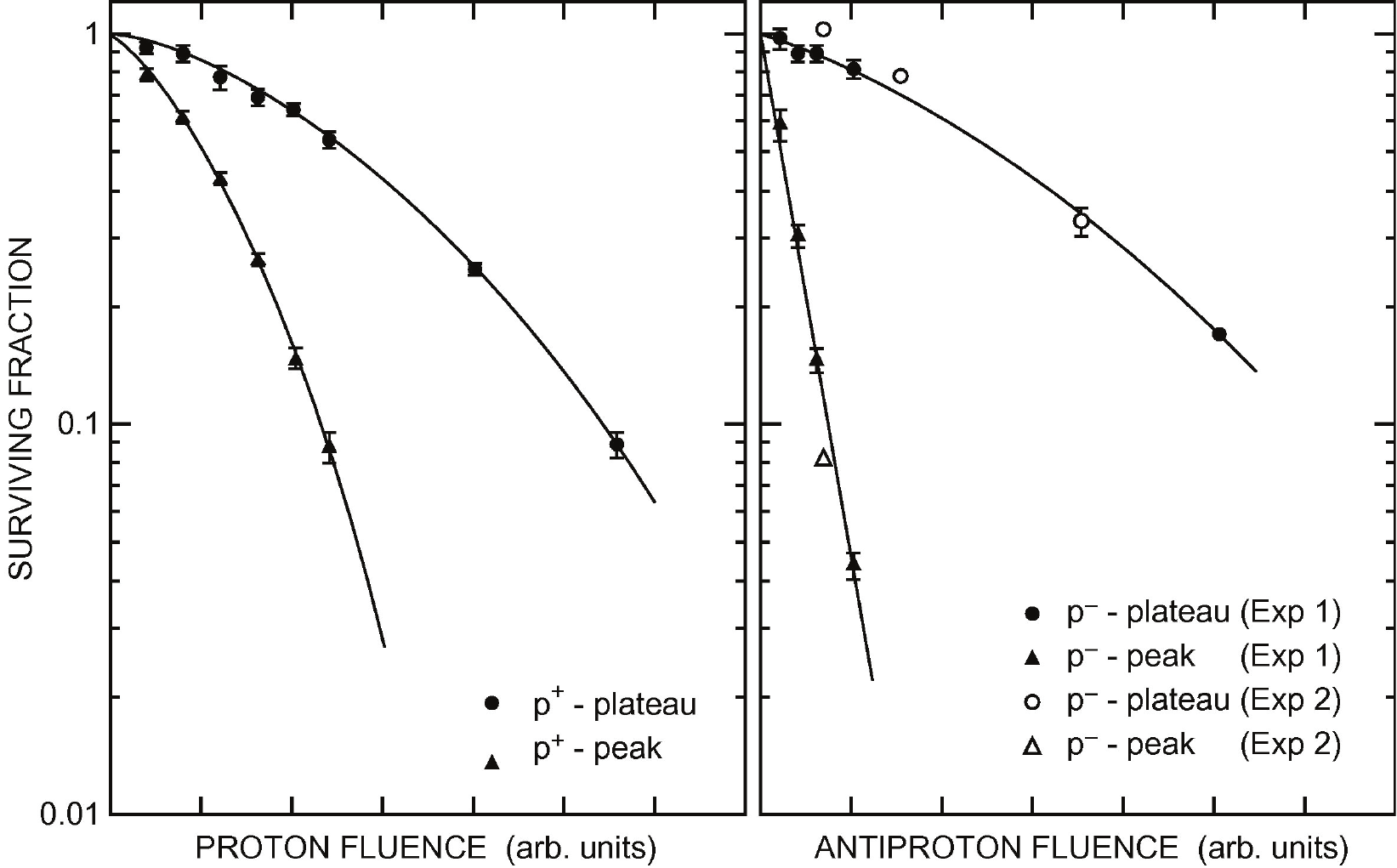,scale=0.8}
\end{minipage}
\begin{minipage}[t]{16.5 cm}
\caption{\label{fig:ace} 
 Left: Measured clonogenical survival fractions of cancer cells irradiated by 50 MeV protons, as a function of the $p$ fluence. Results are shown for cells situated in the spread-out Bragg peak and in the plateau, closer to the target surface, respectively. Right: Corresponding data for 50-MeV $\overline{p}$ irradiation. The curves are drawn to guide the eye. The two $\overline{p}$ data sets were obtained in two different runs. Figures from Ref.~\cite{ace2006}.}
\end{minipage}
\end{center}
\end{figure}

The Antiproton Cancer Experiment (ACE) measured the biological effectiveness of $\overline{p}$ beams
destroying cancer cells \cite{ace2006,knudsen2008a}. 
In conventional radiation therapy involving accelerators, patients are irradiated 
by $e^-$, $p$, or heavy ion beams \cite{suit2003,mazeron2004}
that reach tumors located at typical depths of $\sim 100$--200 mm.
During the passage of these projectiles through cancer cells, free radicals and $e^-$ are created which can interact with 
the biomolecules and destroy them. While single-strand breaks of DNA can be efficiently repaired by the cell, 
double-strand breaks are more permanent and lead to the deactivation of the cell. Ideally, the beam should 
deactivate the cancer cells that lie within a small volume relative to the total size of the tumor, while 
sparing the outside healthy cells. Thus the shape of the energy loss or Bragg curve along the trajectory 
of the penetrating particle is important. Gray and Kalogeropoulos \cite{gray1984}
suggested that $\overline{p}$ may be superior to other particles in this respect, 
since in addition to the energy loss $-dE/dx$ of the $\overline{p}$, its subsequent annihilation in a cancer cell 
deposits some additional 20--30 MeV of energy close to the annihilation point \cite{sullivan1985}. Ions with MeV-energies are 
created either by nuclear recoil or fission following the annihilation, and since the range of these are short 
($<10$$\mu{\rm m}$), biological damage is expected to stay within the proximity of the annihilation.

The ACE collaboration irradiated samples of V79-WNRE Chinese hamster cancer cells with pulsed $\overline{p}$ beams of
energy $E=50$ MeV provided by the AD (Fig.~\ref{fig:acesetup}). The cells were suspended in a gel placed in a plastic tube, and the 
tube in turn was positioned in a bath of glycol-water mixture which mimicked a patient's body. After $\overline{p}$ 
irradiation, the gel was cut in 0.5-mm-thick slices, and the clonogenic survival of the cells in each slice was measured.
Comparative measurements were made with 50-MeV protons and $^{60}$Co gamma rays. 

In Fig.~\ref{fig:ace}, the clonogenic survival fractions of cancer cells measured near the
plateau of the Bragg peak, and the plateau near the target surface, are compared as a function
of the fluence of $p$ and $\overline{p}$ beams in arbitrary units. The 
biological effective dose ratio (BEDR) is defined as the ratio of beam fluences which create a 
certain clonogenic cell survival (for example 20$\%$) in the peak versus the plateau of this 
energy deposition curve. This corresponds to cell survival in respectively the cancer 
tumor versus the healthy tissue along the incoming path. ACE found that the BEDR for $\overline{p}$ 
is $3.75 \pm 0.50$ times larger than for protons.
The authors claim this stems from two contributions: the increase in dose deposition in the Bragg peak 
due to the $\overline{p}$ annihilation, and the fact that the dose is deposited mainly by heavy fragment ions. 
They also pointed out that the BEDR improvement is substantial; Levegr\"un et al. \cite{levegrun2002} 
showed that an increase of dose administered to a prostate cancer from 60 to 90 Gy increases the tumor 
control probability from 15 to $95\%$. 

The authors also found that the cell deactivation outside the Bragg peak is small. For example, for 
a $\overline{p}$ induced dose which resulted in a cell survival of around 15$\%$ in the Bragg 
peak, the cell survival was better than 95$\%$ only 2 mm downstream of the position where the 
$\overline{p}$ stopped. This is due to the fact that the secondary radiation caused by antiproton 
annihilation that penetrates over long distances ($>1$ mm) consists mostly of minimum-ionizing particles.

\section{Future experiments and facilities}

\label{sec:future}
\subsection{\it Extra Low ENergy Antiproton (ELENA) ring}

\begin{figure}[htbp]
\epsfysize=8.0cm
\begin{center}
\begin{minipage}[t]{10 cm}
\epsfig{file=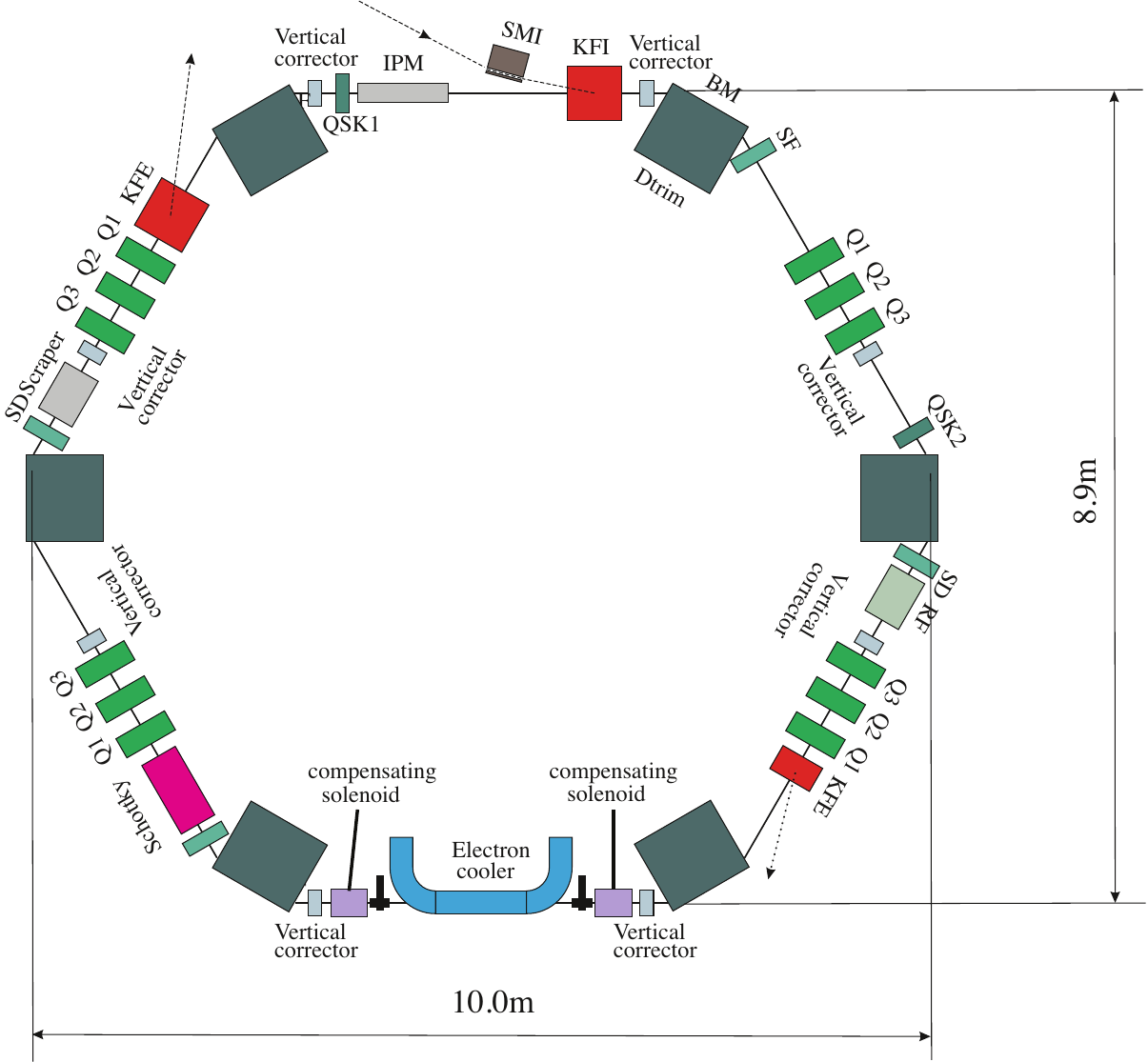,scale=0.7}
\end{minipage}
\begin{minipage}[t]{16.5 cm}
\caption{\label{fig:elena} 
Possible layout of future ELENA storage ring now being designed at CERN.
The 10-m-diameter, hexagonal ring contains an electron cooler,
a magnetic injection kicker (labeled KFI) which allows 5.3 MeV antiprotons to be
injected into the ring, two electrostatic ejection kickers (KFE) to extract 
100-keV antiprotons to experiments located outside the ring. Figure from Refs.~\cite{elena2010,oelert2012}.}
\end{minipage}
\end{center}
\end{figure}

CERN and the AD user community are now constructing the Extra-Low Energy Antiproton (ELENA) facility 
(Fig.~\ref{fig:elena}), which is a magnetic storage ring of circumference $\sim 30$ m located inside the
AD ring \cite{elena2010,oelert2012}. The 5.3-MeV $\overline{p}$ provided by AD will be injected into ELENA, 
where they will be decelerated over a 20-s cycle to $E=100$ keV.  
The magnetic fields of the six dipole magnets in the hexagonal ring (Fig.~\ref{fig:elena}) 
will be decreased from 3000 to 500 Gauss during this deceleration. 
Electron cooling will reduce the 
emittance and momentum spread of the beam to $\varepsilon=4\pi$ mm mrad and 
$\Delta p/p\sim 10^{-4}$. For this the circulating $\overline{p}$ beam at momenta 
35 and 13.7 MeV/c are merged with an $e^-$ beam of respective energies 355 and 55 eV, 
and currents 15 and 2 mA over a 1-m-long interaction region. The $e^-$ beam must have
small transverse (0.1 eV) and longitudinal energy spreads, to avoid heating the $\overline{p}$ beam. 
To prevent the rapid increase in the diameter of the $\overline{p}$ beam due to collisions with 
residual gases, the vacuum inside ELENA must be maintained better than $5\times 10^{-12}$ mbar.
Four 300-ns-long pulses, each containing $\sim 6\times 10^6$ $\overline{p}$, will be extracted
using two electrostatic ejection kickers located inside ELENA. This beam will be delivered to several 
experiments simultaneously. For this $>100$ m of beamlines containing electrostatic 
quadrupoles and dipoles will be installed. 

The low energy and small emittance of the ELENA beam is expected to allow the existing ATRAP, ALPHA, and 
\aegis \ experiments to capture and accumulate $\sim 100$ times more $\overline{p}$ in Penning traps 
per unit time, compared to directly using the 5.3-MeV AD beam. Significant improvements in the atomic
spectroscopy and collision experiments of ASACUSA are expected as well. Additional beamlines will be constructed
for new collaborations, e.g. GBAR. The building and commissioning of ELENA will take place between 
2013--2017.

\subsection{\it Towards antihydrogen laser spectroscopy}

Much of the fascination which drives the present $\overline{\rm H}$ trapping
experiments arise from the prospect to test $CPT$ symmetry at 
unprecedented levels of precision, by carrying out 
ultrahigh-precision laser spectroscopy of $\overline{\rm H}$.

\subsubsection{\it Ultrahigh-resolution laser spectroscopy of hydrogen}

The metastable $2s$ state of H has a long lifetime (122\,ms) because 
the electric-dipole transition to the $1s$ ground state is forbidden by 
the parity selection rule.  The $1s\rightarrow 2s$ transition can be excited
by two counterpropagating ultraviolet
laser photons at wavelength $\lambda=243$ nm, 
which cancel the linear Doppler shift in the resulting resonance profile
 \cite{Haensch:1975}.  These excitations to the $2s$ state can be readily
detected by applying a weak electric field (several V/cm is
enough, see \cite{Bethe:1977}, Sect. 67) which mixes the $2s$
state with the short-lived $2p$ state by Stark effects.  The $2p$ state has a lifetime
of 1.6 ns and decays by emitting a Lyman-$\alpha$ photon of wavelength 122 nm.

Over the last four decades since the pioneering experiment of Ref.~\cite{Haensch:1975} 
which employed a cold atomic H beam, the spectral resolution of the Doppler-free two-photon 
laser spectroscopy has been improved by a factor of $>10^6$.
The observed linewidth of the $1s-2s$ resonance now reaches levels of $<1$ kHz
\cite{Parthey:PRL107:2011,Fischer:2004,Biraben:LaserPhysicsLimits:2002}. 
The results have been used to determine
the Lamb shift in the $1s$ ground state \cite{Weitz:1992}, the Rydberg constant
\cite{Udem:1997}, the H--D isotope shift
\cite{Huber:1998,Parthey:PRL104:2010}, the deuteron structure radius
\cite{Huber:1998}, and a constraint on the variation of the fundamental
constants \cite{Fischer:2004}.  For a review which includes a
discussion of the development of the theory, see Ref.~\cite{Pachucki:1996e}.

The $1s-2s$ transition of H atoms confined in a neutral magnetic trap have also
been studied by laser spectroscopy \cite{Cesar:1996}.  These experiments 
are the culminations of a long effort at MIT and Amsterdam to observe Bose-Einstein 
condensation (BEC) in H, which date much before the introduction of magnetic 
trapping techniques of neutral atoms \cite{Midgall:1985}, see
Ref.~\cite{Greytak:1984} for an account of the early history.  The techniques of
magnetic trapping and evaporative cooling of H \cite{Hess:1986,Ketterle:1996} 
eventually enabled BEC in H, which was detected by measuring the accompanying 
collisional frequency shifts in the $1s-2s$ transition \cite{Killian:1998,Fried:1998}.

\subsubsection{\it Laser cooling of trapped antihydrogen}

\begin{figure}[tbh!]
\begin{center}
\includegraphics[height=10cm]{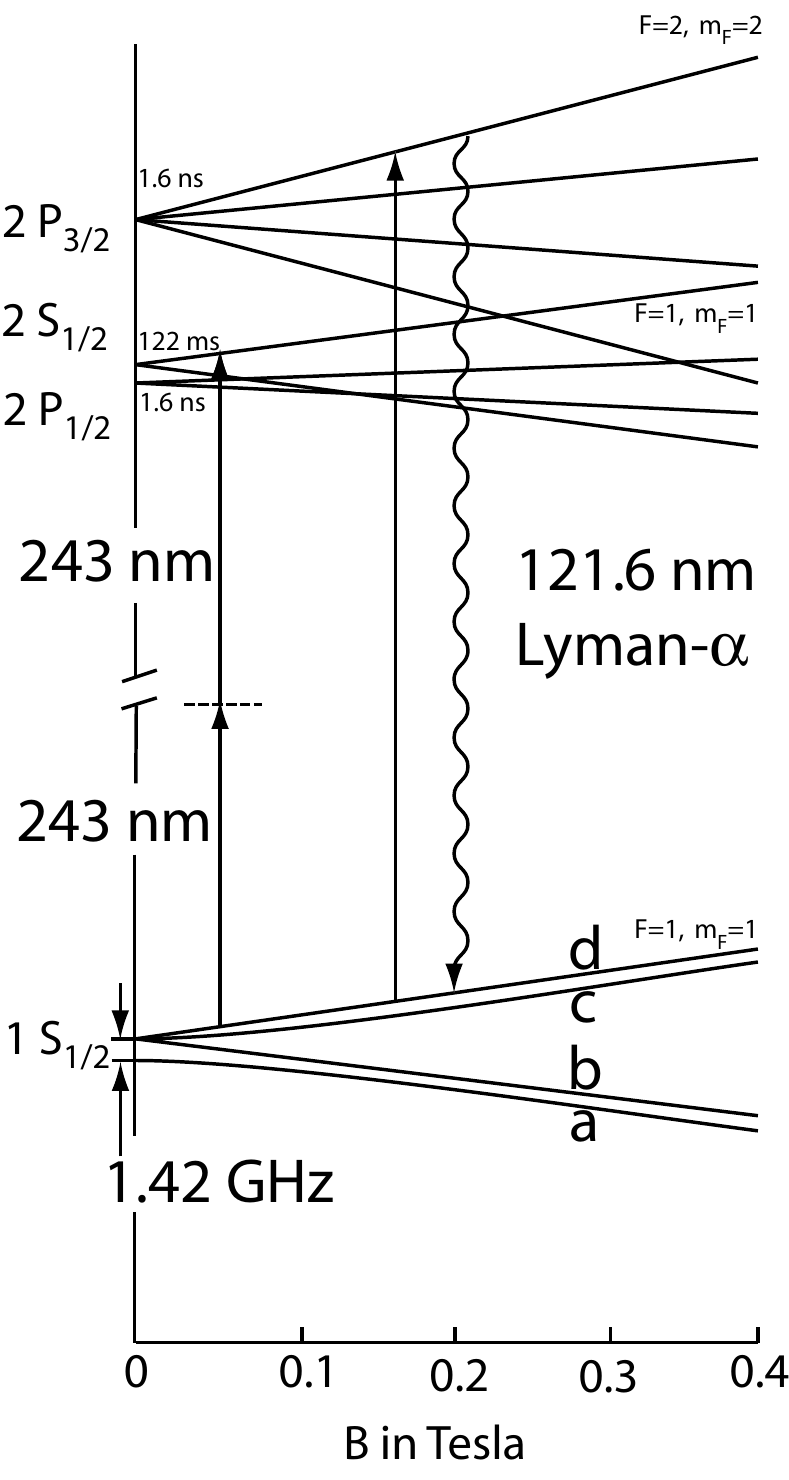}
\begin{minipage}[t]{16.5 cm}
\caption{Energy levels of H or $\overline{\rm H}$ as a function of an
external magnetic field (not to scale).  The hyperfine splitting in the $n=2$ states is
not resolved -- see \cite{Hijmans:1989} for a version with more
detail.  Precision spectroscopy in the microwave range at 1.42 GHz
can be carried out on the hyperfine splitting in the ground state.  The
substates $c$ and $d$ are low-field seeking states
which can be confined in a magnetic trap.  Doppler-free two-photon
laser-spectroscopy can be carried out on the $1s-2s$ transition using
counterpropagating laser beams at a wavelength of 243 nm 
in the vacuum ultraviolet region. Laser-cooling can be done on the strong Lyman-$\alpha$
transition and will reduce residual Zeeman shifts and broadenings on
the $1s-2s$ transition. Figure from Ref.~\cite{Walz:Varenna:2009}.
\label{levels}%
}
\end{minipage}
\end{center}
\end{figure}

Future experiments on high-resolution laser spectroscopy of $\overline{\rm H}$ will likely
utilize trapped atoms \cite{Haensch:1993:47}.  One problem is that external magnetic 
fields shift the $1s-2s$ ($F=1, m_F= 1\rightarrow F=1, m_F=1$) transition frequency 
by an amount 186\,kHz T$^{-1}$. To reduce the magnetically-induced broadening 
\cite{Haensch:1993:47,Cesar:2001:1S2S} of the spectral line, it is important to cool 
$\overline{\rm H}$ atoms and minimize their spatial distribution in the inhomogeneous magnetic 
field of the trap. Although evaporative cooling has been used with great success in the case 
of ordinary H and has enabled BEC, it is unlikely to be of much use in the case of $\overline{\rm H}$ 
due to the small number of atoms that can be trapped \cite{Walraven:1993:205}.

Laser cooling of $\overline{\rm H}$, on the other hand, does not require having high atom
densities or numbers, and can be carried out by utilizing the strong $1s-2s$ electric dipole 
transition at wavelength 122\,nm.  Laser cooling with Lyman-$\alpha$ radiation has been discussed by
several authors \cite{Phillips:1993,Ertmer:1988,Lett:1988}, while alternative 
laser-cooling methods have also been proposed for H \cite{Allegrini:1993,Zehnle:2001}.  
Most encouragingly, laser cooling of H confined in a magnetic trap to temperatures of
a few millikelvin has been demonstrated by using a pulsed Lyman-$\alpha$ source 
\cite{Setija:1993}.

Generating coherent Lyman-$\alpha$ radiation is a challenge, since there are no tunable lasers or nonlinear 
frequency-doubling crystals available for that spectral region.  Sum-frequency generation
of several incident laser beams that utilize the nonlinear susceptibility
of atomic vapors and gases is commonly used.  Four-wave sum-frequency mixing produces the
sum-frequency of three fundamental beams \cite{Vidal:FWM:1992} and has been
employed to generate {\em{pulsed}} laser radiation at Lyman-$\alpha$,
typically using Kr gas
\cite{Mahon:1978,Cotter:1979a,Wallenstein:1980,Cabaret:1987,Marangos:90}.
{\em{Continuous-wave}} (cw) coherent radiation at Lyman-$\alpha$ can have distinct
advantages for laser-cooling of $\overline{\rm H}$, compared to typical 
pulsed sources that have ns-scale pulse lengths.
Since the pulse length is comparable to the lifetime of the $2p$ states of 1.6\,ns, 
only a few excitations can be induced per $\overline{\rm H}$ per laser pulse, and moreover
the rate of laser cooling is limited by the pulse repetition rate.
A cw source on the other hand can provide a larger rate for laser cooling.
Its smaller spectral bandwidth provides 
higher selectivity for magnetic substates of $\overline{\rm H}$ in the trap, thereby 
reducing losses due to spurious optical pumping to the untrapped magnetic 
sublevels $a$ and $b$ in Fig.~\ref{levels}.

The cross section for Lyman-$\alpha$ light being resonantly absorbed by
$\overline{\rm H}$ can be as high as $3 \lambda_{\alpha}^2 / 2 \pi$ \cite{Luiten:1994}.
This implies that the excitation rate for $\overline{\rm H}$ illuminated with a
1 nW Lyman-$\alpha$ laser beam with diameter of 1 mm is 5 s$^{-1}$.  
Suppose that we would like to cool $\overline{\rm H}$ confined in a magnetic 
trap starting with an initial temperature of 1\,K which corresponds to an average 
velocity of 150\,m s$^{-1}$.  The average velocity change per excitation is 3.3 m s$^{-1}$,
and so cooling to temperatures close to the Doppler- and recoil-limits could be completed in 
less than a minute using only 1\,nW of laser power.

An important difference between pulsed and cw Lyman-$\alpha$
generation is that the power levels of cw fundamental beams are 
many orders of magnitude lower than the peak powers typically used
in pulsed Lyman-$\alpha$ generation.  The cw Lyman-$\alpha$ generation
therefore requires the use of resonances and near-resonances in the nonlinear optical
medium. In the case of pulsed sum-frequency mixing, close resonances
are usually avoided as they can cause premature saturation of the VUV yield 
due to step-wise excitation and multi-photon ionization.

Currently, cw Lyman-$\alpha$ laser light can be generated at power levels of 
up to 20\,nW using mercury vapor as a nonlinear optical medium and three 
fundamental laser beams.  The first laser is tuned to the longer wavelength side 
of the transition $6s^1S_0$ $\rightarrow$ $6p^3P_1$ in mercury 
at a wavelength 253.7\,nm.  The second fundamental beam at
408\,nm establishes an exact two-photon resonance with the $7^1S_0$
state.  The wavelength of the third fundamental light field at 545\,nm
is chosen such that the sum-frequency is at Lyman-$\alpha$.  Bound
states in mercury such as $11p^1P_1$ and  $12p^1P_1$ contribute
significantly to the nonlinear susceptibility \cite{Smith:1987a}.
The first cw coherent Lyman-$\alpha$ source
\cite{Eikema:99,Eikema:PRL:01,Pahl:2005} employed up to three
large-frame argon-ion lasers, which limited the
reliability of Lyman-$\alpha$ generation.  A new cw
Lyman-$\alpha$ source has therefore been set up based on
solid-state laser systems
\cite{Scheid:OL:2007,Markert:OE:2007,Scheid:oe:2009}.  It is hoped
that this source will be a reliable basis for laser-cooling of
$\overline{\rm H}$ in a magnetic trap.  The limits for laser-cooling with
Lyman-$\alpha$ are in the millikelvin range \cite{Phillips:1993} which is
cold enough to enable $1s-2s$ spectroscopy at a precision in the
kHz range and below.  This corresponds to a test of $CPT$ symmetry at a precision 
better than a few parts in $10^{12}$.

\subsection{\it Higher-precision microwave spectroscopy of the antihydrogen hyperfine structure}
\label{sec:antihydrogen_hfs}

As described in Sect.~\ref{antihydrogenhyperfine}, the ALPHA collaboration has
recently measured the ground-state hyperfine structure of $\overline{\rm H}$
confined in a magnetic bottle trap with a relative precision of $4\times 10^{-3}$ 
\cite{Alpha:SpinFlips:2012}. The authors note that this was a first proof-of-principle measurement,
and no attempt was made to accurately determine the spectroscopic lineshape. 
Further improvements in the experimental precision can be expected in the nearest future. 
One source of systematic error is caused by the Zeeman shift
in the $\overline{\rm H}$ microwave transition frequencies induced by the 
inhomogeneous magnetic field $B$ in the atom trap. ALPHA intends to reduce this error by
constructing a trap with an extended area near the center with a relatively 
homogeneous field where the spectroscopy can be carried out. They also intend 
\cite{Alpha:SpinFlips:2012,ashkezari2011} 
to measure the nuclear magnetic resonance (NMR) transition between the two energetically lowest 
hyperfine levels (denoted $\left|c\right>$$\leftrightarrow$$\left|d\right>$ in Fig.~\ref{hydrogen_levels}).
Since this transition frequency $f_{c-d}$ has a minimum value 
765483207.7(3) Hz at a magnetic field of $B_0=0.65$ T \cite{hardy1979},
experiments carried out using traps with this field strength will be relatively
free from the effects of inhomogenities in the magnetic fields shifting or broadening the
resonance. In this way, the collaboration expects to achieve an experimental precision
of $10^{-6}$. 

\begin{figure}[tbh!]
\begin{center}
\includegraphics[height=4cm]{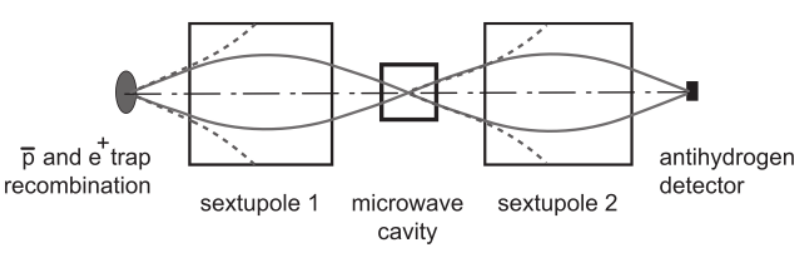}
\begin{minipage}[t]{16.5 cm}
\caption{\label{fig:cusp} 
Schematic layout of a possible method to carry out microwave spectroscopy of the ground-state hyperfine structure of 
$\overline{\rm H}$ using a polarized beam of $\overline{\rm H}$.}
\end{minipage}
\end{center}
\end{figure}

An alternative method to measure the microwave transition is pursued by the ASACUSA 
collaboration \cite{Widmann:LoI:2003} (Fig.~\ref{fig:cusp}). Here a beam of $\overline{\rm H}$ emerging 
from the trap is allowed to pass through a sextupole magnet, which focuses the $\overline{\rm H}$ 
occupying the low-field-seeking states towards the beam axis,
whereas the high-field seekers are defocused. The $\overline{\rm H}$ beam which is spin-polarized 
in this way then traverses a microwave cavity where a spin-flip transition (e.g., from a low-field-seeking
to a high-field seeking state) is induced. The transition is probed by allowing the 
$\overline{\rm H}$ to traverse a second analyzing magnet, which focuses atoms 
occupying the low-field-seeking states onto a detector. The experimental precision is here
determined by the velocity, temperature, and emittance of the $\overline{\rm H}$ beam, and 
may achieve a precision similar to the ALPHA experiment described above. Simulations
indicate that the beam emerging along the axis of the anti-Helmholtz field of the 
cusp trap may be spin polarized, thereby avoiding the necessity of the first sextupole 
magnet \cite{Mohri:2003}. Efforts are currently underway to obtain a cold, high-intensity 
$\overline{\rm H}$ beam.

\subsection{\it Antihydrogen experiments to measure antimatter gravity}
\label{sec:antimatter_gravity}

The {\em{inertial}} mass of $\overline{p}$ has been measured with a precision
of $7\times 10^{-10}$ \cite{hori2011} by combining the results of cyclotron frequency measurements
in Penning traps \cite{gabrielse1999} with laser spectroscopy of $\overline{p}{\rm He}^+$
(see Sect.~\ref{pbarhelium}).
The {\em{gravitational}} mass of $\overline{p}$ (or antiparticles in
general), however, has never been measured; the equivalence
principle, which is at the heart of general relativity, has
never been tested with antimatter.  Interest in such questions is
enhanced by the unknown origin of the acceleration of the expansion of
the universe and by the hypothesis of dark matter, which both suggest
that our understanding of gravitation may be incomplete.  For possible
scenarios of anomalous gravitational behavior and a general overview,
see Refs.~\cite{Scherk:1979,Chardin:HI109:83:1997,Nieto:91,Nieto:92}.

Previous attempts to measure the gravitational acceleration of $e^+$ and
$\overline{p}$ have been unsuccessful, due to the fact that the trajectory of the
charged particles were deflected by stray electric fields in the experimental
apparatus \cite{Darling:1992}. Two collaborations, {\aegis} and GBAR, are now
attempting to measure instead the antimatter gravity of neutral
$\overline{\rm H}$ for the first time using novel methods.
Gravitational experiments using ordinary atoms have produced impressive results for 
quite some time (see e.\,g.~\cite{Peters:1999,Poli:PRL106:2011}), although none of
them have involved H, since this atom is difficult to manipulate.

\subsubsection{\it The {\aegis} antimatter gravity experiment}
\label{sec:aegis}

\begin{figure}[tbh!]
\begin{center}
\includegraphics[height=8.5cm]{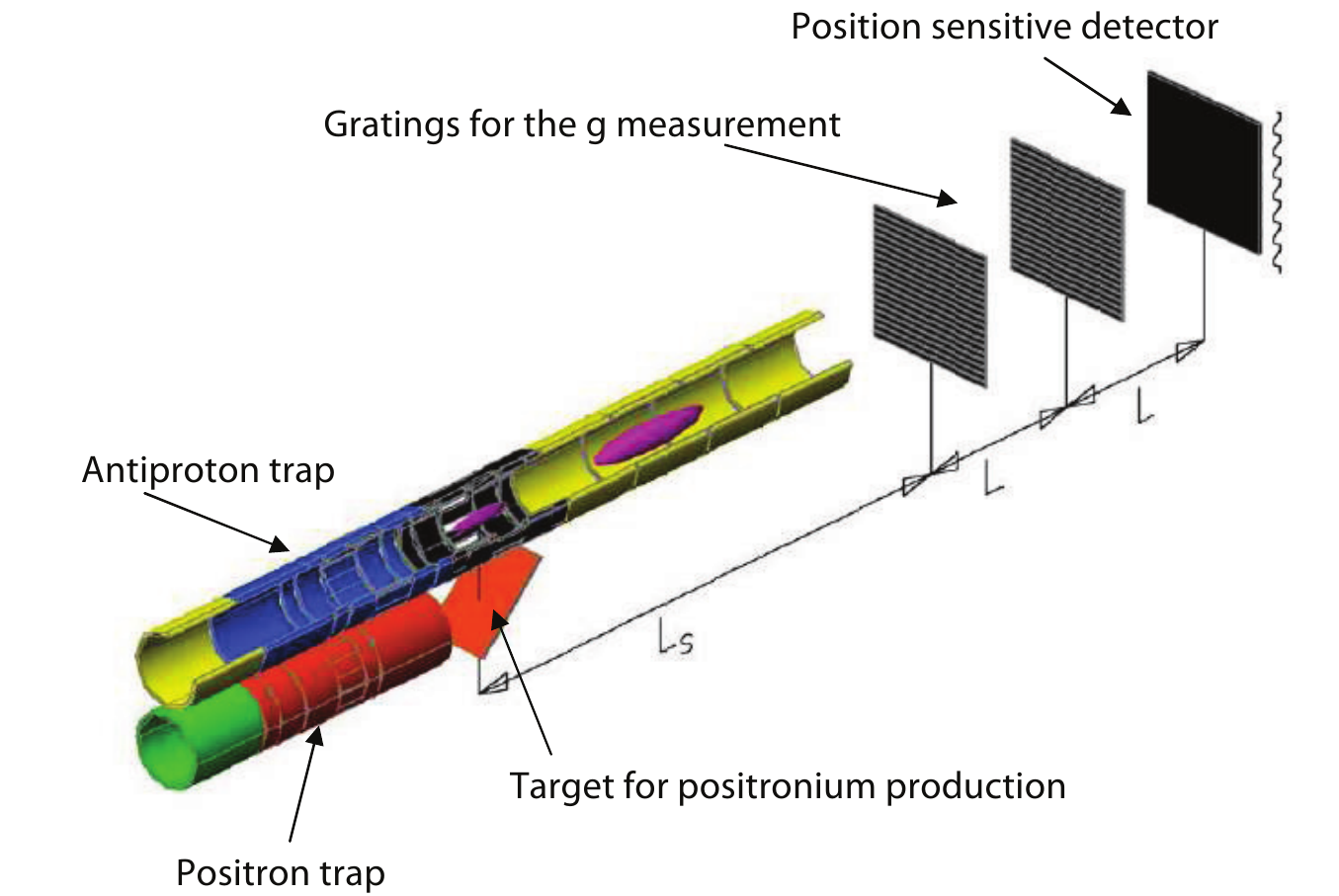}
\begin{minipage}[t]{16.5 cm}
\caption{Schematic layout of the {\aegis} experiment,
involving two parallel Penning-Malmberg traps.
The $\overline{p}$ will be accumulated and stored in the
black region of the upper trap. The $e^{+}$ accumulated
in the lower trap will be allowed to strike a porous target
mounted in front of a window in the $\overline{p}$ trap, thereby producing
$Ps$. Laser pulses will excite the $Ps$ to Rydberg states with $n=20$--30.
The $Ps$ will drift through the $\overline{p}$ cloud, thereby producing
Rydberg $\overline{\rm H}$. Electric fields will accelerate the $\overline{\rm H}$
as shown in the yellow region of the trap. The $\overline{\rm H}$
will traverse the two gratings and reach a position-sensitive detector
where the gravitational deflection of the beam trajectory will be measured.
Figures from Ref.~\cite{AEGIS:AIPCP1037:2008}.
\label{AEGIS:fig}%
}
\end{minipage}
\end{center}
\end{figure}

\begin{figure}[tbh!]
\begin{center}
\includegraphics[height=6cm]{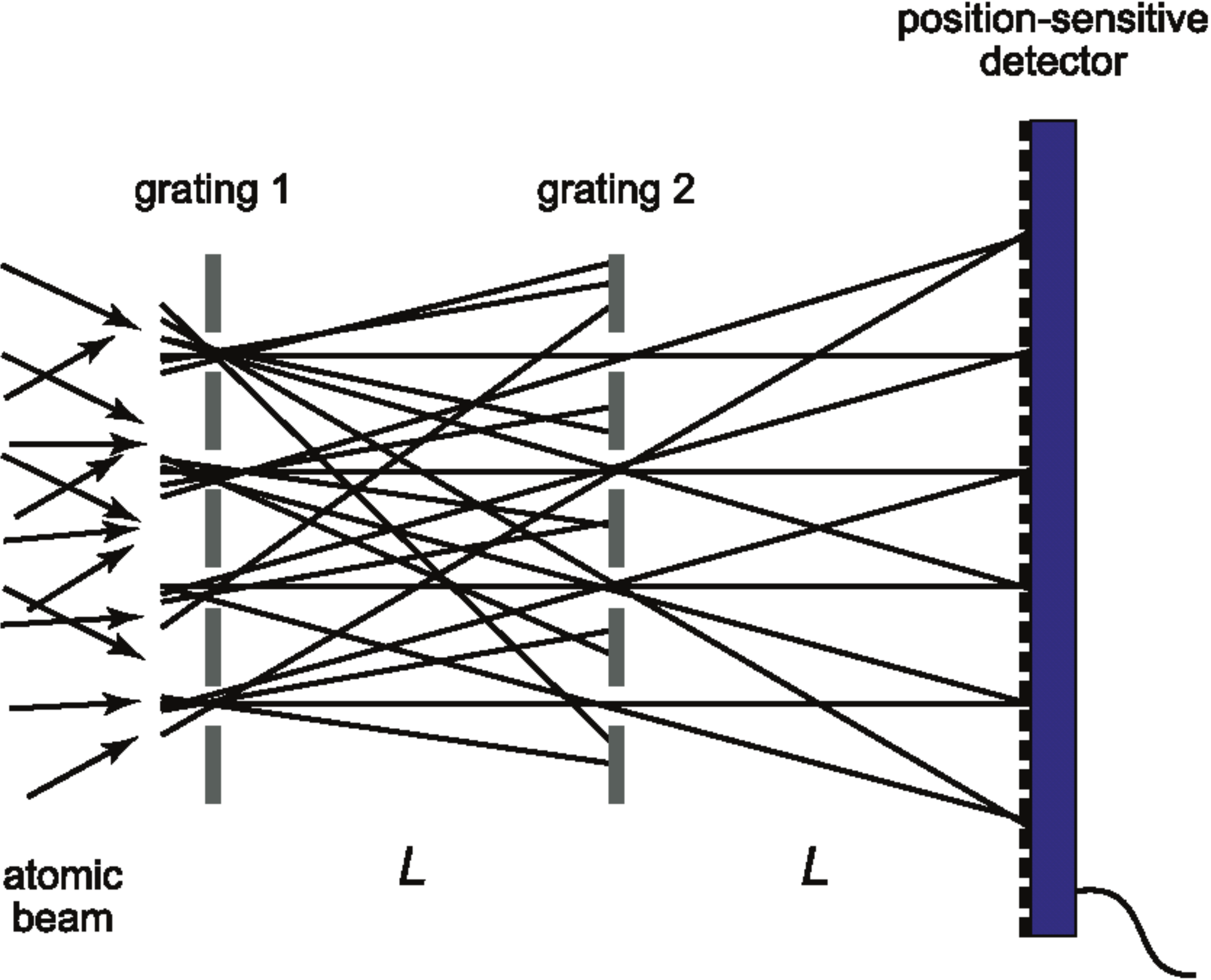}
\begin{minipage}[t]{16.5 cm}
\caption{Schematic of the {\aegis} Moir{\'{e}} deflectometer for
measuring antimatter gravity.  The device consists of two identical
gratings and a position-sensitive detector.  The length $L$ may be
of order 300 mm and the grating period $80\,\mu{\textrm{m}}$.
Figure from Ref.~\cite{AEGIS:NIM:B266:351:2008}.
\label{moire}%
}
\end{minipage}
\end{center}
\end{figure}

The proposed {\aegis} antimatter gravity experiment (Fig.~\ref{AEGIS:fig})
\cite{AEGIS:NIM:B266:351:2008,AEGIS:AIPCP1037:2008,AEGIS:CJP89:2011}
will produce cold $\overline{\rm H}$ in resonant charge-exchange
collisions of Rydberg $Ps^*$ and $\overline{p}$.  The $Ps$ will be produced in 
ultrahigh vacuum by allowing $e^+$ to collide with a porous SiO$_2$ converter.
The resulting $Ps$ will then be excited using two laser beams, the
first at wavelength 205 nm which induces the transition $n=1$$\rightarrow$$n=3$,
and the second one at wavelength 1650--1700 nm which excites the transition from $n=3$
to the final Rydberg state.
The laser pulses needed for this will be produced by parametric generation and
amplification \cite{Cialdi:NIMB269:2011}.  The Rydberg $\overline{\rm H}$
will thus be produced as a pulsed beam. The population distribution of $\overline{\rm H}$
among the Rydberg states will be controlled by adjusting the wavelength of the second 
laser beam.  The Rydberg $\overline{\rm H}$ will then be accelerated along the axis of the trap
by electric fields, so that a beam of temperature 100 mK emerges from the trap.

The free fall of  $\overline{\rm H}$ cannot be directly measured using this beam,
because the radial beam divergence and the distribution of vertical starting
positions will be too large.  These problems can be solved by
utilizing a Moir{\'e} deflectometer, which consists
of two gratings and a position-sensitive detector (Fig.~\ref{moire}).  
The diffraction effect of $\overline{\rm H}$ in this device due to the 
de-Broglie wavelength is negligible, since the grating period is relatively large.
The gratings instead select the propagation direction of the incoming 
$\overline{\rm H}$ beam. The atoms that traverse the two gratings then strike a 
position-sensitive silicon detector, producing a shadow image comprising 
a set of fringes.  The free fall of the atoms induces a shift $- \bar{g} T^2$ in the 
fringe positions, which can be used to determine the gravitational acceleration $\bar{g}$ of
antimatter.  Here $T$ denotes the time-of-flight of $\overline{\rm H}$ traversing the
distance between the two gratings. It can be experimentally determined by measuring 
the time interval between the arrival of the laser pulse which induces $\overline{\rm H}$
recombination, and the detection of  $\overline{\rm H}$ at the position-sensitive silicon detector.
Simulations have shown that a measurement of $\bar{g}$ to 1\,\%
relative precision requires about 10$^5$ $\overline{\rm H}$ atoms with a
temperature of 100\,mK in {\aegis} \cite{AEGIS:NIM:B266:351:2008}.

\subsubsection{\it The GBAR antimatter gravity experiment}
\label{sec:gbar}

\begin{figure}[tbh!]
\begin{center}
\includegraphics[height=14cm]{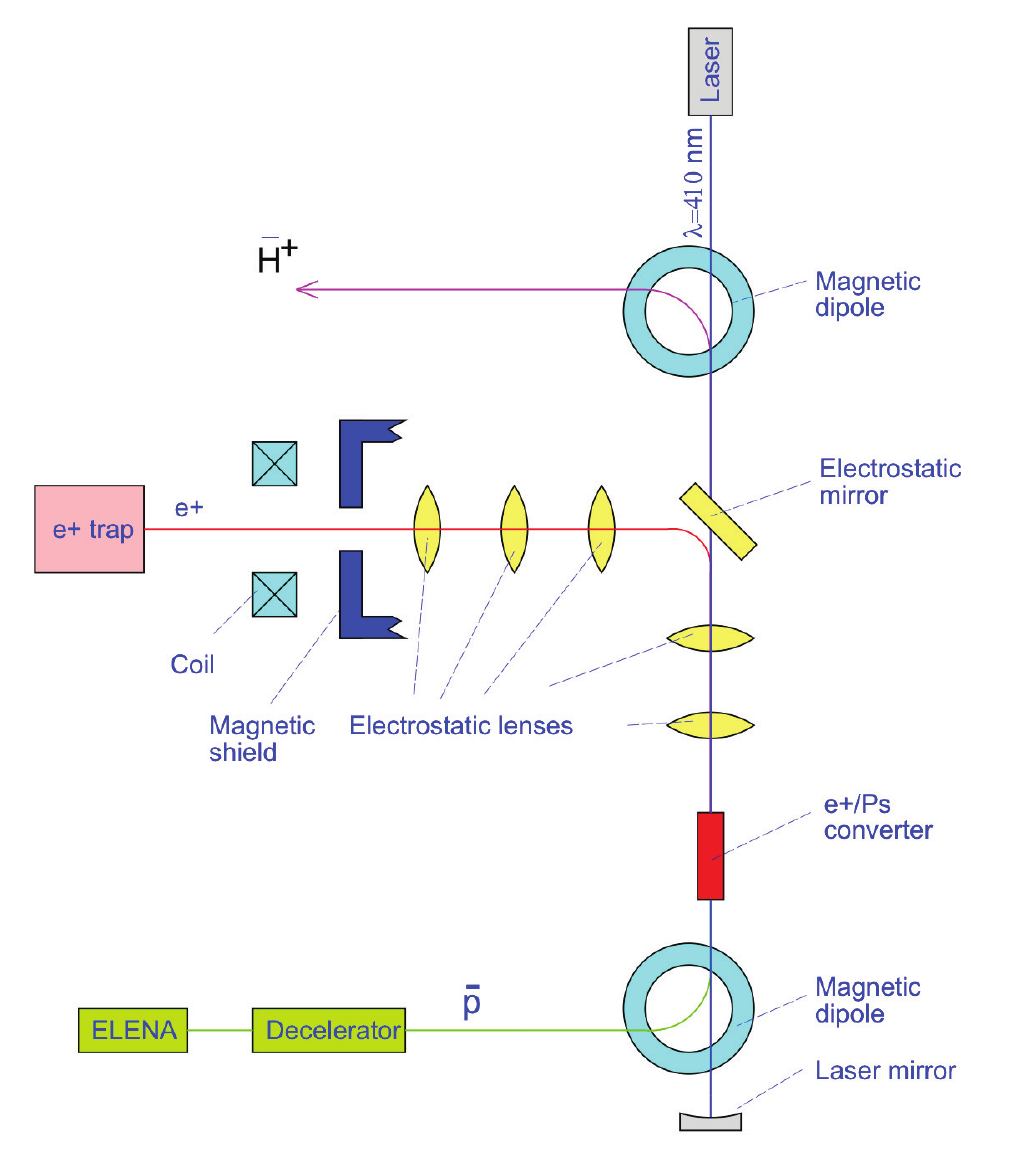}
\begin{minipage}[t]{16.5 cm}
\caption{Proposed layout of the GBAR experiment
to produce beams of $\overline{\rm H}^+$ ions. 
High-intensity pulses of $e^+$ extracted from
a Penning trap are guided onto a mesoporous film,
thereby producing Ps. These are excited to the
metastable 3$d$ state by laser-induced two-photon excitation.
A beam of $\overline{p}$ is allowed to traverse
the cloud of $Ps$, which results in the formation
of $\overline{\rm H}^+$ ions. Figure from Ref.~\cite{gbar2011}.
\label{gbar:fig}
}
\end{minipage}
\end{center}
\end{figure}

\newcommand{\hbarplus}{$\bar{\textrm{H}}^+$}

The GBAR collaboration plans to produce ultracold $\overline{\rm H}$ of even
lower ($\mu\textrm{K}$) temperatures, and directly measure antimatter
gravity in a time-of-flight experiment
\cite{Perez:NIM:A545:2005,gbar2011}.  Laser-cooling with
Lyman-$\alpha$ radiation cannot be used to achieve such ultracold
temperatures, because both the Doppler- and recoil-limits for $\overline{\rm H}$
are in the mK range.  Instead GBAR will first produce positive antihydrogen
ions ({\hbarplus}), confine them in an ion trap together with
ultracold laser-cooled ions (such as Be$^+$), and
utilize sympathetic cooling \cite{Walz:GRG:2004}.  The
ultracold {\hbarplus} can then be irradiated with a laser pulse to photodetach and neutralize it.
This laser pulse can also define the start timing
for a measurement of the time-of-flight of $\overline{\rm H}$ falling 
to an annihilation detector located some 100 mm below the trap.  From this
time-of-flight, the $\overline{\rm H}$ gravitational acceleration $\bar{g}$ can be 
determined. Some 500\,000 {\hbarplus} at a temperature of 20\,$\mu\textrm{K}$
may be sufficient to determine $\bar{g}$ with a relative precision
of $10^{-3}$.  

This scheme involves {\hbarplus} ions, which have not been produced yet.  Whereas 
negative ${\rm H}^-$ ions are commonly produced by
asymmetric dissociation of H$_2$ molecules, this formation method
is excluded in the {\hbarplus} case as there are no
$\overline{\rm H}_2$ molecules available so far. 
The GBAR experiment proposes to use an alternative method involving 
charge-exchange with $Ps$ as shown in Fig.~\ref{gbar:fig},
\begin{equation}
\bar{\rm H} + Ps \rightarrow \bar{\rm H}^+ + e^-.
\end{equation}
This requires high densities of $Ps$ atoms \cite{Perez:CP1037:35:2008}.  For 
this an intense source of slow $e^+$ based on pair production with a beam of
electrons from an industrial linac is being developed
\cite{Perez:NIM:A532:2004,Perez:ASS255:2008}.

\subsection{\it Sub-ppb-scale determination of the antiproton-to-electron mass ratio 
by laser spectroscopy of antiprotonic helium}

The experimental precision on the antiproton-to-electron mass ratio determined by
laser spectroscopy of $\overline{p}{\rm He}^+$ is currently around a factor of $\sim 3$ worse
than the proton-to-electron value recommended by CODATA, obtained by statistically
averaging several experimental results having a lower precision \cite{codata2002}.
Some $\overline{p}{\rm He}^+$ states have 
lifetimes of 1--2 $\mu{\rm s}$ against annihilation, corresponding to a natural width of $\sim 100$ kHz. 
This implies that a laser transition frequency of 1--2 PHz between two such states 
can in principle be measured to a fractional precision of better than $10^{-12}$.

One of the factors which limited the experimental precision to 
$(2.3-5)\times 10^{-9}$ in Ref.~\cite{hori2011} was the relatively small number of $\overline{p}$
that could be stopped in the experimental helium target to synthesize the atoms. The AD provided
a pulsed beam containing $\sim 3\times 10^7$ $\overline{p}$ every 100 s, but after deceleration in the
RFQD, the beam emittance and energy spread 
became so large (100 mm mrad and $>30$ keV) that most of the $\overline{p}$ were not stopped in the 
He target but rather annihilated in the inner walls of the RFQD and beamlines, or the walls of the
experimental apparatus. The $\pi^+$ and $\pi^-$ emerging from all these annihilations caused
a large background in the experimental data.
Monte-Carlo estimations indicate that the higher intensity and lower emittance of the 
electron-cooled $\overline{p}$ beam available from ELENA would increase the number of synthesized
$\overline{p}{\rm He}^+$ by a factor of $\sim 10$, whereas the background annihilations
would be suppressed by an order of magnitude. This would improve both the statistical error and 
signal-to-noise ratio in the experimental data.

The theoretical uncertainty on the transition frequencies is currently limited \cite{korobov2008}
to around $1\times 10^{-9}$ by the uncalculated radiative QED corrections with orders higher than 
$m_ec^2\alpha^6/h$. Recent advances in the techniques of calculating higher-order terms
in the H and He cases is expected to allow the $\overline{p}{\rm He}^+$
transition frequencies to soon be calculated to a precision of $1\times 10^{-10}$.

\subsection{\it Higher-precision determination of the antiproton magnetic moment}

\begin{figure}[tbh!]
\begin{center}
\includegraphics[height=6cm]{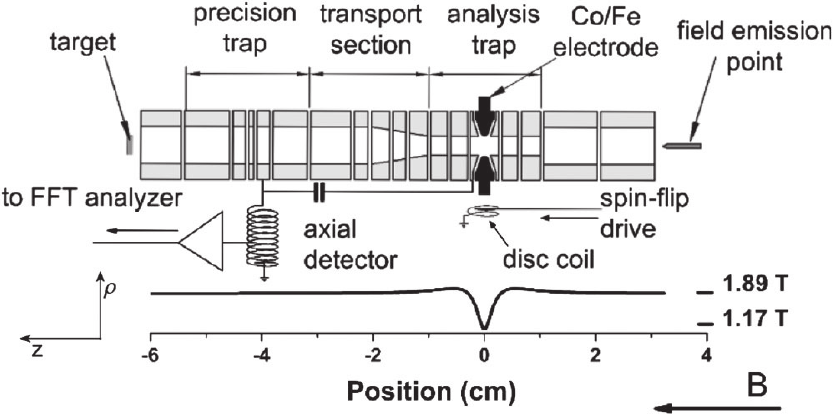}
\begin{minipage}[t]{16.5 cm}
\caption{Schematic layout of the recent experiment carried
out by BASE collaborators, which
measured the magnetic moment of a single proton
confined in a Penning trap. The experiment consisted
of two traps connected by cylindrical transport
electrodes. The central ring in the right analysis trap
 was made of a ferromagnetic material, which created the
 magnetic bottle field needed to detect the $p$ spin-flips.
 The spin-flips were induced by a disc coil. 
 The lower graph shows the magnetic field along the z-axis.
 Figure from Ref.~\cite{Ulmer:ProtonSpinFlips:2011}.
\label{ulmer:fig}
}
\end{minipage}
\end{center}
\end{figure}

The ATRAP collaboration is continuing efforts to improve the experimental precision on
the $\overline{p}$ magnetic moment, beyond their current limit of $4\times 10^{-6}$
by attempting to detect individual spin-flips of a single $\overline{p}$ confined in a Penning trap.
The $\overline{p}$ or $p$ will first be confined in a ``precision" Penning trap with a pure solenoidal 
field, where a spin-flip is induced by an oscillating field applied to one of the electrodes. This spin-flip
is detected by transferring the particle to an adjunct analysis trap which has a quadrupole magnetic 
field superimposed on the solenoid field, and using the continuous Stern-Gerlach effect described in 
Sect.~\ref{sec:magneticmoment}. An additional improvement of 10$^{3}$ on the experimental 
precision of $\mu_{\overline{p}}$ is believed to be possible.
Another collaboration (Baryon Symmetry Experiment, BASE) has recently proposed a similar measurement
(Fig.~\ref{ulmer:fig}).  These direct comparisons of $\mu_{\overline{p}}$ and $\mu_p$ would provide an important
test of the consistency of $CPT$ symmetry. When combined with the hyperfine structure experiments on $\overline{\rm H}$,
they may also help to derive constraints on the internal structure of $\overline{p}$.

\subsection{\it Reaction MIcroscope (ReMI) in ELENA}

\begin{figure}[htbp]
\epsfysize=8.0cm
\begin{center}
\begin{minipage}[t]{10 cm}
\epsfig{file=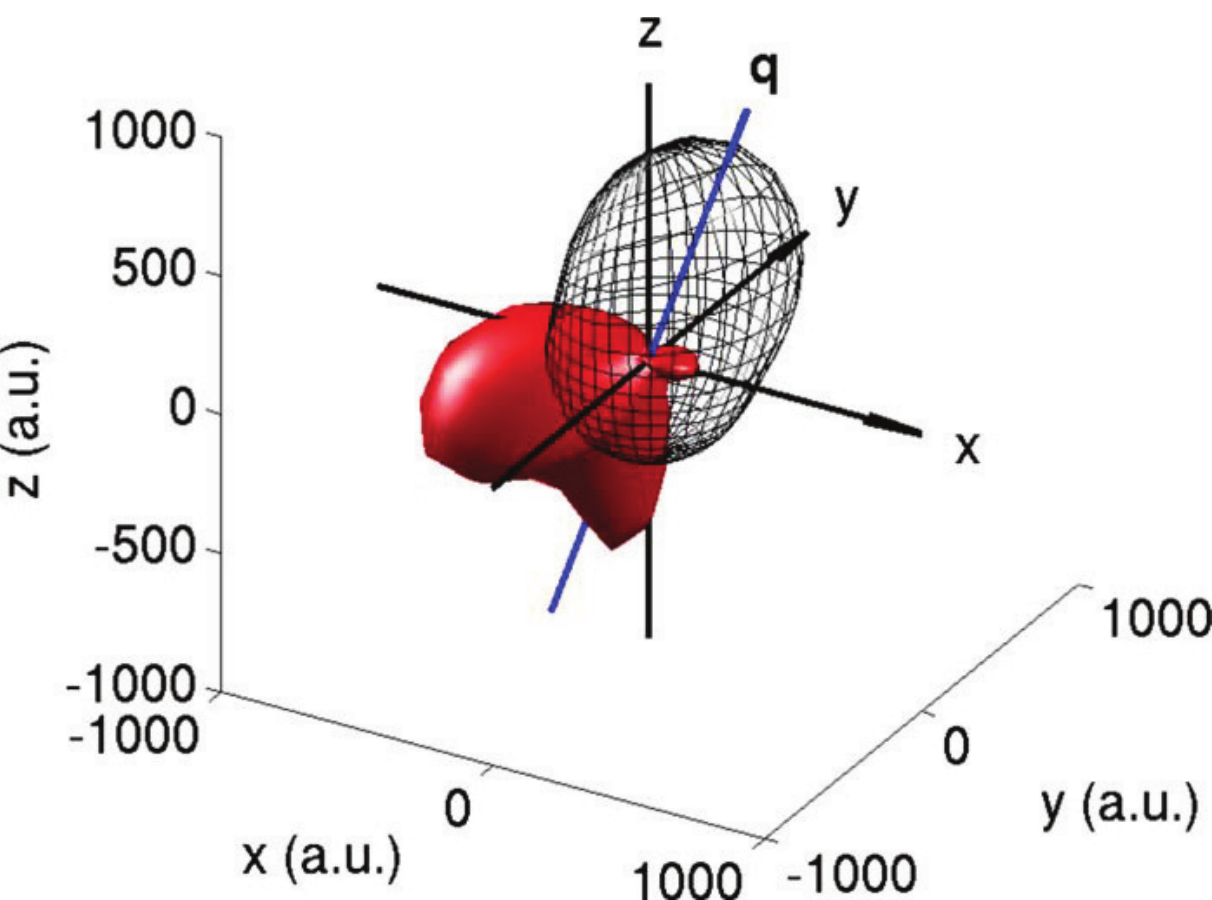,scale=0.6}
\end{minipage}
\begin{minipage}[t]{16.5 cm}
\caption{\label{fig:remi3d} 
Differential cross sections for single ionization of He by 3-kV $\overline{p}$ in the
laboratory frame, calculated by first Born approximation with a frozen core (transparent
meshed area) and coupled pseudostate approximation (filled area). From
Ref.~\cite{mcgovern2009}.}
\end{minipage}
\end{center}
\end{figure}

The {\it total} ionization cross sections of $\overline{p}$ on atomic and molecular gas targets
that were recently measured (Fig.~\ref{fig:knudsen_he}) are in good agreement
with several theoretical models. Future measurements will instead concentrate 
on the {\it differential} cross sections which will impose far stronger constraints on the models.
For example in Fig.~\ref{fig:remi3d}, two sets of differential cross sections for single ionization of He 
by 3 keV $\overline{p}$ are shown, calculated by using first Born approximation (indicated by the transparent 
meshed area) and a more advanced coupled pseudostate approximation (filled red area). Whereas the 
total cross sections calculated by the two methods give similar values, the differential cross sections
differ substantially \cite{mcgovern2009}.

One possible way to carry out this experiment involves placing a reaction microscope \cite{ullrich2003}
within the ELENA ring. The circulating $\overline{p}$ of energy $E\le 100$ keV are
allowed to make multiple passes through a gas jet target. The angular distributions of the
recoiling $e^-$ and ion pairs are analyzed using a magnetic Helmholtz field and a parallel electrostatic
acceleration field superimposed around the gas jet.
This setup will allow a kinematically complete experiment to be carried out, providing a
measurement of the differential cross section similar to the ones shown in Fig.~\ref{fig:remi3d}
over a solid angle of $\sim 4\pi$.

\section{Conclusions and discussions}

The first twelve years of AD operation have seen the first trapping of neutral $\overline{\rm H}$ atoms,
and microwave excitations of magnetic transitions between the ground-state hyperfine sublevels. 
Ppb-scale laser spectroscopy of
$\overline{p}{\rm He}^+$ atoms has yielded a new value for the antiproton-to-electron mass ratio.
The magnetic moment of a single $\overline{p}$ confined in a Penning trap was measured with a fractional
precision of $4\times 10^{-6}$, and
further improvements in the experimental precision may be possible in the near future.
Several atomic and nuclear collision experiments were carried out for the first time at 
kinetic energy ranges of 2--5000 keV, while the biological effectiveness of $\overline{p}$ beams deactivating
cancer cells were measured. 

Higher-precision spectroscopy experiments on $\overline{\rm H}$ and $\overline{p}{\rm He}^+$
appear to be possible in the near future. The magnetic moment of $\overline{p}$ confined in Penning traps
may be measured with $10^{3}$ times higher precision. The first two-photon laser spectroscopy experiments on 
the $1s-2s$ interval of $\overline{\rm H}$ are anticipated. Meanwhile, experiments to measure the gravitational
acceleration of $\overline{\rm H}$ have begun. The new ELENA
facility of CERN will allow experiments to confine 10--100 times more $\overline{p}$ in traps than before.
On the other hand, a new reaction microscope installed in ELENA may allow kinematically 
complete experiments to be carried out on 100-keV $\overline{p}$ ionizing atomic or molecular gasses.

\section{Acknowledgments}
We thank the AD user community, the CERN AD operations group, and ELENA construction
team for the many years of dedicated efforts described in this review article. 
We are indebted to H.~Higaki, V.I.~Korobov, S.~Maury, W.~Oelert, and S.~Ulmer for discussions.
This work was supported by the Bundesministerium f\"{u}r Bildung und Forschung, 
the Deutsche Forschungsgemeinschaft (DFG), the European Science Foundation (EURYI),
and the European Research Council (ERC-Stg).


\begin{thebibliography}{99}
\itemsep -2pt  
\bibitem{Maury:97} S.~Maury, {\it Hyperfine Interact.} 109 (1997) 43.
\bibitem{Pavel:04} P.~Belochitskii, T.~Eriksson, and S.~Maury, {\it Nucl. Instrum. Meth. Phys. Res. A} 214 (2004) 176.
\bibitem{ATHENA:2002} M.~Amoretti {\it et al.}, {\it Nature} 419 (2002) 456.
\bibitem{ATRAP:Hbar:2002} G.~Gabrielse {\it et al.}, {\it Phys. Rev. Lett.} 89 (2002) 213401.
\bibitem{ATRAP:Cesium:2004} C.H.~Storry {\it et al.}, {\it Phys. Rev. Lett.} 93 (2004) 263401.
\bibitem{Enomoto:PRL:2010} Y.~Enomoto {\it et al.} {\it Phys. Rev. Lett.} 105 (2010) 243401.
\bibitem{alpha_cool2010} G.B.~Andresen {\it et al,}, {\it Phys. Rev. Lett.} 105 (2010) 013003.
\bibitem{ALPHA:Nature:2010} G.B.~Andresen {\it et al.}, {\it Nature} 468 (2010) 673.
\bibitem{Alpha:NatPhys:2011} G.B.~Andresen {\it et al.}, {\it Nature Physics} 7 (2011) 558.
\bibitem{ATRAP:TrappedHbar:2012} G.~Gabrielse {\it et al.}, {\it Phys. Rev. Lett.} 108 (2012) 113002.
\bibitem{Alpha:SpinFlips:2012} C.~Amole {\it et al.}, {\it Nature}  483 (2012) 439.
\bibitem{hayanorpp} R.S.~Hayano, M.~Hori, D.~Horv\'ath, E.~Widmann, \Journal{\it Rep. Prog. Phys.}{70}{1995}{2007}.                                                    
\bibitem{mhori2001} M.~Hori {\it et al.}, \Journal{\PRL}{87}{093401}{2001}.    
\bibitem{mhori2003} M.~Hori {\it et al.}, \Journal{\PRL}{91}{123401}{2003}.                                                                               
\bibitem{mhori2006} M.~Hori {\it et al.}, \Journal{\PRL}{96}{243401}{2006}.              
\bibitem{hori2011} M.~Hori {\it et al.}, {\it Nature} 475 (2011) 484. 
\bibitem{korobov2008} V.I.~Korobov {\it et al.}, \Journal{\PRA}{77}{042506}{2008}.
\bibitem{pask2009} T.~Pask {\it et al.}, {\it Phys. Lett.} B 678 (2009) 55.
\bibitem{kreissl} A.~Kreissl {\it et al.}, {\it Z. Phys.} C 37 (1988) 557.
\bibitem{atrap_mag} J.~DiSciacca {\it et al.}, {\it Phys. Rev. Lett.} 110, 130801 (2013).
\bibitem{Moller2002} S.P.~M\o ller {\it et al.} \Journal{\PRL}{88}{193201}{2002}.
\bibitem{Moller2004} S.P.~M\o ller {\it et al.} \Journal{\PRL}{93}{042502}{2004}.
\bibitem{Moller2008} S.P.~M\o ller {\it et al.} {\it Eur. Phys. J.} D 46 (2008) 89.
\bibitem{knudsen2008} H.~Knudsen {\it et al.} \Journal{\PRL}{101}{043201}{2008}.
\bibitem{knudsen2010} H.~Knudsen {\it et al.} \Journal{\PRL}{105}{213201}{2010}.
\bibitem{infn2011} A.~Bianconi {\it et al.} {\it Phys. Lett.} B {704} (2011) 461.
\bibitem{ace2006} M.H.~Holzscheiter {\it et al.} {\it Radiotherapy and Oncology} 81 (2006) 233.
\bibitem{eades} J.~Eades, F.J.~Hartmann, {\it Rev. Mod. Phys.} 71 (1999) 373.                                       
\bibitem{andresencold2010} G.B.~Andresen {\it et al.} \Journal{\PRL}{105}{013003}{2010}.
\bibitem{atrapcold2011} G.~Gabrielse {\it et al.} \Journal{\PRL}{106}{073002}{2011}.
\bibitem{Bylinsky:RFQD:2000}
Y.~Bylinsky, A.~M. Lombardi, and W.~Pirkl, {\em Proceedings XXth International
  Linac Conference, 21-25 August 2000, Monterey, California, USA}.
\bibitem{lombardi2001}
A.~M. Lombardi, W.~Pirkl, and Y.~Bylinsky, 
{\em Proceedings of the 2001 Particle Accelerator Conference} (Chicago, Illinois, 2001).
\bibitem{kuroda2005} N.~Kuroda {\it et al.} \Journal{\PRL}{94}{023401}{2005}.
\bibitem{aegis2010} G.~Bonomi et al., {\it Hyperfine Interact.} 193 (2009) 297.
\bibitem{aegis2011} M.~Doser, {\it J. Phys.: Conf. Ser.} 264 (2011) 012006.
\bibitem{gbar2011} P.~Debu et al., {\it Hyperfine Interact.} 212 (2012) 51.
\bibitem{elena2010} M.-E.~Angoletta {\it et al.}, ELENA - an updated cost and
feasibility study, CERN-BE-2010-029 OP, CERN, Geneva (2010).
\bibitem{pdg2010} J.~Beringer and Particle Data Group, \Journal{\PRD}{86}{010001}{2012}.
\bibitem{dubbers2011} D.~Dubbers and M.G.~Schmidt, {\it Rev. Mod. Phys.} 83 (2011) 1111.
\bibitem{baker2006} C.A.~Baker {\it et al.} \Journal{\PRL}{97}{131801}{2006}.
\bibitem{christenson1964} J.H.~Christenson {\it et al.} \Journal{\PRL}{13}{138}{1964}.
\bibitem{ktev2011} E.~Abouzaid {\it et al.} \Journal{\PRD}{83}{092001}{2011}.
\bibitem{na482007} A.~Lai {\it et al.} {\it Phys. Lett.} B 645 (2007) 26.
\bibitem{babar2001} B.~Aubert {\it et al.} \Journal{\PRL}{87}{091801}{2001}.
\bibitem{belle2001} K.~Abe {\it et al.} \Journal{\PRL}{87}{091802}{2001}.
\bibitem{bs0} R.~Aaij {\it et al.} \Journal{\PRL}{108}{201601}{2012}
\bibitem{aaij2012} R.~Aaij {\it et al.} \Journal{\PRL}{108}{111602}{2012}.
\bibitem{pauli1955} W.~Pauli, L.~Rosenfeld, and V.~Weisskopf, {\it Niels Bohr and the Development of Physics}, Pergamon (1955).
\bibitem{luders1957} G.~L\"{u}ders, {\it Ann. Phys.} 2 (1957) 1.
\bibitem{jost1957} R.~Jost, {\it Helv. Phys. Acta} 30 (1957) 409.
\bibitem{streater1964} R.F.~Streater and A.S.~Wightman, {\it PCT, Spin, Statistics and All That} (Benjamin, New York, 1964) and references therein.
\bibitem{greenberg2002} O.W.~Greenberg, \Journal{\PRL}{89}{231602}{2002}.
\bibitem{chaichian2011} M.~Chaichian, A.D.~Dolgov, V.A.~Novikov, and A.~Tureanu, {\it Phys. Lett.} B 699 (2011) 177.
\bibitem{Colladay:97} D.~Colladay and V.A.~Kosteleck\`y, {\it Phys. Rev.} D 55 (1997) 6760.
\bibitem{Colladay:PRD_58_116002} D.~Colladay and V.A.~Kosteleck\`y, {\it Phys. Rev.} D 58 (1998) 116002.
\bibitem{kostelecky2011} V.A.~Kosteleck\`y and N.~Russell, {\it Rev. Mod. Phys.} 83 (2011) 11.
\bibitem{mavromatos2005} N.E.~Mavromatos, {\it Lecture Notes in Physics} 669 (2005) 245.          
\bibitem{mavromatos2010} N.E.~Mavromatos, {\it Found. Phys.} 40 (2010) 917.
\bibitem{klinkhamer2004} F.R.~Klinkhamer and C.~Rupp, {\it Phys. Rev.} D 70 (2004) 045020.                                                            
\bibitem{murayama2001} H.~Murayama and T.~Yanagida, {\it Phys. Lett.} B 520 (2001) 263.
\bibitem{barenboim2002} G.~Barenboim, L.~Borissov, J.~Lykken, and A.Y.~Smirnov, {\it J. High Energy Phys.} 10 (2002) 1.
\bibitem{chaichian2012} M.~Chaichian, K.~Fujikawa, and A.~Tureanu, {\it Phys. Lett.} B 718 (2013) 1500.
\bibitem{Bamberger} A.~Bamberger {\it et al.}, {Phys. Lett.} 33B (1970) 233.                                    
\bibitem{Hu1975} E.~Hu {\it et al.}, {\it Nucl. Phys.} A 254 (1975) 403.                                               
\bibitem{Roberson} P.~Roberson {\it et al.}, {\it Phys. Rev.} C 16 (1977) 1945.                                       
\bibitem{Roberts} B.L.~Roberts {\it et al.}, {\it Phys. Rev.} D 17 (1978) 358.
\bibitem{gabrielse1986} G.~Gabrielse {\it et al.}, {\it Phys. Rev. Lett.} 57 (1986) 2504.                                                                              
\bibitem{gabrielse1989} G.~Gabrielse {\it et al.}, {\it Phys. Rev. Lett.} 63 (1989) 1360.                                                                              
\bibitem{gabrielse1990} G.~Gabrielse {\it et al.}, {\it Phys. Rev. Lett.} 65 (1990) 1317.                                                                              
\bibitem{gabrielse1995} G.~Gabrielse {\it et al.}, {\it Phys. Rev. Lett.} 74 (1995) 3544.                                                                             
\bibitem{gabrielse1999} G.~Gabrielse {\it et al.}, {\it Phys. Rev. Lett.} 82 (1999) 3198.                                                                             
\bibitem{Brown:1986} L.S.~Brown, G.~Gabrielse, {\it Rev. Mod. Phys.} 58 (1986) 233.                                        
\bibitem{thompson2004} J.K.~Thompson, S.~Rainville, and D.E.~Pritchard, {\it Nature} 430 (2004) 58.                                                     
\bibitem{sno2004} S.N.~Ahmed {\it et al.}, {\it Phys. Rev. Lett.} 92 (2004) 102004. 
\bibitem{bregman1978} M.~Bregman {\it et al.}, {\it Phys. Lett.} 78B (1978) 174.      
\bibitem{bell1979} M.~Bell {\it et al.}, {\it Phys. Lett.} 86B (1979) 215.                                          
\bibitem{autin1990} B.~Autin {\it et al.}, {\it Proceedings of the European Particle Accelerator Conference}, 1990.                                                                 
\bibitem{gear2000} S.~Geer {\it et al.}, {\it Phys. Rev.} D 62 (2000) 052004.                                     
\bibitem{borie} E.~Borie, {\it Phys. Rev.} A 28 (1983) 555.                                                           
\bibitem{bohnert} G.~Bohnert et al., {\it Phys. Lett.} B 174 (1986) 15.                                             
\bibitem{jpcrd} P.J.~Mohr, B.N.~Taylor, and D.B.~Newell, {\it Rev. Mod. Phys.} 84 (2012) 1527. 
\bibitem{Winkler:1972} P.F.~Winkler, D.~Kleppner, T.~Myint, and F.G.~Walther, {\it Phys. Rev.} A {5} (1972) 83.
\bibitem{Karshenboim:PLB566:2003} S.G.~Karshenboim and V.G.~Ivanov, {\it Phys. Lett.} B {566} (2003) 27.
\bibitem{kobayashi1992} M.~Kobayashi and A.I.~Sanda, \Journal{\PRL}{69}{3139}{1992}.
\bibitem{fee93} M.S.~Fee {\it et al.}, {\it Phys. Rev.} A 48 (1993) 192. 
\bibitem{lsnd2001} A.~Aguilar {\it et al.}, {\it Phys. Rev.} D 64 (2001) 112007.
\bibitem{miniboone2010} A.A.~Aguilar-Arevalo {\it et al.}, {\it Phys. Rev. Lett.} 105 (2010) 181801.
\bibitem{minos2011} P.~Adamson {\it et al.}, {\it Phys. Rev. Lett.} 107 (2011) 021801. 
\bibitem{minos2012} P.~Adamson {\it et al.}, {\it Phys. Rev. Lett.} 108 (2012) 191801.
\bibitem{Mohl:97} D.~M{\"o}hl, {\it Hyperfine Interact.} 109 (1997) 33.
\bibitem{Munger:1994} C.T.~Munger, S.J.~Brodsky, and I.~Schmidt, {\it Phys. Rev.} D {49} (1994) 3228.
\bibitem{Bertulani:1998} C.~A. Bertulani and G.~Baur, {\it Phys. Rev.} D {58} (1998) 034005.
\bibitem{Baur:1996} G.~Baur {\it et al.}, {\it Phys. Lett.} B {368} (1996) 251.
\bibitem{Blanford:1998} G.~Blanford {\it et al.}, {\it Phys. Rev. Lett.} {80} (1998) 3037.
\bibitem{Meshkov:1998} I.~N. Meshkov, {\it Phys. At. Nuclei} {61} (1998) 1679.
\bibitem{Blanford:Lamb:1998} G.~Blanford {\it et al.}, {\it Phys. Rev.} D 57 (1998) 6649.
\bibitem{Westig:EPJD57:2010} M.~P. Westig {\it et al.}, {\it Eur. Phys. J.} D {57} (2010) 27.

\bibitem{Gabrielse:IJMSIP:1989a} G.~Gabrielse, L.~Haarsma, and S.~L. Rolston, {\it Int. J. Mass Spectrom. Ion Processes}
 88 (1989) 319.
\bibitem{ONeil:1998} T.~M. O'{N}eil and D.H.E.~Dubin, {\it Phys. Plasmas} 5 (1998) 2163.
\bibitem{Amsler:HydrogenAtom:2001} C.~Amsler {\it et al.}, in S.~G. Karshenboim {\it et al.}, editors, {\it The Hydrogen Atom.
  {P}recision Physics of Simple Atomic Systems\/} (Springer, Berlin, 2001)
\bibitem{Gabrielse:Adv:01} G.~Gabrielse, {\it Adv. At. Mol. Opt. Phys.} 45 (2001) 1.
\bibitem{Kuroda:PRST:2012} N.~Kuroda {\it et al.}, {\it Phys. Rev. ST Accel. Beams} 15 (2012) 024702.
\bibitem{Kuroda:PRL:2008} N.~Kuroda {\it et al.} \Journal{\PRL}{100}{203402}{2008}.
\bibitem{Torii:LEAP:2005} H~A. Torii {\it et al.}, {\it AIP Conf. Proc.} {796} (2005) 413.
\bibitem{mills1986}  A. P.~Mills and E. M.~Gullikson, {\it Appl. Phys. Lett.} {49} (1986) 1121.
\bibitem{Murphy:PRA:1992} T.~J. Murphy and C.~M. Surko, {\it Phys. Rev.} A {46} (1992) 5696.
\bibitem{Surko:1997} C.~M. Surko, R.~G. Greaves, and M.~Charlton, {\it Hyperfine Interact.} {109} (1997) 181.
\bibitem{Surko:PP11:2004} C.~M. Surko and R.~G. Greaves, {\it Phys. Plasmas} {11} (2004) 2333.
\bibitem{Hollmann:2000} E.~M. Hollmann, F.~Anderegg, and C.~F. Driscoll, {\it Phys. Plasmas} {7} (2000) 2776.
\bibitem{ATHENA:Apparatus:2004} M.~Amoretti {\it et al.}, {\it Nucl. Instrum. Meth. Phys. Res.} A {518} (2004) 679.
\bibitem{ALPHA:JPCS262:2011} M.~Charlton {\it et al.}, {\it J. Phys. Conf. Ser.} {262} (2011) 012001.
\bibitem{comeau2012} D.~Comeau {\it et al.}, {\it New J. Phys.} {14} (2012) 045006.
\bibitem{Imao:HI194:2009} H.~Imao {\it et al.}, {\it Hyperfine Interact.} {194} (2009) 71.
\bibitem{Gabrielse:AMOP50:2005} G.~Gabrielse, {\it Adv. At. Mol. Opt. Phys.} {50} (2005) 155.
\bibitem{Neumann:1983} R.~Neumann, H.~Poth, A.~Winnacker, and A.~Wolf, {\it Z. Phys.} A {313} (1983) 253.
\bibitem{Bates:CaseStudies:1974} D.~R. Bates, {\it Case Studies in Atomic Physics} {4} (1975) 57.
\bibitem{Bell:1982} M.~Bell and J.~S. Bell, {\it Particle Accelerators} {12} (1982) 49.
\bibitem{Andersen:1990} L.~H. Andersen and J.~Bolko, {\it Phys. Rev.} A {42} (1990) 1184.
\bibitem{Schuessler:1995a} T.~Sch{\"{u}}ssler {\it et al.}, {\it Phys. Rev. Lett.} {75} (1995) 802.
\bibitem{Yousif:PRL:1991} F.~B. Yousif {\it et al.}, {\it Phys. Rev. Lett.} {67} (1991) 26.
\bibitem{Wolf:NATO:1992} A.~Wolf, in W.~G. Graham {\it et al.}, editors, {\it Recombination of Atomic
  Ions\/}, volume 296 of {\em NATO ASI Series B\/} (Plenum, New York, 1992) 209.
\bibitem{Wolf:97} A.~M{\"u}ller and A.~Wolf, {\it Hyperfine Interact.} {109} (1997) 233.
\bibitem{Noordam:2000} C.~Wesdorp, F.~Robicheaux, and L.~D. Noordam, {\it Phys. Rev. Lett.} {84} (2000) 3799.
\bibitem{Wesdorp:2001} C.~Wesdorp, F.~Robicheaux, and L.~D. Noordam, {\it Phys. Rev.} A {64} (2001) 033414.
\bibitem{Zeibel:2002} J.~G. Zeibel and R.~R. Jones, {\it Phys. Rev. Lett.} {89} (2002) 093204.
\bibitem{Gabrielse:1988b} G.~Gabrielse, S.~L. Rolston, L.~Haarsma, and W.~Kells, {\it Phys. Lett.} A {129} (1988) 38.
\bibitem{Makin:63} B.~Makin and J.~C. Keck, {\it Phys. Rev. Lett.} {11} (1963) 281.
\bibitem{Glinsky:91} M.~E. Glinsky and T.~M. O'Neil, {\it Phys. Fluids} B {3} (1991) 1279.
\bibitem{Menshikov:95} L.~I. Men'shikov and P.~O. Fedichev, {\it JETP} {81} (1995) 78.
\bibitem{Fedichev:1997} P.~O. Fedichev, {\it Phys. Lett.} A {226} (1997) 289.
\bibitem{Robicheaux:PRA73:2006} F.~Robicheaux, {\it Phys. Rev.} A {73} (2006) 033401.
\bibitem{Humberston:1987} J.~W. Humberston, M.~Charlton, F.~M. Jacobsen, and B.~I. Deutch, {\it J. Phys.} B {20} (1987) L25.
\bibitem{Merrison:1997} J.~P. Merrison {\it et al.}, {\it Phys. Rev. Lett.} {78} (1997) 2728.
\bibitem{Charlton:1990} M.~Charlton, {\it Phys. Lett.} A {143} (1990) 143.
\bibitem{Hessels:1998} E.~A. Hessels, D.~M. Homan, and M.~J. Cavagnero, {\it Phys. Rev.} A {57} (1998) 1668.
\bibitem{Speck:2004} A.~Speck, C.H. Storry, E.A. Hessels, and G.~Gabrielse, {\it Phys. Lett.} B {597} (2004) 257.
\bibitem{Athena:LaserStim:2006} M.~Amoretti {\it et al.}, {\it Phys. Rev. Lett.} {97} (2006) 213401.
\bibitem{Hall:1996} D.S.~Hall, G.~Gabrielse, {\it Phys. Rev. Lett.} {77} (1996) 1962.
\bibitem{Ordonez:1997} C.~A. Ordonez, {\it Phys. Plasmas} {4} (1997) 2313.
\bibitem{ATRAP:HbarStates:2002} G.~Gabrielse {\it et al.}, {\it Phys. Rev. Lett.} {89} (2002) 233401.
\bibitem{ALPHA:Autoresonant:2011} G.~B. Andresen {\it et al.}, {\it Phys. Rev. Lett.} {106} (2011) 025002.
\bibitem{Gabrielse:PLB:2001} G.~Gabrielse {\it et al.}, {\it Phys. Lett.} B 507 (2001) 1.
\bibitem{Li:89} G.-Z. Li, {\it Commun. Theor. Phys.} 12 (1989) 355.
\bibitem{Walz:95} J.~Walz {\it et al.}, {\it Phys. Rev. Lett.} {75} (1995) 3257.
\bibitem{Dehmelt:PhysScr:1995} H.~Dehmelt, {\it Physica Scripta} T59 (1995) 423.
\bibitem{Widmann:LoI:2003} E.~Widmann {\it et al.}, CERN/SPSC 2003-009; SPSC-I-226; February 25, 2003
\bibitem{Bluemel:89} R.~Bl{\"{u}}mel, C.~Kappler, W.~Quint, and H.~Walther, {\it Phys. Rev.} A {40} (1989) 808.
\bibitem{ATHENA:Diagnostic:2003} M.~Amoretti {\it et al.}, {\it Phys. Plasmas} {10} (2003) 3056.
\bibitem{ATHENA:PRL:2003} M.~Amoretti {\it et al.}, {\it Phys. Rev. Lett.} {91} (2003) 055001.
\bibitem{ATRAP:ApertureMethod:2004} P.~Oxley {\it et al.} {\it Phys. Lett.} B {595} (2004) 60.
\bibitem{ATHENA:Imaging:2004} M.~C. Fujiwara {\it et al.}, {\it Phys. Rev. Lett.} {92} (2004) 065005.
\bibitem{ATHENA:Dynamics:2004} M.~Amoretti {\it et al.}, {\it Phys. Lett.} B {590} (2004) 133.
\bibitem{ATHENA:HighRate:2004} M.~Amoretti {\it et al.}, {\it Phys. Lett.} B {578} (2004) 23--32
\bibitem{ATHENA:temperature:2004} M.~Amoretti {\it et al.}, {\it Phys. Lett.} B {583} (2004) 59--67
\bibitem{ATHENA:Fujiwara:PRL:2008} M.~C. Fujiwara {\it et al.}, {\it Phys. Rev. Lett.} {101} (2008) 053401.
\bibitem{Robicheaux:JPB41:2008} F.~Robicheaux, {\it J. Phys.} B {41} (2008) 192001.
\bibitem{Jonsell:JPB42:2009} S.~Jonsell, D.~P. van~der Werf, M.~Charlton, and F.~Robicheaux, {\it J. Phys.} B {42} (2009) 215002.
\bibitem{ATRAP:Velocity:2004} G.~Gabrielse {\it et al.}, {\it Phys. Rev. Lett.} {93} (2004) 073401.
\bibitem{Pohl:PRL:2006} T.~Pohl, H.~R. Sadeghpour, and G.~Gabrielse, {\it Phys. Rev. Lett.} {97} (2006) 143401.
\bibitem{zurlo} N.~Zurlo {\it et al.}, {\it Phys. Rev. Lett.} {97} (2006) 153401.
\bibitem{Parthey:PRL107:2011} C.G.~Parthey {\it et al.}, {\it Phys. Rev. Lett.} {107} (2011) 203001.
\bibitem{Walraven:1982} J.~T.~M. Walraven and I.F.~Silvera, {\it Rev. Sci. Instrum.} {53} (1982) 1167.
\bibitem{Bergeman:1987} T.~Bergeman, G.~Erez, and H.J.~Metcalf, {\it Phys. Rev.} A {35} (1987) 1535.
\bibitem{Hess:1987} H.F.~Hess, {\it et al.}, {\it Phys. Rev. Lett.} {59} (1987) 672.
\bibitem{Roijen:1988} R.~van Roijen, J.~J. Berkhout, S.~Jaakkola, and J.~T.~M. Walraven, {\it Phys. Rev. Lett.} {61} (1988) 931.
\bibitem{Fried:1998} D.G.~Fried {\it et al.}, {\it Phys. Rev. Lett.} {81} (1998) 3811.
\bibitem{Mohri:2003} A.~Mohri and Y.~Yamazaki, {\it Europhys. Lett.} {63} (2003) 207.
\bibitem{Raizen:Coilgun:2009} E.~Narevicius {\it et al.} {\it Phys. Rev. Lett.} {100} (2008) 093003.
\bibitem{ONeil:PF23:22161980} T.~M. O'Neil, {\it Phys. Fluids} {23} (1980) 2216.
\bibitem{Gilson:2003} E.~P. Gilson and J.~Fajans, {\it Phys. Rev. Lett.} {90} (2003) 015001.
\bibitem{Fajans:2005} J.~Fajans, W.~Bertsche, K.~Burke, S.~F. Chapman, and D.~P. van~der Werf, {\it Phys. Rev. Lett.} {95} (2005) 155001.
\bibitem{Fajans:2004} J.~Fajans and A.~Schmidt, {\it Nucl. Instrum. Meth. Phys. Res.} A {521} (2004) 318.
\bibitem{Dubin:2001} D.H.E.~Dubin, {\it Phys. Plasmas} {8} (2001) 4331.
\bibitem{Squires:2001} T.M.~Squires, P.~Yesley, and G.~Gabrielse, {\it Phys. Rev. Lett.} {86} (2001) 5266.
\bibitem{ALPHA:2007a} G.~Andresen {\it et al.}, {\it Phys. Rev. Lett.} {98} (2007) 023402.
\bibitem{Gabrielse:PenningIoffe:2007} G.~Gabrielse {\it et al.}, {\it Phys. Rev. Lett.} {98} (2007) 113002.
\bibitem{Andresen:JPB41:2008} G.~B. Andresen {\it et al.}, {\it J. Phys.} B {41} (2008) 011001.
\bibitem{PenningIoffe:PRL:2008} G.~Gabrielse {\it et al.}, {\it Phys. Rev. Lett.} {100} (2008) 113001.
\bibitem{Mohri:AIP793:2005} A.~Mohri, Y.~Kanai, Y.~Nakai, and Y.~Yamazaki, {\it AIP Conf. Proc.} {793} (2005) 147.
\bibitem{Pohl:PRL97:2006} T.~Pohl, H.~R. Sadeghpour, Y.~Nagata, and Y.~Yamazaki, {\it Phys. Rev. Lett.} {97} (2006) 213001.
\bibitem{Taylor:JPB4945:2006} C.~L. Taylor, J.~Zhang, and F.~Robicheaux, {\it J. Phys.} B {39} (2006) 4945.
\bibitem{hellwig1970} H.~Hellwig {\it et al.}, {\it IEEE Trans. Instrum. Meas.} IM-19 (1970) 200. 
\bibitem{essen1971} L.~Essen, M.J.~Donaldson, M.J.~Bangham, and E.G.~Hope, {\it Nature} 229 (1971) 110. 
\bibitem{Brodsky:2005} S.J.~Brodsky, C.E.~Carlson, J.R.~Hiller, and D.S.~Hwang, {\it Phys. Rev. Lett.} {94} (2005) 022001.
\bibitem{cjp:2006:429} C.E.~Carlson, {\it Can. J. Phys.} {85} (2007) 429.
\bibitem{yamazaki2002} T.~Yamazaki {\it et al.}, \Journal{\it Phys. Reports}{366}{183}{2002}.
\bibitem{iwasaki1991} M.~Iwasaki {\it et al.}, {\it Phys. Rev. Lett.} 67 (1991) 1246.
\bibitem{briggs1999} J.S.~Briggs, P.T.~Greenland, E.A.~Solov'ev, {\it Hyperfine Int.} {119} (1999) 235.
\bibitem{cohen2004} J.S.~Cohen, {\it Rep. Prog. Phys.} {67} (2004) 1769.
\bibitem{hesse} M.~Hesse, A.T.~Le, C.D.~Lin, {\it Phys. Rev.} A {69} (2004) 052712.
\bibitem{ovchinnikov} S.Yu.~Ovchinnikov and J.H.~Macek, {\it Phys. Rev.} A {71} (2005) 052717.
\bibitem{tong08} X.M.~Tong, K.~Hino, N.~Toshima, {\it Phys. Rev. Lett.} {101} (2008) 163201.
\bibitem{genkin} M.~Genkin and E.~Lindroth, {\it Eur. Phys. J.} D {51} (2009) 205.
\bibitem{sakimoto2007} K.~Sakimoto, \Journal{\PRA}{76}{042513}{2007}.
\bibitem{sakimoto2009} K.~Sakimoto, \Journal{\PRA}{79}{042508}{2009}.
\bibitem{kor97} V.I.~Korobov and I.~Shimamura, \Journal{\PRA}{56}{4587}{1997}.
\bibitem{revai97} J.~R\'evai and A.T.~Kruppa, \Journal{\PRA}{57}{174}{1998}.
\bibitem{kartavtsev2000} O.I.~Kartavtsev, D.E.~Monakhov, and S.I.~Fedotov, \Journal{\PRA}{61}{062507}{2000}.
\bibitem{yamaguchi2002} H.~Yamaguchi {\it et al.}, \Journal{\PRA}{66}{022504}{2002}.
\bibitem{yamaguchi2004} H.~Yamaguchi {\it et al.}, \Journal{\PRA}{70}{012501}{2004}.
\bibitem{mhori1998} M.~Hori {\it et al.}, \Journal{\PRA}{57}{1698}{1998}; \Journal{\PRA}{58}{1612}{1998}. 
\bibitem{mhori2004} M.~Hori {\it et al.}, \Journal{\PRA}{70}{012504}{2004}.
\bibitem{tokesi} K.~T\"ok\'esi, B.~Juh\'asz, and J.~Burgd\"orfer, {\it J. Phys.} B {38} (2005) S401.
\bibitem{revai06} J.~R\'evai and N.~Shevchenko, {\it Euro. Phys. J.} D {37} (2006) 83.
\bibitem{mhori2002} M.~Hori {\it et al.}, \Journal{\PRL}{89}{093401}{2002}.
\bibitem{Shimamura:92} I.~Shimamura, {\it Phys. Rev.} A {46} (1992) 3776.
\bibitem{Yamazaki:92} T.~Yamazaki and K.~Ohtsuki, {\it Phys. Rev.} A {45} (1992) 7782.
\bibitem{Greenland:93} P.T.~Greenland and R.~Th{\"u}rw{\"a}chter, {\it Hyperfine Interact.} {76} (1993) 355.
\bibitem{korobov1996} V.I.~Korobov, \Journal{\PRA}{54}{R1749}{1996}.
\bibitem{Kartavtsev:96} O.I.~Kartavtsev, {\it Phys. At. Nucl.} {59} (1996) 1483.
\bibitem{Elander:97} N.~Elander and E.~Yarevsky, {\it Phys. Rev.} A {56} (1997) 1855.
\bibitem{Korobov:97a} V.I.~Korobov and D.D.~Bakalov, {\it Phys. Rev.} A {79} (1997) 3379.
\bibitem{andersson1998} S.~Andersson, N.~Elander, and E.~Yarevsky, \Journal{\it J. Phys. {\rm B}}{31}{625}{1998}.
\bibitem{Kino:99} Y.~Kino, M.~Kamimura, and H.~Kudo, {\it Hyperfine Interact.} {119} (1999) 201.
\bibitem{Korobov:99c} V.I.~Korobov, D.~Bakalov, and H.J.~Monkhorst, {\it Phys. Rev.} A {59} (1999) R919.
\bibitem{kino2001hpc} Y.~Kino {\it et al.}, {\it Hyperfine Interact.} {138} (2001) 179. 
\bibitem{Korobov:00} V.I.~Korobov, \Journal{\PRA}{61}{064503}{2000}.
\bibitem{Korobov:03} V.I.~Korobov, \Journal{\PRA}{67}{062501}{2003}.
\bibitem{kino2004} Y.~Kino, M.~Kamimura, and H.~Kudo, \Journal{\it Nucl. Instrum. Meth. Phys. Res. {\rm B}}{214}{84}{2004}.
\bibitem{goldman1998} S.P.~Goldman, \Journal{\PRA}{57}{R677}{1998}.
\bibitem{drake1999} G.W.F.~Drake, {\it Phys. Scr.} {T83} (1999) 83.
\bibitem{korobovprivate} V.I.~Korobov, {\it private communication}.
\bibitem{codata2002} P.J.~Mohr and B.N.~Taylor, {\it Rev. Mod. Phys.} 77 (2005) 1.
\bibitem{hannemann} S.~Hannemann {\it et al.}, \Journal{\PRA}{74}{062514}{2006}.
\bibitem{hilico} L.~Hilico, N.~Billy, B.~Gr\'emaud, and D.~Delande, \Journal{\it J. Phys. {\rm B}}{34}{1}{2001}.
\bibitem{koelemeij} J.C.J.~Koelemeij {\it et al.}, {\it Phys. Rev. Lett.} {98} (2007) 173002.
\bibitem{Mor93} N.~Morita {\it et al.}, {\it Nucl. Instrum. Meth. Phys. Res.} A {330} (1993) 439.
\bibitem{Mor94} N.~Morita {\it et al.}, \Journal{\PRL}{72}{1180}{1994}.
\bibitem{mhorioptics} M.~Hori, R.S.~Hayano, E.~Widmann, and H.A.~Torii, \Journal{\it Opt. Lett.}{28}{2479}{2003}.  
\bibitem{mhori2005} M.~Hori {\it et al.}, \Journal{\PRL}{94}{063401}{2005}.
\bibitem{cherenkov} M.~Hori, \Journal{\it Nucl. Instrum. Meth. Phys. Res. {\rm A}}{496}{102}{2003}.
\bibitem{mhori2010} M.~Hori and V.I.~Korobov, \Journal{\PRA}{81}{062508}{2010}.
\bibitem{chirp2009} M.~Hori and A.~Dax, \Journal{\it Opt. Lett.}{34}{1273}{2009}.
\bibitem{udem} Th.~Udem, R.~Holzwarth, T.W.~H\"ansch, \Journal{\it Nature}{416}{233}{2002}.    
\bibitem{eikema1997} K.S.E.~Eikema, W.~Ubachs, W.~Vassen, and W.~Hogervorst, \Journal{\PRA}{55}{1866}{1997}.
\bibitem{meyer} V.~Meyer {\it et al.}, \Journal{\PRL}{84}{1136}{2000}.                                                             
\bibitem{fendel} P.~Fendel, S.D.~Bergeson, Th.~Udem, T.W.~H\"ansch, \Journal{\it Opt. Lett.}{32}{701}{2007}.
\bibitem{horiphoto} M.~Hori, \Journal{\it Rev. Sci. Instrum.}{76}{113303}{2005}.
\bibitem{parallel} M.~Hori, \Journal{\it Nucl. Instrum. Meth. Phys. Res. {\rm A}}{522}{420}{2004}.
\bibitem{Haas2006} M.~Haas {\it et al.}, \Journal{\PRA}{73}{052501}{2006}.
\bibitem{farnham} D.L.~Farnham, R.S.~Van Dyck Jr., and P.B.~Schwinberg, \Journal{\PRL}{75}{3598}{1995}.
\bibitem{beier} T.~Beier, {\it et al.}, \Journal{\PRL}{88}{011603}{2002}.
\bibitem{verdu} J.~Verd\'{u}, {\it et al.} \Journal{\PRL}{92}{093002}{2004}. 
\bibitem{hughes} R.J.~Hughes, B.I.~Deutch, \Journal{\it Phys. Rev. Lett.}{69}{578}{1992}.                                    
\bibitem{widmann2002} E.~Widmann {\it et al.}, \Journal{\PRL}{89}{243402}{2002}.
\bibitem{sakaguchi2004} J.~Sakaguchi {\it et al.}, \Journal{\it Nucl. Instrum. Meth. Phys. Res. {\rm A}}{533}{598}{2004}. 
\bibitem{pask2008} T.~Pask {\it et al.}, {\it J. Phys.} B 41 (2008) 081008.
\bibitem{bakahfs} D.~Bakalov and V.I.~Korobov, \Journal{\PRA}{57}{1662}{1998}.
\bibitem{yam01} N.~Yamanaka, Y.~Kino, H.~Kudo, and M.~Kamimura, {\it Phys. Rev.} A {63} (2001) 012518.
\bibitem{kinohfs} Y.~Kino, N.~Yamanaka, M.~Kamimura, and H.~Kudo, {\it Hyperfine Interact.} {146/147} (2003) 331.
\bibitem{Kor06} V.I.~Korobov, {\it Phys.\ Rev.} A {73} (2006) 022509.
\bibitem{korobov2009} V.I.~Korobov and Z.X.~Zhong, {\it Phys. Rev.} A {80} (2009) 042506. 
\bibitem{friedreich11} S.~Friedreich {\it et al.}, \Journal{\it Phys. Lett. {\rm B}}{700}{1}{2011}.  
\bibitem{obreshkov2004} B.D.~Obreshkov, D.D.~Bakalov, B.~Lepetit, and K.~Szalewicz, {\it Phys. Rev.} A {69} (2004) 042701.
\bibitem{korenman07} G.Ya.~Korenman and S.N.~Yudin, {\it J. Phys. Conf. Series} {88} (2007) 012060.
\bibitem{juhasz2002} B.~Juhasz {\it et al.}, \Journal{\it Eur. Phys. J. {\rm D}}{18}{261}{2002}.
\bibitem{juhasz2003} B.~Juhasz {\it et al.}, \Journal{\it Chem. Phys. Lett.}{379}{91}{2003}.
\bibitem{juhasz2006} B.~Juhasz {\it et al.}, \Journal{\it Chem. Phys. Lett.}{427}{246}{2006}.

\bibitem{DiSciacca:Proton:2012} J.~Di{S}ciacca and G.~Gabrielse, {\it Phys. Rev. Lett.} {108} (2012) 153001.
\bibitem{Dehmelt:PNAS:1986a} H.~Dehmelt, {\it Proc. Natl. Acad. Sci. USA} {83} (1986) 2291.
\bibitem{odom} B.~Odom, D.~Hanneke, B.D'Urso, and G.~Gabrielse, {\it Phys. Rev. Lett.} {97} (2006) 030801.
\bibitem{hanneke} D.~Hanneke, S.~Fogwell, and G.~Gabrielse, {\it Phys. Rev. Lett.} {100} (2008) 120801.
\bibitem{Ulmer:ProtonSpinFlips:2011} S.~Ulmer {\it et al.},  {\it Phys. Rev. Lett.} {106} (2011) 253001.
\bibitem{Rodegheri:NJP:2012} C.~C. Rodegheri {\it et al.}, {\it New. J. Phys.} {14}  (2012) 063011.


\bibitem{fermi1947} E.~Fermi and E.~Teller, {\it Phys. Rev.} 72 (1947) 399.                              
\bibitem{golser1991} R.~Golser and D.~Semrad, {\it Phys. Rev. Lett.} 66 (1991) 1831.
\bibitem{schief1993} A.~Schieferm\"uller {\it et al.}, {\it Phys. Rev.} A 48 (1993) 4467.
\bibitem{andersen2002} H.H.~Andersen {\it et al.}, {\it Nucl. Instrum. Meth. Phys. Res.} B 194 (2002) 217.
\bibitem{arista2002} N.R.~Arista and A.F.~Lifschitz, {\it Nucl. Instrum. Methods Phys. Res.} B 193 (2002) 8.
\bibitem{sorensen1990} A.H.~S\o rensen, {\it Nucl. Instrum. Methods Phys. Res.} B 48 (1990) 10.
\bibitem{mikkelsen1989} H.H.~Mikkelsen and P.~Sigmund, {\it Phys. Rev.} A 40 (1989) 101.
\bibitem{sigmund2001} P.~Sigmund and A.~Schinner, {\it Eur. Phys. J.} D 15 (2001) 165.
\bibitem{barkas1956} W.H.~Barkas, W.~Birnbaum, and F. M.~Smith, {\it Phys. Rev.} 101 (1956) 778.
\bibitem{eder1997} K.~Eder {\it et al.}, {\it Phys. Rev. Lett.} 79 (1997) 4112.
\bibitem{knudsen2009} H.~Knudsen {\it et al.}, \Journal{\it Nucl. Instrum. Meth. Phys. Res. {\rm B}}{267}{244}{2009}.
\bibitem{knudsen2011} T.~Kirchner and H.~Knudsen, \Journal{\it J. Phys. {\rm B}}{44}{122001}{2011}.
\bibitem{higaki2002} H.~Higaki {\it et al.}, \Journal{\it Phys Rev. {\rm E}}{65}{046410}{2002}.
\bibitem{yoshiki2003} K.~Yoshiki Franzen {\it et al.}, \Journal{\it Rev. Sci. Instrum.}{74}{3305}{2003}.
\bibitem{andersen1990} L.H.~Andersen {\it et al.}, {\it Phys. Rev.} A 41 (1990) 6536.
\bibitem{hvelplund1994} P.~Hvelplund {\it et al.}, {\it J. Phys.} B 27 (1994) 925.
\bibitem{wehrman1996} L. A.~Wehrman {\it et al.}, {\it J. Phys.} B 29 (1996) 5831.
\bibitem{reading1997} J. F.~Reading {\it et al.}, {\it J. Phys.} B 30 (1997) L189.
\bibitem{bent1998} G.~Bent, P.S.~Krsti\'c, and D.R. Schultz, {\it J. Chem. Phys.} 108 (1998) 1459.
\bibitem{lee2000} T. G. Lee, H.C.~Tseng, and C.D.~Lin, {\it Phys. Rev.} A 61 (2000) 062713.
\bibitem{igarashi2000} A.~Igarashi, A.~Ohsaki, and S.~Nakazaki, {\it Phys. Rev.} A 62 (2000) 052722.
\bibitem{igarashi2004} A.~Igarashi, S.~Nakazaki, and A.~Ohsaki, {\it Nucl. Instrum. Meth. Phys. Res.} B 214 (2004) 135.
\bibitem{sahoo2005} S. Sahoo, S.C.~Mukherjee, and H.R.J.~Walters, {\it Nucl. Instrum. Meth. Phys. Res.} B 233 (2005) 318.
\bibitem{schultz2003} D.R.~Schultz and P.S.~Krsti\'c, {\it Phys. Rev.} A 67 (2003) 022712.
\bibitem{tong2002} X.-M.~Tong {\it et al.}, {\it Phys. Rev.} A 66 (2002) 032709.
\bibitem{kirchner2002} T. Kirchner {\it et al.}, {\it J. Phys.} B 35 (2002) 925.
\bibitem{keim2003} M. Keim {\it et al.}, {\it Phys. Rev.} A 67 (2003) 062711.
\bibitem{foster2008} M. Foster, J.~Colgan, M.S.~Pindzola, {\it Phys. Rev. Lett.} 100 (2008) 033201.
\bibitem{luhr2008} A.~L\"uhr and A. Saenz, {\it Phys. Rev.} A 77 (2008) 052713.
\bibitem{cohen1997} J. S.~Cohen, {\it Phys. Rev.} A 56 (1997) 3583.
\bibitem{luhr2010} A.~L\"uhr and A.~Saenz, {\it Phys. Rev.} A 81 (2010) 010701.
\bibitem{iazzi2000} F.~Iazzi {\it et al.}, {\it Phys. Lett.} B 475 (2000) 378.
\bibitem{batty2001} C.J.~Batty, E.~Friedman, A.~Gal, {\it Nucl. Phys.} A 689 (2001) 721.
\bibitem{bruckner1990} W. Br\"uckner {\it et al.}, {\it Z. Phys.} A 335 (1990) 217.
\bibitem{bertin1996} A. Bertin {\it et al.}, {\it Phys. Lett.} B 369 (1996) 77.
\bibitem{benedettini1997} A. Benedettini {\it et al.}, {\it Nucl. Phys. {\rm B} (Proc. Suppl.)} 56 (1997) 58.
\bibitem{zenoni1999} A. Zenoni {\it et al.}, {\it Phys. Lett.} B 461 (1999) 405; {\it Phys. Lett.} B 461 (1999) 413.
\bibitem{kalogeropoulous1980} T.E. Kalogeropoulos and G.S. Tzanakos, {\it Phys. Rev.} D 22 (1980) 2585.
\bibitem{bizzarri1974} R. Bizzarri {\it et al.}, {\it Nuovo Cim.} A 22 (1974) 225.
\bibitem{balestra1989} F. Balestra {\it et al.}, {\it Phys. Lett.} B 230 (1989) 36.
\bibitem{balestra1984} F. Balestra {\it et al.}, {\it Phys. Lett.} B 149 (1984) 69.
\bibitem{bianconi2000} A. Bianconi {\it et al.}, {\it Phys. Lett.} B 492 (2000) 254.
\bibitem{corradini2013} M. Corradini {\it et al.}, {\it Nucl. Instrum. Methods Phys. Res} A (2013).
\bibitem{nakamura1984} K. Nakamura {\it et al.}, {\it Phys. Rev. Lett.} 52 (1984) 731.
\bibitem{bianconi2000a} A. Bianconi {\it et al.}, {\it Phys. Lett.} B 481 (2000) 194.
\bibitem{balestra1986} F. Balestra {\it et al.}, {\it Nucl. Phys.} A 452 (1986) 573.
\bibitem{ashford1985} V. Ashford {\it et al.}, {\it Phys. Rev.} C 31 (1985) 663.
\bibitem{garreta1984} D. Garreta {\it et al.}, {\it Phys. Lett.} B 149 (1984) 64.
\bibitem{todoroki2012} K. Todoroki and M. Hori {\it et al.}, {\it J. Inst.} 7 (2012) C02052.
\bibitem{aghai2012} H.~Aghai-Khozani {\it et al.}, {Eur. Phys. J. Plus} 127 (2012) 125.
\bibitem{knudsen2008a} H.V.~Knudsen {\it et al.}, {\it Nucl. Instrum. Meth. Phys. Res.} B 266 (2008) 530. 
\bibitem{suit2003} H.~Suit {\it et al.}, {\it Acta Oncol.} 42 (2003) 800.
\bibitem{mazeron2004} J.J.~Mazeron {\it et al.}, {\it Radiother. Oncol.} 73 (2004) 50.
\bibitem{gray1984} L.~Gray and T.E.~Kalogeropoulous, {\it Radiat. Res.} 97 (1984) 246.
\bibitem{sullivan1985} A.H.~Sullivan, {\it Phys. Med. Biol.} 30 (1985) 1297.
\bibitem{levegrun2002} S.~Levegr\"un {\it et al.}, {\it Radiother. Oncol.} 63 (2002) 11.
\bibitem{oelert2012} W.~Oelert {\it et al.}, {\it Hyperfine Interact.} 213 (2012) 227.
\bibitem{Haensch:1975} T.~W. H{\"{a}}nsch, S.~A. Lee, R.~Wallenstein, and C.~Wieman, 
{\it Phys. Rev. Lett.} {34} (1975) 307.
\bibitem{Bethe:1977} H.A.~Bethe and E.E.~Salpeter: {\it Quantum Mechnanics of One- and
  Two-Electron Atoms\/}, paperback edition (Plenum, New York, 1977).
\bibitem{Fischer:2004} M.~Fischer {\it et al.}, {\it Phys. Rev. Lett.} {92} (2004) 230802.
\bibitem{Biraben:LaserPhysicsLimits:2002}
F.~Biraben and L.~Julien, in H.~Figger, D.~Meschede, and C.~Zimmermann, editors, {\it Laser
  Physics at the Limits\/} (Springer, Berlin, 2002).
\bibitem{Weitz:1992} M.~Weitz, F.~Schmidt-Kaler, and T.~W. H{\"{a}}nsch, {\it Phys. Rev. Lett.} {68}  (1992) 1120.
\bibitem{Udem:1997} Th. Udem {\it et al.}, {\it Phys. Rev. Lett.} {79} (1997) 2646.
\bibitem{Huber:1998} A.~Huber {\it et al.}, {\it Phys. Rev. Lett.} {80} (1998) 468.
\bibitem{Parthey:PRL104:2010} C.G.~Parthey {\it et al.}, {\it Phys. Rev. Lett.} {104} (2010) 233001.
\bibitem{Pachucki:1996e} K.~Pachucki {\it et al.}, {\it J. Phys.} B {29} (1996) 177.
\bibitem{Cesar:1996} C.L.~Cesar {\it et al.}, {\it Phys. Rev.  Lett.} {77} (1996) 255.
\bibitem{Midgall:1985} A.~L. Midgall {\it et al.}, {\it Phys. Rev.  Lett.} {54} (1985) 2596.
\bibitem{Greytak:1984} T.J.~Greytak and D.~Kleppner, in G.~Grynberg and R.~Stora, editors, {\it Les Houches, Session
  {XXXVIII}, 1982 --- {\it Tendances actuelles en physique atomique / New trends in atomic physics}\/} (Elsevier, 1984).
\bibitem{Hess:1986} H.F.~Hess, {\it Phys. Rev.} B {34} (1986) 3476.
\bibitem{Ketterle:1996} W.~Ketterle and N.~J. Van~Druten, {\it Adv. At. Mol. Opt. Phys.} {37} (1996) 181.
\bibitem{Killian:1998} T.C.~Killian {\it et al.}, {\it Phys. Rev. Lett.} {81} (1998) 3807.
\bibitem{Hijmans:1989} T.~W. Hijmans {\it et al.}, {\it J. Opt. Soc.  Am.} B {6} (1989) 2235.
\bibitem{Walz:Varenna:2009} J.~Walz in A.~Dupasquier, A.~P.  Mills, Jr., and R.~S. Brusa, editors, {\it Proceedings of the International School of Physics ``Enrico Fermi,'' Course CLXXIV. Varenna 7--17 July 2009:
  Physics with Many Positrons\/} (Societ{\'{a}} Italiana di Fisica, Bologna, 2010).
\bibitem{Haensch:1993:47} T.~W. H{\"{a}}nsch and C.~Zimmermann, {\it Hyperfine Interact.} {76} (1993) 47.
\bibitem{Cesar:2001:1S2S} C.L.~Cesar, {\it Phys. Rev.} A {64} (2001) 023418.
\bibitem{Walraven:1993:205} J.~T.~M. Walraven, {\it Hyperfine Interact.}  {76} (1993) 205.
\bibitem{Phillips:1993} W.~D. Phillips {\it et al.}, {\it Hyperfine Interact.} {76} (1993) 265.
\bibitem{Ertmer:1988} W.~Ertmer and H.~Wallis, {\it Hyperfine Interact.} {44} (1988) 319.
\bibitem{Lett:1988} P.~D. Lett, P.~L. Gould, and W.~D. Phillips, {\it Hyperfine Interact.} {44} (1988) 335.
\bibitem{Allegrini:1993} M.~Allegrini and E.~Arimondo, {\it Phys. Lett.} A {172} (1993) 271.
\bibitem{Zehnle:2001} V.~Zehnl{\'{e}} and J.C.~Garreau, {\it Phys. Rev.} A {63} (2001) 021402.
\bibitem{Setija:1993} I.D.~Setija {\it et al.}, {\it Phys. Rev. Lett.} {70} (1993) 2257.
\bibitem{Vidal:FWM:1992} C.R.~Vidal, in L.F.~Mollenauer, J.C.~White, and C.R.~Pollock, editors, {\it Tunable
  Lasers\/}, volume~59 of {\em Topics in Applied Physics\/}, 2 edition, chapter~3 (Springer, Berlin, 1992) 57.
\bibitem{Mahon:1978} R.~Mahon, T.~J. {McI}lrath, and D.W.~Koopman, {\it Appl. Phys. Lett.} {33} (1978) 305.
\bibitem{Cotter:1979a} D.~Cotter, {\it Opt. Commun.} {31} (1979) 397.
\bibitem{Wallenstein:1980} R.~Wallenstein, {\it Opt. Commun.} {33} (1980) 119.
\bibitem{Cabaret:1987} L.~Cabaret, C.~Delsart, and C.~Blondel, {\it Opt. Commun.} {61} (1987) 116.
\bibitem{Marangos:90} J.~P. Marangos {\it et al.}, {\it J. Opt. Soc.  Am.} B {7} (1990) 1254.
\bibitem{Luiten:1994} O.~J. Luiten {\it et al.}, {\it Appl. Phys.} B {59} (1994) 311.
\bibitem{Smith:1987a} A.~V. Smith and W.~J. Alford, {\it J. Opt. Soc. Am.} B {4} (1987) 1765.
\bibitem{Eikema:99} K.~S.~E. Eikema, J.~Walz, and T.~W. H{\"a}nsch, {\it Phys. Rev. Lett.} {83} (1999) 3828.
\bibitem{Eikema:PRL:01} K.~S.~E. Eikema, J.~Walz, and T.~W. H{\"{a}}nsch, {\it Phys. Rev. Lett.} {86} (2001) 5679.
\bibitem{Pahl:2005} A.~Pahl {\it et al.}, {\it Laser Physics} {15} (2005) 46.
\bibitem{Scheid:OL:2007} M.~Scheid {\it et al.}, {\it Opt. Lett.} {32} (2007) 955.
\bibitem{Markert:OE:2007} F.~Markert, M.~Scheid, D.~Kolbe, and J.~Walz, {\it Optics Express} {15} (2007) 14476.
\bibitem{Scheid:oe:2009} M.~Scheid {\it et al.}, {\it Optics Express} {17} (2009) 11276.
\bibitem{ashkezari2011} M.D.~Ashkezari {\it et al.}, {\it Hyperfine Interact.}  212 (2012) 81.
\bibitem{hardy1979} W.N.~Hardy, A.J.~Berlinsky, and L.A.~Whitehead, {\it Phys. Rev. Lett.} 42 (1979) 1042.
\bibitem{Scherk:1979} J.~Scherk, {\it Phys. Lett.} {88B} (1979) 265.
\bibitem{Chardin:HI109:83:1997} G.~Chardin, {\it Hyperfine Interact.} {109} (1997) 83.
\bibitem{Nieto:91} M.~M. Nieto and T.~Goldman, {\it Phys. Rep.} {205} (1991)  221.
\bibitem{Nieto:92} M.~M. Nieto and T.~Goldman, {\it Phys. Rep.} {216} (1992) 343.
\bibitem{Darling:1992} T.~W. Darling, F.~Rossi, G.~I. Opat, and G.~F. Moorhead, {\it Rev. Mod. Phys.} {64} (1992) 237.
\bibitem{Peters:1999} A.~Peters, K.Y.~Chung, and S.~Chu, {\it Nature} {400} (1999) 849.
\bibitem{Poli:PRL106:2011} N.~Poli {\it et al.}, {\it Phys. Rev. Lett.} {106} (2011) 038501.
\bibitem{AEGIS:NIM:B266:351:2008} A.~Kellerbauer {\it et al.}, {\it Nucl. Instrum. Meth. Phys. Res.} B {266} (2008) 351.
\bibitem{AEGIS:AIPCP1037:2008} G.~Testera {\it et al.}, {\it AIP Conf. Proc.} {1037} (2008) 5.
\bibitem{AEGIS:CJP89:2011} R.~Ferragut {\it et al.}, {\it Can. J. Phys.} {89} (2011) 17.
\bibitem{Cialdi:NIMB269:2011} S.~Cialdi {\it et al.}, {\it Nucl. Instrum. Meth. Phys. Res.} B {269} (2011) 1527.
\bibitem{Perez:NIM:A545:2005} P.~P{\'{e}}rez and A.~Rosowsky, {\it Nucl. Instrum. Meth. Phys. Res.} A {545} (2005) 20.
\bibitem{Walz:GRG:2004} J.~Walz and T.W.~H{\"{a}}nsch, {\it Gen. Rel. Grav.} {36} (2004) 561.
\bibitem{Perez:CP1037:35:2008} P.~P{\'{e}}rez {\it et al.}, {\it AIP Conf. Proc.} {1037} (2008) 35.
\bibitem{Perez:NIM:A532:2004} P.~P{\'{e}}rez and A.~Rosowsky,  {\it Nucl. Instrum. Meth. Phys. Res.} A {532} (2004) 523.
\bibitem{Perez:ASS255:2008} P.~P{\'{e}}rez {\it et al.}, {\it Appl. Surface Sci.} {255} (2008) 33.
\bibitem{mcgovern2009} M.~McGovern {\it et al.}, {\it Phys. Rev.} A {79} (2009) 042707.
\bibitem{ullrich2003} J.~Ullrich {\it et al.}, {\it Rep. Prog. Phys.} {66} (2003) 1463.


\end{thebibliography}
\end{document}